\documentclass[12pt]{article}
\pdfoutput=1
\usepackage[]{hyperref}
\usepackage{xcolor}
\usepackage[]{putex}
\usepackage[utf8]{inputenc}
\usepackage{subcaption}

\makeatletter
\DeclareRobustCommand*{\bfseries}{%
  \not@math@alphabet\bfseries\mathbf
  \fontseries\bfdefault\selectfont
  \boldmath
}
\makeatother
\hypersetup{
  colorlinks=true,
  linkcolor=blue,
  citecolor=green!60!black,
  urlcolor=cyan!70!black,
  linktoc=all
  }

\numberwithin{equation}{section}

\input{glyphtounicode}
\usepackage{amsmath,amsthm,amssymb}
\usepackage{physics}
\usepackage{mathrsfs}
\usepackage{tikz}
\usepackage{tikz-cd}
\usetikzlibrary{decorations.markings}
\usetikzlibrary{calc}
\usetikzlibrary{intersections}
\usetikzlibrary{patterns}
\usepackage{intcalc}
\usepackage{mathtools}

\def\H{\mathscr{H}}
\def\C{\mathbb{C}}
\def\Z{\mathbb{Z}}
\def\Q{\mathbb{Q}}
\def\R{\mathbb{R}}
\def\F{\mathbb{F}}
\DeclareMathOperator{\Heis}{Heis}
\DeclareMathOperator{\Fun}{Fun}
\DeclareMathOperator{\Sp}{Sp}

\DeclareMathOperator{\End}{End}
\DeclareMathOperator{\wt}{wt}
\DeclareMathOperator{\Span}{span}

\def\sred{{/\!\!/}}
\def\defeq{\doteq}

\def\cat#1{\ensuremath{\mathsf{#1}}}

\usepackage{xparse}
\NewDocumentCommand \deq { o m }{
\begin{equation}
#2
\IfNoValueF{#1}{\label{#1}}
\end{equation}
}

\newtheoremstyle{iremark}
  {\topsep}   
  {\topsep}   
  {\upshape}  
  {0pt}       
  {\itshape}  
  {.}         
  {5pt plus 1pt minus 1pt} 
  {\thmname{#1}\thmnumber{ \itshape#2}\thmnote{ (#3)}} 

\theoremstyle{plain}
\newtheorem{theorem}{Theorem}
\newtheorem*{thm}{Theorem}

\theoremstyle{definition}
\newtheorem{definition}[theorem]{Definition}

\theoremstyle{iremark}

\DeclareMathOperator{\GL}{GL}
\DeclareMathOperator{\SL}{SL}

\DeclareMathOperator{\PGL}{PGL}
\DeclareMathOperator{\PSL}{PSL}

\DeclareMathOperator{\ord}{ord}

\DeclareMathOperator{\im}{im}

\def\R{\mathbb{R}}
\def\C{\mathbb{C}}
\def\Z{\mathbb{Z}}
\def\Q{\mathbb{Q}}

\def\FF{\mathbb{F}} 

\def\ket#1{\left|#1\right\rangle}
\def\Hilb{\mathscr{H}}

\def\bP{\mathrm{P}}
\def\bZ{\mathbb{Z}}
\def\bC{\mathbb{C}}
\def\bR{\mathbb{R}}

\newcommand{\bQ}{\mathbb{Q}}

\newsavebox{\bigtensor}
\savebox{\bigtensor}{%
\begin{tikzpicture}[scale=0.8]
\tikzstyle{vertex}=[draw,scale=0.4,fill=black,circle]
\tikzstyle{ver2}=[draw,scale=0.6,circle]
\tikzstyle{ver3}=[draw,scale=0.4,circle]
\coordinate (C) at (0,0);
\coordinate (a) at (1,0);
\coordinate (b) at (0,1);
\coordinate (c) at (-1,0);
\coordinate (d) at (0,-1);
\draw (C) -- (a);
\draw (C) -- (b);
\draw (C) -- (c);
\draw (C) -- (d);
\draw (C) node[vertex] {};
\draw (a) node[anchor=west] {$a$};
\draw (b) node[anchor=south] {$b$};
\draw (c) node[anchor=east] {$c$};
\draw (d) node[anchor=north] {$d$};
\end{tikzpicture}%
}

\newsavebox{\bigtensorprime}
\savebox{\bigtensorprime}{%
\begin{tikzpicture}[scale=0.8]
\tikzstyle{vertex}=[draw,scale=0.4,fill=black,circle]
\tikzstyle{ver2}=[draw,scale=0.6,circle]
\tikzstyle{ver3}=[draw,scale=0.4,circle]
\coordinate (C) at (0,0);
\coordinate (a) at (1,0);
\coordinate (b) at (0,1);
\coordinate (c) at (-1,0);
\coordinate (d) at (0,-1);
\draw (C) -- (a);
\draw (C) -- (b);
\draw (C) -- (c);
\draw (C) -- (d);
\draw (C) node[vertex] {};
\draw (a) node[anchor=west] {$a'$};
\draw (b) node[anchor=south] {$b'$};
\draw (c) node[anchor=east] {$c'$};
\draw (d) node[anchor=north] {$d'$};
\end{tikzpicture}%
}

\newsavebox{\bigtree}
\savebox{\bigtree}{%
\begin{tikzpicture}[scale=0.8]
\tikzstyle{vertex}=[draw,scale=0.4,fill=black,circle]
\tikzstyle{ver2}=[draw,scale=0.6,circle]
\tikzstyle{ver3}=[draw,scale=0.4,circle]
\coordinate (C) at (0,0);
\foreach \x in {0,1,2,3} {
	\coordinate (a\x) at (\x*360/4:1);
	\draw (a\x) -- (C);
	\draw (a\x) node[vertex] {};
	};
\foreach \x in {0,1,...,11} {
	\coordinate (b\x) at (\x*360/12 - 30 :2);
	\pgfmathparse{floor(\x/3)}
	\draw (b\x) -- (a\pgfmathresult);
	};
\draw (C) node[vertex] {};
\end{tikzpicture}}

\newsavebox{\contractfour}
\savebox{\contractfour}{%
\begin{tikzpicture}[scale=0.8]
\tikzstyle{vertex}=[draw,scale=0.4,fill=black,circle]
\tikzstyle{ver2}=[draw,scale=0.6,circle]
\tikzstyle{ver3}=[draw,scale=0.4,circle]
\coordinate (A) at (0,0);
\coordinate (B) at (2,0);
\foreach \x in {-60,-20,20,60} {
	\draw (A) to [out =  \x, in =  180 - \x, looseness = 1.3] (B);	
	};
\draw (A) node[vertex] {};
\draw (B) node[vertex] {};
\end{tikzpicture}
}

\newsavebox{\contractthree}
\savebox{\contractthree}{%
\begin{tikzpicture}[scale=0.8]
\tikzstyle{vertex}=[draw,scale=0.4,fill=black,circle]
\tikzstyle{ver2}=[draw,scale=0.6,circle]
\tikzstyle{ver3}=[draw,scale=0.4,circle]
\coordinate (A) at (0,0);
\coordinate (B) at (2,0);
\coordinate (a) at (-1,0);
\coordinate (b) at (3,0);
\foreach \x in {-40,0,40} {
	\draw (A) to [out =  \x, in =  180 - \x, looseness = 1.3] (B);	
	};
\draw (A) node[vertex] {};
\draw (B) node[vertex] {};
\draw (A) -- (a);
\draw (B) -- (b);
\end{tikzpicture}
}

\newsavebox{\contractthreeinline}
\savebox{\contractthreeinline}{%
\begin{tikzpicture}[scale=0.6]
\tikzstyle{vertex}=[draw,scale=0.4,fill=black,circle]
\tikzstyle{ver2}=[draw,scale=0.6,circle]
\tikzstyle{ver3}=[draw,scale=0.4,circle]
\coordinate (A) at (0,0);
\coordinate (B) at (2,0);
\coordinate (a) at (-1,0);
\coordinate (b) at (3,0);
\foreach \x in {-40,0,40} {
	\draw (A) to [out =  \x, in =  180 - \x, looseness = 1.3] (B);	
	};
\draw (A) node[vertex] {};
\draw (B) node[vertex] {};
\draw (A) -- (a);
\draw (B) -- (b);
\end{tikzpicture}
}

\newsavebox{\contracttwo}
\savebox{\contracttwo}{%
\begin{tikzpicture}[scale=0.8]
\tikzstyle{vertex}=[draw,scale=0.4,fill=black,circle]
\tikzstyle{ver2}=[draw,scale=0.6,circle]
\tikzstyle{ver3}=[draw,scale=0.4,circle]
\coordinate (A) at (0,0);
\coordinate (B) at (2,0);
\coordinate (a1) at ($ (A) + (150:1) $);
\coordinate (a2) at ($ (A) + (210:1) $);
\coordinate (b1) at ($ (B) + (30:1) $);
\coordinate (b2) at ($ (B) + (-30:1) $);
\foreach \x in {-30,30} {
	\draw (A) to [out =  \x, in =  180 - \x, looseness = 1.3] (B);	
	};
\draw (A) node[vertex] {};
\draw (B) node[vertex] {};
\draw (A) -- (a1);
\draw (A) -- (a2);
\draw (B) -- (b1);
\draw (B) -- (b2);
\end{tikzpicture}
}

\newsavebox{\contractone}
\savebox{\contractone}{%
\begin{tikzpicture}[scale=0.8]
\tikzstyle{vertex}=[draw,scale=0.4,fill=black,circle]
\tikzstyle{ver2}=[draw,scale=0.6,circle]
\tikzstyle{ver3}=[draw,scale=0.4,circle]
\coordinate (A) at (0,0);
\coordinate (B) at (1.5,0);
\foreach \x in {-1,0,1} {
	\coordinate (a\x) at ($ (A) + (180 + 40*\x:1) $);
	\coordinate (b\x) at ($ (B) + (40*\x:1) $);
	\draw (A) to (a\x);
	\draw (B) to (b\x);	
	};
\draw (A) node[vertex] {};
\draw (B) node[vertex] {};
\draw (A) -- (B);
\end{tikzpicture}}

\newsavebox{\contractsix}
\savebox{\contractsix}{%
\begin{tikzpicture}[scale=0.8]
\tikzstyle{vertex}=[draw,scale=0.4,fill=black,circle]
\tikzstyle{ver2}=[draw,scale=0.6,circle]
\tikzstyle{ver3}=[draw,scale=0.4,circle]
\coordinate (A) at (0,0);
\coordinate (B) at (1.5,0);
\foreach \x in {-5,-3,-1,1,3,5} {
	\coordinate (a\x) at ($ (A) + (180 + 15*\x:1) $);
	\coordinate (b\x) at ($ (B) + (15*\x:1) $);
	\draw[color=blue] (A) to [out =  15*\x, in =  180 - 15*\x, looseness = 1.3] (B);		
	};
\foreach \x in {-1,1} {
	\draw (A) to (a\x);
	\draw (B) to (b\x);
	};
\draw (A) node[vertex] {};
\draw (B) node[vertex] {};
\end{tikzpicture}}

\newsavebox{\fourtofour}
\savebox{\fourtofour}{%
\begin{tikzpicture}[scale=0.8]
\tikzstyle{vertex}=[draw,scale=0.4,fill=black,circle]
\tikzstyle{ver2}=[draw,scale=0.6,circle]
\tikzstyle{ver3}=[draw,scale=0.4,circle]
\coordinate (A) at (0,0);
\coordinate (B) at (1.5,0);
\foreach \x in {-3,-1,1,3} {
	\coordinate (a\x) at ($ (A) + (180 + 15*\x:1) $);
	\coordinate (b\x) at ($ (B) + (15*\x:1) $);
	\draw (A) to (a\x);
	\draw (B) to (b\x);
	\draw[color=blue] (A) to [out =  25*\x, in =  180 - 25*\x, looseness = 1.3] (B);		
	};
\draw (A) node[vertex] {};
\draw (B) node[vertex] {};
\end{tikzpicture}}

\newsavebox{\fourlines}
\savebox{\fourlines}{%
\begin{tikzpicture}[scale=0.8]
\tikzstyle{vertex}=[draw,scale=0.4,fill=black,circle]
\tikzstyle{ver2}=[draw,scale=0.6,circle]
\tikzstyle{ver3}=[draw,scale=0.4,circle]
\coordinate (A) at (0,0);
\coordinate (B) at (3,0);
\foreach \x in {-3,-1,1,3} {
	\coordinate (a\x) at ($ (A) + (0,0.2*\x) $);
	\coordinate (b\x) at ($ (B) + (0,0.2*\x) $);
	\draw (a\x) to (b\x);		
	};
\end{tikzpicture}}

\newsavebox{\twolines}
\savebox{\twolines}{%
\begin{tikzpicture}[scale=0.8]
\tikzstyle{vertex}=[draw,scale=0.4,fill=black,circle]
\tikzstyle{ver2}=[draw,scale=0.6,circle]
\tikzstyle{ver3}=[draw,scale=0.4,circle]
\coordinate (A) at (0,0);
\coordinate (B) at (3,0);
\foreach \x in {-1,1} {
	\coordinate (a\x) at ($ (A) + (0,0.2*\x) $);
	\coordinate (b\x) at ($ (B) + (0,0.2*\x) $);
	\draw (a\x) to (b\x);		
	};
\end{tikzpicture}}

\newsavebox{\smcircle}
\savebox{\smcircle}{%
\begin{tikzpicture}[scale=0.8]
\coordinate (A) at (0,0);
\draw (A) circle (0.5);
\end{tikzpicture}}

\newsavebox{\tinycircle}
\savebox{\tinycircle}{%
\begin{tikzpicture}
\coordinate (A) at (0,0);
\draw (A) circle (1ex);
\end{tikzpicture}}

\newsavebox{\oneintpic}
\savebox{\oneintpic}{%
\begin{tikzpicture}[scale=1.6]
\tikzstyle{vertex}=[draw,scale=0.4,fill=black,circle]
\tikzstyle{bluevert}=[draw,scale=0.6,fill=blue,color=blue,circle]
\tikzstyle{ver2}=[draw,scale=0.6,circle]
\tikzstyle{ver3}=[draw,scale=0.4,circle]
\coordinate (C) at (0,0);
\foreach \x in {0,1,2,3} {
	\coordinate (a\x) at (\x*360/4:1);
	\draw[color=red] (a\x) -- (C);
	};
\foreach \x in {0,1,...,11} {
	\coordinate (b\x) at (\x*360/12 - 30 :2);
	\pgfmathparse{floor(\x/3)}
	\draw[color=red] (b\x) -- (a\pgfmathresult);
	};
\foreach \x in {0,1,...,11} {
	\coordinate (c\x) at (\x*360/12 - 45 :2);
	};
\foreach \x/\y in {0/1,1/2,2/3,3/4,4/5,5/6,6/7,7/8,8/9,9/10,10/11,11/0} {
	\draw[thick] (c\x) to [out =  \x*360/12 + 115, 
		in =  \y*360/12 + 155, looseness = 1.5] (c\y);	
	};
\foreach \x/\y in {0/3,3/6,6/9,9/0} {
	\draw[thick] (c\x) to [out =  \x*360/12 + 135, 
		in =  \y*360/12 + 135, looseness = 1.3] (c\y);	
	};
\foreach \x in {2,3,4,5} {
    \draw (c\x) node[bluevert] {};
    };
\draw[dotted] (0,0) circle[radius=1];
\draw[dotted] (0,0) circle[radius=2];
\draw[color=red] (C) node[below left] {$C$};
\draw (C) node[color=red,fill=red,circle,scale=0.4] {};
\draw[color=red] (b1) node[right] {$x$};
\draw[color=red] (b5) node[above left] {$y$};
\end{tikzpicture}}

\newsavebox{\multiintpic}
\savebox{\multiintpic}{%
\begin{tikzpicture}[scale=1.6]
\tikzstyle{vertex}=[draw,scale=0.4,fill=black,circle]
\tikzstyle{bluevert}=[draw,scale=0.6,fill=blue,color=blue,circle]
\tikzstyle{emptyvert}=[draw,scale=0.6,fill=none,color=black,circle]
\tikzstyle{ver2}=[draw,scale=0.6,circle]
\tikzstyle{ver3}=[draw,scale=0.4,circle]
\coordinate (C) at (0,0);
\foreach \x in {0,1,2,3} {
	\coordinate (a\x) at (\x*360/4:1);
	\draw[color=red] (a\x) -- (C);
	};
\foreach \x in {0,1,...,11} {
	\coordinate (b\x) at (\x*360/12 - 30 :2);
	\pgfmathparse{floor(\x/3)}
	\draw[color=red] (b\x) -- (a\pgfmathresult);
	};
\foreach \x in {0,1,...,11} {
	\coordinate (c\x) at (\x*360/12 - 45 :2);
	};
\foreach \x/\y in {0/1,1/2,2/3,3/4,4/5,5/6,6/7,7/8,8/9,9/10,10/11,11/0} {
	\draw[thick] (c\x) to [out =  \x*360/12 + 115, 
		in =  \y*360/12 + 155, looseness = 1.5] (c\y);	
	};
\foreach \x/\y in {0/3,3/6,6/9,9/0} {
	\draw[thick] (c\x) to [out =  \x*360/12 + 135, 
		in =  \y*360/12 + 135, looseness = 1.3] (c\y);	
	};
\foreach \x in {0,1,3,4,5} {
    \draw (c\x) node[bluevert] {};
    };
\foreach \x in {2,6,7,8,9,10,11} {
    \draw (c\x) node[emptyvert] {};
    };
\draw[dotted] (0,0) circle[radius=1];
\draw[dotted] (0,0) circle[radius=2];
\draw[color=red] (C) node[below left] {$C$};
\draw (C) node[color=red,fill=red,circle,scale=0.4] {};
\draw[color=red] (b1) node[right] {$x_4$};
\draw[color=red] (b2) node[right] {$x_1$};
\draw[color=red] (b5) node[above left] {$x_2$};
\draw[color=red] (b11) node[below right] {$x_3$};
\end{tikzpicture}}

\newsavebox{\computeone}
\savebox{\computeone}{%
\begin{tikzpicture}[scale=1.1]
\tikzstyle{vertex}=[draw,scale=0.4,fill=black,circle]
\tikzstyle{ver2}=[draw,scale=0.6,circle]
\tikzstyle{ver3}=[draw,scale=0.4,circle]
\coordinate (C) at (0,0);
\foreach \x in {0,1,2,3} {
	\coordinate (a\x) at (\x*360/4:1);
	};
\foreach \x in {0,1,...,11} {
	\coordinate (b\x) at (\x*360/12 - 30 :2);
	\pgfmathparse{floor(\x/3)}
	};
\foreach \x in {0,1,...,11} {
	\coordinate (c\x) at (\x*360/12 - 45 :2);
	};
\foreach \x/\y in {1/2,4/5,7/8,10/11,0/1,3/4,6/7,9/10,2/3,5/6,8/9,11/0} {
	\draw[thick] (c\x) to [out =  \x*360/12 + 115, 
		in =  \y*360/12 + 155, looseness = 1.5] (c\y);	
	};
\foreach \x/\y in {0/3,3/6,6/9,9/0} {
	\draw[thick] (c\x) to [out =  \x*360/12 + 135, 
		in =  \y*360/12 + 135, looseness = 1.3] (c\y);	
	};
\end{tikzpicture}
}

\newsavebox{\computetwo}
\savebox{\computetwo}{%
\begin{tikzpicture}[scale=1.1]
\tikzstyle{vertex}=[draw,scale=0.4,fill=black,circle]
\tikzstyle{ver2}=[draw,scale=0.6,circle]
\tikzstyle{ver3}=[draw,scale=0.4,circle]
\coordinate (C) at (0,0);
\foreach \x in {0,1,2,3} {
	\coordinate (a\x) at (\x*360/4:1);
	};
\foreach \x in {0,1,...,11} {
	\coordinate (b\x) at (\x*360/12 - 30 :2);
	\pgfmathparse{floor(\x/3)}
	};
\foreach \x in {0,1,...,11} {
	\coordinate (c\x) at (\x*360/12 - 45 :2);
	};
\foreach \x in {0,1,...,11} {
	\foreach \y in {-1,1} {
		\coordinate (d\x_\y) at (\x*360/12 - 45 + 3*\y :2);
		};
	};
\foreach \x in {0,1,...,11} {
	\foreach \y in {-1,1} {
		\coordinate (e\x_\y) at (\x*360/12 - 45 + 1.2*\y :2);
		};
	};
\foreach \x/\y in {1/2,4/5,7/8,10/11,0/1,3/4,6/7,9/10,2/3,5/6,8/9,11/0} {
	\draw[thick] (d\x_1) to [out =  \x*360/12 + 135, in =  \y*360/12 + 135, looseness = 1.5] (d\y_-1);	
	};
\foreach \x/\y in {0/3,3/6,6/9,9/0} {
	\draw[thick] (e\x_1) to [out =  \x*360/12 + 135, in =  \y*360/12 + 135, looseness = 1.3] (e\y_-1);	
	};
\foreach \x in {1,2,4,5,7,8,10,11} {
	\draw (\x*360/12 - 45 :1.45) node {$r^2$};
	};
\end{tikzpicture}}

\newsavebox{\computethree}
\savebox{\computethree}{%
\begin{tikzpicture}[scale=0.9]
\tikzstyle{vertex}=[draw,scale=0.4,fill=black,circle]
\tikzstyle{ver2}=[draw,scale=0.6,circle]
\tikzstyle{ver3}=[draw,scale=0.4,circle]
\coordinate (C) at (0,0);
\foreach \x in {0,1,2,3} {
	\coordinate (a\x) at (\x*360/4:1);
	};
\foreach \x in {0,1,...,11} {
	\coordinate (b\x) at (\x*360/12 - 30 :2);
	\pgfmathparse{floor(\x/3)}
	};
\foreach \x in {0,1,...,11} {
	\coordinate (c\x) at (\x*360/12 - 45 :2);
	};
\foreach \x in {0,1,...,11} {
	\foreach \y in {-1,1} {
		\coordinate (d\x_\y) at (\x*360/12 - 45 + 3*\y :2);
		};
	};
\foreach \x in {0,1,...,11} {
	\foreach \y in {-1,1} {
		\coordinate (e\x_\y) at (\x*360/12 - 45 + 1.2*\y :2);
		};
	};
\foreach \x/\y in {1/2,4/5,7/8,10/11,0/1,3/4,6/7,9/10,2/3,5/6,8/9,11/0} {
	\draw[thick] (d\x_1) to [out =  \x*360/12 + 135, in =  \y*360/12 + 135, looseness = 1.5] (d\y_-1);	
	\draw[thick] (d\x_1) to [out =  \x*360/12 - 45, in =  \y*360/12 - 45, looseness = 1.5] (d\y_-1);	
	};
\foreach \x/\y in {0/3,3/6,6/9,9/0} {
	\draw[thick] (e\x_1) to [out =  \x*360/12 + 135, in =  \y*360/12 + 135, looseness = 1.3] (e\y_-1);	
	\draw[thick] (e\x_1) to [out =  \x*360/12 - 45, in =  \y*360/12 - 45, looseness = 2.3] (e\y_-1);	
	};
\foreach \x in {1,2,4,5,7,8,10,11} {
	\draw (\x*360/12 - 45 :1.4) node {$r^2$};
	};
\end{tikzpicture}
}

\newsavebox{\numerouscontractions}
\savebox{\numerouscontractions}{%
\begin{tikzpicture}
\tikzstyle{vertex}=[draw,scale=0.4,fill=black,circle]
\tikzstyle{ver2}=[draw,scale=0.6,circle]
\tikzstyle{ver3}=[draw,scale=0.4,circle]
\coordinate (A) at (0,0);
\coordinate (B) at (1.5,0);
\foreach \x in {-5,...,5} {
	\coordinate (a\x) at ($ (A) + (180 + 15*\x:1) $);
	\coordinate (b\x) at ($ (B) + (15*\x:1) $);
	};
\foreach \x in {-5,-3,3,5} {
	\draw (A) to [out =  15*\x, in =  180 - 15*\x, looseness = 1.3] (B);		
	};
\foreach \x in {-2,2} {
	\draw (A) to (a\x);
	\draw (B) to (b\x);
	};
\draw[thick,dotted] (a1) -- (a-1);
\draw[thick,dotted] (b1) -- (b-1);
\draw (0.75,0) node {$n_c$};
\draw (A) node[vertex] {};
\draw (B) node[vertex] {};
\draw (a2) node[anchor=east] {$a_{n_d}$};
\draw (b2) node[anchor=west] {$a_1'$};
\draw (a-2) node[anchor=east] {$a_1$};
\draw (b-2) node[anchor=west] {$a_{n_d}'$};
\end{tikzpicture}
}

\newsavebox{\numerouslines}
\savebox{\numerouslines}{%
\begin{tikzpicture}
\draw (-1,1) node[anchor=east]{$a_1$\,} -- (1,1) node[anchor=west]{$a_1'$};
\draw[thick,dotted] (0,0.2) -- (0,0.8);
\draw (-1,0) node[anchor=east]{$a_{n_d}$} -- (1,0) node[anchor=west]{$a_{n_d}'$};
\end{tikzpicture}}

\newsavebox{\bigcontraction}
\savebox{\bigcontraction}{%
\begin{tikzpicture}[scale=0.7]
\draw [color=white,use as bounding box] (-3.1,-7.3) rectangle (8.1,7.3);
\tikzstyle{vertex}=[draw,scale=0.4,fill=black,circle]
\tikzstyle{ver2}=[draw,scale=0.6,circle]
\tikzstyle{ver3}=[draw,scale=0.4,circle]
\coordinate (C) at (0,0);
\coordinate (D) at (5,0);
\foreach \x in {0,1,2,3} {
	\coordinate (a\x) at (\x*360/4:1);
	\draw[color=red] (a\x) -- (C);
	};
\foreach \x in {0,1,2,3} {
	\coordinate (A\x) at ($ (\x*360/4:1) + (D) $);
	\draw[color=red] (A\x) -- (D);
	};
\foreach \x in {0,1,...,11} {
	\coordinate (b\x) at (\x*360/12 - 30 :2);
	\coordinate (c\x) at (\x*360/12 - 45 :2);
	\pgfmathparse{floor(\x/3)}
	\draw[color=red] (b\x) -- (a\pgfmathresult);
	};
\foreach \x in {0,1,...,11} {
	\coordinate (B\x) at ($ (D) + (\x*360/12 - 30 :2) $);
	\coordinate (C\x) at ($ (D) + (\x*360/12 - 45 :2) $);
	\pgfmathparse{floor(\x/3)}
	\draw[color=red] (B\x) -- (A\pgfmathresult);
	};
\foreach \x/\y in {0/1,1/2,2/3,3/4,4/5,5/6,6/7,7/8,8/9,9/10,10/11,11/0} {
	\draw[thick] (c\x) to [out =  \x*360/12 + 115, in =  \y*360/12 + 155, looseness = 1.5] (c\y);	
	\draw[thick] (C\x) to [out =  \x*360/12 + 115, in =  \y*360/12 + 155, looseness = 1.5] (C\y);	
	};
\foreach \x/\y in {0/3,3/6,6/9,9/0} {
	\draw[thick] (c\x) to [out =  \x*360/12 + 135, in =  \y*360/12 + 135, looseness = 1.3] (c\y);	
	\draw[thick] (C\x) to [out =  \x*360/12 + 135, in =  \y*360/12 + 135, looseness = 1.3] (C\y);	
	};
\foreach \x/\y in {0/9,3/6} {
	\foreach \z in {-3,-1,1,3} {	
		\draw[thick, color=blue] (c\x) to [out =  \x*360/12 + 9*\z - 45, in =  \y*360/12 - 9*\z - 45, looseness = 1.4] (C\y);
		};
	};
\foreach \x/\y in {6/3,9/0} {
	\foreach \z in {-3,-1,1,3} {	
		\draw[thick, color=blue] (c\x) to [out =  \x*360/12 + 4*\z - 45, in =  \y*360/12 - 4*\z - 45, looseness = 3] (C\y);
		};
	};
\foreach \x/\y in {1/8,2/7}
	{
	\foreach \z in {-5,-3,-1,1,3,5} {	
		\draw[thick, color=blue] (c\x) to [out =  \x*360/12 + 7*\z - 45, in =  \y*360/12 - 7*\z - 45, looseness = 1.1] (C\y);
		};
	};
\foreach \x/\y in {4/5,11/10}
	{
	\foreach \z in {-5,-3,-1,1,3,5} {	
		\draw[thick, color=blue] (c\x) to [out =  \x*360/12 + 5*\z - 45, in =  \y*360/12 - 5*\z - 45, looseness = 1.6] (C\y);
		};
	};

\foreach \x/\y in {5/4}
	{
	\foreach \z in {-5,-3,-1,1,3,5} {	
		\draw[thick, color=blue] (c\x) to [out =  \x*360/12 + 5*\z - 10 - 45, in =  \y*360/12 - 5*\z + 10- 45, looseness = 1.4] (C\y);
		};
	};
\foreach \x/\y in {10/11}
	{
	\foreach \z in {-5,-3,-1,1,3,5} {	
		\draw[thick, color=blue] (c\x) to [out =  \x*360/12 + 5*\z + 10 - 45, in =  \y*360/12 - 5*\z - 10- 45, looseness = 1.4] (C\y);
		};
	};
\foreach \x/\y in {7/2,8/1} {
	\foreach \z in {-5,-3,-1,1,3,5} {	
		\coordinate (d\x_\z) at (\x*360/12 + \z - 45 : 2.4);
		\coordinate (D\y_\z) at ($ (D) + (\y*360/12 + \z - 45 : 2.4) $);
		\coordinate (e\x_\z) at (\x*360/12 + \z - 45 : 2.9);
		\coordinate (E\y_\z) at ($ (D) + (\y*360/12 + \z - 45 : 2.9) $);
		\draw[thick, color=blue] (c\x) -- (d\x_\z);
		\draw[thick, color=blue] (C\y) -- (D\y_\z);
		\draw[thick,dotted, color=blue] (d\x_\z) -- (e\x_\z);
		\draw[thick,dotted, color=blue] (D\y_\z) -- (E\y_\z);
		};
	};
\draw[dotted] (C) circle[radius=1];
\draw[dotted] (C) circle[radius=2];
\draw (C) node[color=red,fill=red,circle,scale=0.4] {};
\draw[dotted] (D) circle[radius=1];
\draw[dotted] (D) circle[radius=2];
\draw (D) node[color=red,fill=red,circle,scale=0.4] {};
\end{tikzpicture}}

\newsavebox{\bigspread}
\savebox{\bigspread}{%
\begin{tikzpicture}[scale=0.9]
\tikzstyle{vertex}=[draw,scale=0.4,fill=black,circle]
\tikzstyle{ver2}=[draw,scale=0.6,circle]
\tikzstyle{ver3}=[draw,scale=0.4,circle]
\coordinate (C) at (0,0);
\newcommand\ZSEP{5}
 \draw [color=white, use as bounding box] (-2.5,-5) rectangle (+2.5+\ZSEP,5);
\foreach \z in {0,1} {
\foreach \x in {0,1,2,3} {
	\coordinate (\z-a\x) at ($ (\ZSEP*\z,0) + (\x*360/4:1) $);
	};
\foreach \x in {0,1,...,11} {
	\coordinate (\z-b\x) at ($ (\ZSEP*\z,0) + (\x*360/12 - 30 :2) $);
	};
\foreach \x in {0,1,...,11} {
	\coordinate (\z-c\x) at ($ (\ZSEP*\z,0) + (\x*360/12 - 45 :2) $);
	};
\foreach \x in {0,1,...,11} {
	\foreach \y in {-1,1} {
		\coordinate (\z-d\x_\y) at ($ (\ZSEP*\z,0) + (\x*360/12 - 45 + 3*\y :2) $);
		};
	};
\foreach \x in {0,1,...,11} {
	\foreach \y in {-1,1} {
		\coordinate (\z-e\x_\y) at  ($ (\ZSEP*\z,0) + (\x*360/12 - 45 + 1.2*\y :2) $);
		};
	};
\foreach \x/\y in {1/2,4/5,7/8,10/11,0/1,3/4,6/7,9/10,2/3,5/6,8/9,11/0} {
	\draw[thick] (\z-d\x_1) to [out =  \x*360/12 + 135, in =  \y*360/12 + 135, looseness = 1.5] (\z-d\y_-1);	
	};
\foreach \x/\y in {0/3,3/6,6/9,9/0} {
	\draw[thick] (\z-e\x_1) to [out =  \x*360/12 + 135, in =  \y*360/12 + 135, looseness = 1.3] (\z-e\y_-1);	
	};
\foreach \x in {1,2,4,5,7,8,10,11} {
	\draw  (\x*360/12 - 45 :1.45) node {$r^2$}; 
	};
};
\foreach \x/\y in {0/9,3/6} {	
		\draw[thick] (0-d\x_-1) to [out =  \x*360/12 - 45, in =  \y*360/12 - 45, looseness = 1.4] (1-d\y_1);
		\draw[thick] (0-d\x_1) to [out =  \x*360/12 - 45, in =  \y*360/12 - 45, looseness = 1.4] (1-d\y_-1);
		\draw[thick] (0-e\x_1) to [out =  \x*360/12 - 45, in =  \y*360/12 - 45, looseness = 1.4] (1-e\y_-1);
		\draw[thick] (0-e\x_-1) to [out =  \x*360/12 - 45, in =  \y*360/12  - 45, looseness = 1.4] (1-e\y_1);
	};
\foreach \x/\y in {6/3} {
		\draw[thick] (0-d\x_-1) to [out =  \x*360/12 - 45, in =  \y*360/12 - 45, looseness = 1.7] (1-d\y_1);
		\draw[thick] (0-d\x_1) to [out =  \x*360/12 - 45, in =  \y*360/12 - 45, looseness = 2.0] (1-d\y_-1);
		\draw[thick] (0-e\x_1) to [out =  \x*360/12 - 45, in =  \y*360/12 - 45, looseness = 1.9] (1-e\y_-1);
		\draw[thick] (0-e\x_-1) to [out =  \x*360/12 - 45, in =  \y*360/12  - 45, looseness = 1.8] (1-e\y_1);
	};
\foreach \x/\y in {9/0} {
		\draw[thick] (0-d\x_-1) to [out =  \x*360/12 - 45, in =  \y*360/12 - 45, looseness = 2.0] (1-d\y_1);
		\draw[thick] (0-d\x_1) to [out =  \x*360/12 - 45, in =  \y*360/12 - 45, looseness = 1.7] (1-d\y_-1);
		\draw[thick] (0-e\x_1) to [out =  \x*360/12 - 45, in =  \y*360/12 - 45, looseness = 1.8] (1-e\y_-1);
		\draw[thick] (0-e\x_-1) to [out =  \x*360/12 - 45, in =  \y*360/12  - 45, looseness = 1.9] (1-e\y_1);
	};
\foreach \x/\y in {2/7,4/5,5/4}
	{
		\draw[thick] (0-d\x_1) to [out =  \x*360/12 - 45, in =  \y*360/12 - 45, looseness = 1.2] (1-d\y_-1);
		\draw[thick] (0-d\x_-1) to [out =  \x*360/12 - 45, in =  \y*360/12 - 45, looseness = 1.1] (1-d\y_1);
	};
\foreach \x/\y in {1/8,11/10,10/11}
	{
		\draw[thick] (0-d\x_1) to [out =  \x*360/12 - 45, in =  \y*360/12 - 45, looseness = 1.1] (1-d\y_-1);
		\draw[thick] (0-d\x_-1) to [out =  \x*360/12 - 45, in =  \y*360/12 - 45, looseness = 1.2] (1-d\y_1);
	};
\foreach \x/\y in {7/2,8/1} {
	\foreach \z in {-1,1} {
		\coordinate (0-f\x_\z) at  ($ (0-d\x_\z) + (\x*360/12 - 45 : 0.5) $);
		\draw[thick, dotted] (0-d\x_\z) -- (0-f\x_\z);
		\coordinate (1-f\y_\z) at  ($ (1-d\y_\z) + (\y*360/12 - 45 : 0.5) $);
		\draw[thick, dotted] (1-d\y_\z) -- (1-f\y_\z);
		};
	};
\end{tikzpicture}}

\newsavebox{\dualDrinfeld}
\savebox{\dualDrinfeld}{%
\begin{tikzpicture}[scale=2.0]
\tikzstyle{vertex}=[draw,scale=0.25,fill=black,circle]
\tikzstyle{ver2}=[draw,scale=0.4,fill=red,circle]
\tikzstyle{ver3}=[];

\newcommand{\f}{0.3}
\newcommand{\widthpar}{1.25ex}
\draw[thin, color=red, name path = trunk] (0,0) coordinate (A) -- (3,0) coordinate (B);
\draw[thick, color=blue] ($ ($ (A)! \f/2 !(B) $) !\widthpar!90:(B) $) 
	-- ($ ($ (B)! \f/2 !(A) $) !\widthpar!-90:(A) $) ;
\draw[thick, color=blue] ($ ($ (A)! \f/2 !(B) $) !\widthpar!-90:(B) $) 
	-- ($ ($ (B)! \f/2 !(A) $) !\widthpar!90:(A) $) ;
\draw[thick, color=black, name path = cir1] ($ ($ (A)! 0.5 !(B) $) !\widthpar!90:(B) $)
	to[out = 0, in = 0, looseness = 1.3]
	($ ($ (A)! 0.5 !(B) $) !\widthpar!90:(A) $);
\path[name intersections={of=trunk and cir1,by=p1}];
\draw (p1) node[vertex] {};

\foreach \x in {-1,0,1} {
	\draw[thin,color=red,name path=brA] (A) -- ($ (A) + (\x*75 + 180:1.5) $) coordinate (A\x);
	\draw[thin,color=red,name path=brB] (B) -- ($ (B) + (\x*75:1.5) $) coordinate (B\x);
	\draw[thick,color=blue] ($ (A\x) !\widthpar!90:(A) $) -- ($ ($ (A)! \f !(A\x) $) !\widthpar!90:(A) $) ;
	\draw[thick,color=blue] ($ (A\x) !\widthpar!-90:(A) $) -- ($ ($ (A)! \f !(A\x) $) !\widthpar!-90:(A) $) ;
	\draw[thick,color=blue] ($ (B\x) !\widthpar!90:(B) $) -- ($ ($ (B)! \f !(B\x) $) !\widthpar!90:(B) $) ;
	\draw[thick,color=blue] ($ (B\x) !\widthpar!-90:(B) $) -- ($ ($ (B)! \f !(B\x) $) !\widthpar!-90:(B) $) ;
	\draw[thick,color=black,name path=circA] 
		($ ($ (A\x)! 0.3 !(A) $) !\widthpar!90:(A) $)
		to[in = \x*75,out=\x*75,looseness=1.5]
		($ ($ (A\x)! 0.3 !(A) $) !\widthpar!-90:(A) $);
	\draw[thick,color=black,name path=circB] 
		($ ($ (B\x)! 0.3 !(B) $) !\widthpar!90:(B) $)
		to[in = \x*75+180,out=\x*75+180,looseness=1.5]
		($ ($ (B\x)! 0.3 !(B) $) !\widthpar!-90:(B) $);
	\path[name intersections={of=circA and brA, by=vA}];
	\path[name intersections={of=circB and brB, by=vB}];
	\draw (vA) node[vertex] {};
	\draw (vB) node[vertex] {};
	;
	};
\foreach \x/\y in {-1/0,0/1} {
	\draw[thick,color=blue] ($ ($ (A)! \f !(A\x) $) !\widthpar!-90:(A) $) to[out = \x*75, in=\y*75, looseness = 2] ($ ($ (A)! \f !(A\y) $) !\widthpar!90:(A) $) ;
	\draw[thick,color=blue] ($ ($ (B)! \f !(B\x) $) !\widthpar!-90:(B) $) to[out = \x*75 + 180, in=\y*75 + 180, looseness = 2] ($ ($ (B)! \f !(B\y) $) !\widthpar!90:(B) $) ;
	}; 
\foreach \x in {-1,1} {
	\draw[thick,color=blue] ($ ($ (A)! \f !(A\x) $) !\widthpar!-\x*90:(A) $) to[out = \x*75, in=180, looseness = 2] ($ ($ (A)! \f/2 !(B) $) !\widthpar!-\x*90:(B) $) ;
	\draw[thick,color=blue] ($ ($ (B)! \f !(B\x) $) !\widthpar!-\x*90:(B) $) to[out = \x*75+180, in=0, looseness = 2] ($ ($ (B)! \f/2 !(A) $) !\widthpar!-\x*90:(A) $) ;
	};
\draw[ color=black, dashed]
	($ ($ (A)! 0.5 !(B) $) ! \widthpar !90:(B) $)
	to[out=180,in=0,looseness=1.5]
	($ ($ (A)! 0.1 !(B) $) ! 0.4*\widthpar !90:(B) $)
	to [out = 180, in = - 75, looseness = 1.5]
	($ ($ (A)! 0.2 !(A-1) $) ! 0.4*\widthpar !-90:(A-1) $)
	--
	($ (A-1) ! 0.4*\widthpar !90:(A) $);
\draw[ color=black, dashed]
	($ ($ (A)! 0.5 !(B) $) ! \widthpar !-90:(B) $)
	to[out=180,in=180,looseness=1.3]
	($ ($ (A)! 0.9 !(B) $) ! 0.4*\widthpar !-90:(B) $)
	to [out = 0, in = - 75+180, looseness = 1.5]
	($ ($ (B)! 0.2 !(B-1) $) ! 0.4*\widthpar !-90:(B-1) $)
	--
	($ (B-1) ! 0.4*\widthpar !90:(B) $);
\foreach \x in {-1,0,1} {	
	\draw[color=black,dashed] 
		($ ($ (B\x)! 0.3 !(B) $) !\widthpar!90:(B) $)
		to[out=\x*75, in = \x*75+180,looseness=1.5]
		($ (B\x) ! 0.6*\widthpar !90:(B) $);
	\draw[color=black,dashed] 
		($ ($ (A\x)! 0.3 !(A) $) !\widthpar!90:(A) $)
		to[out=\x*75 + 180, in = \x*75,looseness=1.5]
		($ (A\x) ! 0.6*\widthpar !90:(A) $);
	};
\foreach \x/\y in {-1/0,0/1} {
	\draw[color=black,dashed] 
		($ ($ (B\x)! 0.3 !(B) $) !\widthpar!-90:(B) $)
		to[out=\x*75, in = \x*75,looseness=1.5]
		($ ($ (B\x)!0.8!(B) $) ! 0.2*\widthpar !-90:(B) $)
		to [out = \x*75+180, in = \y*75+180, 
		looseness = 1.5]
		($ ($ (B)! 0.2 !(B\y) $) ! 0.2*\widthpar !-90:(B\y) $)
		--
		($ (B\y) ! 0.2*\widthpar !90:(B) $);
	\draw[color=black,dashed] 
		($ ($ (A\x)! 0.3 !(A) $) !\widthpar!-90:(A) $)
		to[out=\x*75+180, in = \x*75+180,looseness=1.5]
		($ ($ (A\x)!0.8!(A) $) ! 0.2*\widthpar !-90:(A) $)
		to [out = \x*75, in = \y*75, 
		looseness = 1.5]
		($ ($ (A)! 0.2 !(A\y) $) ! 0.2*\widthpar !-90:(A\y) $)
		--
		($ (A\y) ! 0.2*\widthpar !90:(A) $);
	};
\draw[color=black,dashed] 
	($ ($ (A1)! 0.3 !(A) $) !\widthpar!-90:(A) $)
	to[out=75-180,in=75-180,looseness=1.5]
	($ ($ (A1)! 0.8 !(A) $) ! 0.2*\widthpar !-90:(A) $)
	to[out=75,in=180,looseness=1.5]
	($ ($ (A)! 0.1 !(B) $) ! 0.2*\widthpar !-90:(B) $)
	--
	($ ($ (A)! 0.9 !(B) $) ! 0.2*\widthpar !-90:(B) $)
	to[out=0,in=-75-180,looseness=1.5]
	($ ($ (B)! 0.2 !(B-1) $) ! 0.2*\widthpar !-90:(B-1) $)
	--
	($ (B-1) ! 0.2*\widthpar !90:(B) $);
\draw[color=black,dashed] 
	($ ($ (B1)! 0.3 !(B) $) !\widthpar!-90:(B) $)
	to[out=75,in=75,looseness=1.5]
	($ ($ (B1)! 0.8 !(B) $) ! 0.2*\widthpar !-90:(B) $)
	to[out=75+180,in=0,looseness=1.5]
	($ ($ (B)! 0.1 !(A) $) ! 0.2*\widthpar !-90:(A) $)
	--
	($ ($ (B)! 0.9 !(A) $) ! 0.2*\widthpar !-90:(A) $)
	to[out=180,in=-75,looseness=1.5]
	($ ($ (A)! 0.2 !(A-1) $) ! 0.2*\widthpar !-90:(A-1) $)
	--
	($ (A-1) ! 0.2*\widthpar !90:(A) $);
\end{tikzpicture}}

\newsavebox{\quotientline}
\savebox{\quotientline}{%
\begin{tikzpicture}[scale=0.85]
\tikzstyle{vertex}=[draw,scale=0.4,color=red,fill=red,circle]
\tikzstyle{vcirc}=[draw,scale=0.4,color=black,fill=black,circle]
\foreach \x in {1,...,4} {
	\coordinate (a\x) at (4*\x,0);
	\coordinate (B\x) at (4*\x,-1);
	\draw (a\x) node[vertex] {};
	\draw (B\x) node[vcirc] {};
	\foreach \y in {-1,1} {
		\coordinate (b\x\y) at (4*\x + \y,-1);
		\draw (b\x\y) node[vertex] {};
		\draw[color=red] (b\x\y) -- (a\x);
		\foreach \z in {-1,0,1} {
			\coordinate(c\x\y\z) at (4*\x + \y + 0.75*\z,-2);
			\draw (c\x\y\z) node[vertex]{};
			\draw[color=red] (c\x\y\z) -- (b\x\y);
			};
		\foreach \z in {-1,1} {
			\coordinate(C\x\y\z) at (4*\x + \y + 0.4*\z,-2);
			};
		\draw (C\x\y-1) to[out = 40, in = 140, looseness=1.5] (C\x\y1);
		\foreach \z in {-1,1} {
			\draw (C\x\y\z) node[vcirc]{};
			};
		};
	\draw (B\x) to[out=220,in=90,looseness=1.5] (C\x-11);
	\draw (B\x) to[out=320,in=90,looseness=1.5] (C\x1-1);
	};
\draw[color=red]  (3,0) -- (a1)--(a2)--(a3)--(a4) -- (17,0);
\draw[color=red,dashed] (2,0) -- (3,0);
\draw[color=red,dashed] (18,0) -- (17,0);
\foreach \x in {1,...,3} {
	\coordinate (A\x) at (4*\x + 2,-0.5);
	\draw (A\x) node[vcirc] {};
	\draw (A\x) to[out=230,in=90,looseness=1.5] (C\x11);
	\pgfmathparse{int(\x+1)};
	\draw (A\x) to[out=180,in=45,looseness=1] (B\x);
	\draw (A\x) to[out=0,in=135,looseness=1] (B\pgfmathresult);
	\draw (A\x) to[out=310,in=90,looseness=1.5] (C\pgfmathresult-1-1);
	};
\draw (2,-0.5) to[out=0,in=135,looseness=1] (B1);
\draw (18,-0.5) to[out=180,in=45,looseness=1] (B4);
\coordinate (V) at (10,2);
\draw (V) node[vcirc]{} -- (A2);
\draw(V) to[out=200,in=90,looseness=1] (A1);
\draw(V) to[out=-20,in=90,looseness=1] (A3);
\draw (V) node[anchor=south] {$O$};
\draw (a2) node[anchor=south] {$a$};
\end{tikzpicture}}
\newsavebox{\quotientloop}
\savebox{\quotientloop}{%
\begin{tikzpicture}[scale=0.9]
\tikzstyle{vertex}=[draw,scale=0.4,color=red,fill=red,circle]
\tikzstyle{vcirc}=[draw,scale=0.4,color=black,fill=black,circle]
\coordinate (C) at (0,0);
\draw (C) node[vcirc]{};
\foreach \x in {0,...,4} {
	\coordinate (A\x) at (\x*360/5 - 90:2);
	\draw (A\x) node[vcirc]{} -- (C);
	\coordinate (a\x) at (\x*360/5 - 90 + 36:1.5);
	\draw (a\x) node[vertex] {};
	\foreach \y in {-1,1} {
		\coordinate (b\x\y) at ($ (a\x) + (\x*360/5 - 90 + 36 + 36*\y:1) $);
		\draw[color=red] (b\x\y) node[vertex]{} -- (a\x);
		\foreach \z in {-1,0,1} {
			\coordinate (c\x\y\z) at ($ (b\x\y) + (\x*360/5 - 90 + 36 + 30*\y + 45*\z:1) $);
			\draw[color=red] (c\x\y\z) node[vertex] {} -- (b\x\y);
			};
		\foreach \z in {-1,1} {
			\coordinate (C\x\y\z) at ($ (b\x\y) + (\x*360/5 - 90 + 36 + 30*\y + 25*\z:0.6) $);
			\draw (C\x\y\z) node[vcirc] {};
			};
		\draw (C\x\y1) to[bend right=45] (C\x\y-1);
		};
	\coordinate (B\x) at (\x*360/5 - 90 + 36:2.5);
	\draw (C\x1-1) to (B\x);
	\draw (C\x-11) to (B\x);
	\draw (B\x) node[vcirc]{};
	\draw (B\x) to[bend right] (A\x);
	};
\foreach \x in {0,...,4} {
	\pgfmathparse{int(\x + 1 - 5*floor((\x+1)/5))};
	\draw (A\pgfmathresult) to[bend right] (B\x);
	\draw (C\x11) to (A\pgfmathresult);
	\draw (C\x-1-1) to (A\x);
	};
\draw[color=red] (a0) -- (a1) -- (a2) -- (a3) -- (a4) -- (a0);
\draw (C) node[anchor=south] {$O$};
\draw (a0) node[anchor=south east] {$a$};
\end{tikzpicture}}
\newsavebox{\geodesicwrap}
\savebox{\geodesicwrap}{%
\begin{tikzpicture}
\tikzstyle{vertex}=[draw,scale=0.4,color=red,fill=red,circle]
\tikzstyle{vcirc}=[draw,scale=0.4,color=black,fill=black,circle]
\coordinate (C) at (0,0);
\draw (C) node[vcirc]{};
\foreach \x in {0,...,4} {
	\coordinate (A\x) at (\x*360/5 - 90:2);
	\draw (A\x) node[vcirc]{} -- (C);
	\coordinate (a\x) at (\x*360/5 - 90 + 36:1.5);
	\draw (a\x) node[vertex] {};
	\foreach \y in {-1,1} {
		\coordinate (b\x\y) at ($ (a\x) + (\x*360/5 - 90 + 36 + 36*\y:1) $);
		\draw[color=red] (b\x\y) node[vertex]{} -- (a\x);
		\foreach \z in {-1,0,1} {
			\coordinate (c\x\y\z) at ($ (b\x\y) + (\x*360/5 - 90 + 36 + 30*\y + 45*\z:1) $);
			\draw[color=red] (c\x\y\z) node[vertex] {} -- (b\x\y);
			};
		\foreach \z in {-1,1} {
			\coordinate (C\x\y\z) at ($ (b\x\y) + (\x*360/5 - 90 + 36 + 30*\y + 25*\z:0.6) $);
			\draw (C\x\y\z) node[vcirc] {};
			};
		\draw (C\x\y1) to[bend right=45] (C\x\y-1);
		};
	\coordinate (B\x) at (\x*360/5 - 90 + 36:2.5);
	\draw (C\x1-1) to (B\x);
	\draw (C\x-11) to (B\x);
	\draw (B\x) node[vcirc]{};
	\draw (B\x) to[bend right] (A\x);
	};
\foreach \x in {0,...,4} {
	\pgfmathparse{int(\x + 1 - 5*floor((\x+1)/5))};
	\draw (A\pgfmathresult) to[bend right] (B\x);
	\draw (C\x11) to (A\pgfmathresult);
	\draw (C\x-1-1) to (A\x);
	};
\draw[color=red] (a0) -- (a1) -- (a2) -- (a3) -- (a4) -- (a0);
\draw (c4-10) node[anchor=north east] {$x$};
\draw (c11-1) node[anchor=west] {$y$};
\draw (c210) node[anchor=south east] {$y'$};
\draw[color=blue, very thick] (a4)node[vertex]{}  (a4) node[vertex]{} -- (a0) node[vertex]{} -- (a1) node[vertex]{} -- (b11) node[vertex]{} -- (c11-1)node[vertex]{};
\draw[color=green!80!black,very thick] (c4-10) node[vertex]{} -- (b4-1) node[vertex]{} -- (a4) node[vertex]{} -- (a3) node[vertex]{} -- (a2) node[vertex]{} -- (b21) node[vertex]{} -- (c210) node[vertex]{};
\draw[color=blue, very thick, loosely dashed] (c4-10) node[vertex]{} -- (b4-1) node[vertex]{} -- (a4) node[vertex]{};
\end{tikzpicture}}
\newsavebox{\gluehole}
\savebox{\gluehole}{%
\begin{tikzpicture}[scale=0.9]
\tikzstyle{vertex}=[draw,scale=0.4,color=red,fill=red,circle]
\tikzstyle{vcirc}=[draw,scale=0.4,color=black,fill=black,circle]
\foreach \p in {-1,1} {
\foreach \x in {1,...,4} {
	\coordinate (a\p\x) at (4*\x,\p);
	\coordinate (B\p\x) at (4*\x,2*\p);
	\draw (a\p\x) node[vertex] {};
	\draw (B\p\x) node[vcirc] {};
	\foreach \y in {-1,1} {
		\coordinate (b\p\x\y) at (4*\x + \y,2*\p);
		\draw (b\p\x\y) node[vertex] {};
		\draw[color=red] (b\p\x\y) -- (a\p\x);
		};
	};
\draw[color=red]  (3,\p) -- (a\p1)--(a\p2)--(a\p3)--(a\p4) -- (17,\p);
\draw[color=red,dashed] (2,\p) -- (3,\p);
\draw[color=red,dashed] (18,\p) -- (17,\p);
\coordinate (V\p) at (10,0.5*\p);
\draw (V\p) node[vcirc]{};
\foreach \x in {1,...,3} {
	\coordinate (A\p\x) at (4*\x + 2,1.5*\p);
	\draw (A\p\x) node[vcirc] {};
	\pgfmathparse{int(\x+1)};
	\draw (A\p\x) to[out=180,in=-\p*45,looseness=1] (B\p\x);
	\draw (A\p\x) to[out=0,in=180+\p*45,looseness=1] (B\p\pgfmathresult);
	\draw (V\p) to[out= - 90*\p*\x - 90*\p ,in=-90*\p,looseness=0.8] (A\p\x);
	};
};
\foreach \x in {-1,0,1} {
	\draw (V-1) to[in= - 30*\x - 90,out=30*\x + 90,looseness=1.2] (V1);
	};
\end{tikzpicture}}
\newsavebox{\twointplain}
\savebox{\twointplain}{%

\begin{tikzpicture}[scale=1.5]
\tikzstyle{vertex}=[draw,scale=0.4,fill=black,circle]
\tikzstyle{vred}=[draw,scale=0.4,color=red,fill=red,circle]
\tikzstyle{vint}=[draw,scale=0.55,color=blue,fill=blue,circle]
\coordinate (C) at (0,0);
\foreach \x in {0,1,2,3} {
	\coordinate (a\x) at (\x*360/4:1);
	\draw[color=red] (a\x) node[vred]{} -- (C);
	};
\foreach \x in {0,1,...,11} {
	\coordinate (b\x) at (\x*360/12 - 30 :2);
	\pgfmathparse{floor(\x/3)}
	\draw[color=red] (b\x) node[vred]{} -- (a\pgfmathresult);
	};
\foreach \x in {0,1,...,11} {
	\coordinate (c\x) at (\x*360/12 - 45 :2);
	\draw (c\x) node[vertex] {};
	};
\foreach \x/\y in {0/1,1/2,2/3,3/4,4/5,5/6,6/7,7/8,8/9,9/10,10/11,11/0} {
	\draw[thick] (c\x) to [out =  \x*360/12 + 115, 
		in =  \y*360/12 + 155, looseness = 1.5] (c\y);	
	};
\foreach \x/\y in {0/3,3/6,6/9,9/0} {
	\draw[thick] (c\x) to [out =  \x*360/12 + 135, 
		in =  \y*360/12 + 135, looseness = 1.3] (c\y);	
	};
\foreach \x in {3,4,5,10,11} {
	\draw (c\x) node[vint] {};
	};
\draw (b2) node[anchor=south west] {$x_1$};
\draw (b5) node[anchor=south east] {$x_2$};
\draw (b9) node[anchor=north east] {$x_3$};
\draw (b11) node[anchor=north west] {$x_4$};
\draw[dotted] (0,0) circle[radius=1];
\draw[dotted] (0,0) circle[radius=2];
\draw (C) node[below left] {$C$};
\draw (C) node[vred] {};
\end{tikzpicture}}
\newsavebox{\twointcolored}
\savebox{\twointcolored}{%
\begin{tikzpicture}[scale=1.5]
\tikzstyle{vertex}=[draw,scale=0.4,fill=black,circle]
\tikzstyle{vred}=[draw,scale=0.4,color=red,fill=red,circle]
\tikzstyle{vint}=[draw,scale=0.55,color=blue,fill=blue,circle]
\coordinate (C) at (0,0);
\foreach \x in {0,1,2,3} {
	\coordinate (a\x) at (\x*360/4:1);
	\draw[color=red] (a\x) -- (C);
	};
\foreach \x in {0,1,...,11} {
	\coordinate (b\x) at (\x*360/12 - 30 :2);
	\pgfmathparse{floor(\x/3)}
	\draw[color=red] (b\x) -- (a\pgfmathresult);
	};
\foreach \x in {0,1,...,11} {
	\coordinate (c\x) at (\x*360/12 - 45 :2);
	};
\foreach \x/\y in {0/1,1/2,2/3,3/4,4/5,5/6,6/7,7/8,8/9,9/10,10/11,11/0} {
	\draw[thick] (c\x) to [out =  \x*360/12 + 115, 
		in =  \y*360/12 + 155, looseness = 1.5] (c\y);	
	};
\foreach \x/\y in {0/3,3/6,6/9,9/0} {
	\draw[thick] (c\x) to [out =  \x*360/12 + 135, 
		in =  \y*360/12 + 135, looseness = 1.3] (c\y);	
	};
\draw[blue, very thick] (b2) -- (a0) -- (C) -- (a1) -- (b5);
\draw[blue, very thick] (b9) -- (a3) -- (b11);
\foreach \x/\y in {2/3,5/6,9/10,11/0} {
	\draw[green,very thick] (c\x) to [out =  \x*360/12 + 115, 
		in =  \y*360/12 + 155, looseness = 1.5] (c\y);	
	};
\foreach \x/\y in {0/3,3/6} {
	\draw[green, very thick] (c\x) to [out =  \x*360/12 + 135, 
		in =  \y*360/12 + 135, looseness = 1.3] (c\y);	
	};
\foreach \x/\y in {0/1,1/2,6/7,7/8,8/9} {
    \draw[white, very thick] (c\x) to [out =  \x*360/12 + 115, 
		in =  \y*360/12 + 155, looseness = 1.5] (c\y);	
	\draw[black, thick, loosely dashed] (c\x) to [out =  \x*360/12 + 115, 
		in =  \y*360/12 + 155, looseness = 1.5] (c\y);	
	};
\foreach \x/\y in {6/9,9/0} {
	\draw[white, very thick] (c\x) to [out =  \x*360/12 + 135, 
		in =  \y*360/12 + 135, looseness = 1.3] (c\y);	
	\draw[black, thick, loosely dashed] (c\x) to [out =  \x*360/12 + 135, 
		in =  \y*360/12 + 135, looseness = 1.3] (c\y);	
	};
\draw[dotted] (0,0) circle[radius=1];
\draw[dotted] (0,0) circle[radius=2];
\draw (C) node[below left] {$C$};
\draw (C) node[vred] {};
\foreach \x in {0,1,2,3} {
	\draw (a\x) node[vred]{};
	};
\foreach \x in {0,1,...,11} {
	\draw (b\x) node[vred]{};
	\draw (c\x) node[vertex] {};
	};
\draw (b2) node[anchor=south west] {$x_1$};
\draw (b5) node[anchor=south east] {$x_2$};
\draw (b9) node[anchor=north east] {$x_3$};
\draw (b11) node[anchor=north west] {$x_4$};
\foreach \x in {3,4,5,10,11} {
	\draw (c\x) node[vint] {};
	};
\end{tikzpicture}}
\newsavebox{\twointIzero}
\savebox{\twointIzero}{%
\begin{tikzpicture}[scale=1.5]
\tikzstyle{vred}=[draw,scale=0.4,color=red,fill=red,circle]
\foreach \x in {0,1,...,11} {
	\coordinate (b\x) at (\x*360/12 - 30 :2);
	\pgfmathparse{floor(\x/3)}
	};
\draw[thick,blue] (b2) arc (-45:-166:1.6); 	
\draw[thick,blue] (b11) arc (45:135:1.4); 	
\draw[thick] (b2) arc (30:119:2);
\draw[thick] (b9) arc (240:300:2);
\draw[dotted] (0,0) circle[radius=2];
\draw (b2) node[anchor=south west] {$x_1$};
\draw (b5) node[anchor=south east] {$x_2$};
\draw (b9) node[anchor=north east] {$x_3$};
\draw (b11) node[anchor=north west] {$x_4$};
\draw (b4) node[anchor=south] {$A_1$};
\draw (b10) node[anchor=north] {$A_2$};
\end{tikzpicture}}
\newsavebox{\twointIpos}
\savebox{\twointIpos}{%
\begin{tikzpicture}[scale=1.5]
\tikzstyle{vred}=[draw,scale=0.4,color=red,fill=red,circle]
\foreach \x in {0,1,...,11} {
	\coordinate (b\x) at (\x*360/12 - 30 :2);
	\pgfmathparse{floor(\x/3)}
	};
\draw[thick,blue] (b2) arc (-45:-166:1.6); 	
\draw[thick,blue] (b11) arc (45:135:1.4); 	
\draw[thick] (b2) arc (30:-60:2);
\draw[thick] (b5) arc (119:240:2);
\draw[dotted] (0,0) circle[radius=2];
\draw (b2) node[anchor=south west] {$x_1$};
\draw (b5) node[anchor=south east] {$x_2$};
\draw (b9) node[anchor=north east] {$x_3$};
\draw (b11) node[anchor=north west] {$x_4$};
\draw (b7) node[anchor=east] {$A_1$};
\draw (b1) node[anchor=west] {$A_2$};
\end{tikzpicture}}
\newsavebox{\intCRone}
\savebox{\intCRone}{%
\begin{tikzpicture}[scale=1.5]
\tikzstyle{vertex}=[draw,scale=0.4,fill=black,circle]
\tikzstyle{vred}=[draw,scale=0.4,color=red,fill=red,circle]
\tikzstyle{vint}=[draw,scale=0.55,color=blue,fill=blue,circle]
\coordinate (C) at (0,0);
\foreach \x in {0,1,2,3} {
	\coordinate (a\x) at (\x*360/4:1);
	\draw[color=red] (a\x) -- (C);
	};
\foreach \x in {0,1,...,11} {
	\coordinate (b\x) at (\x*360/12 - 30 :2);
	\pgfmathparse{floor(\x/3)}
	\draw[color=red] (b\x) -- (a\pgfmathresult);
	};
\foreach \x in {0,1,...,11} {
	\coordinate (c\x) at (\x*360/12 - 45 :2);
	};
\foreach \x/\y in {0/1,1/2,2/3,3/4,4/5,5/6,6/7,7/8,8/9,9/10,10/11,11/0} {
	\draw[thick] (c\x) to [out =  \x*360/12 + 115, 
		in =  \y*360/12 + 155, looseness = 1.5] (c\y);	
	};
\foreach \x/\y in {0/3,3/6,6/9,9/0} {
	\draw[thick] (c\x) to [out =  \x*360/12 + 135, 
		in =  \y*360/12 + 135, looseness = 1.3] (c\y);	
	};
\draw[blue, very thick] (b2) -- (a0) -- (C) -- (a1) -- (b5);
\draw[blue, very thick] (b8) -- (a2) -- (C) -- (a3) -- (b9);
\foreach \x/\y in {2/3,5/6,8/9,9/10} {
	\draw[green,very thick] (c\x) to [out =  \x*360/12 + 115, 
		in =  \y*360/12 + 155, looseness = 1.5] (c\y);	
	};
\foreach \x/\y in {0/3,3/6,6/9,9/0} {
	\draw[green, very thick] (c\x) to [out =  \x*360/12 + 135, 
		in =  \y*360/12 + 135, looseness = 1.3] (c\y);	
	};
\foreach \x/\y in {10/11,11/0,0/1,1/2,6/7,7/8} {
    \draw[white, very thick] (c\x) to [out =  \x*360/12 + 115, 
		in =  \y*360/12 + 155, looseness = 1.5] (c\y);	
	\draw[black, thick, loosely dashed] (c\x) to [out =  \x*360/12 + 115, 
		in =  \y*360/12 + 155, looseness = 1.5] (c\y);	
	};

\draw[dotted] (0,0) circle[radius=1];
\draw[dotted] (0,0) circle[radius=2];
\draw (C) node[below left] {$C$};
\draw (C) node[vred] {};
\foreach \x in {0,1,2,3} {
	\draw (a\x) node[vred]{};
	};
\foreach \x in {0,1,...,11} {
	\draw (b\x) node[vred]{};
	\draw (c\x) node[vertex] {};
	};
\draw (b2) node[anchor=south west] {$x_1$};
\draw (b5) node[anchor=south east] {$x_2$};
\draw (b9) node[anchor=north east] {$x_4$};
\draw (b8) node[anchor=north east] {$x_3$};
\foreach \x in {3,4,5,9} {
	\draw (c\x) node[vint] {};
	};
\end{tikzpicture}}
\newsavebox{\intCRlarge}
\savebox{\intCRlarge}{%
\begin{tikzpicture}[scale=1.5]
\tikzstyle{vertex}=[draw,scale=0.4,fill=black,circle]
\tikzstyle{vred}=[draw,scale=0.4,color=red,fill=red,circle]
\tikzstyle{vint}=[draw,scale=0.55,color=blue,fill=blue,circle]
\coordinate (C) at (0,0);
\foreach \x in {0,1,2,3} {
	\coordinate (a\x) at (\x*360/4:1);
	\draw[color=red] (a\x) -- (C);
	};
\foreach \x in {0,1,...,11} {
	\coordinate (b\x) at (\x*360/12 - 30 :2);
	\pgfmathparse{floor(\x/3)}
	\draw[color=red] (b\x) -- (a\pgfmathresult);
	};
\foreach \x in {0,1,...,11} {
	\coordinate (c\x) at (\x*360/12 - 45 :2);
	};
\foreach \x/\y in {0/1,1/2,2/3,3/4,4/5,5/6,6/7,7/8,8/9,9/10,10/11,11/0} {
	\draw[thick] (c\x) to [out =  \x*360/12 + 115, 
		in =  \y*360/12 + 155, looseness = 1.5] (c\y);	
	};
\foreach \x/\y in {0/3,3/6,6/9,9/0} {
	\draw[thick] (c\x) to [out =  \x*360/12 + 135, 
		in =  \y*360/12 + 135, looseness = 1.3] (c\y);	
	};
\draw[blue, very thick] (b2) -- (a0) -- (b1);
\draw[blue, very thick] (b5) -- (a1)  -- (C) -- (a3) -- (b11);
\foreach \x/\y in {2/3,5/6,11/0,1/2} {
	\draw[green,very thick] (c\x) to [out =  \x*360/12 + 115, 
		in =  \y*360/12 + 155, looseness = 1.5] (c\y);	
	};
\foreach \x/\y in {3/6,9/0} {
	\draw[green, very thick] (c\x) to [out =  \x*360/12 + 135, 
		in =  \y*360/12 + 135, looseness = 1.3] (c\y);	
	};
\foreach \x/\y in {6/7,7/8,8/9,9/10,10/11} {
    \draw[white, very thick] (c\x) to [out =  \x*360/12 + 115, 
		in =  \y*360/12 + 155, looseness = 1.5] (c\y);	
	\draw[black, thick, loosely dashed] (c\x) to [out =  \x*360/12 + 115, 
		in =  \y*360/12 + 155, looseness = 1.5] (c\y);	
	};
\foreach \x/\y in {6/9} {
	\draw[white, very thick] (c\x) to [out =  \x*360/12 + 135, 
		in =  \y*360/12 + 135, looseness = 1.3] (c\y);	
	\draw[black, thick, loosely dashed] (c\x) to [out =  \x*360/12 + 135, 
		in =  \y*360/12 + 135, looseness = 1.3] (c\y);	
	};
\draw[dotted] (0,0) circle[radius=1];
\draw[dotted] (0,0) circle[radius=2];
\draw (C) node[below left] {$C$};
\draw (C) node[vred] {};
\foreach \x in {0,1,2,3} {
	\draw (a\x) node[vred]{};
	};
\foreach \x in {0,1,...,11} {
	\draw (b\x) node[vred]{};
	\draw (c\x) node[vertex] {};
	};
\draw (b2) node[anchor=south west] {$x_1$};
\draw (b5) node[anchor=south east] {$x_2$};
\draw (b1) node[anchor=north west] {$x_4$};
\draw (b11) node[anchor=north west] {$x_3$};
\foreach \x in {3,4,5,0,1} {
	\draw (c\x) node[vint] {};
	};
\end{tikzpicture}}
\newsavebox{\uconfig}
\savebox{\uconfig}{%
\begin{tikzpicture}[scale=1.5]
\tikzstyle{vred}=[draw,scale=0.4,color=red,fill=red,circle]
\foreach \x in {0,2,6,8} {
	\coordinate (b\x) at (\x*360/12 - 30 :2);
	\pgfmathparse{floor(\x/3)}
	\draw[color=red] (b\x) node[vred]{} -- (a\pgfmathresult);
	};
\foreach \x in {0,1,2,3} {
	\coordinate (a\x) at (\x*360/4:1);
	};
\draw[color=red] (a0) -- (a2);
\draw (b2) node[anchor=south west] {$x_4$};
\draw (b0) node[anchor=north west] {$x_3$};
\draw (b8) node[anchor=north east] {$x_2$};
\draw (b6) node[anchor=south east] {$x_1$};
\draw[dotted] (0,0) circle[radius=2];
\end{tikzpicture}}
\newsavebox{\uvconfig}
\savebox{\uvconfig}{%
\begin{tikzpicture}[scale=1.5]
\tikzstyle{vred}=[draw,scale=0.4,color=red,fill=red,circle]
\foreach \x in {0,2,6,8} {
	\coordinate (b\x) at (\x*360/12 - 30 :2);
	\pgfmathparse{floor(\x/3)}
	\draw[color=red] (b\x) node[vred]{} -- (a\pgfmathresult);
	};
\foreach \x in {4} {
	\coordinate (b\x) at (\x*360/12 - 30 :2);
	\pgfmathparse{floor(\x/3)}
	\draw[color=red] (b\x) node[vred]{} -- (0,0);
	};
\foreach \x in {0,1,2,3} {
	\coordinate (a\x) at (\x*360/4:1);
	};
\draw[color=red] (a0) -- (a2);
\draw (0,0) node[anchor=north] {$C$};
\draw (b4) node[anchor=south] {$x_5$};
\draw (b2) node[anchor=south west] {$x_4$};
\draw (b0) node[anchor=north west] {$x_3$};
\draw (b8) node[anchor=north east] {$x_2$};
\draw (b6) node[anchor=south east] {$x_1$};
\draw[dotted] (0,0) circle[radius=2];
\end{tikzpicture}}

\def\musepic#1{\vcenter{\hbox{\usebox{#1}}}}

\begin{document}
\preprint{CALT-TH-2018-053}

\title{Nonarchimedean Holographic Entropy from Networks of Perfect Tensors}
\authors{
Matthew~Heydeman,$^1$\footnote{mheydema@caltech.edu} Matilde~Marcolli,$^{2,3,4}$\footnote{matilde@caltech.edu} Sarthak~Parikh,$^2$\footnote{sparikh@caltech.edu} \&~Ingmar~Saberi$^5$\footnote{saberi@mathi.uni-heidelberg.de}}
\institution{Caltech}{$^1$Walter Burke Institute for Theoretical Physics,\cr
California Institute of Technology,  Pasadena, CA 91125, USA}
\institution{Caltech}{$^2$Division of Physics, Mathematics and Astronomy,\cr 
California Institute of Technology, Pasadena, CA 91125, USA}
\institution{Toronto}{$^3$Department of Mathematics, University of Toronto, Toronto, ON M5S 2E4, Canada}
\institution{PI}{$^4$Perimeter Institute for Theoretical Physics, Waterloo, ON N2L 2Y5, Canada}
\institution{RKU}{$^5$Mathematisches Institut, Ruprecht-Karls-Universit\"{a}t Heidelberg,\cr
Im Neuenheimer Feld 205, 69120 Heidelberg, Germany}

\abstract{
We consider a class of holographic quantum error-correcting codes, built from perfect tensors in network configurations dual to Bruhat--Tits trees and their quotients by Schottky groups corresponding to BTZ black holes. The resulting holographic states can be constructed in the limit of infinite network size. We obtain a $p$-adic version of entropy which obeys a Ryu--Takayanagi like formula for bipartite entanglement of connected or disconnected regions, in both genus-zero and genus-one $p$-adic backgrounds, along with a Bekenstein--Hawking-type formula for black hole entropy. We prove entropy inequalities obeyed by such tensor networks, such as subadditivity, strong subadditivity, and monogamy of mutual information (which is always saturated). In addition, we construct infinite classes of perfect tensors directly from semiclassical states in phase spaces over finite fields, generalizing the CRSS algorithm, and give Hamiltonians exhibiting these as vacua. 
}

\date{\today}

\maketitle

\setcounter{tocdepth}{2}

{\hypersetup{linkcolor=black}
\tableofcontents
}

\section{Introduction}
\label{INTRODUCTION}

In \cite{Man84} and \cite{Man87}, Manin suggested that physical theories carry arithmetic structures
and that physics as we know it in the real world in fact exists in an adelic form, with $p$-adic 
realizations alongside its real form. These $p$-adic manifestations of physical theories can
be used, by virtue of consistency constraints between the archimedean and the nonarchimedean
places of adelic objects, to determine the real part through knowledge of the $p$-adic side. Since 
some of the objects that exist on the $p$-adic side, such as the Bruhat--Tits trees \cite{BT72}, have a discrete,
combinatorial nature, one can take advantage of this structure to carry out computations in a
more convenient discretized setting. There has been a considerable amount of work over the years concerned
with developing various aspects of physics in a $p$-adic setting, see for instance the reviews \cite{Vlad,Khren,Dragovich:2009hd,Dragovich:2017kge}. In particular, the Bruhat--Tits tree of ${\mathbb Q}_p$ (the $p$-adic numbers) and its
boundary ${\mathbb P}^1({\mathbb Q}_p)$ were used in the setting of $p$-adic string theory~\cite{Freund:1987ck,Brekke:1988dg,Chekhov:1989bg,Zab,Brekke:1993gf}. More recently, a different perspective was taken on $p$-adic numbers in physics based on the observation that there exist natural pairs of ``bulk and boundary'' spaces associated to $p$-adic algebraic curves, with many similarities to the AdS/CFT correspondence.
 
 In AdS$_3$/CFT$_2$, one often realizes the boundary conformal field theory as living on the genus zero Riemann surface $\mathbb{P}^1(\mathbb{C})$ with the hyperbolic bulk AdS space as a coset. Higher genus bulk and boundary spaces can be seen as generalizations of the (Euclidean) BTZ black hole~\cite{Krasnov:2000zq}. The original BTZ
black hole~\cite{Banados:1992wn} corresponds to the genus one case. As curves and cosets, the construction of these spaces is entirely algebraic, and one may find other spaces by changing the underlying field. Based on this analogy, in \cite{Manin:2002hn} it was suggested that certain $p$-adic algebraic curves coming from the replacement $\mathbb{C} \rightarrow \mathbb{Q}_p$ such as $\mathbb{P}^1(\mathbb{Q}_p)$ and the higher genus Mumford curves of \cite{Mum}, are suitable boundary spaces for holography. An attractive feature of this proposal is that the bulk space is still a coset, but now becomes the Bruhat--Tits tree at genus zero. This opens up the possibility of studying certain features of AdS/CFT by passing to a $p$-adic setting where the bulk and boundary geometry are relatively simple.

A detailed theory of the $p$-adic AdS/CFT correspondence was only established much more recently. 
Appropriate boundary and bulk field theories for $p$-adic holography were developed
recently and independently in \cite{Gubser:2016guj, Heydeman:2016ldy}, where certain essential features such as the holographic computation of correlation functions in $p$-adic conformal field theories were established. Many lines of inquiry parallel the situation in real AdS/CFT, but the discrete $p$-adic geometry often makes these models much more solvable.
These models have been explored further in a series of subsequent papers, such as \cite{Gubser:2016htz,Gubser:2017vgc,Gubser:2017tsi,Bhattacharyya:2017aly,Gubser:2017qed,Bhowmick:2018bmn,Qu:2018ned,Jepsen:2018dqp,Gubser:2018cha}. 
Much of this work has focused on exploring analogies between $p$-adic models and ordinary AdS/CFT, and searching for structures familiar from the traditional holographic correspondence in the  discretized or $p$-adic world. Beyond holographic correlators, one may look for structures associated to the bulk geometry directly, including the Ryu--Takayanagi (RT) formula~\cite{Ryu:2006bv,Ryu:2006ef} and its quantum and covariant generalizations~\cite{Faulkner:2013ana,Hubeny:2007xt,Wall:2012uf}. 
One can also ask about other basic properties of the boundary entropy,  such as the strong subadditivity property~\cite{Headrick:2007km} or other entropy inequalities. One may expect these aspects of holographic entropy to have a $p$-adic analog as well.

The purpose of the present paper is to focus on the tensor network approach to
holography, and in particular on holographic states generated by networks of perfect tensors as in~\cite{Pastawski:2015qua}.
Like the Bruhat--Tits tree, tensor networks are generally discrete and provide simplified models to study bulk and boundary entanglement properties.
In fact one is tempted to view the Bruhat--Tits tree itself as a tensor network~\cite{Bhattacharyya:2017aly,Marcolli:2018ohd}, or closely related to a tensor network~\cite{Heydeman:2016ldy}; however these models were unsuccessful in reproducing the RT formula  or various expected entropy inequalities in the $p$-adic setting, so the question of whether entanglement entropy results familiar from the usual real holography hold over the $p$-adics has remained open. 

In this work, we present a simple model of holographic quantum error correction in the $p$-adic setting based on the existence of an infinite class of perfect tensors which can be used to build networks associated with the bulk geometry; either the Bruhat--Tits tree or its black hole variant. 
The tensor networks we use are closely related to those of ~\cite{Pastawski:2015qua} based on hyperbolic tessellations; in the simplest case of vacuum AdS this is described by the Schl\"{a}fli symbol $\{m,n\}$ and the associated tensor network $\{n,m\}$. Heuristically, in our model we study the limit in which the Sch\"{a}fli symbols of the associated  bulk geometry and the dual tensor network tend to $\{\infty,p+1\}$ and $\{p+1,\infty\}$ respectively. 
The subtleties associated with the $p$-adic interpretation of such tensor networks built from perfect tensors of rank tending to infinity are discussed at length in this paper.
We emphasize, however, that the genus 1 black hole tensor networks we consider are fundamentally different from those proposed in~\cite{Pastawski:2015qua}, and are instead obtained via a physically well motivated quotient procedure.

This model addresses a number of shortcomings of previous approaches to tensor network holography and allows for the explicit analytic computation of holographic states, density matrices, and entropies. 
While this model is discrete and preserves a (finite) group of conformal symmetries at finite cutoff, we see a full restoration of conformal symmetry as the cutoff is taken to zero, in the form of the $p$-adic fractional linear transformations $\PGL(2,\mathbb{Q}_p)$, which we interpret as the conformal group acting on a spatial region of the boundary. This group acts by isometries on the Bruhat--Tits tree, thought of as the analog of a time slice of AdS$_3$. The tensor network inherits the symmetries, and is essentially related to minimal geodesics of the Bruhat--Tits tree. Additionally, while the network is defined and manipulated in the bulk space, all quantities we compute can ultimately be defined or described in terms of purely boundary data such as configurations of points and sets on $\mathbb{P}^1(\mathbb{Q}_p)$ or the genus one curve. 

Given a choice of prime number $p$ and choice of bulk IR cut off, there is essentially a single network defined for each bulk space which generates a highly entangled state of boundary qudits, interpreted as the analog of a vacuum state of a boundary conformal field theory. Equipped with the $p$-adic analogs of various quantities and the knowledge of how to manipulate perfect tensors, we are able to explicitly compute nonarchimedean entanglement entropies for connected and disconnected intervals; the results are dual to minimal surfaces in the bulk, as expected from the Ryu--Takayanagi formula.
We also compute the black hole entropy and find that it is proportional to the perimeter 
of the $p$-adic BTZ black hole, as expected according to the Bekenstein--Hawking formula, 
and verify the RT formula, which equates the von Neumann entropy of reduced density matrices obtained from mixed states on the boundary to the lengths of minimal geodesics in the $p$-adic black hole background homologous to boundary regions. 
We also give a holographic derivation of subadditivity, strong subadditivity, and the monogamy of mutual information in the $p$-adic setting. 
In the limit of an infinite network, all of these results can be phrased in terms of conformally invariant information on the boundary, where the ultrametric geometry plays an essential simplifying role. Another interesting feature is our use of graphical tools to perform bulk computations; essentially all entropy quantities can be obtained by geometric operations such as cutting, gluing, and tracing the discrete vertices and bonds of the network. Among other things, this leads to the interpretation of a thermal density as being dual to a two-sided AdS black hole obtained by gluing two bulk regions together.

We construct the network of perfect
tensors associated to the Bruhat--Tits tree as follows. Rather than placing the tensors at the nodes of
the Bruhat--Tits tree as previously suggested, we identify two explicit conditions for the
construction of an appropriate ``dual graph" on which the tensor network lives. While this
at first appears to require an embedding of a nonarchimedean bulk space into an ordinary archimedean
plane, we show that this embedding and the dual graph can also be constructed by remaining entirely 
inside the $p$-adic world, using the Drinfeld $p$-adic plane \cite{Dri} together with the choice of 
a section of its projection to the Bruhat--Tits tree.

For practical purposes, all the computations can be carried out using an embedding in the
ordinary plane. The two main conditions required for the construction of the tensor network are that the sets of edges of the Bruhat--Tits tree and its dual graph are in one-to-one correspondence, with an edge of the dual graph
cutting exactly one edge of the Bruhat--Tits tree and that the arrangements of dual graph
edges around each vertex of the Bruhat--Tits tree form ``plaquettes," i.e., admit a cyclic ordering. Any construction of
a dual graph that satisfies these properties can be used for the purpose of von Neumann
entropy computations.

In trying to capture holographic states, perfect tensors have appeared as a convenient way of generating maximally entangled states. We offer a refined point of view on perfect tensors, which was already 
partially outlined in~\cite{Heydeman:2016ldy}  and~\cite{Marcolli:2018ohd} by some of the present authors. Starting with classical error-correcting codes in the form of Reed--Solomon codes built over projective lines over finite fields~\cite{Tsfa}, one may upgrade these to quantum codes by applying the CRSS algorithm~\cite{CRSS}, which we
show can be generalized to directly obtain perfect tensors from certain self-orthogonal codes. These self-orthogonal codes are Lagrangian subspaces of symplectic vector spaces over finite fields; they can thus be thought of as analogous to semiclassical states, and the theory of the Heisenberg group over finite fields can be used to quantize them, replacing the equations defining the Lagrangian by operator equations (or eigenvalue problems) and producing the corresponding quantum codes.  Our construction both generalizes the families of
perfect tensors used in the construction of holographic codes in~\cite{Pastawski:2015qua}, and gives a physical interpretation of the perfect-tensor condition.
In fact, we also prove more generally that the perfect tensor condition is, in a suitable sense,
``generic" within the CRSS construction of quantum codes, where the generic condition
is described geometrically in terms of the position of the corresponding Lagrangian subspaces or semiclassical states. We work in the setting of the
Gurevich--Hadani functorial quantization of symplectic vector spaces over finite fields~\cite{GH}.

We now give a concise summary of the main results obtained in this paper, followed by a more detailed organization of the paper.

\subsection{Summary of the main results}

\begin{itemize}
\item We show that for static holographic
states built through a  network of perfect tensors dual to the $p$-adic Bruhat--Tits tree (both when the  boundary is the ``infinite line'' $\bQ_p$ and when it is the projective line $\mathbb{P}^1(\bQ_p)$), the bipartite entanglement entropy of a single connected interval as well as disconnected intervals obeys a Ryu--Takayanagi like formula. The 
perfect tensors may be viewed as built via the CRSS algorithm from algebro-geometric codes on projective lines 
over finite fields but this is not crucial to our setup. The entanglement is computed by constructing the holographic state, tracing out 
regions of the tensor network, and explicitly computing the reduced density matrices and the von 
Neumann entropy, which is expressed in terms of (regularized) lengths of minimal geodesics in
the bulk Bruhat--Tits tree.  We also prove subadditivity, strong subadditivity and monogamy of mutual information in this setup.

\item We construct $p$-adic BTZ black holes as quotients of the Bruhat--Tits tree by a rank-one
Schottky group with boundary a Mumford--Tate elliptic curve, and demonstrate that the construction of the
tensor network adapts naturally to this case. Essentially the tensor network is obtained as a quotient of the genus 0 tensor network, paralleling the quotient construction of the geometry. Instead of a pure state at the boundary one has in this case a vertex
behind the horizon that needs to be traced out, which results in a thermal density matrix with 
a Bekenstein--Hawking entropy measured in terms of the length of the horizon (the polygon in
the quotient of the Bruhat--Tits tree). This density matrix can be seen to be dual to a bulk geometry with two asymptotic regions connected by the analog of a two-sided black hole, with the entropy given by the number of tensor bonds suspended between the two sides. We also prove that the entanglement entropy satisfies 
an analog of the Ryu--Takayanagi formula in this geometry in terms of the minimal length of homologous geodesics in the black hole background. 

\item We prove that perfect tensors can be constructed through a general procedure of geometric quantization
from general-position Lagrangians in a symplectic vector space over a finite field. This shows
that perfect tensors are ``generic" for the CRSS algorithm producing quantum codes from classical codes. 
The construction also provides natural Hamiltonians for which the vacuum state is the perfect tensor state.
\end{itemize}

\subsection{Organization of the paper}

In section~\ref{REVIEW} we review some basic background material that we will be using throughout the
paper.  
Section~\ref{ssec:padic}
gives the minimal background on the geometry of the $p$-adic Bruhat--Tits trees, which serve as our
bulk spaces in the rest of the paper. A discussion of quotients of Bruhat--Tits trees by
$p$-adic Schottky groups is given later, in section~\ref{GENUSONE}. In section~\ref{ssec:tensors} we
briefly review networks of perfect tensors and maximally entangled states. Section~\ref{CLASSQUANTCODES}
recalls several facts about classical and quantum codes that we will be using in the rest of the paper,
with particular focus on the CRSS algorithm that promotes classical to quantum error correcting codes,
which we describe in terms of Heisenberg group representations. We review in particular the
classical algebro-geometric codes associated to the projective line over a finite field (the Reed--Solomon
codes), and we show that they can be used to construct, through the CRSS algorithm,
quantum codes given by perfect tensors.  An explicit example is illustrated in appendix \ref{THREEQ}.

Section~\ref{CRSS} focuses on the construction and physical interpretation of perfect tensors, proving some new 
general results about the CRSS algorithm and identifying conditions under which
it can be used to produce perfect tensors in terms of the geometry of semiclassical states. Section~\ref{ssec:HeisGrp} provides more details on
irreducible representations of the Heisenberg group than what was discussed in section~\ref{CLASSQUANTCODES}, in particular discussing their construction from the regular representation as invariant subspaces of the commuting action of an abelian group, corresponding to a choice of  Lagrangian in the symplectic vector space; this provides a choice of polarization
data analogous to the choice of the Hilbert space of wave functions in quantum mechanics. 
Section~\ref{ssec:quant} reviews the Gurevich--Hadani functorial quantization of \cite{GH} from
a category of symplectic vector spaces and isomorphisms over a finite field to complex vector spaces
and isomorphisms, which assign canonical models of Weil representations, in a way that is
monoidal and compatible with symplectic reduction. Section~\ref{ssec:PT} presents our
general construction of perfect tensors, from the data of a symplectic vector space over
a finite field with a Darboux basis and a Lagrangian subspace in general position with
respect to the splitting determined by the Darboux basis. This gives a simple physical interpretation
of the CRSS algorithm as canonical quantization, in the stronger sense of~\cite{GH}, and translates properties of perfect tensors naturally into properties of the corresponding semiclassical states (Lagrangian subspaces). We show how our
construction works in a simple explicit example in section~\ref{ssec:example}.  In section~\ref{ssec:Ham}
we show how to write Hamiltonians (in a form similar to random walk/discretized Laplacian
operators) that have the perfect tensor state as ground state.

In section~\ref{GENUSZERO} we present the main results on our construction of 
a quantum error-correcting tensor network, built using perfect tensors, associated to
the $p$-adic Bruhat--Tits trees via a ``dual graph'' construction, and we establish
the $p$-adic analog of the Ryu--Takayanagi formula. Section~\ref{DUALGRAPH} describes
the construction of the ``dual graph'' tensor network associated to the 
$p$-adic Bruhat--Tits trees by identifying two axiomatic properties that
characterize the network in relation to the tree. As discussed more in detail
in section~\ref{TNPROPS}, different choices satisfying these properties are possible,
which can be characterized in terms of different choices of embeddings.
For the purpose of the entanglement entropy computation any such choice
of a ``dual graph'' will achieve the desired result. We also describe the perfect
tensors associated to the nodes of the dual graph for a finite cutoff of the
infinite Bruhat--Tits tree, the number of dangling (uncontracted) legs at the vertices and the resulting
boundary wavefunction. The rank of the perfect tensors is related to the cutoff
on the tree and goes to infinity in the limit of the infinite tree.
In section~\ref{GENUSZERORESULTS} we summarize the main technical results in the genus 0 background, that the dual graph tensor network satisfies a 
Ryu--Takayanagi like formula, where instead of ``intervals" we specify the boundary datum in
terms of configurations of points (though we still use the terminology ``connected interval"
or ``disconnected interval").  We show that the 
von~Neumann entropy computation matches what is expected for CFT$_2$
and that it is naturally expressed in terms of the $p$-adic norm, which  
leads to the expected bulk interpretation (consistent with 
the minimal cut rule obeyed by perfect tensor networks~\cite{Pastawski:2015qua}) as the length of the minimal geodesic joining the entangling surfaces
determined by the chosen configuration of boundary points. We also comment on
the disconnected  interval (four points) entropy case, the dependence of the mutual information on the cross-ratio and entropy inequalities such as subadditivity, strong subadditivity and monogamy of mutual information, where the ultrametric property plays a direct, simplifying role, the details of which are found in sections \ref{GENUS0DOUBLE}-\ref{ssec:Inequalities}.

Section~\ref{GENUSONE} deals with the $p$-adic BTZ black hole, described in
terms of Mumford--Tate elliptic curves as boundary and with bulk space a quotient
of the $p$-adic Bruhat--Tits tree by a rank one Schottky group. In section~\ref{ssec:Schottky}
we review the $p$-adic geometry of Mumford curves of genus one and the associated
bulk spaces, comparing it with the case of complex elliptic curves with Tate uniformization
by the multiplicative group. We also explain how to adapt our construction of the tensor network
as dual graph of the Bruhat--Tits tree to a network similarly dual to the homologically non-trivial
quotient of the Bruhat--Tits tree in the genus one case. In particular, the tensor network obtained
in this way has a vertex beyond the black hole horizon that does not correspond to
boundary degrees of freedom. In section~\ref{BTZResults} we come to our main results in
the BTZ black hole case. Computing 
on the tensor network the thermal entropy of the boundary density matrix 
obtained by tracing out this special vertex gives the black hole horizon perimeter, 
which can be seen as a Bekenstein--Hawking formula for the $p$-adic BTZ black hole.
In section~\ref{RTBTZresults} we discuss the Ryu--Takayanagi formula in genus one backgrounds, with the
boundary entanglement entropy of a single interval corresponding to the length of a 
minimal geodesic in the bulk black hole geometry. Unlike the genus zero case, 
the entropy of a  boundary region and its complement are not necessarily the same,
corresponding to the fact that the boundary state is no longer pure, and in the bulk geometry a geodesic may wrap around the
loop of the quotient graph (the black hole horizon).

Section~\ref{COMPUTATION} contains all the detailed explicit computations
used in section~\ref{GENUSZERO} and section~\ref{GENUSONE} for obtaining
the von Neumann entropy via the density matrices determined by the tensor
network.  
Section~\ref{SIMPLESTATES} illustrates the rules for the computation of states and
reduced density matrices from perfect tensors, and the graphical calculus used to keep track
of contractions, and a convenient representation of the resulting density matrices 
in block-diagonal form.
We describe how to obtain the reduced density matrices corresponding to tracing out regions 
determined by sets of vertices of the tensor network, and we compute the associated 
von Neumann entropy.
In section~\ref{SPLITSCYCLES} we discuss the computation of the inner product of the
holographic state with itself. The computation method is described in terms of
certain  
graphical contraction rules (``splits''), decomposing the network into disjoint simple curves;
each resulting closed cycle then determines an overall multiplicative factor. 
Section~\ref{NORM} then contains the computation of the norm of the
holographic state obtained from our tensor network dual to the
Bruhat--Tits tree, with an assigned cutoff on the infinite tree. This
depends on different types of vertices (in terms of number of dangling legs)
and the corresponding multiplicities and the application of the ``splits and cycles"
method. 
In section~\ref{GENUS0SINGLE} we then show how a Ryu--Takayanagi like
formula is obeyed exactly in the single connected interval (``two point'') case. The entanglement of a boundary region with its
complement is computed by computing the density matrix of the full holographic state
produced by the tensor network and then computing a partial trace using the computational
techniques developed in the previous subsections. The result is then compared with the
(regulated) geodesic length in the bulk Bruhat--Tits tree.

 The disconnected interval case (in particular the ``four point'' case)  is discussed in section~\ref{GENUS0DOUBLE}  in terms of overlapping or non-overlapping geodesics in the bulk, depending on the sign of the logarithm of the cross-ratio, and the corresponding properties of the mutual information.
 We show  subadditivity (both Araki--Lieb inequality as well as non-negativity of mutual information) and give the exact dependence of mutual information on the cross-ratio constructed from the boundary points. 
 We show that a Ryu--Takayanagi like formula is satisfied exactly for  disconnected intervals, and in section~\ref{ssec:Inequalities} we proceed to prove strong subadditivity and monogamy of mutual information. In fact we show that in this tensor network mutual information is exactly extensive. 
 
Section~\ref{GENUS1BTZ} contains the computation of the black hole entropy as well as 
the Ryu--Takayanagi formula for
the minimal geodesics in the black hole background. Using tools from the previous sections,
the norm of the black hole boundary state is computed in terms of types of vertices and multiplicities, and the density matrix and corresponding von Neumann entropy is determined. This entropy
is seen to be proportional to the length of the horizon or the length of the minimal geodesic homologous to the boundary interval, where the homologous condition, an important feature of the RT formula, is obeyed automatically by the tensor network.

Section~\ref{GEOMETRIC} further discusses some of the geometric aspects of
our tensor network construction. In section~\ref{TNPROPS} we discuss more in
detail the symmetry properties of the tensor networks with respect to the
global symmetries of the Bruhat--Tits tree, showing how the properties needed
for the construction of a suitable ``dual graph" reduce the symmetries and
obtaining in this way a characterization of all the possible choices of dual graph.
In section~\ref{ssec:Drinfeld} we show that the construction of the dual graph and
the tensor network can be done entirely within the $p$-adic world, by embedding
it in the Drinfeld $p$-adic plane, using a choice of lifts of the projection from
the Drinfeld plane to the Bruhat--Tits tree. This is illustrated in section~\ref{ssec:toymodel}
in a toy model given by the tubular neighborhood of an infinite tree, and
adapted in section~\ref{ssec:pplane} and section~\ref{ssec:dualpplane} to the $p$-adic plane. 
Similarly, in section~\ref{ssec:geometryTATE}, the tensor network for the genus one $p$-adic
BTZ black hole is embedded in the quotient of the Drinfeld $p$-adic plane by the
uniformizing $p$-adic Schottky group. In the same section we also discuss the
construction of measures on the Mumford--Tate curve induced from the Patterson--Sullivan
measure on the $p$-adic projective line. Finally, in section~\ref{AFALGEBRA} we show how
to interpret the density matrices in the limit of the infinite Bruhat--Tits tree, as states
on an approximately finite dimensional $C^*$-algebra.

Finally, a list of possible further directions of investigation and open
questions is given in section~\ref{DISCUSSION}. 

Since several sections of the paper are quite independent, it may be useful for readers to know the shortest path required to reach a given result. In the leitfaden below,  arrows indicate logical dependence between sections. For example, in order to read section~6, one should read all of its predecessors in the diagram: sections 1, 2, and~4.
\[
\begin{tikzcd}[row sep = 1 em, column sep = 0.4 em]
& \parbox[c]{2cm}{\centering \S\ref{INTRODUCTION}\\ Introduction} \ar[d] & & &   \\ 
& \parbox[c]{2cm}{\centering \S\ref{REVIEW}\\ Background}\ar[dr] \ar[dl]  & & & \\ 
\parbox[c]{3cm}{\centering \S\ref{CRSS}\\ Perfect~tensors} & & \parbox[c]{2cm}{\centering \S\ref{GENUSZERO}\\ Empty~AdS} \ar[dr] \ar[dl] & & \\ 
 &   \parbox[c]{2.5cm}{\centering \S\ref{COMPUTATION}\\ Details} & & \parbox[c]{2cm}{\centering \S\ref{GENUSONE} \\ Black~hole}\ar[dl] \ar[dr] & \\
 & &   \parbox[c]{2cm}{\centering \S\ref{DISCUSSION}\\ Discussion} & & \parbox[c]{4cm}{\centering \S\ref{GEOMETRIC}\\ Geometric~properties} 
\end{tikzcd}
\]
However, readers wishing to directly view the entanglement entropy results in the $p$-adic setting may consider skipping directly ahead to sections \ref{GENUSZERO} and \ref{GENUSONE}, and refer back to section \ref{REVIEW} when needed.

\section{Background}
\label{REVIEW}

\subsection{The Bruhat--Tits tree as a $p$-adic bulk space}
\label{ssec:padic}

 Let us briefly recall the setup of $p$-adic AdS/CFT~\cite{Gubser:2016guj,Heydeman:2016ldy}. 
In the simplest formulation, the bulk geometry is described by an infinite $(p+1)$-regular graph (without cycles), called the Bruhat--Tits tree,  and its asymptotic boundary is given by the projective line over the $p$-adic numbers, $\mathbb{P}^1(\mathbb{Q}_p)$. 
 (For an introduction to the theory of $p$-adic numbers, see~\cite{Koblitz,Gouvea}; a shorter discussion in the physics literature can be found in e.g.~\cite{Brekke:1993gf}.)
The Bruhat--Tits tree ${\cal T}_p$ is a discrete, maximally symmetric space
of constant negative curvature, which plays the role of (Euclidean) AdS space. One can study perturbative bulk dynamics on the tree by considering lattice actions defined on its nodes (or bonds)~\cite{Gubser:2016guj,Heydeman:2016ldy,Gubser:2016htz}. A central result of these works is that semiclassical dynamics of these lattice models in the bulk ${\cal T}_p$ compute correlation functions in a dual conformal field theory defined on $\mathbb{P}^1(\mathbb{Q}_p)$.

In the following, we will view the Bruhat--Tits tree as a constant-time spatial slice of a higher-dimensional $p$-adic analog of Lorentzian AdS space. 
To make the analogy with the real setup more concrete, we view the Bruhat--Tits tree as the $p$-adic analog of the Poincar\'{e} disk (or equivalently the hyperbolic plane $\mathbb{H}^2$), arising as a constant-time slice of an appropriate higher-dimensional building describing ($p$-adic) Lorentzian AdS$_3$.\footnote{In this paper we remain agnostic about the appropriate higher dimensional origins of the Bruhat--Tits tree (such as hyperbolic buildings), and will only be interested in studying entanglement entropy in the static (time symmetric) case.} 
This analogy is motivated by the fact that the real hyperbolic plane $\mathbb{H}^2$ is a symmetric, homogeneous space of constant negative curvature, and arises algebraically as the quotient space $\mathbb{H}^2= \SL(2,\mathbb{R})/SO(2,\mathbb{R})$, where $\SL(2,\mathbb{R})$ is the isometry group of $\mathbb{H}^2$ and $SO(2,\mathbb{R})$ is its maximal compact subgroup. Similarly, the Bruhat--Tits tree is a symmetric, homogeneous space which can be viewed as the quotient ${\cal T}_p = \PGL(2,\mathbb{Q}_p)/\PGL(2,\mathbb{Z}_p)$. Here, the group $\PGL(2,\mathbb{Q}_p)$ acts on ${\cal T}_p$ by isometries, and $\PGL(2,\mathbb{Z}_p)$ is its maximal compact subgroup. However, while the hyperbolic plane is a two-dimensional manifold, the Bruhat--Tits tree is described by a discrete (but infinite) collection of points (as seen later in figure~\ref{fig:clopen}). In section~\ref{GENUSZERO}, we use the ``dual'' of this discrete tree to define a tensor network, and the entanglement properties of the boundary state are described geometrically in terms of the Bruhat--Tits tree.

We now describe the action of~$G = \PGL(2,\mathbb{Q}_p)$ on the Bruhat--Tits tree in more detail. 
Let $H=\PGL(2,\mathbb{Z}_p) < G$ denote a maximal compact subgroup. 
Choose representatives $g_i \in G$ of the left cosets of~$H$. In other words,
$G = \bigcup_{i=0}^\infty g_i H$, where the cosets $g_i H, g_j H$ are pairwise disjoint for $i\neq j$. 
The cosets $g_i H$ are in bijective correspondence with the equivalence classes of $\mathbb{Z}_p$-lattices in $\mathbb{Q}_p \times \mathbb{Q}_p$,
as well as with the nodes on the Bruhat--Tits tree (see e.g.\ \cite{Brekke:1993gf}).
The group $G$ has a natural action on equivalence classes of lattices $(r,s)$ 
by matrix multiplication:
\begin{equation}
    g \cdot (r,s) \equiv g \cdot \left\{ \begin{pmatrix} a r_1 + b s_1 \\ a r_2 + bs_2 \end{pmatrix}: a,b \in \mathbb{Z}_p \right\} = \begin{pmatrix} A & B \\ C & D \end{pmatrix} \left\{ \begin{pmatrix} a r_1 + b s_1 \\ a r_2 + bs_2 \end{pmatrix}: a,b \in \mathbb{Z}_p  \right\},
\end{equation}
for $r = (r_1,r_2)^T, s=(s_1,s_2)^T \in \mathbb{Q}_p\times \mathbb{Q}_p$, $g \in G$. 

Equivalently, $G$ acts on the space of cosets $G/H$, by the rule $g_i H \mapsto g g_i H$. Each equivalence class (or equivalently, coset) is stabilized by a conjugate of $H$; the coset $g_i H$ is stabilized by $g_i H g_i^{-1} < G$. Either of these descriptions gives the action of~$G$ on the nodes of~${\cal T}_p$.

Two nodes on the tree are defined to be adjacent when the relation 
\begin{equation}
p\Lambda \subset \Lambda^\prime \subset \Lambda
\end{equation}
holds between the corresponding $\bZ_p$-lattices $\Lambda$ and~$\Lambda'$. This relation is reflexive, so that the previous inclusion holds if and only if $p\Lambda^\prime \subset \Lambda \subset \Lambda^\prime$.
The action of the group $G$ on the nodes of the Bruhat--Tits tree 
preserves these incidence relations. In other words, the group $G$ acts by isometries of the Bruhat--Tits tree, preserving the graph distance between any pair of nodes. Intuitively, $G$ acts by translations and rotations on the (infinite) nodes of the tree in analogy to the ordinary isometries of AdS; additionally for any given vertex we may find a stabilizer subgroup, which  is always a conjugate of $H = \PGL(2,\Z_p)$, which rotates the entire tree around this point. 
 
As is well known in AdS/CFT, the isometry group of the bulk acts as conformal transformations on the boundary. In $p$-adic AdS/CFT, we have $\partial {\cal T}_p = \mathbb{P}^1(\mathbb{Q}_p)$, with $G$ acting as fractional linear transformations:
\eqn{}{
\mathbb{P}^1(\mathbb{Q}_p) \ni z \mapsto g \cdot z = {A z + B \over C z + D} \qquad   g = \begin{pmatrix} A & B \\ C & D \end{pmatrix} \in G=\PGL(2,\mathbb{Q}_p)\,.
}
These are  interpreted as the global ($p$-adic) conformal transformations acting on the dual theory defined on the boundary $\partial {\cal T}_p$.

In analogy with static AdS$_3$, it is possible to obtain black hole like bulk geometries algebraically. One may quotient the Bruhat--Tits tree by an abelian discrete subgroup $\Gamma \in PGL(2,\mathbb{Q}_p)$ to obtain an analog of the BTZ black hole. The bulk and boundary properties of this construction are explored in section~\ref{GENUSONE}, where we will describe why this is a good model of a $p$-adic black hole geometry and compute the entropy via the tensor network proposed in this work.

\subsection{Tensor networks}
\label{ssec:tensors}

In the recent literature, there has been much interest in so-called \emph{tensor network} models, which describe a state or family of states in a Hilbert space that is the tensor product of many qubits or local Hilbert spaces of fixed rank. Such states are built by considering concatenations of many tensors, each operating on a finite number of qubits in the manner of a quantum circuit. Early proposals~\cite{Vidal:2007hda,EvenblyVidal12} showed that such setups could be used to construct states whose entanglement structure mimics that of the vacuum state of a conformal field theory~\cite{Evenbly:2007hxg,EvenblyVidal08,EvenblyVidal10}. Subsequently, it was proposed~\cite{Swingle:2009bg,Swingle:2012wq} that the geometry of the tensor network could be thought of as a discrete analogue of an AdS bulk space, and various models have been developed to try and exhibit this correspondence more precisely~\cite{Pastawski:2015qua,Hayden:2016cfa,Almheiri:2014lwa,Harlow:2016vwg,Harlow:2018fse,Osborne:2017woa}.

In particular, in the proposal of~\cite{Pastawski:2015qua}, tensor networks were associated to uniform tilings of hyperbolic two-dimensional space by $k$-gons, by placing a tensor with $m$ indices on each polygon and contracting indices across each adjacent face. The residual $m-k$ indices represent ``logical'' inputs in the bulk. (Of course, $m\geq k$; equality is not necessary, but $2m-k$ should not be too large. Furthermore, $m$ is taken to be even. See~\cite{Pastawski:2015qua} for a discussion of the precise conditions.)
Using this construction, analogues of the Ryu--Takayanagi formula for the entanglement entropy were proved.

This formula follows from a key property of the $m$-index tensors that are used on each plaquette:
\begin{definition}
\label{def:PT}
Let $T\in V^{\otimes m}$ be an $m$-index tensor, where each index labels an identical tensor factor or ``qubit'' $V\cong \C^q$. $V$ is equipped with a Hilbert space structure, so that we can raise and lower indices using the metric. Let $I\subseteq M = \{1,\ldots, m\}$ be any subset of the index set, and $J$ its complement; without loss of generality, we can take $\# I \leq \# J$. $T$ is said to be \emph{perfect} if, for every such bipartition of the indices, 
\deq{
T_I^J: V^I \rightarrow V^J
}
is an isometric map of Hilbert spaces. Here we are using the notation that $T_I^J$ means $T$ with the indices in the set $I$ lowered, and those in the set $J$ raised. (In particular, $T_I^J$ is injective, so that one can think of the condition as asking that $T$ have the largest possible rank for any such tensor decomposition.)
\end{definition}

The parity of~$m$ is not important to the above definition, but the applications in~\cite{Pastawski:2015qua} make use of perfect tensors for which $m$ is even. It is then shown that requiring $T_I^J$ to define a unitary map for every bipartition with $\# I = \# J$ is sufficient to imply perfection in the sense of definition~\ref{def:PT}.

The connection to maximally entangled states should hopefully be apparent: Recall that a state is said to be maximally entangled between two subsystems if the reduced density matrix, obtained by tracing out one subsystem, is ``as mixed as possible.'' At the level of density matrices, this means ``proportional to the identity matrix'' (recall that pure states correspond one-to-one to density matrices of rank one). So, for a state defined by a perfect tensor, we can write
\deq{
\rho = |\alpha|^2 \, T_M T^M,
}
where $\alpha$ is a normalization constant required so that $\rho$ has unit trace (equivalently, so that the state $T_M \in V^{\otimes m}$ is normalized). 
Note that no Einstein summation convention applies; $M$ denotes a set of indices, rather than an index. To reduce the density matrix, though, we \emph{do} contract along the indices in the set $J$:
\deq{
\qty(\rho_\text{red})_I^I
= |\alpha|^2 \, T_I^J \circ T_J^I.
}
In the case that $m$ is even and $\# I = \# J$, this shows that $\rho_\text{red}$ is the composition of two unitary maps; as a consequence, it is full-rank. Indeed, by unitarity, the two maps $T_I^J$ and~$T_J^I$ are inverses of one another, so that the reduced density matrix is proportional to the identity. From the condition of unit trace, it follows that the normalization constant must be taken to be
\deq{
|\alpha|^2 = p^{-m/2}.
}
For more details on computations with perfect tensors, look forward to section~\ref{COMPUTATION}.

\subsection{Classical and quantum codes}
\label{CLASSQUANTCODES}

There are many close relations between perfect tensors, maximally entangled states, and quantum error-correcting codes. We have outlined some of the connections between the first two ideas above; in this subsection, we will discuss the third, which will give us a way to produce examples of perfect tensors. The key construction will be the CRSS algorithm~\cite{CRSS}, which produces quantum error-correcting codes from a particular class of classical codes. In turn, the CRSS algorithm makes use of a particular complete set of matrices acting on qubits, which come from the theory of Heisenberg groups; these groups generalize the familiar theory of the canonical commutation relations to variables which are discrete ($\FF_q$-valued) rather than continuous. As such, the CRSS procedure can be seen as perfectly analogous to canonical quantization problems of a familiar sort. We review the CRSS algorithm and the necessary theory of Heisenberg groups here; in~section~ \ref{CRSS}, we will develop this analogy further, and show that it provides a natural way to write down Hamiltonians whose vacuum states arise from perfect tensors.


\subsubsection{Heisenberg groups}

The simplest example of a finite Heisenberg group can be presented as follows:
\deq[eq:simplestHeis]{
G = \langle X,Z,c : ZX = c XZ, ~ X^p = Z^p = 1,  ~ c \text{ central}\rangle.
}
(It follows from these relations that $c^p = 1$ as well.) 
The center of the group is a copy of $\F_p$, generated by~$c$, and the quotient by the center is also abelian, so that the group fits into a short exact sequence
\deq[eq:cseq]{
0 \rightarrow \F_p \rightarrow G \rightarrow \F_p^2 \rightarrow 0
}
exhibiting it as a central extension of one abelian group by another. Despite this, $G$ itself is nonabelian. 

Representations of this group are also easy to understand; each representation can be restricted to $Z(G) = \F_p$, and defines a character $\chi$ of that group, called the central character. In the case at hand, a central character is just a choice of $p$-th root of unity, corresponding to~$\chi(c)$.

Given a choice of representation, a corresponding representation can be constructed on the vector space $\H=\C^p$. In a particular basis, the generators of~$V$ act according to the rule
\deq{
X\ket{a} = \ket{a+1}, \quad Z\ket{a} = \chi(c)^a \ket{a}.
}
For a nontrivial central character, this representation is irreducible. An analogue of the Stone--von~Neumann theorem shows that this is in fact the unique irreducible representation with central character $\chi$. Furthermore, the representation matrices form an additive basis (over~$\C$) for the matrix algebra $M_{p\times p}(\C)$. 

On the other hand, when the central character is trivial, any representation factors through the quotient map to~$\F_p^2$; since that (additive) group is abelian, there are $p^2$ different one-dimensional representations.
As such, we have understood the complete representation theory of $G$. A quick check reveals that we've found the whole character table: there are $(p-1)$ nontrivial central characters, each with a representation of dimension~$p$, together with $p^2$ abelian representations. 
This makes a total of $p^2 + p - 1$ irreps, which corresponds to the number of conjugacy classes: these are the powers of~$c$, together with the nonzero powers $X^i Z^j$. One can also double-check that
\deq{
\sum_\rho (\dim \rho)^2 = (p-1)p^2 + p^2 = p^3 = \# G.
}

This simple example already contains most of the structural features, and motivates the following definition. In what follows, $k$ will be an arbitrary field, though the cases that will be relevant will be when $k$ is locally compact (i.e., $k$ is a local field or a finite field). In fact, we will only really consider the cases where $k=\R$ or~$\F_q$, although some amount of the discussion even continues to make sense over an arbitrary commutative ring, for instance~$\Z$.

\begin{definition}
Let $V$ be a symplectic vector space over~$k$, with symplectic form~$\omega$. The \emph{Heisenberg group} associated to this data, denoted $\Heis(V)$, is the central extension of the additive abelian group $V$ by the cocycle 
\deq{
\omega: V^2 \rightarrow k.
}
\end{definition}
Note that, since $\Heis(V,\omega)$ is a central extension, there is a natural short exact sequence of abelian groups
\deq[eq:seq]{
0 \rightarrow k \rightarrow \Heis(V) \rightarrow V \rightarrow 0,
}
generalizing the sequence~\eqref{eq:cseq}.
Furthermore, the image of~$k$ is the center of the group. Our previous example arises in the case $k= \F_p$, $V = \F_p^2$, and $\omega$ the standard Darboux symplectic form on a two-dimensional vector space.  

In speaking of $\omega$ as a group cocycle, we are thinking of the inhomogeneous group cochains of~$(V,+)$. The cocycle condition is obeyed because
\begin{align}
d\omega(u,v,w) &\defeq \omega(v,w) - \omega(u+v,w) + \omega(u,v+w) - \omega(u,v) \nonumber \\
&= 0.
\end{align}
An analogue of the Stone--von~Neumann theorem also holds in this more general case, so that $\Heis(V)$ admits a unique irreducible representation for any choice of central character.
We will discuss this theorem further below.
Furthermore, just as in our example above, it is true for more general Heisenberg groups that the representation matrices form an additive basis for~$M_{p^n\times p^n}(\C)$, where $n = \dim(V)/2$.

An explicit construction of that unique irreducible representation can be given as follows. Let ${\H} = \Fun(\F_q,\C) \cong \mathbb{C}^q $ be the Hilbert space of a single $q$-ary qubit. 
An orthonormal basis of ${\H}$ is labeled by states $|a\rangle$ where $a \in \mathbb{F}_q$. 
Quantum error-correcting spaces are subspaces of ${\H}^{\otimes n}$ which are error-correcting for a certain number of qubits. 
All errors can be constructed from the error operators $E = E_1 \otimes \cdots \otimes E_n$ which are the representation matrices of the Heisenberg group. Each of the $E_i$ can be thought of as a particular combination of  bit-flip and phase-flip operators, which we now describe.

Define the $p \times p$ matrices
\eqn{TRDefs}{
T = \begin{pmatrix} 0 & 1 & 0 & \cdots & 0 \\ 0 & 0 & 1 & \cdots & 0 \\ & \vdots & & \ddots & \vdots \\ 0 & 0 & 0 & \cdots & 1 \\1 & 0 & 0 & \cdots & 0\end{pmatrix} \qquad R = \begin{pmatrix} 1 & & & & \\ & \xi & & & \\ & & \xi^2 & & \\ & & & \ddots & \\ & & & & \xi^{p-1} \end{pmatrix}\,,
}
where $\xi = e^{2\pi i/ p}$ is a (nontrivial) $p$-th root of unity. If $q=p$, $\xi$ is precisely a choice of central character, and it is easy to check that these matrices define the representation of the simplest Heisenberg group~\eqref{eq:simplestHeis} with that central character. However, in the case $q=p^r$, 
we must do slightly more work.

Let $\{\gamma_j: 1 \leq j \leq r\}$ be a basis of $\mathbb{F}_q$ as an $\mathbb{F}_p$-vector space; see e.g.~\eno{abComponents}. Then we can write $a = \sum_{j=1}^r a_j \gamma_j$ for any $a \in \mathbb{F}_q$, where $a_i \in \mathbb{F}_p$. 
This also defines a tensor product basis in~$\Hilb=\Fun(\F_q,\C)$,
such that $|a\rangle = |a_1 \rangle \otimes \cdots \otimes |a_r \rangle$ when $a$ is decomposed as a direct sum, as above. 
Then the error operators act on individual copies of $\mathbb{C}^p$ as follows:
\eqn{TRCp}{
T^{b_j} |a_j\rangle = |a_j + b_j \rangle \qquad R^{b_j} |a_j \rangle = \xi^{\Tr ( a_j b_j )} |a_j \rangle \qquad a_j,b_j \in \mathbb{F}_p\,.
}
Here, the trace function $\Tr_{q:p}: \mathbb{F}_q \to \mathbb{F}_p$ (with $q=p^r$) is defined as
\eqn{TrDef}{
\Tr_{q:p}(a) = \sum_{i=0}^{r-1} a^{p^i} \qquad a \in \mathbb{F}_q \,.
}
It is easy to see that this is precisely the trace of the endomorphism of~$\F_q$ that is multiplication by the element $a$, regarded as an $n\times n$ matrix over~$\F_p$.

It is now simple to define the bit- and phase-flip operators acting on single $q$-ary qubits; they are the $q \times q$ matrices
\eqn{TbRbDefs}{
T_b = T^{b_1} \otimes \cdots \otimes T^{b_r} \qquad R_b = R^{b_1} \otimes \cdots \otimes R^{b_r}\,.
}
These operators act on a single $q$-ary qubit via
\eqn{TRsingle}{
T_{b} |a \rangle = |a + b \rangle \qquad R_{b} |a \rangle = \xi^{\Tr (\langle a, b \rangle)} |a \rangle\,,
}
where $a = \sum_{j=1}^r a_j \gamma_j\in \mathbb{F}_q$, and
\eqn{aSingleState}{
|a \rangle = \otimes_{j=1}^r |a_j\rangle\,.
}
As emphasized above, the operators $T_a R_b$ form an orthonormal basis for $M_{q \times q}(\mathbb{C})$ under the inner product $\langle A,B \rangle = q^{-1} \Tr(A^\dagger B)$, and thus generate all possible errors on $\Hilb$.
We can further construct error operators which act on~$\Hilb^{\otimes n}$ as follows. Given $a = (a_1, \ldots, a_n), b=(b_1, \ldots , b_n) \in \mathbb{F}_q^n$, define
\eqn{EabDef}{
E_{a,b} = T_a R_b = (T_{a_1} \otimes \cdots \otimes T_{a_n}) (R_{b_1} \otimes \cdots \otimes R_{b_n})\,.
}
It is straightforward to check that $E_{a,b}^p = 1$, and that they obey the following commutation and composition laws:
\eqn{EabComLaws}{
E_{a,b}E_{a^\prime,b^\prime} = \xi^{\langle a,b^\prime \rangle - \langle a^\prime,b \rangle} E_{a^\prime,b^\prime} E_{a,b} \qquad E_{a,b}E_{a^\prime,b^\prime} = \xi^{-\langle b,a^\prime \rangle} E_{a+a^\prime, b+b^\prime}\,.
}
Here, we have made use of an $\F_p$-valued pairing, 
\eqn{CompInProd}{
\langle a, b \rangle = \sum_{i=1}^n \langle a_i, b_i \rangle = \sum_{i=1}^n \sum_{j=1}^r a_{i,j} b_{i,j} \qquad a,b \in \mathbb{F}_q^n, 
}
where 
the elements $a_i, b_i \in \mathbb{F}_q$ are expanded in terms of an $\F_p$-basis as
\eqn{abComponents}{
a_i = \sum_{j=1}^r \gamma_j a_{i,j} \qquad 
b_i = \sum_{j=1}^r \gamma_j b_{i,j}\qquad a_{i,j}, b_{i,j} \in \mathbb{F}_p\,.
}
\eqref{EabDef} therefore produces the explicit representation  matrices of the Heisenberg group of~$\F_q^{2n}$, corresponding to a particular (nontrivial) choice of central character.


\subsubsection{Classical algebrogeometric codes}
\label{ALGEBCODES}

In this section, we give a few general remarks about classical codes over finite  fields. 
The next section will review the CRSS algorithm, which associates a quantum error-correcting code to each such self-orthogonal classical code. Placing the two together, one can demonstrate that perfect tensors with arbitrarily many indices can be constructed, which we will require for the models of~section~ \ref{DUALGRAPH}; an additional ingredient is an appropriate family of classical codes, an example of which is given in appendix~\ref{THREEQ}. Section~\ref{CRSS} gives a new perspective on the CRSS algorithm, showing that perfect tensors arise naturally in the context of quantization of symplectic vector spaces over finite fields.

A (classical) \emph{linear code} is nothing more than a linear subspace of a vector space over a finite field. In a basis, it is defined by an injective map
\deq{
i : \F_q^k \hookrightarrow \F_q^n,
}
which can be thought of as encoding $k$ bits of information (each bit being of size $q$) into $n$ bits of information. 

The \emph{Hamming weight} is defined to be the function
\deq{
\wt: \F_q^n \rightarrow \mathbb{N}, \quad
c \mapsto \#\{ i : c_i \neq 0 \}.
}
Note that this is a basis-dependent definition! The \emph{minimum weight} of a code is simply the minimum Hamming weight of all nonzero elements of the code subspace; one often uses the notation ``$[n,k,d]_q$ code'' to speak of a code with the given parameters. We may sometimes omit the weight parameter $d$ from this list; no confusion should arise.

Equipping $\F_q^n$ with an inner product or more generally a bilinear form, one can classify codes according to the properties of the code subspace. In particular, a code is said to be \emph{self-orthogonal} when the code subspace is isotropic with respect to the bilinear form, i.e., contained in its orthogonal complement: $\im(i) \subseteq \im(i)^\perp$.\footnote{\label{fn:DualCode} The superscript $^\perp$ denotes the dual (orthogonal) code. The dual code is defined as follows: If $C$ is a classical code over $\mathbb{F}_q$ of size $n$, then $C^\perp = \{ v \in \mathbb{F}_q^n : a * v = 0 \  \forall a \in C\}$. }  

The CRSS algorithm produces a quantum error-correcting code from classical self-orthogonal codes associated to symplectic vector spaces over finite fields. Such codes are generally of the form $[2n,\ell]_q$, where $\ell \leq n$ is the dimension of the isotropic subspace, and the inner product on~$\F_q^{2n}$ may, without loss of generality, be taken to have the standard Darboux form. We review the construction in the following subsection. For certain choices of the code parameters, CRSS quantum codes may then be used in turn to produce perfect tensors; in fact, as explained in~section~ \ref{CRSS}, a generalization of the CRSS algorithm relates perfect tensors to Lagrangian subspaces in general position in~$\F_q^{2n}$. 

Let us also remark that isotropic subspaces in symplectic vector spaces may be constructed from other types of classical codes. For example, let $D$ be a classical self-orthogonal $[n,k,d]_{q^2}$ code over $\mathbb{F}_{q^2}$, 
where the self-orthogonality is established with respect to the Hermitian inner product
\eqn{HermInProd}{
v * w = \sum_{i=1}^n v_i w_i^q \qquad v, w \in \mathbb{F}_{q^2}^n\,.
}
By Theorem~4 of~\cite{AshKni:2001}, there exists a classical code $C$ of length $2n$ and size $2k$ over $\mathbb{F}_q$ which is self-orthogonal with respect to the inner product,
\eqn{TrInProd}{
(a,b) * (a^\prime,b^\prime) = \Tr \left(\langle a,b^\prime\rangle_* - \langle a^\prime,b\rangle_*\right) \qquad (a,b), (a^\prime, b^\prime) \in \mathbb{F}_q^{2n}\,,
}
where the Euclidean inner product $\langle \cdot, \cdot \rangle_*$ is defined to be
\eqn{EucInProd}{
\langle a, b \rangle_* = \sum_{i=1}^n a_i b_i \qquad a,b \in \mathbb{F}_q^n \quad a_i, b_i \in \mathbb{F}_q\,.
}
This is of course precisely the standard Darboux symplectic form on~$\F_q^{2n}$.

The inner product given in \eno{TrInProd} has an equivalent description in terms of the inner product of \eno{CompInProd}, as follows (see \cite{AshKni:2001}):
\eqn{TrInProdEqui}{
(a,b) * (a^\prime,b^\prime) = \langle a, \varphi(b^\prime) \rangle - \langle a^\prime, \varphi(b) \rangle\,,
}
where\footnote{More generally, $\varphi$ is an automorphism of the vector space $\mathbb{F}_p^r$, but for convenience we will restrict our focus to the particular choice of $\varphi$ described here.}
\eqn{phiAction}{
\varphi(a) = \left(\varphi(a_1),\ldots,\varphi(a_n)\right) \qquad a \in \mathbb{F}_q^n \quad a_i \in \mathbb{F}_q,
}
and the action of $\varphi$ on elements of $\mathbb{F}_q$ is given by matrix multiplication, where $\varphi$ acts as an $r \times r$ matrix $M$ on the elements of $\mathbb{F}_q$, with
\eqn{phiMatrix}{
M_{ij} = \Tr (\gamma_i \gamma_j) \qquad i,j = 1,\ldots r\,.
}

\subsubsection{CRSS algorithm}

We briefly review the CRSS algorithm~\cite{CRSS}, which produces a quantum error-correcting code from an appropriately chosen classical code. We emphasize the perspective that CRSS is intimately related to the formalism of canonical quantization, albeit for Heisenberg groups over~$\F_p$ rather than~$\R$. For further discussion, the reader is referred to~\cite{Marcolli:2018ohd,CRSS,GBR}.

As mentioned above, the CRSS algorithm starts with a symplectic vector space~$V$ of dimension~$2n$ over a finite field. We let  $\H(V)$ denote the ``quantization'' of this symplectic space, i.e., the unique irreducible representation of~$\Heis(V)$ with central character $\chi$. In fact, by results of~\cite{GH}, there is a canonical model for~$\H(V)$; we review these results in~section~ \ref{CRSS}. Now, $\H(V)$ is isomorphic to the tensor product of $n$ $p$-dimensional Hilbert spaces, one for each ``qubit'' or discrete degree of freedom. Such a tensor product decomposition corresponds to a choice of Darboux basis for~$V$, which splits it as the direct sum of standard  two-dimensional symplectic spaces. As noted above, the representation matrices of~$\Heis(V)$ additively span the space $\End(\H(V))$ of all operators over~$\C$.

Now, consider any maximal isotropic subspace $L$ of~$V$; every such subspace defines a maximal abelian subgroup of $\Heis(V)$. 
Mutually diagonalizing the action of the operators representing~$L$ splits~$\H$ as a direct sum of one-dimensional eigenspaces.

Then, consider a (necessarily isotropic) subspace $C \subset L$, whose dimension is $i < n$. $C$ is to be thought of as the classical code subspace. 
The mutual eigenspaces of the abelian group associated to~$C$ define a decomposition of~$\H$ into $\#C = q^i$ eigenspaces, each of dimension $q^{n-i}$. Each of these is further split as a sum of one-dimensional eigenspaces of~$L$.
Now, one can define the quantum code space to be the invariant subspace of $C$, $\H(V)^C$, which is a Hilbert space of $(n-i)$ qubits, isomorphic to $\qty(\C^q)^{\otimes(n-i)}$. (One could equivalently have chosen any of the joint eigenspaces of~$C$.) Choosing an identification of this space with a standard set of $n-i$ qubits
gives an encoding of $n-i$ qubits to $n$ qubits; the ``code words'' can be thought of as the natural basis in the code space consisting of eigenspaces of~$L$.

To think of the code as a perfect tensor, we'd like to view the isometric injection of the $(n-i)$-qubit code space into the $n$-qubit encoding space as arising from a partitioning of the indices of a $(2n-i)$-index  tensor. In other words, we should consider the larger space consisting of $(n-i)$ degrees of freedom to be encoded, together with $n$ degrees of freedom for the encoding space. Note that the number of indices of the perfect tensor will be even precisely when $i = \dim C$ is even, as is the case for the codes of the previous section.
 The error-correction properties of such a code are discussed in~\cite{CRSS}; we note that quantization of a self-orthogonal $[2n,2k]_q$ code, such as those discussed above,  produces a quantum code with parameters $[\![n, n-2k, d_Q]\!]_q$. That is, one encodes $n-2k$ qubits in $n$ qubits in a manner that protects against $d_Q$ errors, where $d_Q = \min \{ \wt(a,b): (a,b) \in C^\perp \smallsetminus C\}$.
 In order to produce a perfect tensor, we will need $d_Q = n - k$. 
 
In appendix \ref{THREEQ} we consider an explicit example of a particular classical code, one of the Reed--Solomon codes, and construct the associated quantum Reed--Solomon code. 
The classical Reed--Solomon codes have parameters $[n,k,n-k+1]_q$, and are constructed using a set of points $X \subseteq \mathbb{P}^1(\mathbb{F}_q)$ with $|X| = n \leq q+1$, and homogeneous polynomials $f \in \mathbb{F}_q[u,v]$ where $x = [u:v] \in X$. For an input $k$-tuple of $q$-ary bits, $a = (a_0,\ldots,a_{k-1}) \in \mathbb{F}_q^k$, the homogeneous polynomial is chosen to be
\eqn{homPol}{
f_a(u,v) = \sum_{i=0}^{k-1} a_i u^i v^{k-1-i}\,,
}
and the resulting code takes the form
\eqn{RSCode}{
C = \{( f_a(u_1,v_1), \ldots, f_a(u_n,v_n): a \in \mathbb{F}_q^k, [u_i:v_i] \in X\}\,.
}
This family of Reed--Solomon codes can be used to construct
quantum error-correcting Reed--Solomon codes $[\![n,n-2k,k+1]\!]_q$~\cite{Marcolli:2018ohd}. The case of perfect tensors is obtained by setting $n=q$ and $k+1=n-k$, which leads to a $[\![q,1,(q+1)/2]\!]_q$ code describing
perfect tensors with $q+1$ indices and bond dimension $q$, where we can take the prime $q$ to be large as required for our later applications. In this paper we will consider precisely this code to construct holographic tensor networks; however our results are applicable more generally to tensor networks built out of any error-correcting code with the ``perfectness'' property in definition \ref{def:PT}.

\smallskip
{\bf Notational remark:} In this and the following section,  as well as in appendix~\ref{THREEQ}, we  set $q=p^r$ where $p$ is a prime and $r$ is a positive integer. Later in section \ref{GENUSZERO} onward, we will reserve the letter $p$ to parametrize the bulk geometry of the Bruhat--Tits tree of valence $p+1$, and will set up on this geometry the quantum Reed--Solomon code  $[\![r,1,(r+1)/2]\!]_r$ where $r$ will be an independent prime number.
It's also worth emphasizing that, in addition to being a prime power, $q$ will be used for the parameter of a multiplicative normalization of an elliptic curve (i.e. a representation as $\C^\times/q^\bZ$, or $\Q_p^\times/q^\bZ$ in the $p$-adic case). Both notations are standard, but the context should always be sufficient to determine which usage is intended.

\section{Perfect Tensors Associated to Semiclassical States} 
\label{CRSS}

The reader will have noticed that we have chosen to emphasize the perspective of canonical quantization in our exposition of the CRSS algorithm. We have done this, in part, to prepare for the discussion in this section, in which we will demonstrate that there is a natural generalization of the CRSS algorithm that produces perfect tensors \emph{directly}, without any intermediate reference to the theory of quantum codes. 

This perspective on perfect tensors has several advantages: First off, it shows that they are naturally associated by quantization to a particular class of semiclassical states, i.e., Lagrangian subspaces of a symplectic vector space $V$ over~$\F_q$. The condition of maximal rank on the perfect tensor, with respect to a decomposition of the Hilbert space into groups of qubits, translates naturally into a general-position requirement on the Lagrangian, asking that the dimensions of its intersections with a symplectic splitting of~$V$ be generic (as small as possible). As such, one is led to the conclusion that perfect tensors---rather than just being an \emph{ad~hoc} choice, adopted for calculational convenience in tensor network models---arise naturally in a way which bears a precise relationship to standard physical constructions.

As a consequence of this perspective, we are able to write down a natural class of Hamiltonians, which are closely related to standard Hamiltonians for discrete degrees of freedom, for which the vacuum state is precisely the perfect-tensor state. With a bit of additional work, related to understanding gluing of perfect tensors, it should be possible to use these ideas to write down concrete spin systems whose vacuum states are computed by networks of perfect tensors. We look forward to returning to this question in future work.

Our constructions make use of results of Gurevich and Hadani~\cite{GH}, who demonstrated the existence of a ``canonical quantization functor'' for symplectic vector spaces over finite fields. Given a choice of central character, this functor associates a finite-dimensional Hilbert space to each such symplectic vector space. Of course, this Hilbert space is just isomorphic to the unique representation of the corresponding Heisenberg group with given central character; however, constructing that representation normally requires a choice of auxiliary data, taking the form of a Lagrangian subspace of~$V$ and playing the role of a choice of polarization in geometric quantization. (The reader should imagine, for example, the choice between the position and momentum representations in constructing the quantum-mechanical Hilbert space of a particle.) Rather than a canonical Hilbert space, one therefore normally gets a family of Hilbert spaces over the oriented Lagrangian Grassmannian of~$V$.
Gurevich and Hadani demonstrate the existence of a collection of intertwining morphisms that naturally identify all of the fibers of this family; the reader should think of equipping this bundle with a natural flat connection (with trivial monodromy). The canonical model of the Hilbert space is then given by horizontal sections of the family.

Using these intertwining morphisms, one can therefore use any model one chooses to study the irrep of the Heisenberg group. 
For the perfect tensor associated to a Lagrangian $L\subset V$, it is natural to choose the polarization to be either $L$ or~$L^\vee$; by making this choice, one obtains a state that looks, roughly speaking, either like a delta function or like a constant function. The reader can imagine that translation-invariant states in $L^2(\R)$ are constant functions in the position representation, or delta functions in the momentum representation. 
Of course, since~$L$ is in general position, the basis for~$\H$ arising from this choice is as far as possible from being a tensor-product basis. To change to such a basis, one must instead take the polarization data to be a Lagrangian $\Lambda\subset V$ which is a direct sum of one-dimensional Lagrangians, one in each of the symplectic direct summands of~$V$. ($\Lambda$ is thus maximally decomposable, i.e., as \emph{far} as possible from being in general position.) Applying the intertwining morphism of~\cite{GH} then gives an explicit formula for the perfect-tensor state, in the tensor product basis.
 
The organization of this section is as follows: We will begin by giving a few more details on the construction of models for the irreducible representation of~$\Heis(V)$, and then continue by reviewing some of the results of~\cite{GH}. From there, we will go on to give the relation between perfect tensors and Lagrangians in general position. 

\subsection{Irreducible representations of Heisenberg groups}
\label{ssec:HeisGrp}

We reviewed some basic facts about Heisenberg groups above; here, we give some more details about the construction of irreducible representations. Any symplectic vector space $(V,\omega)$ can be written in some basis in the form
\deq[eq:darboux]{
V = \Lambda \oplus \Lambda^\vee,\quad
\omega: (x_1,z_1;x_2,z_2) \mapsto z_1(x_2) - z_2(x_1),
}
i.e., in a standard set of Darboux-type coordinates as the direct sum of two Lagrangian subspaces, which are placed in duality by~$\omega$. So we can simply write $\Heis(n,q)$ for the Heisenberg group where $\Lambda \cong \F_q^n$. Here, a basis of $\Lambda$ also determines a splitting of~$V$ as a direct sum of two-dimensional symplectic spaces; with respect to this splitting, $\Lambda$ is of course maximally decomposable, in the sense mentioned above.

If we choose a section of the projection map in the exact sequence~\eqref{eq:seq}, which is equivalent to choosing a normal-ordering prescription, we can begin to write familiar-looking explicit formulas. For example, given a Darboux basis of the form~\eqref{eq:darboux}, we can pick the section of the projection map defined by the condition that all operators from~$L^\vee$ appear to the right of those coming from~$L$, and thereby identify~$\Heis(V)$ with $\F_q \times V$.
In other words, by an obvious Poincar\'e--Birkhoff--Witt-type property, we can write each element of the group uniquely in the form
\deq{
c^\ell X^I Z^J,
}
where the multi-index $I$ is an element of~$L$, $J$ of~$L^\vee$, and~$\ell$ of~$\F_q$.
(Clearly, $X^I$ means $X_1^{i_1}\cdots X_n^{i_n}$, corresponding to a chosen basis of~$L$, and so on. Since $L$ is abelian, there is no further ordering ambiguity.) In what follows, we use the notation $[v]$ to mean the element of~$\Heis(V)$ corresponding to~$v\in V$ under the above prescription.
 
The commutator of two such elements is then determined by the standard symplectic form:
\deq{
[v][w] = c^{\omega(v,w)} [w][v].
}
In a basis, we could write
\deq{
X^I Z^J X^{I'} Z^{J'} = c^{(I',J) - (I,J')} X^{I'} Z^{J'} X^I Z^J,
}
where  the pairing $(I,J) = i_1j_1 + \cdots + i_nj_n$, and the computation is carried out in~$\F_q$. This is  the dual pairing between~$L$ and~$L^\vee$, and as such we have just rewritten the standard symplectic form~\eqref{eq:darboux}. The reader should compare this with~\eqref{EabComLaws}; note, however, that a specific choice of central character has been made, whereas here we have not done this yet. 

Now, consider the regular representation of~$\Heis(V)$, which is just on the space of complex-valued functions on the group itself. This space admits an action of~$\Heis(V)$ by both left and right translations, and furthermore has the natural $L^2$ Hermitian inner product, which is translation invariant.
Furthermore, 
it breaks up into a direct sum over the set of central characters; we will denote by $F_\chi$ the subspace of functions where $Z(\Heis(V))\cong \F_q$ acts through the character~$\chi$. After choosing a normal-ordering prescription as above, $F_\chi$ can be thought of as identified with~$\Fun(V)\cong (\C^{p})^{\otimes 2n}$; In the presence of this additional data, it is precisely the space 
\begin{equation}
\begin{tikzcd}
\Fun(V) \otimes \chi \arrow[hookrightarrow]{d}{} \arrow{r}{\sim} & F_\chi \arrow[hookrightarrow]{d}{} \\
\Fun(V) \otimes \Fun(\F_q) \arrow{r}{\sim} &\Fun(\Heis(V)).
\end{tikzcd}
\end{equation}
Note, however, that $F_\chi$ it is defined independent of a normal ordering, even though the above identification is not. Thus, the horizontal isomorphisms in the above diagram are not canonical. Moreover, it is apparent that, with respect to the $L^2$ inner product, 
\deq{
(F_\chi)^\vee = F_{\bar\chi},
}
where $\bar\chi = \chi^{-1}$ is the inverse (or complex conjugate) character.

Now, by the Stone--von~Neumann theorem, $F_\chi$ must decompose as the direct sum of $p$ irreducibles, each isomorphic to the unique (Heisenberg) representation with central character $\chi$. This decomposition can be seen as follows: Given a choice of Lagrangian $\Lambda$ and the corresponding decomposition $V = \Lambda \oplus \Lambda^\vee$ as above, one can take the \emph{right} $\Lambda^\vee$-invariants inside of~$\Fun_\chi(V)$. Specifically, we mean functions $f \in F_\chi$ such that 
\deq{
f( g [z]) = f(g), \quad \forall g \in \Heis(V),\ z \in \Lambda^\vee.
}
This is then an invariant subspace (in fact, an irreducible representation) with respect to the action of~$\Heis(V)$ by \emph{left} translations; indeed, the entire right action of~$\Lambda^\vee$ commutes with the left action of~$\Heis(V)$, so that the eigenspaces of that action are $\Heis(V)$-invariant. They are precisely the $p$ irreducible factors of the isotypic component $F_\chi$, and can be labeled by a set of characters of~$L$. 

With respect to a choice of normal ordering as above, we can identify $(F_\chi)^{\Lambda^\vee}$ in the obvious way with~$\Fun(\Lambda)$, and may write $f(x)$ for an element using this identification. 
In formulas, one has
\deq{
[x'] \cdot f(x) = f(x + x'),
}
as well as 
\deq{
[z] \cdot f(x) = f([z][x]) = f(c^{z(x)}[x+z]) = \chi(c^{z(x)}) f(x),
}
after using right $\Lambda^\vee$-invariance. This is the model of the Heisenberg representation corresponding to the polarization data $\Lambda^\vee$, and we will sometimes denote it $\Hilb_{\Lambda^\vee}$.

It should be apparent that this construction is precisely analogous to the construction of the Hilbert space in quantum mechanics, or (in more sophisticated terms) to the role of a ``choice of polarization'' in geometric quantization. For example, functions on phase space would be ``wavefunctions'' $\psi(x,p)$, but the Hilbert space in fact consists of wavefunctions $\psi(x)$---i.e., that subset of functions on phase space that are invariant under translation in the momentum directions. The choice of central character corresponds to a choice of numerical value for~$\hbar$.

\subsection{Functorial quantization}
\label{ssec:quant}

In this section, we give a brief review of the main results of~\cite{GH}, which constructed a canonical model of the Heisenberg representation. As mentioned above, this was done by constructing a family of intertwining morphisms
\deq{
\phi_{\Lambda}^{\Lambda'}: \H_\Lambda \rightarrow \H_{\Lambda'},
}
trivializing the family over the space of oriented Lagrangians in~$V$. (The morphisms $\phi$---although not the representations $\Hilb_\Lambda$---depend on the additional data of orientations on~$\Lambda$ and~$\Lambda'$; an orientation is just a choice of nonzero vector in the top exterior power of~$\Lambda$. We suppress this from the notation for the sake of simplicity;  in fact, the orientation only enters $\phi$ in the form of a normalization constant, which will play no role in our considerations.)

Let us begin by stating the theorem:
\begin{thm}[\cite{GH}]
To each nontrivial central character $\chi$, there is associated a canonical ``quantization functor''
\deq{
\H: \cat{symp}^\text{\rm iso}(\F_q) \rightarrow \cat{vect}^\text{\rm iso}(\C),
}
where the first category is that of finite-dimensional symplectic vector spaces over~$\F_q$ (with arrows being isomorphisms of such) and the second is that of finite-dimensional complex vector spaces together with their isomorphisms. (We may occasionally write $\H^\chi$ for clarity.)
By the action on arrows, one obtains a natural group homomorphism
\deq[eq:Wrep]{
\Sp(V) \rightarrow GL\qty(\H(V)),
}
for every symplectic vector space~$V$; this gives a canonical model of the Weil representation of~$\Sp(V)$. Moreover, $\H(V)$ carries a natural action of~$\Heis(V)$, isomorphic to the Heisenberg representation with the corresponding central character.

The functor $\H$ is monoidal, carrying the Cartesian product in~$\cat{symp}$ to the tensor product in~$\cat{vect}$. 
It is compatible with symplectic duality, meaning that 
\deq{
\H(\overline{V}) = \H(V)^\vee.
}
Here $\overline{V} = (V,-\omega)$ is the ``symplectic dual'' of~$(V,\omega)$.

Moreover, $\H$ is compatible with symplectic reduction, in the following sense: Let $I \subset V$ be an isotropic subspace. Then there is a natural isomorphism
\deq{
\H(V)^I \cong \H(V\sred I).
}
The left-hand side is the $I$-invariants in the quantization of~$V$, whereas the right-hand side is the quantization of the linear symplectic reduction
\deq[eq:sreddef]{
V\sred I \defeq I^\perp / I
}
of~$V$ along~$I$. 
\end{thm}
A few remarks on this result: First off, as mentioned above, one should think of the choice of central character as the value of~$\hbar$. As such, this is not erroneous extra data, but conceptually essential to the idea of a quantization functor.
Second, the compatibility with symplectic reduction is an example of a result of the form ``quantization commutes with reduction'' (i.e., the Guillemin--Sternberg conjecture) in the context of finite fields. 

As mentioned above, the proof takes the form of giving a natural family $\phi$, trivializing the dependence of~$\Hilb_\Lambda$ on additional data. This means that, to check a given property, it is enough to establish it for each such model, and further check that it is compatible with the trivialization maps. For example, the action of the functor on morphisms is quite simple to see: for a symplectomorphism $f_V\rightarrow V'$, it is just the pullback of $\Hilb_{\Lambda'} \subset F_\chi'$ to~$V$, which lands in~$\Hilb_{f^{-1}(\Lambda')}$. That is, one identifies polarization data in the two cases using the natural map induced from the symplectomorphism between the respective spaces of Lagrangians.

Furthermore, the family of maps $\phi$ are relatively simple to describe: For two Lagrangians $L$ and~$L'$, in general position with respect to  one another (in the sense that $L \cap L' = \{0\}$), the map is simply given by averaging:
\deq{
\phi(f)(h) \propto \sum_{m \in L'} f(h\cdot[m]).
}
The result is obviously right $L'$-invariant; as mentioned above, the proportionality constant depends on a choice of orientation on each Lagrangian. When $L\cap L' = I$ is nontrivial, one averages over a set of representatives for the cosets $L'/I$. For more detailed discussion, the reader is referred to~\cite{GH}.

\subsection{Perfect tensors from symplectic vector spaces}
\label{ssec:PT}

We're now at the point where we can formulate our central result. Take~$V$ to be a symplectic vector space of dimension $4n$ (as always, over a finite field~$\FF_q$), and choose a Darboux basis as above, defining a splitting of~$V$ into classical degrees of freedom (two-dimensional symplectic subspaces). 
We will show that a choice of Lagrangian subspace $L$ in~$V$, which is in generic position with respect to the splitting, defines a perfect tensor upon application of the quantization functor~$\H$.

To start, let's make the notion of general position a bit more precise. The splitting of~$V$, coming from a Darboux basis, writes it as a symplectic direct sum
\deq[eq:decomp]{
V = \bigoplus_{i=1}^{2n} V_i
}
of copies of~$\FF_q^2$. Since the functor is monoidal, this is correspondingly a tensor product decomposition
\deq{
\Hilb(V) = \bigotimes_{i=1}^{2n} \Hilb(V_i) \cong (\C^q)^{\otimes 2n}
}
of the corresponding Hilbert space into qubits.

Partition the index set $\{1,\ldots,2n\}$ into disjoint sets $K$ and~$K'$, with $k = \# K \leq n$, and denote the corresponding splitting~$V=W\oplus W'$. It is then simple to check that 
\deq[eq:dims]{
\dim(L \cap W) \geq 2(k - n), \quad
\dim(L \cap W') \geq 2(n - k). 
}
The first condition, though, is obviously vacuous, since $n - k \geq 0$. In particular, if $k=n$, $L$ will generically have zero-dimensional intersection with both $W$ and~$W'$. 
We will say that $L$ is in \emph{strongly general position} with regard to the decomposition~\eqref{eq:decomp} if it has zero-dimensional intersection with all such $W$ and $W'$, i.e., for any partition of the indices into two equal sets.

Now, it is straightforward to check the following simple result:
Given such a splitting, a Lagrangian $L$ in general position is precisely equivalent to a choice of symplectomorphism
\deq{
\psi: \overline{W} \rightarrow {W}',
}
where $\overline{W}$ is the symplectic dual of~$W$.
Indeed, an element $v\in L$ is a pair of elements $(v_1,v_2)$ of~$W \times W'$, and the symplectic form is of direct product type, so that
\deq[eq:symplecto]{
0 = \omega(u,v) = \omega(u_1,v_1) + \omega(u_2,v_2).
}
Define the map $\psi$ by the rule $\psi(v_1) = v_2$; this is clearly linear. Moreover, it is unambiguous, since the existence of more than one $v_2$ such that $(v_1,v_2) \in L$ would contradict the assumption that $L \cap W = 0$. There must exist at least one such $v_2$, though, since $L$ has maximal dimension. By~\eqref{eq:symplecto}, $\psi^*\omega = - \omega$, so that $\psi$ is a symplectomorphism between $W'$ and the symplectic dual of~$W$. The converse construction is obvious; just take $L$ to be the graph of the map $\psi$.

It is now easy to see that the $L$-invariant element defines a linear isomorphism. We can interpret $L$ as a symplectomorphism, which will be carried by the quantization to a linear map
\deq{
\H(\psi): \H(W)^\vee \rightarrow  \H(W').
}
This map is equivalently an element in~$\H(W) \otimes \H(W') = \H(V)$, which is the perfect tensor.
Of course, it remains to show that the element associated to the symplectomorphism $\psi$ is, in fact, identical to the invariant element under the abelian subgroup associated to the graph of~$\psi$. But this is not difficult to see;
for a given basis vector in~$L$, which can be written as a sum over the Darboux basis, the invariants of that basis vector will be those tensor products of eigenvectors of the Darboux generators whose eigenvalues multiply to unity.  Upon splitting the generators into two subsets, this means that the product of eigenvalues from the first set is equal to (the inverse of) the product of eigenvalues from the second.

We can furthermore make the following observation: The dimensions of the isotropic spaces $J = L\cap W$ and~$J' = L\cap W'$, must, in fact, be equal. To see that this is true, suppose that $J$ has dimension $j$. Then consider the symplectic reduction of~$V$ along~$J$; because $W' \subset J^\perp$, this is naturally isomorphic to the direct sum of~$W'$ with the symplectic reduction of~$W$ along~$J$, and $L/J$ is a Lagrangian subspace there (as one can see by checking dimensions). But then it follows from dimension formulas analogous to~\eqref{eq:dims} that 
\deq{
j' = \dim J' \geq j.
}
By reversing $W$ and~$W'$, the equality $j=j'$ follows.

It further follows that the rank of the $L$-invariant element $T$, with respect to the tensor-product decomposition $\H(W) \otimes \H(W')$, is in fact always given by the formula
\deq{
\rank T = p^{n-j},
}
where, as above, $j = \dim(L\cap W) = \dim(L  \cap W')$. This is a simple consequence of compatibility with symplectic reduction, as defined in~\eqref{eq:sreddef}: after reducing along $J \oplus J'$, one obtains a Lagrangian in general position, of the form
\deq{
L/(J \oplus J') \subset (W\sred J) \oplus (W'\sred J').
}
The formula then follows by recalling the above considerations in the case of general position.

In fact, nothing in our arguments requires that $V$ consist of an even number of degrees of freedom. One is free to consider the more general case of a decomposition, where
\deq{
L^m \subset V^{2m} = W^{2k} \oplus (W')^{2(m-k)}.
}
Here, superscripts denote dimension, and $2k\leq m$ without loss of generality; with regard to our previous notation, $m = 2n$, although $m$ may of course now be odd. It is trivial to check, as above, that 
\deq{
j = \dim(L\cap W) \geq 0, \quad
j' = \dim(L \cap W') \geq m - 2k.
}
But it is further true that 
\deq{
j' = m - 2k + j,
}
as may be seen by considering symplectic reduction along~$J$ and~$J'$ in turn, and counting dimension. Moreover, the rank of the quantization of~$L$ will be precisely $p^{k-j}$. It is thus true that a Lagrangian in general position with respect to a Darboux basis gives rise to a perfect tensor, in the sense of our above definition, independent of the number of indices under consideration.

A further remark on this result: It is simple to see that, for each splitting of~$V$ as a symplectic direct sum, generic position is an open condition in the Lagrangian Grassmannian. Since there are only finitely many splittings to check, we impose only finitely many open conditions by insisting that $L$ is in strongly general position. As such, this condition is ``generic'' among all Lagrangians in~$V$. Of course, this is subject to the caveat that (over $\F_p$) the Lagrangian Grassmannian itself consists of only finitely many points. As such, for a given choice of $q$ and~$\dim V$, there may not be \emph{any} such Lagrangian---even though the condition is open! However, for large enough $q$, one expects perfect tensor states to be generic in this precise sense among all semiclassical states.

\subsection{A simple example}
\label{ssec:example}

Just for concreteness, let's consider the simplest example of the above considerations: a two-qubit space $V\cong \FF_p^4$, with Darboux basis $\{x_1,x_2;z_1,z_2\}$ over $\F_p$. An example of a Lagrangian in general position is
\deq{L = \Span(x_1 + z_2, x_2 + z_1);
}
this is the graph of the symplectomorphism
\deq{
\psi: V_1 \rightarrow \overline{V}_2,
\qquad x_1 \mapsto z_2,\quad z_1 \mapsto x_2.
}
It is now trivial to see that, in the representation where the explicit matrices are
\deq{
x_i \ket{a} = \ket{a+1}, \quad z_i \ket{a} = e^{2 \pi i a / p} \ket{a},
}
a set of invariant eigenvectors for the first basis element of~$L$ consists of
\deq{
\psi(a) =  \sum_{b \in \F_p} e^{-2 \pi i a b / p} \ket{a,b},
}
since this element is just the basis vector $\ket{a}$ of~$x_1$ tensored with the eigenvector of $z_2$ with eigenvalue $a^{-1}$.
Similarly, for the second basis element,
\deq{
\psi'(b) = \sum_{a \in \F_p} e^{-2 \pi i a b / p} \ket{a,b},
}
from which it is trivial to see that the perfect tensor associated to~$L$ (the intersection of the above two subspaces) is just
\deq{
T = \sum_{a,b \in \F_p} e^{-2 \pi i a b / p} \ket{a,b}.
}
It is, of course, obvious that this defines a full-rank matrix after lowering an index, since the matrix elements are just 
\deq{
T_{ab} = e^{-2 \pi i a b / p}.
}
Furthermore, it is clearly unitary, up to a normalization by~$1/\sqrt{p}$ that we should have included all along (to make use of the \emph{normalized} eigenvectors of~$z_i$). And we should expect it to be the isomorphism that identifies the eigensystem of~$x_1$ in the first copy of~$\C^p$ with the eigensystem of~$z_2$ in the second copy; a quick look at the $a$-th column of~$T_{ab}$ demonstrates that this is indeed the case. (This is, of course, closely connected to the kernel representation of the Fourier transform over~$\F_p$.)

An equally simple example can be constructed by thinking of the Lagrangian
\deq{
L = \Span(x_1+x_2,z_1-z_2);
}
the reader will find it pleasant to check that the matrix $T_{ab}$, in this case, reduces exactly to the identity matrix.

\subsection{Hamiltonians with perfect tensor vacua}
\label{ssec:Ham}

We've worked hard to develop the idea that it is profitable to think about the systems of qubits arising in CRSS-type algorithms as quantizations of the canonical commutation relations on a discrete and periodic degree of freedom (i.e., representations of Heisenberg groups over finite fields). This makes a lot of analogies with usual quantum mechanics---which, after all, is about representations of Heisenberg groups over~$\R$---apparent.

As a simple application of these ideas, we'd like to show that it's possible to write a physically natural Hamiltonian, starting with the semiclassical state (i.e. Lagrangian subspace) whose quantization is the perfect tensor. This Hamiltonian has the property that its spectrum is positive semidefinite, and the unique vacuum state is precisely the perfect tensor state.

Of course, the elements of~$\Heis(n,q)$ are represented as \emph{unitary} operators, rather than Hermitian ones---the typical commutation relation,
\deq{ZX = cXZ,} is after all an analogue of the \emph{exponentiated} canonical commutation relations.
This seems like an obstruction to writing down interesting Hamiltonians, especially since $\Heis(n,q)$ is a discrete group and one cannot simply ask about its Lie algebra! 

However, it is straightforward to see that the situation is analogous to that of writing discrete derivative operators in quantum mechanics. Here, the only natural operators are the shift operators, which of course are unitary: the adjoint of a shift is the inverse shift. But it's therefore straightforward to see that combinations like
\deq{
 x + x^{-1}, ~ z + z^{-1}
 }
are Hermitian, with spectra that look like (twice) the real parts of the roots of unity. And these Hermitian operators are obviously diagonal in the same basis that the unitaries themselves are.

In fact, there's also the option to simply write
\deq{
h(x) = 2 - x - x^{-1},
}
which is still obviously Hermitian. Recalling that $x$ is an elementary shift operator, this is precisely analogous to a standard quadratic kinetic term in a lattice model: a discrete analogue of the operator $-\nabla^2$.
Moreover, since $X$ has eigenvalue $e^{2 \pi i a / p}$, the spectrum of $h(X)$ is
\deq{
h(X) = 2 - 2 \cos\qty(\frac{2\pi i a}{p}) = 4 \sin^2 \qty(\frac{\pi i a}{p}).
}

The construction of a suitable Hamiltonian is now clear: Choose any basis $b_i$ for the Lagrangian~$L$ whose quantization is the perfect tensor state, and then write
\deq{
h = \sum_i h(b_i) = \sum_i \qty( 2 - b_i - b_i^{-1} ).
}
Since $L$ is Lagrangian, the $b_i$ (and therefore also the $h(b_i)$) are mutually commuting, and $h$ can be diagonalized by a set of mutual eigenvectors of the $b_i$---which is precisely an eigenbasis of~$L$.

These Hamiltonians provide a natural set of candidates for constructing physical Hamiltonians, analogous to spin systems, whose sets of vacua are precisely the states obtained by networks of perfect tensors. 
What remains to be developed is an understanding of how the operation of gluing perfect tensors together lifts to the construction of a glued Hamiltonian, whose vacuum is the glued state. 
It seems plausible that this could be done in a way that has a natural semiclassical interpretation; whether the resulting model would have a Hamiltonian of commuting-projector type is not obvious. We look forward to returning to this question in future work.

\section{Entanglement in $p$-adic AdS/CFT} 
\label{GENUSZERO} 

In this section we build on section \ref{REVIEW} to initiate the study of holographic entanglement entropy in $p$-adic AdS/CFT, via a quantum error-correcting tensor network construction built using perfect tensors. We begin by discussing the framework for the vacuum ($p$-adic) AdS geometry, culminating in the verification  of a Ryu--Takayanagi like formula in this purely $p$-adic setting, and in the next section proceed to discuss entanglement in a genus 1 ($p$-adic) black hole geometry.

\subsection{The dual graph tensor network}
\label{DUALGRAPH}

The states we want to focus on in this paper are a subset of all possible states which can be constructed using contractions of perfect tensors, each of which can be referred to as a ``tensor network''. The basic idea of this construction involves the contraction of many tensors in a ``bulk'' space to produce a complicated entangled state at the boundary of the network. One may interpret this boundary state as an analog of the ground state of a boundary conformal field theory,
and there are many proposals in the literature on how this may be realized.
The details of the particular tensor network proposed here are along the lines of~\cite{Pastawski:2015qua} and are important to the overall conclusions and generalizations. We will see the tensor network is closely associated with $p$-adic AdS/CFT. 

To construct a holographic state $|\psi \rangle$ in the boundary Hilbert space, we consider a tensor network given by what we call the ``dual graph'' of the holographically dual bulk geometry. 
 For instance, if the boundary is $\mathbb{P}^1(\mathbb{Q}_p)$, then we consider the ``dual graph'' of the Bruhat--Tits tree in the bulk. 
If we are interested in states dual to the $p$-adic analog of the BTZ black hole, we must consider the corresponding ``dual graph'' of the genus $1$ Schottky uniformization of the Bruhat--Tits tree. 
In this section we focus on the tensor network associated with the Bruhat--Tits tree (which as mentioned in section \ref{ssec:padic} is to be thought of as the $p$-adic analog of a time slice of vacuum AdS$_3$). 
In the following, the introduction of this dual graph to the Bruhat--Tits tree may at first sight appear to be an additional structure beyond what is needed to study bulk dynamics $p$-adic AdS/CFT, but it will turn out to be crucial to our investigation of the relationship between bulk geometries and boundary entanglement. 

We recall from section~\ref{ssec:padic} that every edge on the Bruhat--Tits tree can be uniquely specified by specifying its two end-points (either as a pair of adjacent lattice equivalence classes or as a pair of cosets) and for every node on the Bruhat--Tits tree, there are $p+1$ edges incident on it. We define the dual graph as follows:

\begin{definition}
\label{def:dualgraph}
A {\it dual graph} of the Bruhat--Tits tree is a graph which satisfies the following two properties:\footnote{These properties are motivated from the ``minimum cut rule'' which provides a discrete analog of the Ryu--Takayanagi formula for connected regions in  networks of perfect tensors~\cite{Pastawski:2015qua}; however we do not assume this in the following. In fact in our setup the ``minimum cut rule'' applies more generally, for instance in the evaluation of the bipartite entanglement of a disconnected region as discussed in section \ref{GENUS0DOUBLE}.}
\begin{itemize}
\item There  exists  a bijective correspondence between bonds on the dual graph and edges on the Bruhat--Tits tree. (Both ``edge'' and ``bond'' refer to the same object in graph theory -- links between nodes on the graph, but for clarity we reserve the term ``edge'' for the Bruhat--Tits tree and ``bond'' for the dual graph.) Consequently, each bond on the dual graph is identified by specifying the corresponding edge on the Bruhat--Tits tree. 

\item The incidence relations of the set of bonds in bijective correspondence with those edges on the Bruhat--Tits tree  incident at a particular node, are such that they form a cycle graph. 
We refer to such cycle graphs as ``plaquettes''. Thus there is a bijective correspondence between nodes on the Bruhat--Tits tree and plaquettes on the dual graph.
\end{itemize}
In fact, the dual graph in the $p$-adic black hole geometry also satisfies the same properties.
\end{definition}

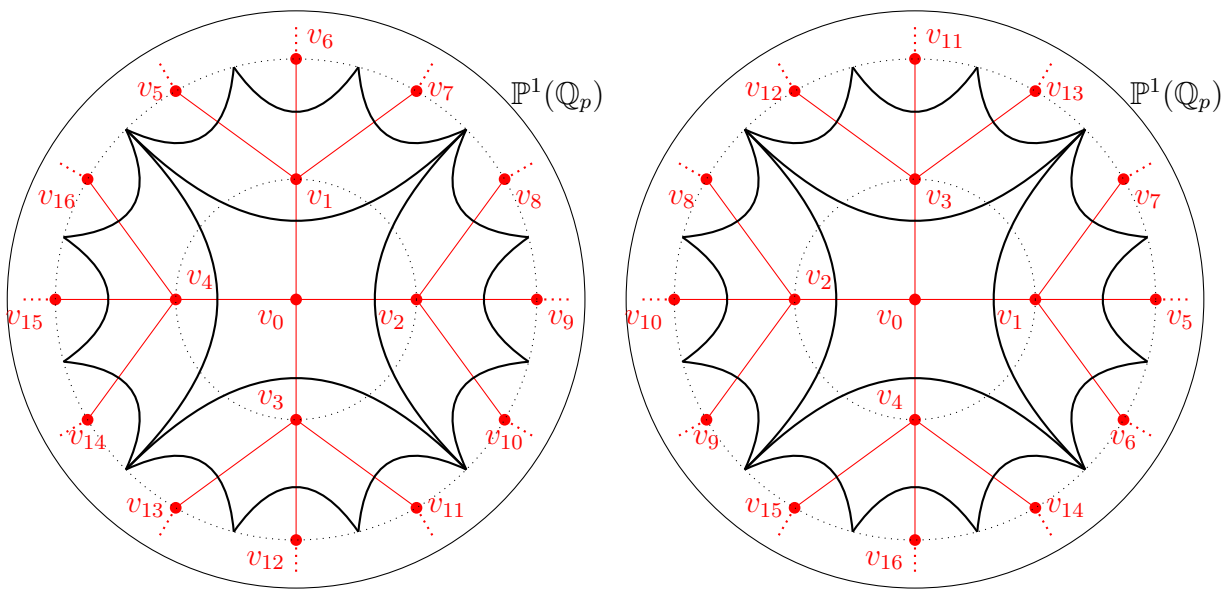
\begin{figure}[!th]
\centering
\begin{subfigure}[]{0.49\textwidth}
  \centering
\begin{tikzpicture}[scale=1.6]
\tikzstyle{vertex}=[draw,scale=0.4,fill=black,circle]
\tikzstyle{ver2}=[draw,scale=0.6,circle]
\tikzstyle{ver3}=[draw,scale=0.4,circle]
\coordinate (C) at (0,0);
\foreach \x in {0,1,2,3} {
	\coordinate (a\x) at (\x*360/4:1);
	\draw (a\x) node[color=red,fill=red,circle,scale=0.4] {};
	\draw[color=red] (a\x) -- (C);
	};
\foreach \x in {0,1,...,11} {
	\coordinate (b\x) at (\x*360/12 - 30 :2);
	\pgfmathparse{floor(\x/3)}
	\draw (b\x) node[color=red,fill=red,circle,scale=0.4] {};
	\draw[color=red] (b\x) -- (a\pgfmathresult);
	\draw[color=red, dotted, thick] (b\x) -- (\x*360/12 - 30 :2.3);
	};
\foreach \x in {0,1,...,11} {
	\coordinate (c\x) at (\x*360/12 - 45 :2);
	};
\foreach \x/\y in {0/1,1/2,2/3,3/4,4/5,5/6,6/7,7/8,8/9,9/10,10/11,11/0} {
	\draw[thick] (c\x) to [out =  \x*360/12 + 115, 
		in =  \y*360/12 + 155, looseness = 1.5] (c\y);	
	};
\foreach \x/\y in {0/3,3/6,6/9,9/0} {
	\draw[thick] (c\x) to [out =  \x*360/12 + 135, 
		in =  \y*360/12 + 135, looseness = 1.3] (c\y);	
	};
\draw[dotted] (0,0) circle[radius=1];
\draw[dotted] (0,0) circle[radius=2];
\draw (0,0) circle[radius=2.4];
\draw (1.697,1.697) node[right] {$\mathbb{P}^1(\mathbb{Q}_p)$};
\draw[color=red] (C) node[below left] {$v_0$};
\draw (C) node[color=red,fill=red,circle,scale=0.4] {};
\draw[color=red] (a0) node[below left] {$v_2$};
\draw[color=red] (a1) node[below right] {$v_1$};
\draw[color=red] (a2) node[above right] {$v_4$};
\draw[color=red] (a3) node[above left] {$v_3$};
\draw[color=red] (b0) node[below] {$v_{10}$};
\draw[color=red] (b1) node[below right] {$v_9$};
\draw[color=red] (b2) node[below right] {$v_8$};
\draw[color=red] (b3) node[right] {$v_7$};
\draw[color=red] (b4) node[above right] {$v_6$};
\draw[color=red] (b5) node[left] {$v_5$};
\draw[color=red] (b6) node[below left] {$v_{16}$};
\draw[color=red] (b7) node[below left] {$v_{15}$};
\draw[color=red] (b8) node[below] {$v_{14}$};
\draw[color=red] (b9) node[left] {$v_{13}$};
\draw[color=red] (b10) node[below left] {$v_{12}$};
\draw[color=red] (b11) node[right] {$v_{11}$};
\end{tikzpicture}
        \caption{A choice of a planar embedding for the dual graph}
        \label{fig:planar1}
\end{subfigure}
\begin{subfigure}[]{0.49\textwidth}
  \centering
\begin{tikzpicture}[scale=1.6]
\tikzstyle{vertex}=[draw,scale=0.4,fill=black,circle]
\tikzstyle{ver2}=[draw,scale=0.6,circle]
\tikzstyle{ver3}=[draw,scale=0.4,circle]
\coordinate (C) at (0,0);
\foreach \x in {0,1,2,3} {
	\coordinate (a\x) at (\x*360/4:1);
	\draw (a\x) node[color=red,fill=red,circle,scale=0.4] {};
	\draw[color=red] (a\x) -- (C);
	};
\foreach \x in {0,1,...,11} {
	\coordinate (b\x) at (\x*360/12 - 30 :2);
	\pgfmathparse{floor(\x/3)}
	\draw (b\x) node[color=red,fill=red,circle,scale=0.4] {};
	\draw[color=red] (b\x) -- (a\pgfmathresult);
	\draw[color=red, dotted, thick] (b\x) -- (\x*360/12 - 30 :2.3);
	};
\foreach \x in {0,1,...,11} {
	\coordinate (c\x) at (\x*360/12 - 45 :2);
	};
\foreach \x/\y in {0/1,1/2,2/3,3/4,4/5,5/6,6/7,7/8,8/9,9/10,10/11,11/0} {
	\draw[thick] (c\x) to [out =  \x*360/12 + 115, 
		in =  \y*360/12 + 155, looseness = 1.5] (c\y);	
	};
\foreach \x/\y in {0/3,3/6,6/9,9/0} {
	\draw[thick] (c\x) to [out =  \x*360/12 + 135, 
		in =  \y*360/12 + 135, looseness = 1.3] (c\y);	
	};
\draw[dotted] (0,0) circle[radius=1];
\draw[dotted] (0,0) circle[radius=2];
\draw (0,0) circle[radius=2.4];
\draw (1.697,1.697) node[right] {$\mathbb{P}^1(\mathbb{Q}_p)$};
\draw[color=red] (C) node[below left] {$v_0$};
\draw (C) node[color=red,fill=red,circle,scale=0.4] {};
\draw[color=red] (a0) node[below left] {$v_1$};
\draw[color=red] (a1) node[below right] {$v_3$};
\draw[color=red] (a2) node[above right] {$v_2$};
\draw[color=red] (a3) node[above left] {$v_4$};
\draw[color=red] (b0) node[below] {$v_{6}$};
\draw[color=red] (b1) node[below right] {$v_5$};
\draw[color=red] (b2) node[below right] {$v_7$};
\draw[color=red] (b3) node[right] {$v_{13}$};
\draw[color=red] (b4) node[above right] {$v_{11}$};
\draw[color=red] (b5) node[left] {$v_{12}$};
\draw[color=red] (b6) node[below left] {$v_{8}$};
\draw[color=red] (b7) node[below left] {$v_{10}$};
\draw[color=red] (b8) node[below] {$v_{9}$};
\draw[color=red] (b9) node[left] {$v_{15}$};
\draw[color=red] (b10) node[below left] {$v_{16}$};
\draw[color=red] (b11) node[right] {$v_{14}$};
\end{tikzpicture}
        \caption{A different planar embedding for the dual graph}
		\label{fig:planar2}
\end{subfigure}
\caption{A finite part of the infinite Bruhat--Tits tree is shown in red, with the nodes labelled $v_i$. The graphical representation of the dual graph for the finite red subgraph is shown in black. The Bruhat--Tits tree in (b) is obtained by acting on the Bruhat--Tits tree in (a) with a $\PGL(2,\mathbb{Q}_p)$ transformation fixing the vertex $v_0$. Equivalently, the dual graphs in (a) and (b) correspond to two different choices of incidence relations for bonds on the dual graph, subject to the two requirements mentioned in definition \ref{def:dualgraph}. In the infinite graph limit, the geometry of the Bruhat--Tits tree is represented by the Schl\"{a}fli symbol $\{\infty, p+1\}$, while the dual graph is given by the Schl\"{a}fli symbol $\{p+1,\infty\}$.}
\label{fig:planar}
\end{figure}

{\it Any} valid dual graph must satisfy the definition above; however, the definition does not uniquely specify {\it a particular} dual graph. 
By construction, a $\PGL(2,\mathbb{Q}_p)$ transformation acts simultaneously on both the Bruhat--Tits tree and its dual graph as an isometry. 
The choice of picking a particular valid dual graph (which corresponds to making a particular choice on the connectivity of the plaquettes at each node) corresponds to a choice of ``planar embedding'' of the Bruhat--Tits tree as we explain in section \ref{TNPROPS}. See figure \ref{fig:planar} for an example.  The construction of the dual graph may appear sensitive to the existence and choice of a planar embedding of the Bruhat--Tits tree.  However, we show  that physical quantities do not depend on this choice, and in  section \ref{GEOMETRIC} we explain this construction entirely in the context of the $p$-adic Drinfeld upper half plane without assuming an embedding in the ordinary (real) upper half plane.

\begin{figure}
\centering
\begin{subfigure}[]
    {0.48\textwidth}
  \centering
\begin{tikzpicture}[scale=1.6]
\tikzstyle{vertex}=[draw,scale=0.4,fill=black,circle]
\tikzstyle{ver2}=[draw,scale=0.6,circle]
\tikzstyle{ver3}=[draw,scale=0.4,circle]
\coordinate (C) at (0,0);
\foreach \x in {0,1,2,3} {
	\coordinate (a\x) at (\x*360/4:1);
	\draw[color=red] (a\x) -- (C);
	};
\foreach \x in {0,1,...,11} {
	\coordinate (b\x) at (\x*360/12 - 30 :2);
	\pgfmathparse{floor(\x/3)}
	\draw[color=red] (b\x) -- (a\pgfmathresult);
	};
\foreach \x in {0,1,...,11} {
	\coordinate (c\x) at (\x*360/12 - 45 :2);
	};
\foreach \x/\y in {0/1,1/2,2/3,3/4,4/5,5/6,6/7,7/8,8/9,9/10,10/11,11/0} {
	\draw[thick] (c\x) to [out =  \x*360/12 + 115, 
		in =  \y*360/12 + 155, looseness = 1.5] (c\y);	
	};
\foreach \x/\y in {0/3,3/6,6/9,9/0} {
	\draw[thick] (c\x) to [out =  \x*360/12 + 135, 
		in =  \y*360/12 + 135, looseness = 1.3] (c\y);	
	};
\draw[dotted] (0,0) circle[radius=1];
\draw[dotted] (0,0) circle[radius=2];
\draw[color=red] (C) node[below left] {$C$};
\draw (C) node[color=red,fill=red,circle,scale=0.4] {};
\end{tikzpicture}

        \caption{Dual graph to a finite tree}
		\label{fig:SimpleDual}
\end{subfigure}
\begin{subfigure}[]
      {0.48\textwidth}
      \centering
\begin{tikzpicture}[scale=1.6]
\tikzstyle{vertex}=[draw,scale=0.4,fill=black,circle]
\tikzstyle{ver2}=[draw,scale=0.6,circle]
\tikzstyle{ver3}=[draw,scale=0.4,circle]
\coordinate (C) at (0,0);
\foreach \x in {0,1,2,3} {
	\coordinate (a\x) at (\x*360/4:1);
	\draw[color=red] (a\x) -- (C);
	};
\foreach \x in {0,1,...,11} {
	\coordinate (b\x) at (\x*360/12 - 30 :2);
	\pgfmathparse{floor(\x/3)}
	\draw[color=red] (b\x) -- (a\pgfmathresult);
	};
\foreach \x in {0,1,...,11} {
	\coordinate (c\x) at (\x*360/12 - 45 :2);
	};
\foreach \x/\y in {0/1,1/2,2/3,3/4,4/5,5/6,6/7,7/8,8/9,9/10,10/11,11/0} {
	\draw[thick] (c\x) to [out =  \x*360/12 + 115, 
		in =  \y*360/12 + 155, looseness = 1.5] (c\y);	
	};
\foreach \x/\y in {0/3,3/6,6/9,9/0} {
	\draw[thick] (c\x) to [out =  \x*360/12 + 135, 
		in =  \y*360/12 + 135, looseness = 1.3] (c\y);	
	};
\foreach \x in {0,3,6,9} {
	\foreach \y in {-3,-1,1,3} {	
		\coordinate (d\x_\y) at (\x*360/12 + \y - 45 : 2.2);
		\draw[thick] (c\x) -- (d\x_\y);
		};
	\draw (\x*360/12 - 45 : 2.35) node {$4$};
	};
\foreach \x in {1,2,4,5,7,8,10,11} {
	\foreach \y in {-5,-3,-1,1,3,5} {	
		\coordinate (d\x_\y) at (\x*360/12 + \y - 45 : 2.2);
		\draw[thick] (c\x) -- (d\x_\y);
		};
	\draw (\x*360/12 - 45 : 2.35) node {$6$};
	};
\draw[dotted] (0,0) circle[radius=1];
\draw[dotted] (0,0) circle[radius=2];
\draw[color=red] (C) node[below left] {$C$};
\draw (C) node[color=red,fill=red,circle,scale=0.4] {};
\end{tikzpicture}
        \caption{Holographic state with free dangling legs}
        \label{fig:SimplePsi}
\end{subfigure}
\caption{The construction of the holographic tensor network for the cut-off Bruhat--Tits tree. In this example, we have used perfect tensors of rank eight. The construction shown here corresponds to the parameters $p=3$, $\Lambda = 2$, and~$r=7$. 
Hereafter we will suppress showing the free uncontracted/dangling legs displayed in (b).
}
\label{fig:TNexample}
\end{figure}
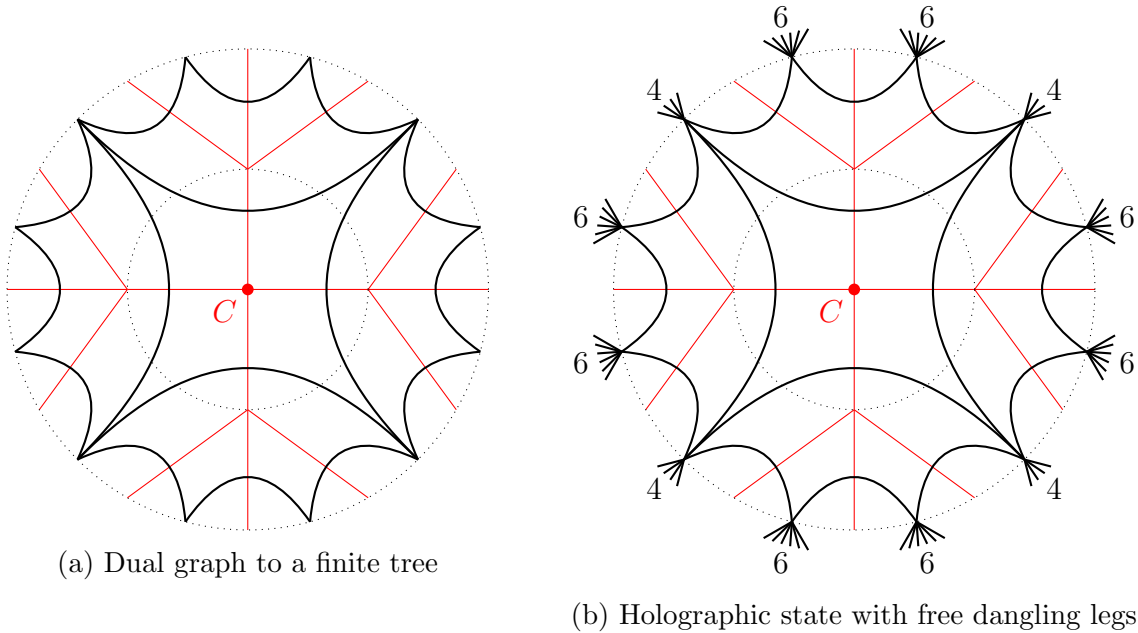

The dual graph will describe a tensor network. Each node on the dual graph will represent a rank-$(r+1)$ perfect tensor for some chosen $r$, with the bonds on the dual graph specifying how tensor indices are contracted among themselves.
In this paper we restrict ourselves  to the study of the so-called holographic states rather than holographic codes~\cite{Pastawski:2015qua}. 
Thus there are no ``bulk logical inputs'' in our setup. Interestingly, the Bruhat--Tits tree and (any choice of) its dual graph may be obtained as the asymptotic limit of the simplest holographic states considered in \cite{Pastawski:2015qua} -- where the geometry is described by a regular hyperbolic tiling using $q$-gons with $p+1$ $q$-gons incident at each vertex (which is represented by the Schl\"{a}fli symbol $\{q, p+1\}$) and the corresponding perfect tensor network with Schl\"{a}fli symbol $\{p+1, q\}$ -- in the limit $q \to \infty$. 
Viewed as such a limit, we observe that in fact all nodes of the dual graph tensor network may be interpreted as having been ``pushed to the boundary'' leaving no ``bulk nodes'' on the tensor network (see figure \ref{fig:SimplePsi} for an example with a finite tree).\footnote{Since we restrict our attention in this paper to only holographic states~\cite{Pastawski:2015qua}, the results  do not depend on this curious feature of the $p$-adic tensor network. Thus we do not comment further on the physical interpretation of this observation.} As mentioned in section \ref{REVIEW} the  perfect tensors themselves originate from quantum Reed-Solomon codes, particularly  the $[\![r,1,(r+1)/2]\!]_r$-code, where $r$ is prime, although in the following the only thing we will explicitly use is the perfectness property of the quantum code and the fact that the tensors have rank and bond dimension $r$. Depending on the chosen rank of the perfect tensor, there will be a varying number of free (uncontracted or ``dangling'') legs at each node on the dual graph. (We have suppressed such ``dangling'' legs in figure \ref{fig:planar}.) As with other tensor network models, we interpret the ``boundary wavefunction'' as a complicated entangled state in the tensor product Hilbert space of these dangling legs.

In practice, we always work with a boundary UV cutoff, so that we only consider finite graphs in the bulk. 
Thus in constructing a dual graph tensor network which describes a holographic state, in addition to the prime $p$ (which parametrizes the bulk geometry ${\cal T}_p$), we need to specify two other integral parameters: the UV ``cut-off parameter'' $\Lambda$ and the rank of the perfect tensors $(r+1)$. 
\begin{definition}
\label{def:cutoff}
The {\it cut-off parameter} $\Lambda$  is defined to be one-half the length of the longest geodesic on the radially truncated (i.e.\ cut-off) Bruhat--Tits tree.
\end{definition}
See figure \ref{fig:TNexample}.
 Eventually, we will take the number of tensors and $r$ to be large; the resulting boundary holographic states will have entanglement properties which are geometrized by this bulk network. 
 
Cutting off ordinary (real) or $p$-adic AdS at a finite radius provides an IR regulator from the bulk point of view, as the length of boundary anchored geodesics formally diverges for an infinite tree. 
In the usual picture, this IR regulator of minimal surfaces is dual to the UV cut-off of the conformal field theory; in this model it is the finiteness of the  number of boundary tensors. For each choice of $p$ and $\Lambda$, the endpoints of the cut-off (truncated) Bruhat-Tits tree form the space $\mathbb{P}^1(\mathbb{Z}/p^{\Lambda} \mathbb{Z}),$\footnote{As usual, points in $\mathbb{P}^1(\mathbb{Z}/p^{\Lambda} \mathbb{Z})$ are obtained by considering pairs in the ring $\mathbb{Z}/p^{\Lambda} \mathbb{Z}$ modulo scaling. For $\Lambda > 1$, this is not a field and there are zero divisors without multiplicative inverses. When forming the projective line, one finds there are multiple ``points at infinity'' beyond the usual inverse of $0$. Perhaps surprisingly, the set of base points and points at infinity are in one to one correspondence with the boundary of the tree cut off at finite distance.} and the tensor network is the dual graph to the tree of this space. As we remove the cutoff, the boundary approaches $\mathbb{P}^1(\mathbb{Q}_p)$, and we will show that various quantities such as the length of geodesics and the entanglement entropy will have a logarithmic UV divergences as expected in a two-dimensional quantum field theory.

Further, we impose a constraint on the rank $(r+1)$. We require $(r+1)/2 \geq 2\Lambda$. Thus in the limit $\Lambda \to \infty$, the rank of the perfect tensor also goes to infinity.
We will return to a detailed conceptual and technical analysis  of such a limit in section \ref{AFALGEBRA}.
If for a chosen vertex $v$ on the dual graph, the number of contracted legs of the tensor is denoted $v_c$, while the number of uncontracted legs is denoted $v_d$, then for cut-off $\Lambda$, all vertices on the dual graph tensor network satisfy $2\Lambda \geq v_c$. 
Since $v_c + v_d = r+1$, the requirement above implies $v_c \leq v_d$ at all vertices of the tensor network. 
Thus this condition ensures that the number of free dangling legs at any vertex on the dual graph is greater than or equal to the number of contractions at the vertex. This requirement may seem arbitrary, but plays an important role in our setup. Without this constraint the minimal cut rule obeyed by perfect tensors may lead to the cuts being made at the uncontracted dangling legs of the tensor network rather than along the contractions in the bulk of the network, which will be necessary for recovering the appropriate RT surface. 

The holographic state so constructed is a pure state which is a ground state of the dual toy model CFT.
We comment on the construction of the Hilbert space to which such states belong in section~\ref{AFALGEBRA}.
We show that the dual graph tensor network satisfies  a Ryu--Takayanagi like formula and is independent of the choice of the planar embedding.   We will prove this in general for the bipartite entanglement of ``connected'' 
and ``disconnected regions'' (which we will be  define more precisely shortly).
In this section, we restrict to discussing the results; all detailed  computations can be found in section \ref{COMPUTATION}.

\subsection{Entanglement in genus zero $p$-adic background}
\label{GENUSZERORESULTS}

\begin{figure}[t]
\begin{subfigure}[]{0.49\textwidth}
\[
\begin{tikzpicture}[scale=1.5]
\tikzstyle{vertex}=[draw,scale=0.4,fill=black,circle]
\foreach \x in {0,...,4} {
	\coordinate (a\x) at (0,\x);
	\coordinate (b\x) at (-1,\x);
	\draw (b\x) node[vertex]{} -- (a\x) node[vertex]{};
	\foreach \y in {-1,1} {
		\coordinate (c\x\y) at (-2,\x+0.3*\y);
		\draw (c\x\y)  -- (b\x);
		\foreach \z in {-1,1} {
			\coordinate (d\x\y\z) at (-2.25,\x+0.3*\y + 0.075*\z);
			\draw[very thick, dotted] (d\x\y\z) -- (c\x\y);
			};
		};
	};
\draw (-2.3,2) node[anchor=east] {$\mathbb{U}_p$};
\draw (-2.3,1) node[anchor=east] {$p \mathbb{U}_p$};
\draw (-2.3,3) node[anchor=east] {$p^{-1} \mathbb{U}_p$};
\draw (0,-0.3) node[anchor=north]{$0$ }-- (0,4.3) node[anchor=south]{$\infty$};
\end{tikzpicture}
\]
\caption{The multiplicative group $\mathbb{Q}_p^\times$.}
   \label{fig:clopenQp} 
\end{subfigure}
\begin{subfigure}[]{0.49\textwidth}
\[
\begin{tikzpicture}[scale=0.9]
\tikzstyle{vertex}=[draw,scale=0.4,fill=black,circle]
\tikzstyle{ver2}=[draw,scale=0.6,circle]
\tikzstyle{ver3}=[draw,scale=0.4,circle]
\coordinate (C) at (0,0);
\foreach \x in {0,1,2} {
	\coordinate (a\x) at (\x*360/3 + 30:1);
	\draw (a\x) -- (C);
	};
\foreach \x in {0,1,...,5} {
	\coordinate (b\x) at (\x*360/6:2);
	\pgfmathparse{floor(\x/2)}
	\draw (b\x) -- (a\pgfmathresult);
	};
\foreach \x in {0,1,...,11} {
	\coordinate (c\x) at (\x*360/12 - 15:3);
	\pgfmathparse{floor(\x/2)}
	\draw (c\x) -- (b\pgfmathresult);
	};
\foreach \x in {0,1,...,23} {
	\coordinate (d\x) at (\x*360/24 - 22.5:4);
	\pgfmathparse{floor(\x/2)}
	\draw (d\x) -- (c\pgfmathresult);
	};
\draw[dotted] (0,0) circle[radius=1];
\draw[dotted] (0,0) circle[radius=2];
\draw[dotted] (0,0) circle[radius=3];
\draw[dotted] (0,0) circle[radius=4];
\draw (C) node[below left] {$C$};
\draw (C) node[vertex] {};
\draw (c1) node[vertex] {};
\draw[very thick] (d2) -- (c1);
\draw[very thick] (d3) -- (c1);

\draw (b4) node[vertex] {};
\draw[very thick] (b4) -- (c8);
\draw[very thick] (b4) -- (c9);
\draw[very thick] (c8) -- (d16);
\draw[very thick] (c8) -- (d17);
\draw[very thick] (c9) -- (d18);
\draw[very thick] (c9) -- (d19);

\draw (d2) node[right] {$y_2$};
\draw (d3) node[right] {$y_1$};
\end{tikzpicture}
\]
\caption{Clopen balls in~$\mathbb{P}^1(\bQ_p)$.}
\label{fig:clopenP1Qp} 
\end{subfigure}
\caption{Two pictures of  the Bruhat--Tits tree for~$P^1(\bQ_2)$, emphasizing either the action of the multiplicative group $\bQ_2^\times$ or the Patterson--Sullivan measure.}
     \label{fig:clopen}
\end{figure}
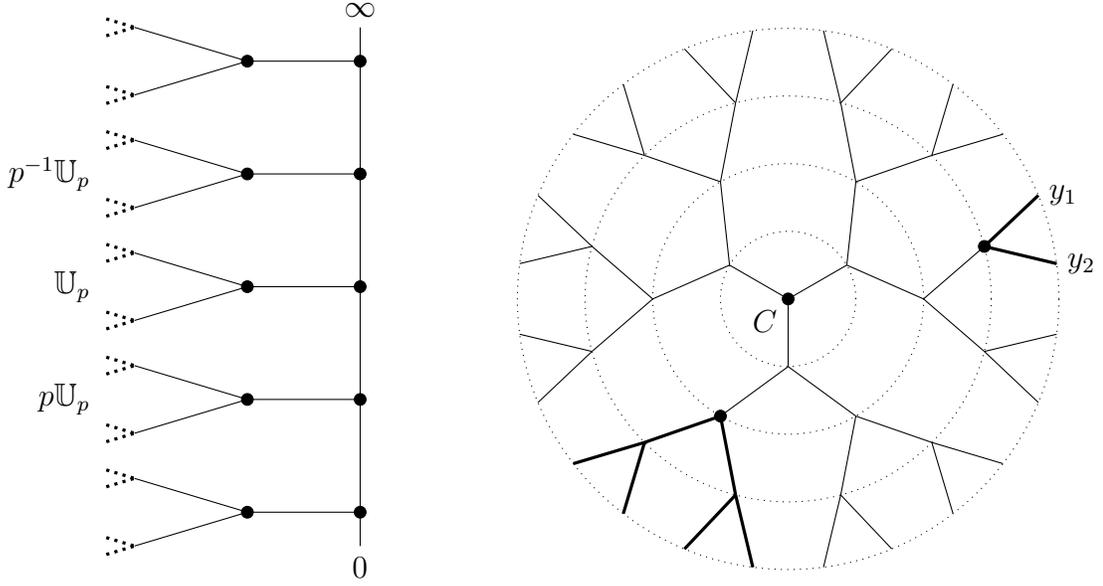

The $p$-adic numbers  have a totally disconnected topology, so that open balls (in fact, all balls are clopen, i.e.\ closed and open) are either fully disjoint or contained one inside another. 
Clopen balls are defined as ${\cal B}_v(x) \equiv \{y \in \mathbb{Q}_p: |x-y|_p \leq p^v\}$ for any integer $v$. The set of non-zero $p$-adic numbers itself can be written as the disjoint union of the clopen balls $\bQ_p \smallsetminus{\{0\}}= \bigcup_{m=-\infty}^\infty p^m \mathbb{U}_p$, where $\mathbb{U}_p = \bZ_p \smallsetminus p\bZ_p = \{x \in \bQ_p : |x|_p =1\}$. 
 Thus although $p$-adic numbers are not ordered, they admit   a partial sense of ordering with respect to the $p$-adic norm.
 This partial ordering is captured by the Bruhat--Tits tree. Using conformal transformation, set any two points on the projective line $\mathbb{P}^1(\bQ_p)$, the boundary of the Bruhat--Tits tree, to $0$ and $\infty$. Then the particular clopen balls of $\bQ_p^\times = \bQ_p \smallsetminus{\{0\}}$  of the form $p^m \mathbb{U}_p$ arrange themselves as shown in figure \ref{fig:clopenQp}.  In the Poincar\'{e} disk picture,  any clopen ball of $\mathbb{P}^1(\bQ_p)$ can be obtained by cutting the Bruhat--Tits tree along one of its edges -- the terminus of the disconnected branches of the tree represent the (mutually complimentary) clopen sets whose union is the whole of $\mathbb{P}^1(\bQ_p)$ (see figure \ref{fig:clopenP1Qp}). More general clopen sets are obtained as finite union of balls.

Now recall from standard results in holography that in a CFT$_2$, the entanglement of a connected region $A$ with its complement on a one-dimensional spatial slice admits an interpretation as the length of the minimal geodesic(s) in AdS$_3$ homologous to the region $A$, i.e.\ one minimizes over the length of the geodesic(s) such that there exists a bulk region $r$ whose boundary is the union of the minimal geodesic(s) and the boundary region $A$. In this case the boundary of $A$ is simply a pair of points (which together comprise the ``entangling surface''). This presents an obvious obstruction over the $p$-adic formulation, since $\bQ_p$ is not ordered, and it is clear one cannot define regions by specifying end points in $\bQ_p$. How, then, is entanglement to be interpreted in a $p$-adic theory over one (spatial) dimension? There are at least two physically motivated points of view:

\begin{itemize}
    \item  One possibility is to study entanglement of clopen sets on the projective line with their complementary clopen sets. However, due to ultrametricity, every point in a clopen  ball is the center of the ball. Thus the notion of ``boundary'' of the ball is ill-defined (more generally, the ``boundary'' of a clopen set is ill-defined), nor is there an analog of entangling surfaces. One can however specify the {\it smallest} clopen ball containing a given pair of points. The size of such a clopen ball is given by the Patterson-Sullivan measure~\cite{Sullivan,CornKool}, and is directly related to the (regulated) length of the boundary-anchored bulk geodesic joining the given pair of points.
    
    \item A second related possibility, but closer in spirit to the real formalism, is to motivate entanglement directly in terms of the ``entangling surface'' on the spatial $p$-adic boundary, namely, in terms of pairs of points on the boundary $\mathbb{P}^1(\bQ_p)$.    We note that the (regulated) geodesic distance between any two chosen points on the boundary is invariant under any automorphism of the Bruhat--Tits tree, and thus  is independent of the planar embedding, which  is an essential feature of the setup. The freedom in the choice of planar embedding reflects the fact that $p$-adic numbers (living on the boundary of the Bruhat--Tits tree) do not form an ordered field and thus admit all possible planar embeddings equally. 
 \end{itemize}  
 
 We  adopt the latter point of view here (although some aspects of the former point of view will also inevitably feature in our discussion of the $p$-adic results due to the inherent ultrametric nature of $p$-adic numbers), as it represents a generalization which is applicable to both the real and the $p$-adic formulations. We  comment further on this point in section~\ref{DISCUSSION}.
    
We emphasize that the geometry, which is given by the Bruhat--Tits tree has a strong nonarchimedean flavor owing to the direct connection with $p$-adic numbers.\footnote{The nonarchimedean property of $p$-adic numbers is as follows: Given two $p$-adic numbers $a,b \in \bQ_p$, such that $|a|_p < |b|_p$, then for all $n \in \mathbb{Z}$, $|na|_a < |b|_p$. We will also use the term ``ultrametricity'' in the context of the $p$-adic norm. Ultrametricity refers to the stronger form of the triangle inequality obeyed by the $p$-adic norm: Given $a,b \in \bQ_p$, $|a+b|_p \leq \sup\{|a|_p,|b|_p\}$. The nonarchimedean property follows from ultrametricity.}
The dual graph tensor network is however closer in spirit to the usual tensor networks framework over the reals. The network still encodes a maximally entangled ground state of the CFT, and the perfect tensors from which it is made provide the quantum-error correction properties. 
In this setup, we will compute entanglement in the usual way: by tracing out ``boundary regions'' of the tensor network, specified by sets of nodes on the dual graph (more precisely the collection of uncontracted tensor legs at those nodes), and then explicitly compute the reduced density matrix, and from it the von Neumann entropy. However, we will argue in the $p$-adic setting that the specification of intervals is not as fundamental as the specification of the entangling surfaces.
    
\subsubsection{Regions in the bulk and boundary of tensor networks}    
\label{sssec:intervals}

 In a static slice of the boundary theory, the standard way of specifying a connected region at the terminus of  a holographic tensor network is by picking a pair of points on the spatial slice of the two-dimensional CFT. This naturally defines a pair of complimentary intervals at the boundary of the tensor network, and provides a factorization of the Hilbert space, ${\cal H} = {\cal H}_1 \otimes {\cal H}_2$, where ${\cal H}_i$ are the Hilbert spaces associated with the individual regions. Starting with a pure state in ${\cal H}$ given by the density matrix $\rho$, the bipartite von Neumann entropy of region $1$ is computed by tracing out the states associated with ${\cal H}_2$, producing the reduced density matrix $\rho_1 = \Tr_{{\cal H}_2} \rho$. The von Neumann entropy, in this case called the entanglement entropy, is then given by $S_1 = -\Tr (\rho_1 \log \rho_1)$.
 
  However, the situation is different  in the $p$-adic setting in an important way, namely the specification of intervals on the boundary. As mentioned above, the notion of an interval with end-points is  ill-defined when the CFT lives on a (spatial) $p$-adic  slice  $\bQ_p$ (or the projective line over $\bQ_p$);
 however one can {\it still} define a corresponding pair of complimentary connected regions at the boundary  on the {\it tensor network}, separated by two boundary points as we now explain.
 
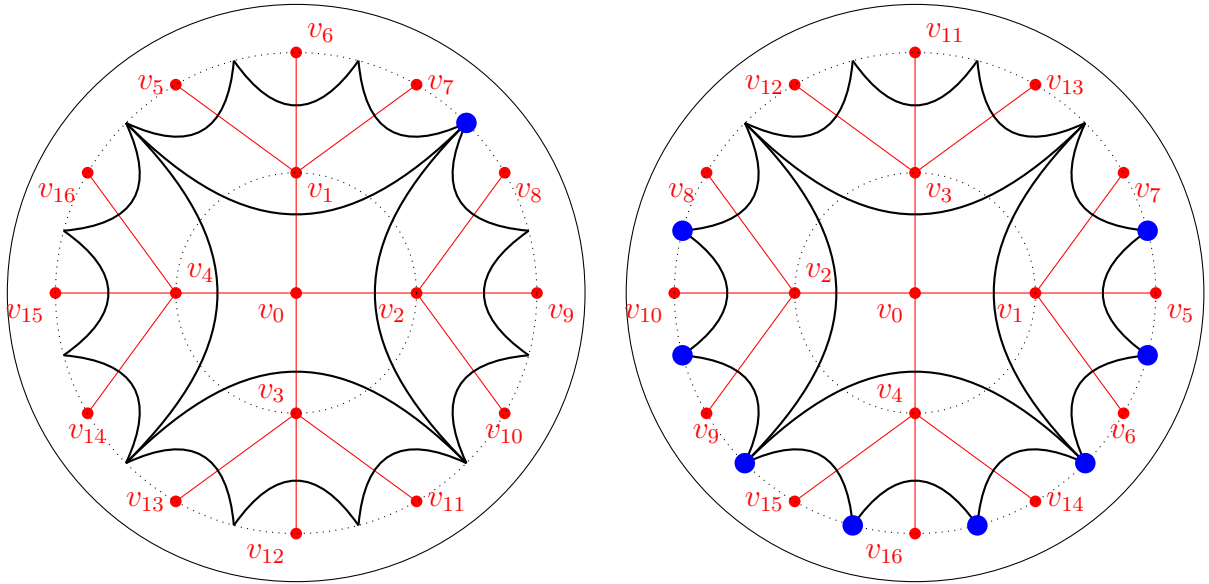
\begin{figure}[!th]
\centering
\begin{subfigure}[]{0.49\textwidth}
  \centering
\begin{tikzpicture}[scale=1.6]
\tikzstyle{vertex}=[draw,scale=0.4,fill=black,circle]
\tikzstyle{ver2}=[draw,scale=0.6,circle]
\tikzstyle{ver3}=[draw,scale=0.4,circle]
\coordinate (C) at (0,0);
\foreach \x in {0,1,2,3} {
	\coordinate (a\x) at (\x*360/4:1);
	\draw (a\x) node[color=red,fill=red,circle,scale=0.4] {};
	\draw[color=red] (a\x) -- (C);
	};
\foreach \x in {0,1,...,11} {
	\coordinate (b\x) at (\x*360/12 - 30 :2);
	\pgfmathparse{floor(\x/3)}
	\draw (b\x) node[color=red,fill=red,circle,scale=0.4] {};
	\draw[color=red] (b\x) -- (a\pgfmathresult);
	};
\foreach \x in {0,1,...,11} {
	\coordinate (c\x) at (\x*360/12 - 45 :2);
	};
\foreach \x/\y in {0/1,1/2,2/3,3/4,4/5,5/6,6/7,7/8,8/9,9/10,10/11,11/0} {
	\draw[thick] (c\x) to [out =  \x*360/12 + 115, 
		in =  \y*360/12 + 155, looseness = 1.5] (c\y);	
	};
\foreach \x/\y in {0/3,3/6,6/9,9/0} {
	\draw[thick] (c\x) to [out =  \x*360/12 + 135, 
		in =  \y*360/12 + 135, looseness = 1.3] (c\y);	
	};
\draw[dotted] (0,0) circle[radius=1];
\draw[dotted] (0,0) circle[radius=2];
\draw (0,0) circle[radius=2.4];
\draw[color=red] (C) node[below left] {$v_0$};
\draw (C) node[color=red,fill=red,circle,scale=0.4] {};
\draw[color=red] (a0) node[below left] {$v_2$};
\draw[color=red] (a1) node[below right] {$v_1$};
\draw[color=red] (a2) node[above right] {$v_4$};
\draw[color=red] (a3) node[above left] {$v_3$};
\draw[color=red] (b0) node[below] {$v_{10}$};
\draw[color=red] (b1) node[below right] {$v_9$};
\draw[color=red] (b2) node[below right] {$v_8$};
\draw[color=red] (b3) node[right] {$v_7$};
\draw[color=red] (b4) node[above right] {$v_6$};
\draw[color=red] (b5) node[left] {$v_5$};
\draw[color=red] (b6) node[below left] {$v_{16}$};
\draw[color=red] (b7) node[below left] {$v_{15}$};
\draw[color=red] (b8) node[below] {$v_{14}$};
\draw[color=red] (b9) node[left] {$v_{13}$};
\draw[color=red] (b10) node[below left] {$v_{12}$};
\draw[color=red] (b11) node[right] {$v_{11}$};

\draw (c3) node[color=blue,fill=blue,circle,scale=0.7] {};

\end{tikzpicture}
        \caption{A choice of a planar embedding for the dual graph}
        \label{fig:interval1}
\end{subfigure}
\begin{subfigure}[]{0.49\textwidth}
  \centering
\begin{tikzpicture}[scale=1.6]
\tikzstyle{vertex}=[draw,scale=0.4,fill=black,circle]
\tikzstyle{ver2}=[draw,scale=0.6,circle]
\tikzstyle{ver3}=[draw,scale=0.4,circle]
\coordinate (C) at (0,0);
\foreach \x in {0,1,2,3} {
	\coordinate (a\x) at (\x*360/4:1);
	\draw (a\x) node[color=red,fill=red,circle,scale=0.4] {};
	\draw[color=red] (a\x) -- (C);
	};
\foreach \x in {0,1,...,11} {
	\coordinate (b\x) at (\x*360/12 - 30 :2);
	\pgfmathparse{floor(\x/3)}
	\draw (b\x) node[color=red,fill=red,circle,scale=0.4] {};
	\draw[color=red] (b\x) -- (a\pgfmathresult);
	};
\foreach \x in {0,1,...,11} {
	\coordinate (c\x) at (\x*360/12 - 45 :2);
	};
\foreach \x/\y in {0/1,1/2,2/3,3/4,4/5,5/6,6/7,7/8,8/9,9/10,10/11,11/0} {
	\draw[thick] (c\x) to [out =  \x*360/12 + 115, 
		in =  \y*360/12 + 155, looseness = 1.5] (c\y);	
	};
\foreach \x/\y in {0/3,3/6,6/9,9/0} {
	\draw[thick] (c\x) to [out =  \x*360/12 + 135, 
		in =  \y*360/12 + 135, looseness = 1.3] (c\y);	
	};
\draw[dotted] (0,0) circle[radius=1];
\draw[dotted] (0,0) circle[radius=2];
\draw (0,0) circle[radius=2.4];
\draw[color=red] (C) node[below left] {$v_0$};
\draw (C) node[color=red,fill=red,circle,scale=0.4] {};
\draw[color=red] (a0) node[below left] {$v_1$};
\draw[color=red] (a1) node[below right] {$v_3$};
\draw[color=red] (a2) node[above right] {$v_2$};
\draw[color=red] (a3) node[above left] {$v_4$};
\draw[color=red] (b0) node[below] {$v_{6}$};
\draw[color=red] (b1) node[below right] {$v_5$};
\draw[color=red] (b2) node[below right] {$v_7$};
\draw[color=red] (b3) node[right] {$v_{13}$};
\draw[color=red] (b4) node[above right] {$v_{11}$};
\draw[color=red] (b5) node[left] {$v_{12}$};
\draw[color=red] (b6) node[below left] {$v_{8}$};
\draw[color=red] (b7) node[below left] {$v_{10}$};
\draw[color=red] (b8) node[below] {$v_{9}$};
\draw[color=red] (b9) node[left] {$v_{15}$};
\draw[color=red] (b10) node[below left] {$v_{16}$};
\draw[color=red] (b11) node[right] {$v_{14}$};

\draw (c2) node[color=blue,fill=blue,circle,scale=0.7] {};
\draw (c1) node[color=blue,fill=blue,circle,scale=0.7] {};
\draw (c7) node[color=blue,fill=blue,circle,scale=0.7] {};
\draw (c8) node[color=blue,fill=blue,circle,scale=0.7] {};
\draw (c9) node[color=blue,fill=blue,circle,scale=0.7] {};
\draw (c10) node[color=blue,fill=blue,circle,scale=0.7] {};
\draw (c11) node[color=blue,fill=blue,circle,scale=0.7] {};
\draw (c0) node[color=blue,fill=blue,circle,scale=0.7] {};

\end{tikzpicture}
        \caption{A different planar embedding for the dual graph}
		\label{fig:interval2}
\end{subfigure}
\caption{The region at the boundary of the dual graph ``in between'' boundary points $v_8$ and $v_7$ for two choices of planar embedding for the dual graph, depicted in blue.}
\label{fig:interval}
\end{figure}

 In our setup, the ``connected region'' of interest on the tensor network will be specified by a set of nodes on the tensor network which lie ``in between'' the given boundary points $x$ and $y$, which themselves lie at the terminus of the cutoff Bruhat--Tits tree. 
 We will explain the terminology ``in between'' shortly, but essentially it corresponds to selecting a set of vertices at the boundary of the tensor network in between the chosen end points, in a given planar embedding. 
 The exact region to be traced out will depend on the choice of planar embedding (i.e.\ the choice of the dual graph). See figure \ref{fig:interval} for an example. The ambiguity in picking a region or its complement is fixed by assigning an orientation, such as an anti-clockwise orientation.
 We stress that we are not assuming any ordering of the $p$-adic numbers. Once a planar embedding is chosen, the region ``in between'' $x$ and $y$ is $\PGL(2,\bQ_p)$ ``covariant'', which follows from the transformation properties of the dual graph explained earlier.\footnote{By covariance, we mean that the region is always given by the set of nodes on the tensor network ``in between'' the given boundary points. The boundary points will in general transform under $\PGL(2,\bQ_p)$ to a new set of points, and accordingly, the region will transform to one between the transformed pair of points.} The final result for the von Neumann entropy for the connected region is independent of this choice of the planar embedding. 

 Let us make this more precise.
\begin{definition}
\label{def:shortestbond}
 The {\it shortest bonds} on the tensor network comprise the subset of bonds (contractions) on the tensor network which are in bijective correspondence with the set of edges on the Bruhat--Tits tree situated at the cutoff boundary. 
\end{definition}
\begin{definition}
\label{def:connectedregion}
A {\it connected region} on the tensor network is defined to be a set of vertices on the tensor network which are ``path-connected'' to each other (by which we mean one can jump, solely via the shortest bonds on the tensor network, from any vertex in the set to any other without landing on a vertex which is not in the set). 
We define a {\it disconnected region} to be a region which is not connected.

We will also interchangeably mean the (connected or disconnected) region to stand for the uncontracted tensor legs situated at the vertices in the specified region.
\end{definition} 
As noted earlier, we specify the connected region (on the dual tensor network) by specifying two boundary points on the Bruhat--Tits tree and considering all vertices on the dual graph which lie ``in between'' the boundary points (after making a choice of orientation), which we now define. 
 \begin{figure}[thb]
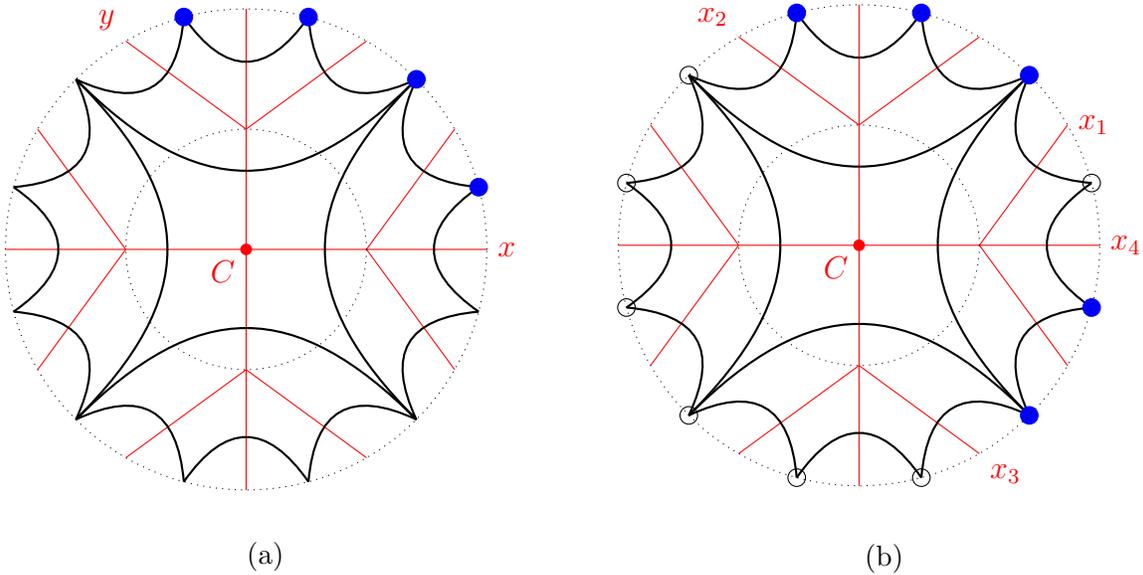

\begin{subfigure}[]{0.49\textwidth}
 \centering
\[\musepic{\oneintpic} \]
\caption{}\label{fig:connectedA}
\end{subfigure}
\begin{subfigure}[]{0.49\textwidth}
 \centering
\[ \musepic{\multiintpic}\]
 \caption{}\label{fig:connectedB}
\end{subfigure}
\caption{(a) The set of vertices on the dual graph marked in blue form a ``connected region''. (b) The set of blue vertices form a ``disconnected region'' which can be specified by a set of four boundary points. Note that for the given set of boundary points and a chosen planar embedding, another choice of disconnected regions, shown in black circles, is possible.}
\label{fig:connected}
\end{figure}
\begin{definition}
\label{def:inbetween}
 Given two points $x$ and $y$ in $\partial T_{p}$ and a choice of a planar embedding, we define the connected region on the tensor network {\it in between} $x$ and $y$ (up to a choice of orientation) as the set of nodes on the tensor network path-connected to each other via the ``shortest bonds'' starting at the ``start bond'' and ending at the ``end bond'', without backtracking. The ``start'' and ``end'' bonds correspond to bonds on the tensor network dual to the cutoff edges on the Bruhat--Tits tree ending at $x$ and $y$.
\end{definition}
 
 For instance, in figure \ref{fig:connectedA}, the chosen connected region lies in between the boundary points $x$ and $y$. 
 By contrast, in figure \ref{fig:connectedB} we show an example of a ``disconnected region'', associated to a given set of {\it four} boundary points. 
The von Neumann entropy of the chosen region in such a case will depend on the von Neumann entropy of its disjoint connected parts and the mutual information shared between them. We will discuss the case of disconnected regions in more detail in sections \ref{GENUS0DOUBLE}-\ref{ssec:Inequalities}.
 
 Looking ahead, we make two more definitions.
 \begin{definition}
 \label{def:bulkregion}
 Given a planar embedding (i.e.\ a choice of a dual graph tensor network) and a geodesic $\gamma$ that separates the tensor network into two connected components, called {\it bulk regions} $P$ and $Q$, the {\it boundary} of each bulk region is a collection of uncontracted tensor legs, which includes the tensor legs which were originally contracted across $\gamma$.
 \end{definition}
 \begin{definition}
 \label{def:homologous}
  Given a boundary interval $A$ specified by a set of nodes on the tensor network (where more precisely, by $A$ we mean the collection of uncontracted tensor legs at the specified set of nodes), we say a geodesic $\gamma$ is {\it homologous} to $A$ if there exists a bulk region on the tensor network, part of whose boundary is given by (the full set of uncontracted tensor legs at) $A$ while the remaining uncontracted legs forming the boundary of the bulk region are in bijective correspondence with the edges of the geodesic $\gamma$. 
 \end{definition}
  This definition applies to both connected and disconnected regions $A$, and has an obvious extension to the setting with multiple geodesics. This notion of homologous geodesics is a natural adaptation of the notion in continuum case to tensor networks, and makes a natural appearance in our tensor network setup.

\subsubsection{Results}    
\label{sssec:results}

 We now summarize the entanglement entropy results for the tensor network described above; the detailed calculations can be found in sections \ref{SIMPLESTATES}-\ref{ssec:Inequalities}.
 In the case of a connected region $A$, parametrized by the points $x, y \in \bQ_p$ as described previously, we prove in section~\ref{GENUS0SINGLE} that the vacuum von Neumann entropy is given by
\eqn{RTgenus0}{
S(x,y) = (2 \log r)\log_p \left| {x-y \over \epsilon} \right|_p,
}
where $\epsilon = p^{\Lambda} \in \bQ_p$ is the UV cutoff, and the tensor network is built out of perfect tensors of rank $r+1$. The overall factor of $\log r$ can absorbed  by taking a logarithm in base $r$ when computing von Neumann entropy via $S = -\Tr \rho \log \rho$.
We use the notation $S(x,y)$ to emphasize that the entropy is independent of the choice of planar embedding and only a function of $p$-adic coordinates $x$ and $y$. 
The $p$-adic norm $|\cdot|_p$ appears naturally in this tensor network setup, and the expression \eno{RTgenus0} makes sense in the limit $|\epsilon|_p \to 0$ when the finite cutoff tree approaches the infinite Bruhat--Tits tree.\footnote{For a finite tree bulk geometry, the expression \eno{RTgenus0} continues to make sense if one views the boundary points $x,y$, which are now elements of the ring $\bZ/p^\Lambda \bZ$, as $p$-adic numbers with a truncated power series expansion.} 
The quantity $|x-y|_p$ has a natural intepretation as the measure of the smallest clopen set in $\bQ_p$ containing both $x$ and $y$ and is the analog of the ``length of an interval'' over the reals.
The result \eno{RTgenus0} is obtained by explicitly computing the reduced density matrix and diagonalizing it.\footnote{Some issues associated with infinite dimensional density matrices in the limit when the cutoff $\Lambda \to \infty$ are discussed in section \ref{AFALGEBRA}.} 
The entanglement entropy obtained is in direct analogy with the corresponding classic result for the entanglement of a connected interval (of size $|x-y|$ over the real line) with its complement, on a static spatial slice of a massless CFT$_2$ with UV cutoff $\epsilon$~\cite{Holzhey,Calabrese:2004eu}.

The tensor network also affords a bulk interpretation for \eno{RTgenus0}. We show in section \ref{GENUS0SINGLE} that the von Neumann entropy $S(x,y)$ is precisely equal to the length of the minimal geodesic in the bulk homologous to the connected region $A$, consistent with the Ryu--Takayanagi formula. On a static slice of CFT$_2$, the minimal geodesic joins the end-points of $A$, given by the ``entangling surfaces'' $x, y \in \bQ_p$. Indeed, we show that
\eqn{RTgenus0bulk}{
S(x,y) = {{\rm length}(\gamma_{xy}) \over \ell} \log r\,,
}
where $\gamma_{xy}$ is the minimal geodesic on the cutoff Bruhat--Tits tree joining boundary points $x$ and $y$, and $\ell$, the length of each edge on the tree, is proportional to the AdS radius. 
As remarked earlier, this result is consistent with the minimum-cut rule obeyed by networks of perfect tensors in the case of bipartite entropy of connected regions~\cite{Pastawski:2015qua}, although we do not assume it in our setup. Essentially, the dual graph tensor network proposed in this paper has the property that the minimal number of tensor contractions on the tensor network which must be ``cut''  across to completely separate  the connected region from the rest of the network, precisely equals the length of the minimal geodesic on the Bruhat--Tits tree. Indeed, we show in section \ref{COMPUTATION} that these cuts  trace out precisely the path of the minimal geodesic joining the entangling surfaces.\footnote{In fact the geodesic is homologous to the specified region. This condition is especially important in black hole geometries where there can exist shorter geodesics not homologous to the given region. The tensor network always picks the one which is homologous. This will discussed in more detail in the next section.}

Moreover, we have the bulk formula
\eqn{lengthxy}{
{\rm length} (\gamma_{xy})/\ell = \delta(C\to x,C\to x)+\delta(C\to y,C\to y) - 2\delta(C\to x,C\to y)\,,
}
where $C$ is an arbitrary node on the Bruhat--Tits tree (or its boundary), and $\delta(\cdot,\cdot)$ is the signed-overlap between the two directed paths in its argument~\cite{Zab}.\footnote{For example, the signed-overlap of a path with itself is simply the length of the path, while the signed-overlap with the same path but with opposite orientation is {\it negative} the total length of the path. The signed overlap vanishes for paths which do not share any edges.} Equations \eno{RTgenus0bulk}-\eno{lengthxy} are applicable for connected regions in the genus 1 geometry as well, which is discussed in the next section. 

We also show in section \ref{COMPUTATION} that the result \eno{RTgenus0bulk}-\eno{lengthxy} continues to apply to a (massless) CFT defined over a circle, more precisely the projective line $\mathbb{P}^1(\mathbb{Q}_p)$. Let's first recall the result in the real case. Over the reals, the entropy formula
\eqn{Sline}{
S(x,y) \propto \log {L \over \epsilon}\,,
}
where $L=|x-y|$ is the size of the connected interval $A=[x,y]$ and $\epsilon$ the UV cutoff in the CFT, gets replaced by~\cite{Calabrese:2004eu}
\eqn{Scircle}{
S(x,y) \propto \log \left( {2 R \over \epsilon} \sin{ L \over 2 R}\right),
}
where $R$, the IR cutoff  parametrizing the total size of the spatial boundary, is the radius of the Poincar\'{e} disk, and $L=R |\arg x-\arg y| $ is the arc length of the interval with $x,y \in \mathbb{P}^1(\mathbb{R})=S^1$.
In the limit $L \ll R$, \eno{Scircle} reduces to \eno{Sline}. 

 When the spatial slice at the boundary is the projective line over $\bQ_p$, the results in the $p$-adic tensor network setup continue to be analogous to the real case.
The measure of the smallest clopen set in $\bQ_p$ containing $x, y \in \bQ_p$, given by $|x-y|_p$ in \eno{RTgenus0} gets replaced by the Patterson-Sullivan measure of the  smallest clopen set in $\mathbb{P}^1(\bQ_p)$ containing $x,y \in \mathbb{P}^1(\bQ_p)$~\cite{Sullivan}, so we have
\eqn{RTgenus0circle}{
S(x,y) = (2 \log r)\log_p  {|{\mathfrak B}(x,y)|_{\rm PS} \over |\epsilon|_p}  \,.
}
Explicitly, choosing $C$ to be the radial center of the cutoff Bruhat--Tits tree in the Poincar\'{e} disk picture (recall figure \ref{fig:clopenP1Qp}), the Patterson-Sullivan measure is given by $|{\mathfrak B}(x,y)|_{\rm PS} \equiv p^{-d(C,{\rm anc}(x,y)}$, where $d(\cdot,\cdot)$ is the graph distance between the nodes in its argument, and ${\rm anc}(x,y)$ is the unique vertex on the Bruhat--Tits tree at which the geodesics from $x,y$ and $C$ simultaneously intersect. 

When $d(C,{\rm anc}(x,y)) \ll d(C,x) = d(C,y)$, the Patterson-Sullivan measure is approximated by $|x-y|_p$, thus recovering the formula \eno{RTgenus0} from \eno{RTgenus0circle}.\footnote{Here we are being loose about the distinction between $x \in \mathbb{P}^1(\bQ_p)$ and $x \in \bQ_p$. 
The precise statement is that when the radial center $C$ is sent to a boundary point, say $\infty \in \mathbb{P}^1(\bQ_p)$, the remaining boundary of the Bruhat--Tits tree is described precisely by the $p$-adic numbers $\bQ_p$ and in this case, the Patterson-Sullivan measure on $\mathbb{P}^1(\bQ_p)$, $|{\mathfrak B}(x,y)|_{\rm PS} = p^{-d(C,{\rm anc}(x,y)}$ reduces exactly to the Haar measure on $\bQ_p$, given by the $p$-adic norm $|x-y|_p$.}
Moreover, the Patterson-Sullivan measure rises, attains a maxima, and then falls as the boundary points $x$ and $y$ are moved away from each other, similar to the sine function in \eno{Scircle} which rises, reaches a maxima, and then falls as $L$ is increased. 
This can be seen from the explicit form of the Patterson-Sullivan measure quoted above, by fixing one of the boundary points while moving the other ``away'' from it.\footnote{The Patterson-Sullivan measure rises, attains a maximum, and then eventually falls in discrete steps in contrast to the smoothly varying sine function in \eno{Scircle}, as one fixes one of the nodes but moves the other ``away'' in the sense of increasing the path length along the ``shortest bonds'' between the two nodes for a chosen orientation.} 

We remark that as is clear from \eno{RTgenus0} and \eno{RTgenus0circle}, the measure of the clopen set is more fundamental than the number of boundary vertices falling within a connected region in a chosen planar embedding.  Fixing a planar embedding, a given pair of end-points $x,y$ may simply contain ``in between'' themselves a single vertex on the tensor network but may still correspond to a clopen set with a larger measure than that of another pair of points which carry ``in between'' themselves a larger number of vertices. This essentially is due to the inherent ultrametric nature of the setup. 

In section \ref{GENUS0DOUBLE} we extend the results for a single connected interval to the case of a disconnected interval. 
Instead of two boundary points specifying a single connected interval, we now have four boundary points parametrizing the disconnected case, with the full interval written as a union of its two connected components. 
We show that the entropy is independent of the planar embedding and obeys an RT-like formula exactly (see, for instance, the discussion around \eno{Sdisconnect}). 
We also verify the non-negativity of mutual information as well as the Araki--Lieb inequality, and in fact in \eno{MIbdyGen} write down an explicit expression for mutual information in terms of the conformal cross-ratio constructed from the boundary points. 
We then provide a dual bulk interpretation of mutual information in terms of the overlap of the minimal geodesics of the individual components. 
Finally, in section \ref{ssec:Inequalities} we give a simple holographic proof of strong subadditivity in the $p$-adic setting, demonstrating its relation to ultrametricity, and a proof for monogamy of mutual information. In fact we find that mutual information is extensive, that is, the tripartite entropy is identically zero.
We refer to these sections for more details.

\section{$p$-adic BTZ Black Hole}
\label{GENUSONE}

In this section, we continue the study of the proposed tensor network in bulk geometries which can be considered the $p$-adic analog of black hole or thermal states. We will first summarize the construction of these $p$-adic geometries $(\mathbb{Q}_p)$ in analogy with the complex case of Euclidean AdS$_3$ $(\mathbb{C})$ or a time slice of the Lorentzian counterpart. This \emph{uniformization} procedure is an algebraic way to obtain black hole geometries from empty AdS, and there is a natural way to apply this construction to the perfect tensor network of the previous section. Instead of a pure state on the boundary in this toy model, one finds degrees of freedom behind the `horizon' which must be traced out. The result is a thermal density matrix, and following~\cite{Maldacena:1998bw} we interpret this as the thermal state at the conformal boundary with entropy analogous to the entropy first observed by Bekenstein and Hawking~\cite{Bekenstein:1973ur, Hawking:1974sw}. In this discrete $p$-adic model, using computational tools described in the next section we find precise agreement between the perimeter of the black hole horizon and the thermal entropy of the boundary density matrix. We postpone the details of this calculation until section~\ref{COMPUTATION}, and here we will focus on the setup and results.

One can further study entanglement entropy in these genus $1$ backgrounds by tracing out regions of qubits at the boundary. The resulting entanglement entropy has a dual interpretation in the bulk as the lengths of minimal geodesics homologous to the boundary intervals in the black hole background, the analog of the Ryu--Takayanagi formula in this geometry~\cite{Ryu:2006bv, Ryu:2006ef}. In our model, the boundary anchored geodesics wrap non-trivially around the horizon to minimize the total length, and one might have expected this minimization property of tensor networks from the minimal cut rule of~\cite{Pastawski:2015qua}; however we emphasize that the genus 1 tensor network is fundamentally different from the one considered in~\cite{Pastawski:2015qua} and is obtained instead as a quotient of the genus 0 construction. We have verified this agreement between the boundary entropy and the bulk geodesic length by direct computation, and conjecture that this gives a holographic derivation of entanglement entropy in $p$-adic AdS/CFT in thermal backgrounds. 

\subsection{Genus $1$ curves and Schottky uniformization}\label{ssec:Schottky}
 In analogy with the complex case of $\text{AdS}_3/\text{CFT}_2$ where the  Ba\~nados, Teitelboim, and Zanelli~\cite{Banados:1992wn} (BTZ) black hole boundary in Euclidean signature is a $T^2$, the boundary picture of these $p$-adic BTZ black holes can be understood as genus $g=1$ curves over nonarchimedean fields. These curves were originally described in the classical work of Tate for $g=1$ and Mumford for $g>1$~\cite{Mum}, and while we focus on the genus $1$ case we will often refer to the boundary curve as a Tate-Mumford curve. The applications of the boundary curve/bulk graph to $p$-adic AdS/CFT are described in~\cite{Heydeman:2016ldy,Manin:2002hn}. The bulk spaces are then given by quotients of the $p$-adic Bruhat--Tits tree (the analog of empty AdS) by the action by isometries of a discrete group. For a general introduction to Mumford curves and their associated bulk spaces see \cite{Man,GevdP}.

In the complex case of a torus boundary, a familiar realization is a uniformization of the elliptic curve $E(\mathbb{C})$ by the complex plane $\mathbb{C}$. If the modular parameter is $\tau$ with $\text{Im}(\tau) >0$, one may construct a $\mathbb{C}$ lattice $\Lambda = \mathbb{Z} \oplus \tau \mathbb{Z}$ and describe the curve as the quotient
\begin{equation}
T^2 \simeq E(\mathbb{C}) \simeq \mathbb{C}/\Lambda \,.
\end{equation}
This is the familiar procedure of identifying opposite sides of a parallelogram. However, a direct $p$-adic analog using a lattice turns out to not be possible. An alternative approach due to Tate is essentially to consider the exponentiated map. Defining the standard Fourier parameter $q = e^{2 \pi i \tau}$ which satisfies $|q| < 1$ since $\text{Im}(\tau) >0$, we may instead consider a quotient of the multiplicative group $\mathbb{C}^{\times}$:
\begin{equation}
\label{complexuniformization}
T^2 \simeq E(\mathbb{C}) \simeq \mathbb{C}^{\times}/q^{\mathbb{Z}} \, ,
\end{equation}
where $q^{\mathbb{Z}}$ for the integers $\mathbb{Z}$ form a discrete abelian group. This construction of the elliptic curve is an example of complex \emph{Schottky uniformization} of genus $1$ curves, which can be generalized to $g>1$ curves and has a natural $p$-adic analog. Schottky uniformization is the uniformization of an elliptic curve by quotienting the projective line by a chosen discrete subgroup of its M\"{o}bius transformations. More precisely, we must first remove a certain limit set of the projective line where the Schottky group acts poorly; in the present $g=1$ case these can be chosen to be the two points $\{0, \infty \}$, which explains the $\mathbb{C}^{\times}$ used above. At higher genus the limit set is much more complicated, see section~\cite{Heydeman:2016ldy} for details.

At genus $1$ we can be even more explicit. Recall in the complex case that M\"{o}bius transformations form the group $G = \PSL(2,\bC)$ which acts on $\bP^1(\bC)$ with complex coordinate $z$ by fractional linear transformations,
\begin{equation}
 g \in G =\begin{pmatrix} a & b \\ c & d \end{pmatrix}: \, \, \, z \mapsto \frac{az+b}{cz+d} \, .
 \end{equation}
Removing the aforementioned limit points, we may now pick a discrete subgroup $\Gamma \in \PSL(2,\bC)$ with which to perform the quotient. For genus $g=1$ with the abelian group $\Gamma=q^\bZ$, a generator $\gamma$ acts on the domain by 
\begin{equation}
\label{complexschottky}  \gamma \in \Gamma = \begin{pmatrix} q^{1/2} & 0 \\ 0 & q^{-1/2} \end{pmatrix}: \, \, \, z \mapsto qz \, ,
\end{equation}
and we may obtain a torus by dividing by this action; explicitly points in the plane are identified under this scaling. 

\begin{figure}[th]
\includegraphics[width=11cm]{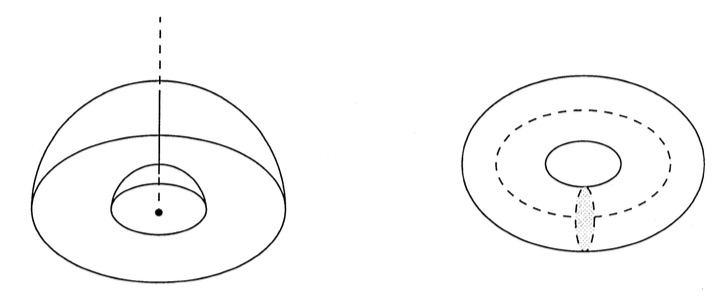} 
\centering
\caption{Left: Fundamental domain and quotient for the Euclidean BTZ black hole~\cite{Manin:2002hn}. The two circles on the plane are identified under $z \rightarrow qz$, while the two domes in the bulk are identified. Right: The resulting quotient gives a torus boundary and a solid torus with AdS metric in the bulk, the BTZ black hole in Euclidean signature.
\label{EuclideanBTZFig}}
\end{figure}

In Euclidean signature, this action uplifts to the 3-dimensional hyperbolic upper half plane which has the complex projective line as its boundary. Here we view $\PSL(2,\bC)$ as the group of isometries of Euclidean AdS$_3$ with the scaling extending into the bulk direction. In the bulk, this Schottky generator $q$ acts by scaling of geodesic surfaces, and in particular it acts on the unique geodesic connecting $\{0,\infty \}$ as translations by $\log q$. Taking a quotient of the bulk by this action gives the Euclidean BTZ black hole, presented as a solid torus with the desired elliptic curve $E(\mathbb{C})$ at the conformal boundary. This is illustrated in figure \ref{EuclideanBTZFig}. (In fact, one may obtain a family of black hole and thermal AdS solutions by acting with modular transformations~\cite{Maldacena:1998bw, Maloney:2007ud}.)

It is possible to repeat the above uniformization in Lorentzian AdS$_3$. In this case, the isometry group is $SO(2,2)$, the connected part of which is isomorphic to $(\SL(2,\mathbb{R}) \times \SL(2,\mathbb{R}))/\mathbb{Z}_2$ (in fact, there is a subtlety in the choice of covering group \cite{Witten:2007kt}, which will not concern us here.) As before, quotienting with discrete abelian subgroups can be used to find BTZ black hole spacetimes in Lorentzian signature, possibly with angular momentum. In our case in analogy with genus $0$, we would like to interpret the $p$-adic tensor network as describing a time slice of a static black hole. To this end, one may pick a discrete subgroup of the diagonal $\PSL(2,\mathbb{R})$ acting on the $t=0$ slice, which is now a copy of the upper half plane $\mathbb{H}^2 \sim \bR \times \bR^+$. For $q\in \mathbb{R}$, the Schottky group is again $\Gamma=q^\bZ$ with matrix form (\ref{complexschottky}), acting by fractional linear transformation, but now on $\mathbb{H}^2$ (rather than the complex boundary coordinate as in the Euclidean case.) The result of this quotient is the $t=0$ slice of the non-rotating BTZ black hole in Lorentzian signature. In the bulk, this is two-sided and has one non-contractible cycle-- the black hole horizon. 

In the conventions of \cite{Carlip:1995qv} with unit cosmological constant, this black hole with mass $M = r_+$ and angular momentum $J=0$ is generated by the Schottky element
\begin{equation}
\gamma = \begin{pmatrix} e^{\pi r_+} & 0 \\ 0 & e^{-\pi r_+} \end{pmatrix}.
\end{equation}
From this one my find the horizon perimeter and compute the Bekenstein Hawking entropy. In these units it is simply $S_{BH} = 4 \pi r_+ = 2 \log q$ in terms of the $q$ parameter.

When moving from $\bR$ or $\bC$ to $\bQ_p$, the topology and geometry of both the bulk and boundary change dramatically. The goal of the present work is not to define or classify the possible choices in this $p$-adic setting (which might involve even more exotic structures such as buildings), but rather to provide a discrete toy model of holography in which aspects of the bulk can be computed exactly. For this reason, in discussing genus $1$ black holes we will assume the simplest interpretation of a static black hole where we work on a spatial slice; this is the situation where our formulas have qualitative similarity to real AdS/CFT in $3$-dimensions. There may be other interpretations of our results, and we will remain agnostic about more general signatures and situations such as rotating black holes.

We now proceed to describe the uniformization of the Tate-Mumford curve by the $p$-adic multiplicative group $\Q_p^\times$. This describes the boundary geometry for the black hole, and there is a natural extension to the bulk Bruhat-Tits tree. As explained above, we will not rely on a lattice, but rather identification of points under a Schottky generator; now a discrete subgroup of the $p$-adic conformal group. Mathematically, we will mimic the above construction with the substitutions $\PSL(2, \bR) \rightarrow \PGL(2, \bQ_p)$ along with the discrete Schottky generator $\Gamma \subset \PGL(2,\bQ_p)$.
Physically as we have explained, we interpret the result in the bulk as a time slice of a static black hole. Asymptotically, this geometry looks like the Bruhat-Tits tree, but the center contains a non-contractible cycle of integer length. In a different context, this uniformization was used in the context of open $p$-adic string theory to compute multi-loop amplitudes in~\cite{Chekhov:1989bg}, which viewed the Bruhat-Tits tree and its higher genus generalizations as worldsheets.

We seek a $p$-adic version of equation~(\ref{complexuniformization}), which is the desired Tate-Mumford curve at the boundary. Beginning with the usual boundary $z \in \bP^1(\bQ_p)$, the M\"{o}bius transformations now form the group $\PSL(2, \bQ_p)$ acting on $z$ by fractional linear transformations, though below we will use $G = \PGL(2, \bQ_p)$ which is the isometry group of the tree.\footnote{We have passed from the special linear group to the general linear group because this more properly accounts for the isometries of the Bruhat-Tits tree. One may see that the $\SL(2)$ matrix in (\ref{complexschottky}) requires us to take a square root in $\bC$; this is in general not possible for $q \in \bQ_p$ without extensions. Among other things, restricting to $\SL(2)$ thus excludes translations on the tree by non-square elements, while a $\GL(2)$ matrix allows one to act with isometries of this type.} We choose the abelian Schottky group $\Gamma = q^{\bZ} \in \PGL(2, \bQ_p)$, with $q \in \bQ_p^{\times}$, $|q|_p <1$. The $\PGL$ matrix which generates the Schottky group can be chosen to be
\begin{equation}
\label{padicschottky}
\gamma \in \Gamma =  \begin{pmatrix} q & 0 \\ 0 & 1 \end{pmatrix}: \, \, \, z \mapsto qz \, .
\end{equation}
Performing the quotient, which identifies $p$-adic numbers related by this scaling, we obtain a curve of genus $1$ at the conformal boundary, which is the elliptic curve uniformized by the $p$-adic multiplicative group:
\begin{equation}
\label{padicuniformization}
E(\mathbb{Q}_p) \simeq \mathbb{Q}_p^{\times}/q^{\mathbb{Z}} \, .
\end{equation}
This in principle completes the description of the boundary curve for the black hole geometry which might be interpreted as a thermal state of a conformal field theory at a fixed time slice. An important technical caveat is that not all elliptic curves over $\bQ_p$ can be uniformized in this way, only those with split multiplicative reduction. However, it is precisely these Tate-Mumford curves which have a natural extension to the Bruhat-Tits tree ${\cal T}_p$, so we will only consider these in this work. 

\begin{figure}[t]
\centering
\begin{tikzpicture}[scale=1.3]
\tikzstyle{vertex}=[draw,scale=0.4,color=red,fill=red,circle]
\tikzstyle{ver2}=[draw,scale=0.4,color=red,fill=red,circle]
\tikzstyle{ver3}=[];

\draw (0*60:1) node[vertex] (a0) {};
\draw (1*60:1) node[vertex] (a1) {};
\draw (2*60:1) node[vertex] (a2) {};
\draw (3*60:1) node[vertex] (a3) {};
\draw (4*60:1) node[vertex] (a4) {};
\draw (5*60:1) node[vertex] (a5) {};

\draw[thick,color=red] (a0) -- (a1) -- (a2) -- (a3) -- (a4) -- (a5) -- (a0);

\draw (a0) + (0*60 + 30:1) node[ver2] (b0) {};
\draw (a0) + (0*60 - 30:1) node[ver2] (c0) {};
\draw (a1) + (1*60 + 30:1) node[ver2] (b1) {};
\draw (a1) + (1*60 - 30:1) node[ver2] (c1) {};
\draw (a2) + (2*60 + 30:1) node[ver2] (b2) {};
\draw (a2) + (2*60 - 30:1) node[ver2] (c2) {};
\draw (a3) + (3*60 + 30:1) node[ver2] (b3) {};
\draw (a3) + (3*60 - 30:1) node[ver2] (c3) {};
\draw (a4) + (4*60 + 30:1) node[ver2] (b4) {};
\draw (a4) + (4*60 - 30:1) node[ver2] (c4) {};
\draw (a5) + (5*60 + 30:1) node[ver2] (b5) {};
\draw (a5) + (5*60 - 30:1) node[ver2] (c5) {};

\foreach \x in {-1,0,1} {
	\foreach \n in {0, 1, 2, 3, 4, 5} {
		\draw (b\n) + (\n*60 + 20 + \x*20:1) node[ver3] (d\n\x) {};
		\draw (c\n) + (\n*60 - 20 + \x*20:1) node[ver3] (e\n\x) {};
		\draw (b\n) + (\n*60 + 20 + \x*20:1.5) node[ver3] (D\n\x) {};
		\draw (c\n) + (\n*60 - 20 + \x*20:1.5) node[ver3] (E\n\x) {};
		\draw[thick,color=red] (d\n\x) -- (b\n);
		\draw[thick,color=red] (e\n\x) -- (c\n);
		\draw[very thick, color=red, loosely dotted] (d\n\x) -- (D\n\x) ;
		\draw[very thick, color=red, loosely dotted] (e\n\x) -- (E\n\x) ;
	};
};

\draw[thick,color=red] (b0) -- (a0) -- (c0);
\draw[thick,color=red] (b1) -- (a1) -- (c1);
\draw[thick,color=red] (b2) -- (a2) -- (c2);
\draw[thick,color=red] (b3) -- (a3) -- (c3);
\draw[thick,color=red] (b4) -- (a4) -- (c4);
\draw[thick,color=red] (b5) -- (a5) -- (c5);
\draw[very thick, loosely dotted] (0,0) circle (3.5);
\end{tikzpicture}
\caption{The $p$-adic BTZ black hole, obtained here for $p =3$ by a quotient of the Bruhat-Tits tree by a Schottky generator with $\log_p |q|_p^{-1} = 6$. The geometry is locally indistinguishable from the Bruhat-Tits tree, but the presence of the horizon signifies the boundary interpretation will be very different. In the tensor network analysis we will find the boundary state of this geometry will have a thermal entropy proportional to $6 \log p$.
\label{fig:padicBTZ}}
\end{figure}
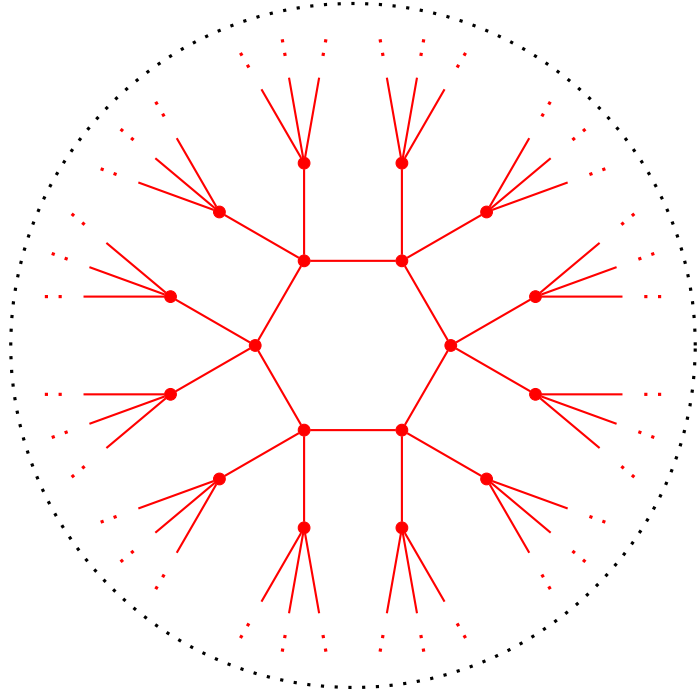

The situation so far over the $p$-adics may be somewhat abstract, but a very intuitive picture resembling a black hole emerges when we consider the quotient of the Bruhat-Tits tree itself by the above Schottky generator. 
Algebraically, one again removes the boundary points $\{0, \infty \}$ of ${\cal T}_p$ and identifies vertices and edges of the tree related by the action of the Schottky generator; we can express this as $({\cal T}_p \smallsetminus \{0,\infty\})/q^\mathbb{Z}$. In analogy with the real case, an explicit form of this generator is a $\PGL(2, \bQ_p)$ element which translates along the $\{ 0, \infty \}$ by $\log_p|q|_p^{-1}$. 
Pictorially (and more formally), the geometry after the quotient is obtained by taking the entire tree and identifying points which are related by translation by $\ord_p(q)=\log_p |q|_p^{-1}$ steps along this main geodesic.
The condition on the norm of $q$ means this translation is always an integer number of steps, and the result is a central ring of length $\ord_p(q)$ with branches which asymptotically look like ${\cal T}_p$. It is a motivating result that the boundary of this ring geometry can be identified with the Tate-Mumford curve, and mathematically it is guaranteed by our uniformization procedure.  This is illustrated in figure \ref{fig:padicBTZ}, where by analogy with the real case we interpret this as a time slice of a static BTZ black hole.

Our final task of this section is to explain how to extend this $p$-adic uniformization to the tensor network living on the dual graph of the the Bruhat-Tits tree. The most natural way to do this is to simply perform the identifications under the Schottky generator on both the tree and the dual graph simultaneously, and one can see graphically that this will introduce a special vertex behind the `horizon'. This vertex is most naturally traced out (as in a two-sided black hole geometry,) and we will later show by explicit computation that this choice produces a mixed density matrix for the boundary state. The thermal entropy of this density matrix is proportional to the perimeter of the $p$-adic BTZ black hole, giving agreement with Bekenstein-Hawking formula up to an overall constant. In this toy model, the interpretation of these microstates are those legs (namely contracted bonds) of the tensor network which stretch across the horizon.

\begin{figure}[t]
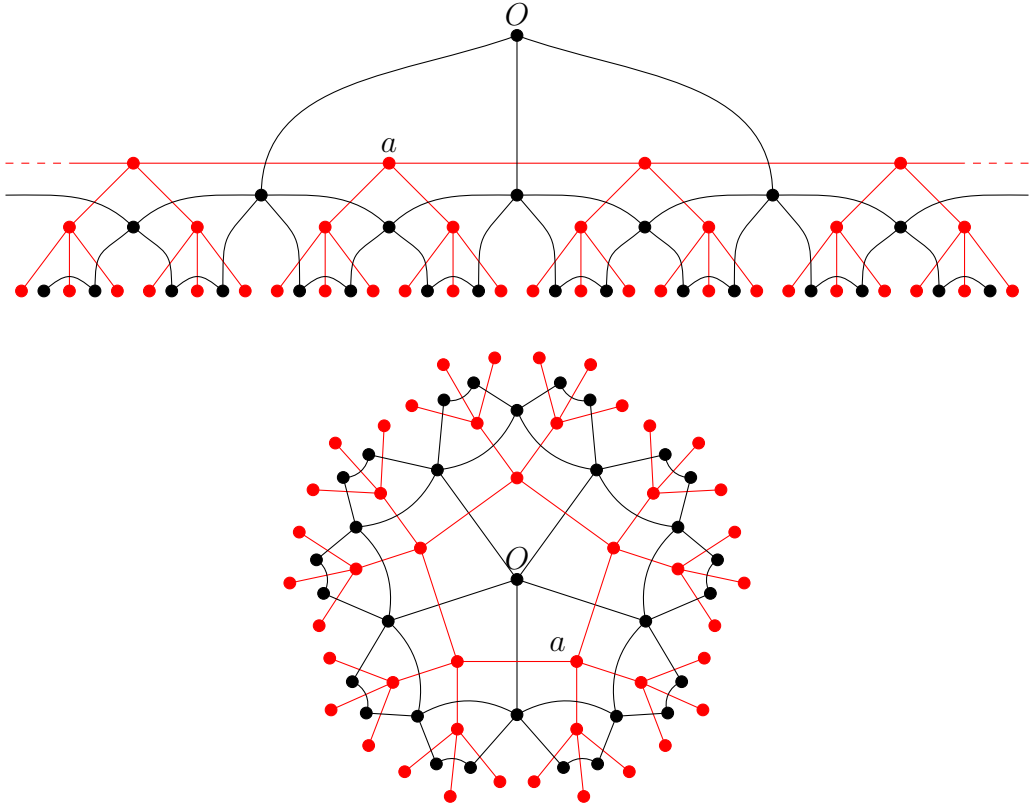

    \[
    \musepic{\quotientline}
    \]
    \bigskip
    \[
    \musepic{\quotientloop}
    \]
    \caption{The quotient construction of the dual graph tensor network (in black). As pictured, $p=3$ and $\ord_p(q)=5$.}
    \label{fig:BTZtensornetwork}
\end{figure}

As usual, the identification and resulting BTZ black hole tensor network are best done with the aid of a figure. We first redraw the tree in a form that is `flattened out' along the preferred $\{ 0\to \infty \}$ geodesic, as shown in the top sub-figure of figure \ref{fig:BTZtensornetwork}, where we have explicitly chosen $p=3$ and $\log_p|q|_p^{-1}= 5$ as an example. 
While we can only display a small portion of the tree, one should think of this geodesic as stretching infinitely, with branches coming off and continuing to the conformal boundary. This is nothing but a relabeling of figure \ref{fig:SimplePsi}, but we have now labeled a special vertex $O$ on the tensor network which sits above the central geodesic as well as a special vertex $a$ on the tree. After the quotient, $O$ will be in the black hole interior and $a$ will be identified with its image under $a \rightarrow a - \log_p|q|_p$. 
Recall also that all the vertices on this dual graph have dangling legs and represent degrees of freedom on the boundary, but we have not yet specified the rank of these tensors due to a subtlety explained below. 

After taking the quotient, we can redraw the tree and its dual graph in a form that has rotational symmetry as seen in the second figure \ref{fig:BTZtensornetwork}. The infinite geodesic has now become the central cycle or horizon of the black hole, and the genus $1$ boundary is now the boundary of the infinite branches coming off of this cycle. The center point $O$, which before the quotient was just another vertex on the boundary, has now moved behind the horizon. The number of internal bonds connected to $O$ is determined by the $p$-adic norm of our Schottky parameter $q$, which can be any integer greater than $1$.

This is an intuitive picture of how one might make a black hole tensor network state which agrees with our uniformization procedure for the tree. A formal recipe for constructing different choices of dual graph is discussed in section~\ref{ssec:geometryTATE}, and here as before we make one convenient choice. Even so, it is now necessary to explain both the subtleties of how the dual graph is defined as well as the cut-off prescription. Recall in the genus $0$ picture, the bulk IR regulator $\Lambda$ was defined as half the length of the longest geodesic- this meant truncating both the tree and the dual graph $\Lambda$ steps from the central vertex in all directions. For the black hole case, we would like the analogous statement to be that we truncate the tree and network $\Lambda$ steps from the horizon. However, in order to achieve this we had to use a different cutoff before identifying points related by the Schottky group. This should be clear from examining the `flat' picture of the tree, which for finite cut-off would not correspond to something radially symmetric. Conceptually, we could treat the genus $0$ and genus $1$ on an equal footing if we worked with the infinite tree and network and only apply the cut-off prescription after taking the quotient. 

One can note that the two criterion expressed in section~\ref{DUALGRAPH} continue to hold in the black hole background; every edge of the tree has exactly one bond  of the dual graph which ``cuts'' it, and every vertex of the tree is surrounded by a plaquette with $p+1$ sides. These facts do not guarantee that the dual graph after taking the quotient is uniquely defined, but later we will discuss the boundary measure associated to the Mumford curve and the dual graph and find a canonical choice. Recall that in the genus 0 case, the non-uniqueness could be interpreted as the lack of ordering of $p$-adic numbers at the boundary. 

A further technical point concerns black holes that are large compared to the cutoff scale. When working with the genus $0$ network with a finite cutoff, we observed that our uniform use of a single kind of perfect tensor of rank $r+1$ required certain conditions on the rank and the cutoff in order for the Ryu--Takayanagi formula to hold. Roughly speaking, the number of bulk contracted legs at any tensor could not exceed the number of boundary uncontracted legs.
Increasing the cutoff thus meant increasing the rank, corresponding to a large number of UV degrees of freedom at the boundary. Similarly, the perimeter of the black hole horizon at genus $1$ also constrains the minimum rank of the tensor, as now the center point may have a larger number of bulk bonds than other points in the network. This becomes an issue for black holes that are large compared to the cutoff, but using a sufficiently large rank perfect tensor will always produce the correct answers for the black hole entropy.

There is one final point to address, which is the curious case of a horizon with length $\log_p |q|_p^{-1} = 1$. 
This is the minimal Tate-Mumford curve allowed by the uniformization; the corresponding quotient of the Bruhat--Tits tree contains a self-looping edge, and it correspondingly leads to a degenerate configuration of the network. One of the plaquettes of the dual graph network collapses. Nonetheless, the entropy computed from this degenerate network still leads to the expected result.

\subsection{Black hole entropy}
\label{BTZResults}
In the previous subsection, we described the construction of BTZ black holes in $p$-adic AdS/CFT via the algebraic process of Schottky uniformization. We also explained how this naturally extended to the dual graph tensor network, essentially by identifying all nodes and bonds related by a translation by the horizon length. The result is a graph with a cycle and a dual graph tensor network with many desirable properties; crucially there is one special vertex behind the horizon which cannot be identified with any boundary degrees of freedom. In this section, we present the results of a computation on the tensor network for the thermal entropy of the boundary density matrix obtained by tracing out this vertex, explained in more detail in section \ref{GENUS1BTZ}. We find perfect agreement between the thermal entropy and the black hole horizon perimeter, as predicted by an analog of the Bekenstein-Hawking formula. 

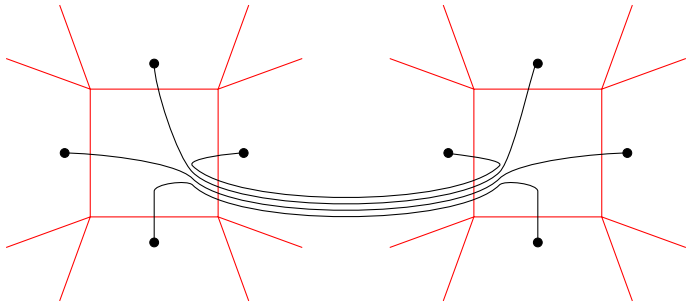
\begin{figure}[t]
 \[
 \begin{tikzpicture}[scale=1.7]
\tikzstyle{vertex}=[draw,scale=0.3,fill=black,circle]
\tikzstyle{ver2}=[draw,scale=0.6,circle]
\tikzstyle{ver3}=[draw,scale=0.4,circle]

\draw[color=red] (0,0) -- (0,1) -- (1,1) -- (1,0) -- cycle;
\draw[color=red] (3,0) -- (3,1) -- (4,1) -- (4,0) -- cycle;

\draw[color=red] ($ (0,0) - (20:0.7) $) -- (0,0) -- ($ (0,0) - (70:0.7) $);
\draw[color=red] ($ (0,1) - (20-90:0.7) $) -- (0,1) -- ($ (0,1) - (70-90:0.7) $);
\draw[color=red] ($ (1,1) - (20-180:0.7) $) -- (1,1) -- ($ (1,1) - (70-180:0.7) $);
\draw[color=red] ($ (1,0) - (20-270:0.7) $) -- (1,0) -- ($ (1,0) - (70-270:0.7) $);

\draw[color=red] ($ (3,0) - (20:0.7) $) -- (3,0) -- ($ (3,0) - (70:0.7) $);
\draw[color=red] ($ (3,1) - (20-90:0.7) $) -- (3,1) -- ($ (3,1) - (70-90:0.7) $);
\draw[color=red] ($ (4,1) - (20-180:0.7) $) -- (4,1) -- ($ (4,1) - (70-180:0.7) $);
\draw[color=red] ($ (4,0) - (20-270:0.7) $) -- (4,0) -- ($ (4,0) - (70-270:0.7) $);

\draw (0.5,-0.2) node[vertex]{};
\draw (0.5,1.2) node[vertex]{};
\draw (-0.2,0.5) node[vertex]{};
\draw (1.2,0.5) node[vertex]{};

\draw (3.5,-0.2) node[vertex]{};
\draw (3.5,1.2) node[vertex]{};
\draw (2.8,0.5) node[vertex]{};
\draw (4.2,0.5) node[vertex]{};

\draw (0.5,-0.2) -- (0.5,0.2)  to[out=90,in=135,looseness=0.5] (0.8,0.25) to[in=225,out=-45,looseness=0.5] (3.2,0.25) to[out=45,in=90,looseness=0.5] (3.5,0.2) -- (3.5,-0.2);
\draw (-0.2,0.5) to[out=0,in=135,looseness=0.5] (0.8,0.3) to[in=225,out=-45,looseness=0.5] (3.2,0.3) to[out=45,in=180,looseness=0.5] (4.2,0.5);
\draw (0.5,1.2) to[out=-90,in=135,looseness=0.5] (0.8,0.35) to[in=225,out=-45,looseness=0.5] (3.2,0.35) to[out=45,in=90,looseness=0.5] (3.5,1.2);
\draw (1.2,0.5) to[out=180,in=135,looseness=0.5] (0.8,0.4) to[in=225,out=-45,looseness=0.5] (3.2,0.4) to[out=45,in=180,looseness=0.5] (2.8,0.5);

\end{tikzpicture}
\]
    \caption{The genus $1$ tensor network after tracing out the vertex behind the horizon. One should imagine the two sides corresponding to $| \psi \rangle$ and $\langle \psi |$, and the total state being the mixed density matrix which has von Neumann entropy proportional to the number of shared bonds. We postpone the explanation of the graphical rules and the computation of the entropy until section~\ref{COMPUTATION}, but the reader might be reminded of the 2-sided BTZ black hole.}
    \label{fig:BTZ2sided}
\end{figure}

Tracing out the center vertex in our tensor network amounts to constructing a mixed density matrix reminiscent of a two-sided BTZ black hole. This is depicted in figure~\ref{fig:BTZ2sided} and follows from our graphical rules for computing density matrices from tensor networks, explained in section~\ref{COMPUTATION}. Defining the perimeter to be $\tau = \log_p|q|_p^{-1}$, we find by detailed computation the von Neumann entropy of the boundary state to be proportional to the perimeter, which is the same as the number of bonds stretched across the two sides. The result is surprisingly simple and analogous to the BTZ black hole entropy discussed in the previous section:
\eqn{BTZEntropysummary}{
S_{\rm BH} = \tau \log r \,.
}

\subsection{Ryu--Takayanagi formula in the black hole background}
\label{RTBTZresults}

Here we will briefly summarize our results which combine the main ideas of sections~\ref{GENUSZERORESULTS} and \ref{BTZResults}. This involves computing the boundary von Neumann entropy of a single connected interval in the thermal background, holographically found to be dual to the length of a minimal geodesic in the black hole geometry homologous to the interval. This presents further computational challenges which are discussed in section~\ref{GENUS1BTZ}. As is often the case with quantities that can be computed in the dual picture, the bulk result is easier to state than derive, but we find agreement in all cases considered. A schematic depiction of the behavior of the minimal surface is shown later in figure~\ref{fig:realBHInt}. It is a surprising and nontrivial fact that the tensor network proposed here automatically captures the three topologically distinct cases for the surface. We take the success of this tensor network proposal as a conjecture for the entanglement entropy of a connected interval in $p$-adic field theory at finite temperature.

\begin{figure}[t]
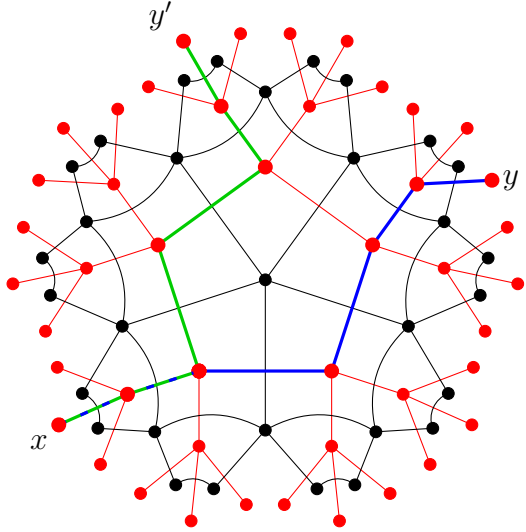

\[
\musepic{\geodesicwrap}
\]
    \caption{Geodesics in the BTZ black hole geometry. Moving $y$ to~$y'$ leads to a ``jump'' of the minimal geodesic to a path wrapping the other side of the horizon.}
    \label{fig:BTZMinSurface}
\end{figure}

A key conceptual difference from the genus $0$ Ryu--Takayanagi formula is that the entropy of a  boundary region and its complement are not equal, since the holographic state generated by the network is no longer a pure state. The bulk interpretation of this is the presence of the black hole horizon which minimal surfaces may wrap around. Varying the ($p$-adic) size of the boundary region, a minimal geodesic might jump from crossing one side of the horizon to the other, and one observes this behavior in the boundary von Neumann entropy as well. This is a feature that is desirable in principle and in practice, though care must be taken in the precise definition of the boundary measure and dual graphs. We chose to parametrize the size of the boundary `intervals' using the measure for the covering space (before taking the quotient), and this is explained in greater mathematical detail in section~\ref{ssec:geometryTATE}. However, after making a choice of a planar embedding for the tensor network, the intuitive picture of the genus $1$ minimal geodesic behavior is easy to see in figure~\ref{fig:BTZMinSurface}.

The structure of entanglement in this $p$-adic black hole setting has a particular novel feature not present in the usual picture of AdS$_3$. In that case, small interval sizes or low temperatures will have entanglement entropy very nearly equal to the flat space result \cite{Calabrese:2004eu}, which can be seen by Taylor expansion or noting the minimal surfaces do not approach the BTZ horizon. In contrast, for $p$-adic AdS/CFT the transition would seem to be much more sharp. If one considers boundary regions that are small enough in the measure described above, one may see that the minimal geodesic never reaches the horizon, and the length and thus the entropy will be precisely equal to the genus $0$ case. This is an interesting prediction for entanglement in thermal $p$-adic field theories, where the indication is that low enough temperatures have exactly zero effect on the short distance physics (rather than parametrically small effect.) This is ultimately due to the nonarchimedean or ultrametric nature of $p$-adic numbers.

The various features of the minimal geodesics in the black hole background can be described using a distance measure in the bulk.
Given a planar embedding, the entropy of the connected region $(x,y)$ is given by a minimal geodesic homologous to the given region (recall definition \ref{def:homologous} of the homologous condition).
In the $p$-adic black hole background, given two boundary points, there are two possible boundary anchored geodesics to choose from, only one of which will be homologous to the given region.
The geodesic homologous to the complimentary region will be  given by the other path. 
This can be unified together into the formula
\eqn{lengthBTZ0}{
S(x,y) = {{\rm length}(\gamma_{xy}) \over \ell } \log r\,,
}
where
\eqn{lengthxybtzCombine}{
{\rm length} (\gamma_{xy})/\ell = \delta(C\to x,C\to x)+\delta(C\to y,C\to y) - 2\delta(C\to x,C\to y)\,.
}
Here $\ell$ is the constant length of each edge of the tree, and $C$ is an arbitrary reference point on the genus $1$ graph. 
Recall that $\delta(\cdot,\cdot)$ is an integer which counts the signed overlap between the two paths in its arguments.
Equation \eno{lengthxybtzCombine} does not depend on the choice of $C$, but the two choices of paths for $C \to x$ (as well as $C\to y$) in \eno{lengthxybtzCombine} correspond  in all to the two possible values of $S(x,y)$, corresponding to the interval $(x,y)$ and its complement on the boundary (such that the geodesics are appropriately homologous).
Define $\epsilon \in \mathbb{Q}_p$ to be the cutoff 
\eqn{}{
\epsilon \equiv p^{\Lambda}\,,
}
which goes to zero $p$-adically, i.e.\ $|\epsilon|_p \to 0$ as $\Lambda \to \infty$. If $x,y \in E(\mathbb{Q}_p) \simeq \mathbb{Q}_p^{\times}/q^{\mathbb{Z}}$, then 
\eqn{lengthBTZ}{
{\rm length} (\gamma_{xy})/\ell = 2\log_p { |\mathfrak{B}(\{x,y\})|_{g=1} \over |\epsilon|_p}\,,
}
where $|\mathfrak{B}(\{x,y\})|_{g=1}$ is the measure of the set containing $x,y \in E(\mathbb{Q}_p)$. 
On the covering space geometry, there are infinitely many sets which contain $x$ and $y$ because there is an infinite set of image points which correspond to these boundary points. From the point of view of the fundamental domain, there are two minimal sets which correspond to the two ways to wrap around the horizon. The measure above corresponds to choosing one of the two depending on which choice of minimal surface(s) is homologous to the boundary region. 
This measure is further explained in section~\ref{ssec:geometryTATE}, and the explanation from tensor network contractions is outlined in section~\ref{GENUS1BTZ}. Here we comment that up to an overall constant factor, the entanglement entropy for the mixed states is equal to these geodesic lengths, and this is encapsulated by this measure. One can see this as a kind of prediction for the single interval entanglement entropies for thermal states in $p$-adic AdS/CFT.

\section{von Neumann Entropy and Inequalities}
\label{COMPUTATION}

In this section we present the detailed computations leading to the results summarized in sections \ref{GENUSZERO} and \ref{GENUSONE}, as well as proofs of various entropy inequalities in sections \ref{GENUS0DOUBLE}-\ref{ssec:Inequalities}.

\subsection{Perfect tensors and density matrices}
\label{SIMPLESTATES}

Before getting to calculations in the holographic setup, we point out some of the basic ingredients and properties which will be useful later using simpler toy examples.
We focus on simple states (not necessarily holographic) constructed using rank-$(r+1)$ perfect tensors; the indices of such tensors will label bases of fixed finite-dimensional Hilbert spaces, which we interchangeably call ``spins,'' ``qubits,'' or ``qudits.'' For example, for $r=3$, consider
\eqn{psiSimplest}{
|\psi \rangle = T_{abcd}\, |abcd \rangle \qquad a,b,c,d \in \{0,1,2\}\,,
}
where repeated indices are summed over, and $|abcd\rangle = |a \rangle \otimes |b\rangle \otimes |c \rangle \otimes |d\rangle$ is a product state of four qutrits. 
$T_{abcd}$ is the rank-4 perfect tensor, and we normalize it so that all its non-zero components are unity. Graphically, we may represent \eno{psiSimplest} as
\eqn{psiSimplestGraph}{
|\psi \rangle = \musepic{\bigtensor}.
}
Since $T$ is a perfect tensor, specifying half of its indices uniquely fixes the remaining half. For instance, we may choose 
\eqn{Trank4}{
T_{0000} = T_{0111} = T_{0222} = T_{1012} = T_{1120} = T_{1201} = T_{2021} = T_{2210} = T_{2102} = 1\,,
}
with all other components vanishing. One may check that for any choice of two indices, there is a unique non-vanishing component of $T$, so the other two indices are also determined. 
The ``perfectness property'' is useful in writing out the full state $|\psi\rangle$ starting from \eno{psiSimplestGraph}. This $|\psi \rangle$ as constructed is a very special entangled superposition of $9$ of the $3^4 = 81$ possible basis states. In this state, any choice of $2$ spins are maximally entangled with the remaining two. This is a general feature of perfect states.

One can construct more complicated states by contracting multiple copies of the perfect tensor $T$ in different ways. For example, the following graph represents a new state,
\eqn{psiComplic}{
|\psi^\prime \rangle = 
\musepic{\bigtree} \,\, ,
}
where we have suppressed the index labels. In \eno{psiComplic} and future graphical representations, a shared edge between two vertices will denote that the corresponding index is to be summed over. Thus, explicitly, \eno{psiComplic} represents the state
\eqn{psiComplicAgain}{
|\psi^\prime \rangle = T_{a_0 b_0 c_0 d_0} T_{a_0 b_1 c_1 d_1} T_{b_0 b_2 c_2 d_2} T_{c_0 b_3 c_3 d_3} T_{d_0 b_4 c_4 d_4} |b_1 c_1 d_1 b_2 c_2 d_2 b_3 c_3 d_3 b_4 c_4 d_4 \rangle\,. 
}
In this example, the internal lines (denoted by indices with a $0$ subscript) appear traced over in the tensors but do not label the basis of boundary states.

The (normalized) density matrix corresponding to the state $|\psi\rangle$ is given by 
\eqn{DenMat}{
\rho ={1 \over \langle \psi | \psi \rangle} |\psi \rangle \langle \psi |\,.
}
For example, for the state in \eno{psiSimplest},
\eqn{DenMatSimplest}{
\rho  = {1 \over 9} T_{abcd} T_{a^\prime b^\prime c^\prime d^\prime}  |abcd \rangle \langle a^\prime b^\prime c^\prime d^\prime |\,,
}
where we used
\eqn{psiSimplestNorm}{
 \langle \psi | \psi \rangle = T_{abcd} T_{abcd} = 9\,,
}
which follows from the perfect tensor property of $T$ and the fact that $a,b,c,d \in \{0,1,2\}$. 

Just as states built from perfect tensor contractions had a convenient graphical representation, we will sometimes also write the density matrix in the same way. Because the density matrix is a product of the perfect state vector and the dual, graphically we can write \eno{DenMatSimplest} as 
\eqn{DenMatSimplestGraph}{
\rho = {1\over 9} \musepic{\bigtensor} \musepic{\bigtensorprime}
\,,
}
with the understanding that \eno{DenMatSimplestGraph} represents a density matrix, with the matrix elements given by specifying the external indices and performing the tensor contractions.

The normalization in \eno{psiSimplestNorm} is a contraction on all indices, and we can represent this contraction by connecting the lines of \eno{DenMatSimplestGraph}, producing:
\eqn{psiSimplestNormGraph}{
\langle \psi | \psi \rangle = 
\musepic{\contractfour}
\,.
}
This has no external legs, so it is a pure number. It evaluates to $9$ as this is the number of non-vanishing components of $T$, equivalently the number of allowed assignments of the internal legs.

Taking partial traces leads to reduced density matrices. 
For example, for the $\rho$ given in \eno{DenMatSimplest}, if $Tr_2$ denotes tracing out the second factor of the direct product state $|\psi \rangle = |abcd \rangle$, then 
\eqn{rho1}{
\rho_1 \equiv \Tr_2\, \rho = \sum_{b^{\prime\prime}} \langle b^{\prime\prime}| \rho | b^{\prime\prime} \rangle = {1 \over 9}\, T_{ab^{\prime\prime}cd} T_{a^\prime b^{\prime\prime} c^\prime d^\prime} |acd \rangle \langle a^\prime c^\prime d^\prime |\,.
}
Graphically, we represent this as
\eqn{rho1Graph}{
\rho_1 = \frac{1}{9} \, \, 
\musepic{\contractone}
\,.
}
We have suppressed index labels on the graph. There is a slight abuse of notation by representing both states and reduced density matrices using the same kinds of pictures, even though the corresponding equations are unambiguous. We will be careful to distinguish between states and matrices in more complicated examples later; a rule of thumb is that the reduced density matrix is mirrored across the contracted lines.

The diagram in \eno{rho1Graph} is identical to that of a reduced density matrix where instead of the second factor, we traced out any of the other single qutrits in $|\psi \rangle = |abcd \rangle$. This follows from the permutation symmetry of the legs.

If instead we trace out two sites, say the first two, we obtain 
\eqn{rho2}{
\rho_2 \equiv \Tr_{12} \rho = \sum_{a^{\prime \prime} b^{\prime\prime}} \langle a^{\prime \prime} b^{\prime\prime}| \rho | a^{\prime \prime} b^{\prime\prime} \rangle = {1\over 9}\, T_{a^{\prime \prime} b^{\prime\prime}cd} T_{a^{\prime \prime} b^{\prime\prime} c^\prime d^\prime} |cd \rangle \langle  c^\prime d^\prime |\,.
}
Graphically, we write
\eqn{rho2Graph}{
\rho_2 = {1 \over 9} \,\,
\musepic{\contracttwo}
\,.
}
Similarly, tracing out three sites leads to the following representation,
\eqn{rho3Graph}{
\rho_3 = {1 \over 9} \,\,
\musepic{\contractthree}
\,.
}

Explicitly evaluating expressions such as \eno{rho1} and \eno{rho2}, and in fact the norm of a given state constructed out of perfect tensor contractions can become cumbersome for more complicated states. 
It would be useful to have a set of graphical rules which can be used to simplify and evaluate reduced density matrices without resorting to tedious (though straightforward) algebra. 
In the end, our goal is to evaluate the von Neumann entropy of the reduced density matrix $\rho_A = \Tr_{A^c} \rho$,
\eqn{vNent}{
S = - \Tr \rho_A \log \rho_A\,,
}
which for pure $\rho$ corresponds to a measure of quantum entanglement between the traced out region $A^c$ and its complement $(A^c)^c = A$. 
With this in mind, we present some  useful (diagrammatic) rules and techniques.

For two rank-$(r+1)$ perfect tensors $T$ which have $n_c$ indices contracted between them, with $n_d$ free indices each, so that $n_c + n_d = r+1$ we have
\eqn{ContractRuleEqn}{
T_{a_1 \ldots a_{n_d} b_1 \ldots b_{n_c}} T_{a^\prime_1 \ldots a^\prime_{n_d}  b_1 \ldots b_{n_c}} = \delta_{a_1 a^\prime_1} \cdots \delta_{a_{n_d} a^\prime_{n_d}} \times r^{(n_c-n_d)/2} \qquad n_c \geq n_d\,.
}
This easily follows from the ``perfectness property'' of perfect tensors and crucially assumes $n_c \geq n_d$. In this situation, we are tracing out at least half of the available indices on each tensor; once we have specified half the indices (for each term in the sum), the rest are uniquely determined. Tracing more than half the indices means we are performing a free sum on the remaining indices, this introduces the multiplicity given by $r^{(n_c-n_d)/2}$. In fact in \eno{ContractRuleEqn}, we need not have contracted precisely the final $n_c$ indices, but some other subset of $n_c$ indices to obtain the same form by symmetry. The Kronecker delta functions between  ``dangling'' (uncontracted) indices in that case would still be between indices at matching positions. 

Graphically, we write this contraction identity as 
\eqn{ContractRule}{
\musepic{\numerouscontractions}
= 
\musepic{\numerouslines}
\times 
\qty(\musepic{\smcircle})^{(n_c-n_d)/2},
}
which is valid whenever~$n_c \geq n_d$.
The horizontal line-segments correspond to the delta function factors in~\eno{ContractRuleEqn}, with the understanding that precisely the free (dangling) indices at matching positions in the two tensors share a delta function. After doing the contraction, we see that we have ``split'' open the tensor to obtain the delta functions, and we refer to this operation as a ``split''. The meaning of each disconnected loop represents a factor of $r$ coming from the free sum. To see this, note that contracting a delta function $\delta_{ab}$ with another $\delta_{ab}$ results in
\eqn{loopEqn}{
\delta_{ab} \delta_{ab} = \delta_{aa} = r\,,
}
which diagrammatically is represented as joining together the end-points of a line-segment, turning it into a (disconnected) loop. One checks that using rule \eno{ContractRule}, the the inner product in \eno{psiSimplestNormGraph} evaluates to $\left(\musepic{\tinycircle}\right)^2 = r^2=3^2$, which was the number of allowed terms in the free sum.

We may also interpret the l.h.s.\ of \eno{ContractRule} as representing the partial trace of a pure (unnormalized) density matrix, leading to a diagonal reduced density matrix, as long as we remember to include appropriate ket and bra state factors in \eno{ContractRuleEqn} when translating \eno{ContractRule} back to equations. 
More precisely, starting with the pure (unnormalized) density matrix
\eqn{TwoTrho}{
\rho = T_{a_1 \ldots a_{n_d} b_1 \ldots b_{n_c}}  T_{a^\prime_1 \ldots a^\prime_{n_d}  b_1^\prime \ldots b_{n_c}^\prime} |a_1 \ldots a_{n_d} b_1 \ldots b_{n_c} \rangle  \langle a^\prime_1 \ldots a^\prime_{n_d}  b_1^\prime \ldots b_{n_c}^\prime|\,,
}
where $n_c + n_d = r+1$, and tracing out, say, the final $n_c$ sites (indices), one obtains a reduced density matrix $\rho_r$ which is diagonal in the basis ${\cal B} = \{ |a_1 \ldots a_{n_d} \rangle : a_i=0,1,\ldots r-1\}$:
\eqn{}{
 \langle a_1 \ldots a_{n_d} | \rho_r   |a_1^\prime \ldots a_{n_d}^\prime  \rangle = r^{(n_c-n_d)/2}\,  \delta_{a_1 a^\prime_1} \cdots \delta_{a_{n_d} a^\prime_{n_d}} \,,
}
{\it as long as} $n_c \geq n_d$ (this is the case where the sum over internal contracted lines fixes the values of the external legs.) Thus $\rho_r$ is a $r^{n_d} \times r^{n_d}$ diagonal density matrix. To start with a normalized $\rho$ such that $\Tr \rho=1$, we normalize $\rho$ in \eno{TwoTrho} by multiplying with an overall factor of $1/\langle \psi|\psi \rangle = r^{-(r+1)/2} = r^{-(n_c+n_d)/2}$. Then we see that the $\Tr \rho_r = 1$ condition is satisfied automatically, and in fact $\rho_r$ has $r^{n_d}$ eigenvalues each equaling $r^{-n_d}$. The von Neumann entropy associated with $\rho_r$ is then 
\eqn{SingleTensorEnt1}{
S = - \Tr \rho_r \log \rho_r = n_d \log r \hspace{5mm} \textrm{(perfect state with $n_c \geq n_d$ spins traced out)\,.}
}

If $n_c < n_d$, then the reduced density matrix will no longer be diagonal in the previously chosen basis. 
In fact, since in this case $n_d > (r+1)/2$, not all possible combinations for the string ``$a_1 \ldots a_{n_d}$'' are  permissible any more, as some combinations will lead to a vanishing tensor component $T_{a_1\ldots a_{n_d} b_1\ldots b_{n_c}}$. 
Thus the basis of states in which we may represent the reduced density matrix is  no longer $r^{n_d}$-dimensional, but in fact an $r^{(r+1)/2}$-dimensional subset ${\cal V} \subset {\cal B}$ (the exponent is $(r+1)/2$ because that is precisely the maximum number of $a_i$ indices one needs to specify before fully determining the tensor component $T_{a_1\ldots a_{n_d} b_1\ldots b_{n_c}}$ uniquely).

The reason that the reduced density matrix is not diagonal in ${\cal V}$ is because  in the l.h.s.\ of \eno{ContractRuleEqn} (or \eno{ContractRule}) knowledge about all the contracted indices no longer uniquely fixes the free dangling indices, since $n_c < (r+1)/2$. In such cases, with an eye on computing the von Neumann entropy \eno{vNent} -- which comes down to finding the eigenvalues of the reduced density matrix -- we  use a convenient parametrization of ${\cal V}$ in which the reduced density matrix assumes a Jordan block-diagonal form, with each diagonal block a matrix with all elements equal to 1. 

To arrive at this convenient block diagonal form, enumerate the basis states $|a_1 \ldots a_{n_d} \rangle$ such that the first $r^{(r+1)/2-n_c}$ states correspond to the set of $\{ a_1,\ldots, a_{n_d}\}$ such that given a particular numerical combination of the string ``$b_1 \ldots b_{n_c}$'', the tensor $T_{a_1 \ldots a_{n_d} b_1 \ldots b_{n_c}}$ is non-zero. 
There are  $r^{n_c}$ distinct combinations possible for  ``$b_1 \ldots b_{n_c}$'', and for each single combination, the set of allowed $\{ a_1,\ldots, a_{n_d}\}$ such that $T_{a_1 \ldots a_{n_d} b_1 \ldots b_{n_c}}$ is non-zero has cardinality $r^{(r+1)/2-n_c}$. The next $r^{(r+1)/2-n_c}$ correspond to a different choice of ``$b_1 \ldots b_{n_c}$'', and so on. 
In this way of enumerating the basis, the reduced density matrix assumes a block-diagonal form, with $r^{n_c}$ blocks along the diagonal, each of size $r^{(r+1)/2-n_c} \times r^{(r+1)/2-n_c}$. Thus the total size of $\rho_r$ is $r^{(r+1)/2} \times r^{(r+1)/2}$, as expected (since $\dim {\cal V} = r^{(r+1)/2}$). 

So for $n_c < n_d$, by abuse of notation, we write graphically
\eqn{}{
\musepic{\numerouscontractions}
= 
 \begin{pmatrix}
 	J_{r^{(r+1)/2-n_c}} \\ & \ddots \\ && J_{r^{(r+1)/2-n_c}}
 \end{pmatrix}_{r^{(r+1)/2} \times r^{(r+1)/2}} \qquad n_c < n_d\,,
}
where $J_{n}$ is the $n \times n$ matrix of all ones and all other entries are $0$. We note that this argument holds regardless of precisely which $n_c$ indices in \eno{TwoTrho} were traced out. 
If we normalize $\rho$ with an overall factor of $1/\langle \psi|\psi\rangle = r^{-(r+1)/2}$ so that $\Tr \rho=1$, the reduced density matrix will automatically have $\Tr \rho_r =1$. 
Further, it will have $r^{n_c}$ non-zero eigenvalues, each equaling $\lambda_i = r^{-n_c}$ (this follows from a standard result on block diagonalization of these matrices.) 
From the eigenvalues, it follows that the von Neumann entropy of the (normalized) reduced density matrix will be 
\eqn{SingleTensorEnt2}{
S= -\sum_i \lambda_i \log \lambda_i =  n_c \log r \hspace{5mm} \textrm{(perfect state with $n_c < n_d$ spins traced out)\,.}
}
Since $n_c$ counts the number of contractions between the two copies of the state $\psi$ in the reduced density matrix (and consequently the number of diagonal blocks in the matrix representation of $\rho_r$, which is $r^{n_c}$), we conclude in this case that for $n_c < n_d$, the von Neumann entropy is proportional to precisely the number of such contractions (more precisely, equal to the logarithm of the number of blocks in the Jordan block-diagonal matrix representation of $\rho_r$). 

It is instructive to apply what we have learned so far to the previous examples of \eno{rho1}-\eno{rho3Graph}. We conclude that the latter two examples satisfy $n_c \geq n_d$ and the reduced density matrices have the explicit diagonal form $\rho_2 = {1\over 3^2} I_{3^2}$, $\rho_3 = {1\over 3} I_3$. The first example has $n_c < n_d$, and in an appropriate basis, $\rho_1 = {1\over 3^2} \diag (J_3, J_3, J_3)$. Correspondingly, the von Neumann entropies are $S_2 = 2\log 3$, $S_3= \log 3$ and $S_1 = \log 3$, which is consistent with the expectation that the von Neumann entropy in tensor networks built out of perfect tensors is proportional to the minimal number of cuts needed to separate out the traced out part  of the tensor network from the rest.

We have so far focused on the simplest explicit examples, but the reasoning we have used for the state given by \eno{TwoTrho} works for {\it any} state constructed out of any number of copies of the perfect tensor $T$ of fixed rank-$(r+1)$, with the proviso that we have already applied the rule \eno{ContractRule} wherever possible to reduce the reduced density matrix to its ``simplest'' form. With even only a few tensors of modest rank, one can quickly construct enormous density matrices due to the doubly exponential power scaling of various quantities. However, armed with \eno{ContractRule} and the properties of perfect tensors, it becomes possible to determine these density matrices analytically.

The dimension of the basis ${\cal V}$ in more complicated examples will differ from the case of a single tensor determined above, but one can still parametrize the basis such that the matrix representation of the reduced density matrix $\rho_r$ (for $n_c < n_d$) is in Jordan form. The size of each diagonal block will also depend on the details of the original state $\rho$ and the choice of $n_c, n_d$ (even with $n_c + n_d$ no longer equaling $r+1$), but the number of blocks will still equal $r^{n_c}$, where by definition, $n_c$ is the number of contractions between the two copies of the state $\psi$ in the ``simplified'' reduced density matrix as described in the previous paragraph. Thus the von Neumann entropy (for normalized states) will evaluate to $S=n_c \log r$ as long as $n_c < n_d$. As an example, if the reduced density matrix was obtained by tracing out the bottom three legs of the state in \eno{psiComplic}, we first apply rule \eno{ContractRule} to replace the contraction of the form 
$\cdots 
\musepic{\contractthreeinline}
\cdots$ 
by a single horizontal line (times a constant factor) to obtain the simplified reduced density matrix, which now has simply one contraction ($n_c = 1$) between the two copies of the state $\psi$. Thus $S = \log r$. This was also expected from the ``minimal number of cuts'' intuition, as we only needed one cut to separate the three sites to be traced out from the rest of the tensor network. We will return to similar calculations for holographic states next.

\subsection{Efficient techniques: splits and cycles}
\label{SPLITSCYCLES}

 Before discussing the partial trace of density matrices associated with the dual graph tensor network, let's warm up with a simpler computation: the inner product of the holographic state with itself. As described in section \ref{DUALGRAPH}, a holographic state in the vacuum AdS cut-off  tree geometry is specified by a prime $p$ which labels the Bruhat--Tits tree ${\cal T}_p$, a prime $r$ which relates to the rank of the perfect tensors forming the tensor network, and the bulk IR cut-off parameter $\Lambda$. In fact, before discussing this in full generality for any $p, r$ and $\Lambda$, let's work it out in the case of the state shown in figure \ref{fig:SimplePsi} and in the process introduce some more useful tools and techniques (the state in figure \ref{fig:SimplePsi} corresponds to the choice $p=3, r=7$, and $\Lambda = 2$).

Diagrammatically, computing $\langle \psi | \psi \rangle$, which is the same as contracting all free dangling legs of $\psi$ with another copy of $\psi$, is represented as
\eqn{psiNormStep1}{
\langle \psi|\psi \rangle = 
\musepic{\bigcontraction}
\ .
}
We have chosen to omit the lines representing contraction of dangling legs at the outermost vertices, but no confusion should arise. 
To evaluate the inner product, one must contract the dangling legs at {\it all} vertices. 
The contractions between the two copies of $\psi$, shown in blue in~\eno{psiNormStep1} to guide the eye, take precisely the form of the contraction rule \eno{ContractRule}. 
Note that the number of dangling legs at the vertices of $|\psi \rangle$, which we denoted $v_d$ in section \ref{DUALGRAPH}, is now reinterpreted as the number of contractions $n_c$ between pairs of vertices from the two copies of $\psi$. 
On the other hand, the number of contractions at a vertex within each $| \psi \rangle$, denoted $v_c$ in section \ref{DUALGRAPH}, can be reinterpreted as the number of ``dangling legs'' $n_d$ in each of the contractions between the two copies of $\psi$. 
Since $v_c \leq v_d$ at any vertex, which we recall follows from the requirement $(r+1)/2 \geq 2\Lambda$, we now have $n_c \geq n_d$ for the contracted pair of vertices.

Thus we can apply rule \eno{ContractRule}. For example,
\eqn{psi4Rule}{
\musepic{\fourtofour} \ = \ \musepic{\fourlines} \  ,
}
at each of the four pairs of vertex contractions between the dangling legs of vertices of $\psi$ which originally had four dangling legs each. 
The second kind of contraction between the two copies of $\psi$ depicted in \eno{psiNormStep1}, which is between vertices carrying six dangling legs each, also admits a simplification. Since $n_c \geq n_d$ here as well, we may simplify this contraction using rule \eno{ContractRule},
\eqn{psi6Rule}{
\musepic{\contractsix} 
= \qty(\musepic{\smcircle} )^2 \times \ \musepic{\twolines}
= r^2 \times\  \musepic{\twolines}\ .
}
The result of applying the contraction rule at all vertex contractions is,
\eqn{psiNormStep4}{
\langle \psi|\psi \rangle =  
\musepic{\bigspread}
\,.
}
We note the creation of new ``cycles'' (or disconnected loops) in \eno{psiNormStep4}. Each such cycle contributes a factor of $r$, as explained in \eno{loopEqn}. The factors of $r^2$ explicitly shown in \eno{psiNormStep4} originate from the application of \eno{psi6Rule}.
One counts $16$ new cycles and eight factors of $r^2$, thus $\langle \psi | \psi \rangle = r^{16} \times (r^2)^8 = r^{32}$, where $r=7$.\footnote{The normalization, as well as several other quantities as computed by these kinds of diagrams, are reminiscent of similar calculations in topological quantum field theory. In that setting, for example, the norm of the state determined by a nullbordism of a particular manifold is computed by gluing two copies of the nullbordism along their common boundary, corresponding to the inner product in Hilbert space.}

Clearly this method is cumbersome to implement. We now propose a shorthand procedure which makes the computation of the norm much more efficient (see figure \ref{fig:proc} for reference):
\begin{figure}
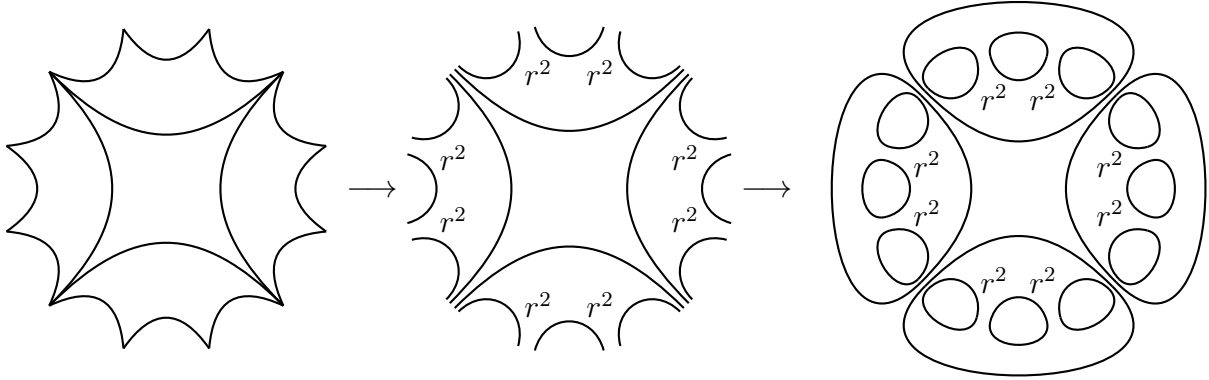

\begin{equation*}
\musepic{\computeone}
\longrightarrow 
\musepic{\computetwo}
\longrightarrow
\musepic{\computethree}
\end{equation*}
\caption{Computation of $\langle \psi | \psi \rangle$.}
\label{fig:proc}
\end{figure}

\begin{enumerate}
\item We start with the holographic state $|\psi \rangle$ depicted in figure \ref{fig:SimplePsi} constructed from rank-$8$ tensors, but suppress drawing the dangling legs. The number of dangling legs at each vertex can be reconstructed from the knowledge of the rank of the tensor.

\item {\bf (Splits.)} We wish to contract all vertices of $\psi$ with itself. Since the number of dangling legs at each vertex of $\psi$ is greater than or equal to half the rank of the perfect tensor, we can apply the contraction rule \eno{ContractRule} at each of the vertices. 
However, we draw just one-half, say the left half, of the entire diagram, and label the vertices with an appropriate power of $r$ wherever the contraction rule \eno{ContractRule} prescribes factors of $r$. 
In practice, as explained below \eno{ContractRule}, this comes down to ``splitting'' open each vertex of the tensor network as shown in the second step in figure \ref{fig:proc}, and assigning any prescribed powers of $r$ at the corresponding vertex, coming from the application of the contraction rule \eno{ContractRule}. Such powers of $r$ will be referred to as ``splits''.

\item {\bf (Cycles.)} Finally, we would like to count the number of new cycles created upon performing all the ``splits'' in step 2. Focusing on the left-half of the diagram in \eno{psiNormStep4}, it is clear that each disconnected bond (line-segment) in step 2 above will end up in a ``cycle'' (disconnected loop). So we simply join together the end-points of each line-segment creating as many cycles as disconnected line-segments. 

\end{enumerate}
Following this procedure to compute $\langle \psi|\psi \rangle$, as depicted in figure \ref{fig:proc}, we verify that we have created $16$ new loops (coming from 16 cycles) and introduced $8$ factors of $r^2$ as before, yielding $\langle \psi|\psi \rangle = r^{16} \times (r^2)^8 = r^{32}$.

\subsection{Norm of a holographic state}
\label{NORM}

In this subsection we work out the norm of a general holographic state dual to the $(p+1)$-regular Bruhat--Tits tree geometry, with cutoff $\Lambda$ and which is constructed as a dual graph tensor network made from perfect tensors of rank-$(r+1)$, where we assume $(r+1)/2 \geq 2\Lambda$. This ground state normalization is necessary for any computation involving the vacuum state or density matrix at the $p$-adic boundary. The black hole and more general backgrounds require an analogous normalization constant which will be determined later.

Let us begin by tabulating the ``type'' of vertices which make up the tensor network corresponding to a general holographic state. Two vertices are of the same ``type'' if the tensors at the respective vertices have the same number of legs (indices) contracted with other tensors. We refer to this number as $v_c$.
Since the total number of legs (contracted and uncontracted) is constant (and equals $r+1$),  vertices of the same type also have identical number of dangling (uncontracted) legs (which we refer to as $v_d$). Thus each type of vertex may appear with a non-trivial multiplicity in the holographic state. 
Some reflection immediately leads to the conclusion that for a fixed cutoff $\Lambda$, all vertices with the number of contracted legs $v_c$ in the set $\{2, 4, \ldots, 2\Lambda\}$ will appear in the tensor network. The multiplicity of each type of vertex is straightforward to work out thanks to the highly symmetric nature of the dual graph. We tabulate the results in table \ref{tb:vertices}.
\begin{table}[h!]
\centering 
\begin{tabular}{|c|c|c|}
\hline
$v_c$ & $v_d$ & multiplicity $M$\\
\hline \hline 
$2\Lambda$ & $(r+1) - 2\Lambda$ & $p+1$ \\
$2\Lambda - 2$ & $(r+1) - (2\Lambda -2)$ & $(p+1)(p^1- p^0)$\\
$2\Lambda - 4$ & $(r+1) - (2\Lambda -4)$ & $(p+1)(p^2- p^1)$\\
\vdots & \vdots & \vdots \\
4 & $(r+1) - 4$ & $(p+1)(p^{\Lambda-2}- p^{\Lambda-3})$\\
2 & $(r+1) - 2$ & $(p+1)(p^{\Lambda-1}- p^{\Lambda-2})$ \\
\hline
\end{tabular}
\caption{Types of vertices in a general holographic state $|\psi \rangle$.}
\label{tb:vertices}
\end{table}
The multiplicities in the table add up to give $\sum_i M^{(i)} = (p+1) p^{\Lambda-1}$, where $M^{(i)}$ is the multiplicity of the vertex type $i$ and we sum over all vertex types. 
This precisely equals the total number of vertices in the tensor network, and consequently the total number of vertices at the boundary of the (cutoff) Bruhat--Tits tree, which is $\mathbb{P}^1(\mathbb{Z}/p^{\Lambda} \mathbb{Z})$.

While computing the norm of $|\psi\rangle$, we argued that we may apply the contraction rule \eno{ContractRule} at each vertex, where the role of $v_d$ gets mapped to $n_c$, the number of legs contracted within  each vertex pair coming from the two copies of $\psi$, while that of $v_c$ is mapped to $n_d$, the number of ``dangling'' legs. 
We put dangling in quotes because in fact these legs are still contracted with other vertices within the {\it same} copy of $\psi$, but for the purposes of applying the contraction rule \eno{ContractRule}, we may treat them as dangling. 
We are able to apply \eno{ContractRule} because $v_d \geq v_c \Rightarrow n_c \geq n_d$. 
As outlined in the previous subsection, we begin by performing ``splits''. 
At each vertex, we pick up a factor of $r^{(n_c-n_d)/2} = r^{(v_d-v_c)/2}$ as prescribed by \eno{ContractRule}.
Since each vertex type comes with a certain multiplicity, we really pick up a factor of $r^{M(v_d-v_c)/2}$ for each vertex {\it type}. 
Multiplying together factors from each vertex type, we obtain that ``splitting'' leads to an overall factor of $r^{N_{\rm splits}}$, where
\eqn{psiNormSplits}{
N_{\rm splits} \equiv \sum_{{\rm type\ }i} M^{(i)} \left({v_d^{(i)} - v_c^{(i)}\over 2}\right) = {p+1 \over 2(p-1)}\left( p^\Lambda(r-3)  - p^{\Lambda-1}(r+1) +4 \right).
}
In the previous example, we specialized to $p=3, \Lambda=2, r=7$, in which case $N_{\rm splits} = {4 \over 2 \cdot 2} \left( 3^2 \cdot 4 - 3^1 \cdot 8 + 4\right) = 16$ as we found earlier.

After the ``splitting'' we proceed to counting the number of new cycles created. Each new cycle contributes a factor of $r$.
Now the number of cycles is equal to the number of disconnected bonds obtained after the splitting. 
This number can be obtained by summing up the number of contracted legs $v_c$ in a single copy of $\psi$ at each vertex, and dividing by half to compensate for the over-counting. This gives, 
\eqn{psiNormLoops}{
N_{\rm cycles} \equiv {1\over 2} \sum_{{\rm type\ }i} M^{(i)} v_c^{(i)} = {p+1 \over p-1} \left(p^\Lambda -1 \right).
}
In our previous example, we had $N_{\rm cycles} = {4\over 2}(3^2-1) = 16$, as expected. 

Combining the results, we obtain
\eqn{psiNorm}{
\log \langle \psi | \psi \rangle = \left(N_{\rm splits} + N_{\rm cycles}\right)\log r =  {p+1 \over 2(p-1)} \left(  p^{\Lambda}(r-1) - p^{\Lambda-1}(r+1) +2 \right) \log r\,.
}
For $\Lambda \gg 1 \Rightarrow r \gg 1$ and fixed $p$, the norm takes the asymptotic form
\eqn{psiNormAsymp}{
\log \langle \psi | \psi \rangle \to  {p+1 \over 2} p^{\Lambda} r \log r\,.
}
One may see from this and other considerations that the dimension of the boundary Hilbert space grows very rapidly. Still, it is possible to make sense of quantum information theoretical quantities such as density matrices in this limit, see section~\ref{AFALGEBRA}.

\subsection{Bipartite entanglement and the Ryu--Takayanagi formula}
\label{GENUS0SINGLE}

As explained in section \ref{GENUSZERORESULTS}, fixing a planar embedding, two given boundary points $x$ and $y$ define a unique (up to the choice of the complimentary set which can be eliminated by specifying the orientation) connected interval on the tensor network, which we denote by $A$. Particularly, $A$ is given as a set of nodes on the tensor network each of which has a number of contracted and uncontracted legs attached to it. Our goal is to compute the entanglement of $A$ with its compliment $B=A^c$. 
We proceed by writing down the pure density matrix for the full holographic state $|\psi\rangle$, 
\eqn{}{
\rho = {1 \over \langle \psi | \psi \rangle } |\psi \rangle\langle \psi |\,,
}
and computing the reduced density matrix obtained by tracing out the region $B$,
\eqn{}{
\rho_A = \Tr_B \rho = {1 \over \langle \psi | \psi \rangle } \Tr_B |\psi \rangle\langle \psi |\,.
}
The trace over region $B$ is performed exactly in the manner we described previously. Graphically, one may represent this trace by taking two copies of $|\psi \rangle$, then ``gluing'' the vertices along $B$. Just as in the computation of the normalization, this set of contractions implements the trace of the density matrix, now only over the qudits in $B$.

The first step is the application of the contraction rule \eno{ContractRule} wherever possible to reduce the density matrix to its simplest form.
At this point, like in the previous section, we parametrize the basis of states ${\cal V}$ such that the reduced density matrix in this basis assumes a Jordan block-diagonal form with all diagonal blocks simply matrices of ones. 
Then the calculation of the von Neumann entropy reduces to the computation of the number of blocks, since as discussed earlier $S = \log N_{\rm blocks}$.

We begin by simplifying the reduced density matrix using the contraction rule \eno{ContractRule} at all vertices in $B$. Like in the previous subsection, we would like keep track of the number of splits and the number of new cycles generated in the process, as these factors not only affect the overall normalization of $\rho_A$ but also dictate the form of the simplified reduced density matrix. We have 
\eqn{}{
N_{\rm splits} &= \sum_{v \in B}  {v_d - v_c \over 2} \cr 
N_{\rm cycles} &= {1 \over 2} \left(\sum_{v \in B} v_c - C_{AB} \right), 
}
where $C_{AB}$ is the number of tensor legs which extend between $A$ and $B$ (equivalently, the number of tensor leg contractions between vertices in $A$ and $B$), and thus are precisely the number of contractions between the two copies of $\psi$ in the diagrammatic representation of the reduced density matrix. For each of the $C_{AB}$ tensor legs, we sum over the possible values in $0,1, \ldots r-1$, giving $r^{C_{AB}}$ terms. In fact, from our discussion in the previous section, it follows that the number of blocks in the block-diagonal representation of $\rho_A$ will be precisely $r^{C_{AB}}$, where each term of the internal sum implies a certain block of non-vanishing matrix elements. 

With the choice of basis explained in the previous subsection, the density matrix becomes block diagonal with identical blocks of all $1$'s. The size of each block can be explicitly determined using the properties of perfect tensors, essentially by counting the number of allowed configurations of external legs for a given assignment of indices on the $C_{AB}$ legs. This is always a power of $r$ which can be determined in terms of other quantities by demanding the usual condition that the trace of the reduced density matrix is $1$. If the size of each block is $r^\sigma \times r^\sigma$, then the total size of the density matrix  is $r^{\sigma +C_{AB}} \times r^{\sigma +C_{AB}}$ (or equivalently  $\dim {\cal V} = r^{\sigma +C_{AB}}$). Thus the explicit form of the reduced density matrix for any single interval can always be written as:
\eqn{RhoAInterval}{
 \rho_A = { r^{N_{\rm splits} + N_{\rm cycles}} \over \langle \psi|\psi \rangle} \begin{pmatrix}
 	\begin{bmatrix} 1 & \cdots & 1 \\
    \vdots & \ddots & \vdots \\
    1 & \cdots & 1
    \end{bmatrix} r^{\sigma} &
     \vspace{-1 mm} \\  \hspace{-3.5 mm} r^{\sigma} \\
    & \ddots &  \\
    & & \vspace{-1mm} \hspace{3.5 mm} r^{\sigma} \\  &  & r^{\sigma} \begin{bmatrix} 1 & \cdots & 1 \\
    \vdots & \ddots & \vdots \\
    1 & \cdots & 1
    \end{bmatrix}
 \end{pmatrix}_{r^{\sigma +C_{AB}} \times r^{\sigma +C_{AB}}}
}
where the number of blocks is $r^{C_{AB}}$. This is a generalization of the earlier examples, where the various parameters depend on $\Lambda$ and the specific interval chosen.

Now we compute the trace
\eqn{}{
\Tr \rho_A = { r^{N_{\rm splits} + N_{\rm cycles}} \over \langle \psi|\psi \rangle} r^{\sigma +C_{AB}}\,,
}
and we recall that we computed the norm of the holographic state $| \psi \rangle$ in the previous subsection. However, for the purposes of simplifying the trace, we note that we can write the norm of $\psi$ as an independent sum over vertices in $A$ and $B$:
\eqn{}{
\log_r \langle \psi|\psi \rangle =  \left( \sum_{v \in A}  {v_d-v_c \over 2}  + \sum_{v \in B}  {v_d-v_c \over 2}  \right) + \left( {1\over 2} \sum_{v \in A} v_c + {1\over 2} \sum_{v \in B} v_c \right). 
}
where the first parenthesis corresponds to the total power of $r$ originating from the contraction rule \eno{ContractRule} upon ``splitting'' all vertices of $\psi$, while the second parenthesis  counts the total number of new loops created after ``splitting'' all vertices of $\psi$. Combining all the results, we obtain an expression which only depends on quantities in region $A$ and the number of bonds which connect $A$ to $B$:
\eqn{TraceCond}{
\log_r \Tr \rho_A = -\sum_{v \in A}  {v_d  \over 2}  + \sigma + {C_{AB} \over 2}\,.
}
For any (normalized) reduced density matrix, the trace is always unity, so the logarithm on the left hand side vanishes. This determines the size of the blocks in terms of other quantities which depend only on the chosen region to be traced out.

Returning back to the calculation of the von Neumann entropy, we have 
\eqn{SACAB}{
S_A = C_{AB} \log r\,.
}
This follows by direct diagonalization of the block diagonal reduced density matrix, and gives an explicit connection between the von Neumann entropy and the number of tensor bonds extending between $A$ and $B$. This kind of behavior for single intervals in perfect tensor networks was an attractive feature of ~\cite{Pastawski:2015qua}, where a general argument was given for this behavior based on the properties of perfect tensors. We will explain in this section that the network proposed here for $p$-adic AdS/CFT gives analogous results for a broader class of physical situations such as black hole backgrounds and multiple intervals.

A key motivation for this specific (i.e.\ ``dual graph'') tensor network is the relationship between $S_A$, thought of as the boundary entanglement entropy between $A$ and $B$, and the bulk geometry of geodesics on the Bruhat-Tits tree. Recall that by construction, $C_{AB}$ is the number of edges in the dual graph tensor network which originate from a vertex in $A$ and end on a vertex outside $A$. 
For example, in the single interval example in figure \ref{fig:connected}, $C_{AB} = 4$, where $A$ is the connected region on the tensor network in between boundary points $x$ and $y$. 
In other words, $C_{AB}$ is the number of edges which start on a tensor network vertex belonging to the region $A$ ``in between'' $x$ and $y$, but end outside this region. 
In the case when the boundary is $\mathbb{P}^1(\mathbb{Q}_p)$ (although this argument also works when the boundary is the ``infinite line'' $\mathbb{Q}_p$) the outside of $A$ is precisely the region ``between'' $x$ and $y$ which is {\it complimentary} to $A$. 
Thus by construction, the edges on the tensor network constituting $C_{AB}$ cut across those edges (geodesics) on the the Bruhat--Tits tree which separate the boundary points $x$ and $y$ into two disconnected parts of the tree, if we imagine cutting the tree at precisely those  edges (geodesics). 
In fact, on a tree geometry, there are precisely as many such edges as the length of the boundary anchored geodesic joining $x$ and $y$ (if we normalize the length of each edge to unity).
Thus we conclude that 
\eqn{CABlength}{
C_{AB} = {\rm length} (\gamma_{xy})/\ell\,,
}
where $\gamma_{xy}$ is the (regulated) geodesic on the Bruhat--Tits tree joining $x$ to $y$ and $\ell$ is the length of each edge on the tree. 
If the boundary is $\mathbb{Q}_p$ so that $x,y \in \mathbb{Q}_p$, then 
\eqn{lengthQp}{
{\rm length} (\gamma_{xy})/\ell &=  d(C,x)+ d(C,y) - 2d(C, {\rm anc}(x,y)) \cr 
  &= 2\Lambda + 2\log_p |x-y|_p \cr 
  &= 2\log_p \left| {x-y \over \epsilon} \right|_p \cr 
}
where $C$ is the any point on the Bruhat--Tits tree, ${\rm anc}(x,y)$ is the unique vertex on the Bruhat--Tits tree where geodesics from $C, x$ and $y$ meet, and $d(\cdot,\cdot)$ measures the graph distance between two points.
We have defined $\epsilon \in \mathbb{Q}_p$ to be the cutoff 
\eqn{}{
\epsilon \equiv p^{\Lambda}\,,
}
which goes to zero $p$-adically, i.e.\ $|\epsilon|_p \to 0$ as $\Lambda \to \infty$. If $x,y \in \mathbb{P}^1(\mathbb{Q}_p)$, then
\eqn{lengthP1Qp}{
{\rm length} (\gamma_{xy})/\ell &= d(C,x)+ d(C,y) - 2d(C, {\rm anc}(x,y)) \cr 
  &= 2\log_p { |{\mathfrak B}(x,y)|_{\rm PS} \over |\epsilon|_p},
}
where we take $C$ to be the ``radial center'' of the Bruhat--Tits tree (so that $d(C,x) = d(C,y) = \Lambda$), and as explained around \eno{RTgenus0circle}, $|{\mathfrak B}(x,y)|_{\rm PS}$ is the Patterson-Sullivan measure of the smallest clopen ball in $\mathbb{P}^1(\bQ_p)$ containing both $x$ and $y$.\footnote{In fact, we can interpret the ``interval size'' $|x-y|_p = |{\mathfrak B}(x,y)|_{\rm Haar}$ in \eno{lengthQp} as the Haar measure of the smallest clopen ball in $\bQ_p$ containing both $x$ and $y$.} As discussed in section \ref{sssec:results}, equations \eno{SACAB}-\eno{lengthP1Qp} correspond to the $p$-adic analog of the RT formula.

We end this discussion by remarking that it is straightforward to see that the lengths in \eno{lengthP1Qp} and \eno{lengthQp} can be re-expressed using the signed-overlap function $\delta(\cdot,\cdot)$ described in section \ref{GENUSZERORESULTS} (see \eno{lengthxy}). This is more convenient because $\delta(\cdot,\cdot)$ admits a choice of alternate paths, which is important in the genus $1$ case (see, for example, sections \ref{RTBTZresults} and \ref{GENUS1BTZ}) where there are always two choices.

\subsection{Bipartite entanglement for a disconnected region and subadditivity}
\label{GENUS0DOUBLE}

So far we have discussed bipartite entanglement entropy only in the case of a connected region, where we are given a pair of points on $\bQ_p$ (or $\mathbb{P}^1(\bQ_p)$). We now extend the discussion to include the case of a ``disconnected region'' (recall definition \ref{def:connectedregion}). 

To specify a disconnected region built from two connected subregions, we must specify a set of four distinct points on the projective line on which the spacial slice of the CFT resides, and which constitute the entangling surface. 
Given a set of four boundary points and a choice of planar embedding for the tensor network, there are two different choices for constructing a (complementary pair of) disconnected region on the tensor network, as previously illustrated in figure \ref{fig:connectedB}. 
Each ``path-disjoint'' piece of a disconnected region is specified by the set of vertices at the boundary of the tensor network ``in between'' a chosen pair of boundary points, just like for connected regions in the previous subsection (recall definition \ref{def:inbetween} for the notion of ``in between''). We define ``path-disjoint'' as follows:
\begin{definition}
Two regions (i.e.\ sets of vertices) on the dual graph tensor network are {\it path-disjoint} if they do not share any common vertices, {\it and}  the tensors located on the vertices in one set are not contracted with the tensors in the other set via any of the ``shortest bonds''.\footnote{Recall that the ``shortest bonds'' on the tensor network are the bonds in bijective correspondence with the UV edges (equivalently the boundary edges) of the cut off Bruhat--Tits tree (see definition \ref{def:shortestbond}). For a choice of planar embedding, the notion of ``shortest bonds'' is $\PGL(2,\bQ_p)$ invariant.}
\end{definition}

 Particularly, we use  the notation $A = (x,y)$ to specify the connected region $A$ in the tensor network which corresponds to vertices on the network between boundary points $x,y$ on the tree. 
 The ordering inside the parenthesis is used to tell $A$ apart from its complement. 
 We will often use the convention that given a planar embedding, the region $A = (x,y)$ is given by the set of vertices ``in between'' $x$ and $y$ going counter-clockwise from $x$ to $y$. 
 Then the two choices for the complimentary pairs of path-disjoint subregions in figure \ref{fig:connectedB} correspond to $A_1 = (x_1,x_2), A_2=(x_3,x_4)$ and $A_3=(x_4,x_1), A_4=(x_2,x_3)$. The disconnected region $A = A_1 \cup A_2$ is the complement of $A^c=A_3 \cup A_4$. In vacuum, we expect $S(A) = S(A^c)$, and indeed this is borne out in our setup. A different choice of planar embedding will lead to a different pair of path-disjoint intervals $A_1=(x_i,x_j), A_2=(x_k,x_\ell)$ for distinct $i,j,k,\ell \in \{1,2,3,4\}$; however the von Neumann entropy $S(A_1A_2)$ will be independent of the choice of embedding, for the same reasons as in the case of the single-interval setup -- in this case it will depend solely on the specified boundary points $x_1, x_2, x_3, x_4$ via a conformally invariant cross-ratio constructed from them.
 For this reason, we  make the following definition: 
 \begin{definition}
 We define the {\it bipartite entanglement entropy of a disconnected region} constructed using boundary points $x_1,x_2,x_3$ and $x_4$, and denoted $S^{\rm disc.}(x_1,x_2,x_3,x_4)$, as the von Neumann entropy of the union $A= A_1 \cup A_2$ of connected regions $A_1 = (x_i, x_j)$ and $A_2=(x_k,x_\ell)$ in any chosen planar embedding, where the distinct indices $i,j,k,\ell \in \{1,2,3,4\}$ are selected such that $A_1$ and $A_2$ are path-disjoint.
 \end{definition}

The calculation of bipartite entanglement in the case where $A = A_1 \cup A_2$ with $A_1, A_2$ appropriately chosen, proceeds almost identically to the single interval calculation described above. 
We begin by applying the contraction rule \eno{ContractRule} on all vertices in $B=A^c$.  
In this case, after applying the diagrammatic methods from the previous subsection, two things may happen. 
Simplifying using the contraction rule, one either ends up with two disjoint pieces for the simplified form of the reduced density matrix (here by disjoint, we mean the diagrammatic representation of the reduced density matrix splits into two pieces which do not share {\it any} edges), or a single connected diagram. 
In either case, the bipartite entanglement for the disconnected region follows the Ryu--Takayanagi formula.

\begin{figure}[thb]
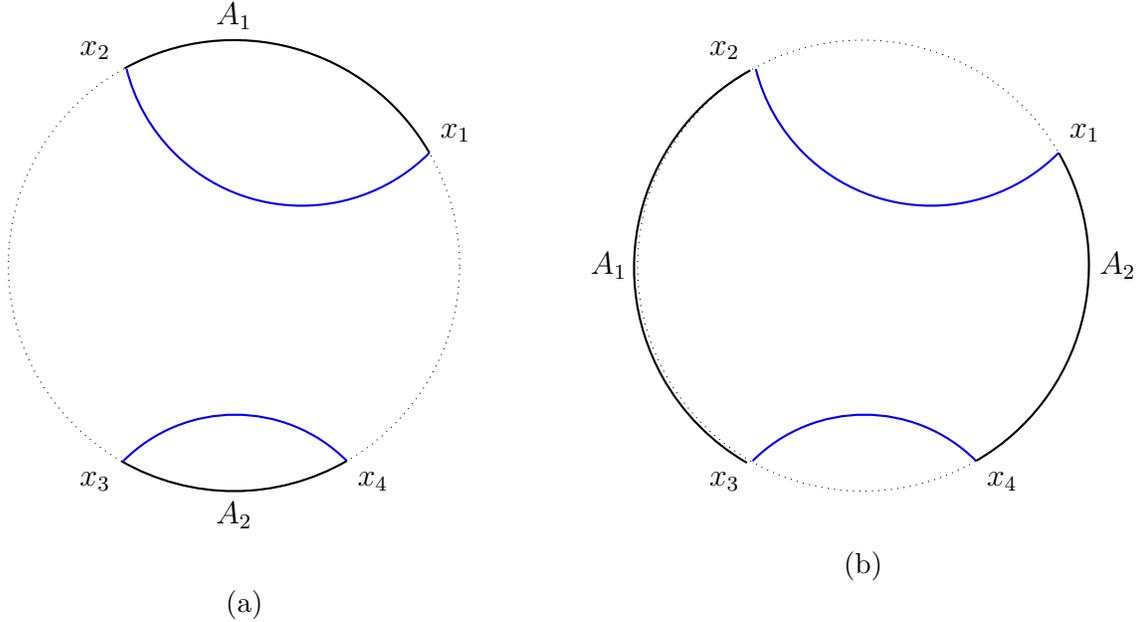

\begin{subfigure}[]{0.49\textwidth}
    \centering
\[
\musepic{\twointIzero}
\]
    \caption{}
    \label{fig:realTwoIntA}
\end{subfigure}
\begin{subfigure}[]{0.49\textwidth}
    \centering
\[
\musepic{\twointIpos}
\]
    \caption{}
    \label{fig:realTwoIntB}
\end{subfigure}
\caption{A schematic representation of the bipartite entanglement entropy calculation for the cases shown in figures \ref{fig:A1A4} and \ref{fig:A1A2}, here (a) and (b) respectively,  to emphasize the parallel with the usual story over the reals. The disconnected region $A= A_1 \cup A_2$ is shown in black, while the minimal geodesics homologous to $A$ is shown in blue.}
\label{fig:realTwoInt}
\end{figure}

The case of the disjoint  reduced density matrix  is interpreted as a direct product reduced state, $\rho_A = \rho_{A_1} \otimes \rho_{A_2}$. 
Each of the disjoint reduced density matrix pieces can be evaluated using the method described in the previous subsection. 
This case is precisely when $S(A)=S(A_1 A_2) = S(A_1) + S(A_2)$, that is, $A_1$ and $A_2$ share no mutual information, defined to be
\eqn{MI}{
I(A_1:A_2) \equiv S(A_1) + S(A_2) - S(A_1 A_2)\,.
}
This general result follows from elementary results on diagonalization of tensor products of density matrices. 
We also write this in the alternate, more suitable notation
\eqn{MIx}{
I(x_i,x_j,x_k,x_\ell) \equiv S(x_i,x_j) + S(x_k,x_\ell) - S^{\rm disc.}(x_1,x_2,x_3,x_4)\,,
}
where $i,j,k,\ell \in \{1,2,3,4\}$ are distinct labels chosen so that $(x_i,x_j)$ and $(x_k,x_\ell)$ are {\it any} connected subregions path-disjoint from each other, and $S^{\rm disc.}(x_1,\ldots,x_4)$ is the bipartite entanglement of the union $(x_i,x_j) \cup (x_k,x_\ell)$.\footnote{In fact as we stressed previously, $S^{\rm disc.}(x_1,\ldots,x_4)$ equals the bipartite entanglement of {\it any} disconnected region constructed from boundary points $x_1,\ldots,x_4$.}
This situation is depicted schematically in figure \ref{fig:realTwoIntA}.

In the other case, where we obtain a reduced density matrix given by a single diagram (as opposed to two disjoint diagrammatic pieces), we can again employ the method of the previous subsection to calculate the reduced density matrix as well as the entanglement entropy.  
In this case $S(A_1 A_2) \leq S(A_1) + S(A_2)$, i.e. the mutual information $I(A_1:A_2)$
is non-negative. This situation is depicted schematically in figure \ref{fig:realTwoIntB}.
In this case the entanglement entropy of the disconnected region $A = A_1 \cup A_2$, $S(A)$ is still given by the logarithm of the number of blocks in the Jordan block-diagonal representation of the reduced density matrix. 
Just like in the case of the single interval, the number of blocks is $r^{C_{AB}}$, where $C_{AB}$ is the number of edges on the tensor network which originate on a vertex in $A = A_1 \cup A_2$ but end outside $A$ (i.e.\ in $B = A^c$). 

Let us now describe these cases in more detail.
The two possible scenarios discussed above can be classified in terms of the entangling surface consisting the boundary points specifying the disconnected region $A$.   If $A_1 = (x_1, x_2)$, and $A_2 = (x_3,x_4)$ are two path-disjoint intervals on the tensor network (with $x_i \neq x_j\,, \forall i,j$) with $A=A_1 \cup A_2$, then the sign of the  logarithm of the cross-ratio 
\eqn{uCrossRatio}{
u(x_1,x_2,x_3,x_4) \equiv \left|{x_{12} x_{34} \over x_{14}x_{23}}\right|_p,
}
 dictates which of the scenarios depicted in figure \ref{fig:realTwoInt} will occur. 

If the logarithm is non-positive, then the pairwise boundary anchored geodesics between $x_1$ \& $x_2$ and $x_3$ \& $x_4$ do not overlap (i.e. they intersect at most at a single vertex) on the Bruhat--Tits tree, and in fact constitute the minimal surfaces homologous to $A_1$ and $A_2$ respectively.\footnote{Recall definition \ref{def:homologous} for the homologous condition.} 
Thus $S^{\rm disc.}(x_1,x_2,x_3,x_4)=S(A_1A_2) = S(A_1) + S(A_2)$.

\begin{figure}[t]
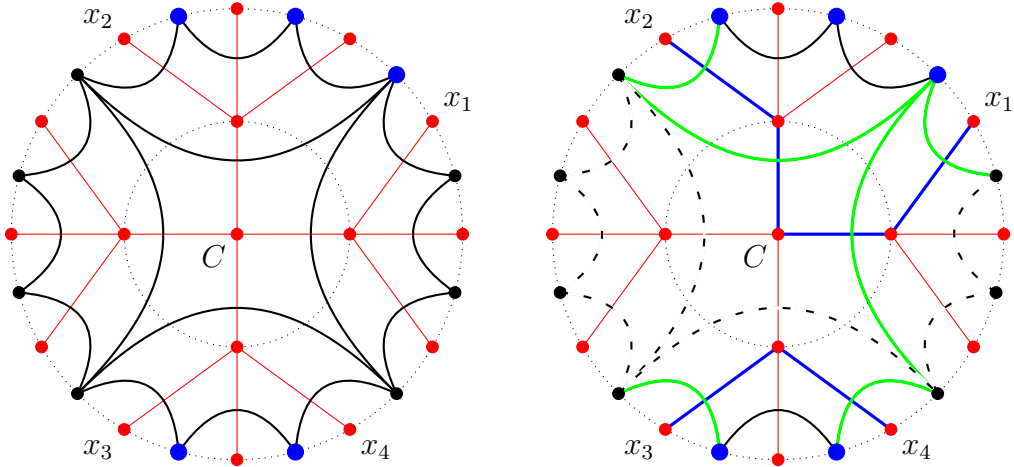

\centering
\[
\musepic{\twointplain}
\qquad
\musepic{\twointcolored}
\]
\caption{The setup for the bipartite entanglement calculation for a disconnected region, in the case when the cross-ratio $u(x_1,x_2,x_3,x_4)$ defined in \eno{uCrossRatio} is strictly less than one. Left: The holographic state with the region marked in blue. Right: Computation of the reduced density matrix, and the depiction of the minimal surface homologous to the boundary region. See the main text for the detailed explanation of the figure.}
\label{fig:A1A4}
\end{figure}

\begin{figure}[t]
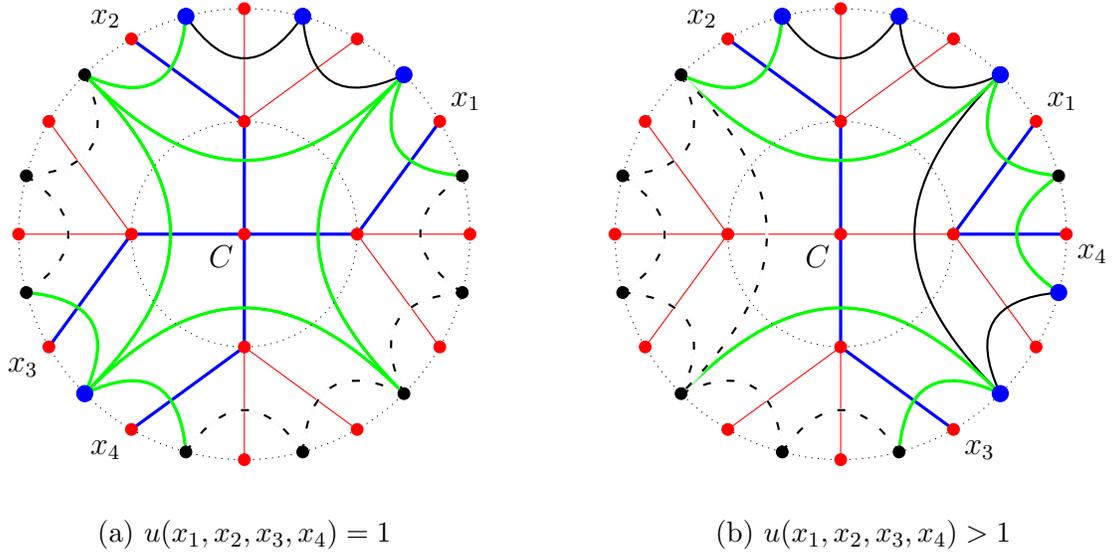

\begin{subfigure}[]{0.49\textwidth}
    \centering
\[
\musepic{\intCRone}
\]
    \caption{$u(x_1,x_2,x_3,x_4)=1$}
    \label{fig:A1A3}
\end{subfigure}
\begin{subfigure}[]{0.49\textwidth}
    \centering
\[
\musepic{\intCRlarge}
\]
    \caption{$u(x_1,x_2,x_3,x_4)>1$}
    \label{fig:A1A2}
\end{subfigure}
\caption{The case of a disconnected region with cross-ratio $u(x_1,x_2,x_3,x_4)$ defined in \eno{uCrossRatio} greater than or equal to 1.}
\end{figure}

Figure~\ref{fig:A1A4} shows an example of a disconnected region setup for $A_1=(x_1,x_2)$, $A_2=(x_3,x_4)$ with $u(x_1,x_2,x_3,x_4)<1$.\footnote{In figure \ref{fig:A1A4}, $u(x_1,x_2,x_3,x_4) = p^{-1}$. See the discussion around \eno{ugraph} for the explanation.}
As usual, the tensor network is shown in red with regions marked by blue vertices, while the Bruhat--Tits tree geometry is shown in black. 
Applying the contraction rule \eno{ContractRule} to the trace out the complement of $A = A_1 \cup A_2$ in the state shown in the left subfigure ``splits'' open vertices on the tensor network marked in black (in the manner described in the previous subsections), while the dashed bonds on the tensor network turn into ``cycles''. 
The bonds marked in green connect vertices in $A$ to vertices in the complement of $A$.
The simplified reduced density matrix is given by the subdiagram containing blue vertices, along with black and green colored bonds, where we remember to ``split open'' all the black vertices so that all bonds originally coincident on such vertices no longer meet. 
Thus the reduced density matrix manifestly decomposes into two disjoint pieces, and can be written as the direct product of the density matrices for the individual sub-intervals.
The number of green bonds in each individual piece correspond to the (base-$r$ logarithm of the) number of blocks in the Jordan block form of the individual reduced density matrices.
It is clear from counting the green bonds that $S^{\rm disc.}(x_1,\ldots,x_4)=S(A_1  A_2) = S(A_1)+S(A_2)$.\footnote{Particularly, applying the methods from the previous subsection, we have $S(A_1)=4, S(A_2)=2$ and $S(A)=6$, which can be confirmed visually in figure \ref{fig:A1A4} by counting the length of the corresponding minimal geodesics homologous to various regions.}  
The edges on the Bruhat--Tits tree corresponding to the green bonds are highlighted in blue. Together they correspond to the minimal length boundary-anchored geodesics homologous to $A$. 
Thus the RT formula is satisfied.
Schematically, this case is the $p$-adic analog of the configuration depicted in figure \ref{fig:realTwoIntA}, where the minimal surface homologous to a disconnected region is the union of the minimal surfaces homologous to each disjoint piece of the region separately.
Figure~\ref{fig:A1A3} shows the disconnected region setup for $A_1 = (x_1,x_2)$, $A_2 = (x_3,x_4)$ with $u(x_1,x_2,x_3,x_4) = 1$. The analysis in this case proceeds identically to the one above, so we do not repeat it here.

In figure \ref{fig:A1A4}, we could have considered the alternate choice of path-disjoint intervals, $A_1^\prime=(x_4,x_1)$ and $A_2^\prime=(x_2,x_3)$ with the full disconnected region given by $A^\prime = A_1^\prime \cup A_2^\prime$. Then the cross-ratio of interest \eno{uCrossRatio} would become
\eqn{}{
u(x_4,x_1,x_2,x_3) = \left|{x_{41}x_{23} \over x_{43}x_{12}}\right|_p = {1 \over u(x_1,x_2,x_3,x_4)} > 1\,.
}
We discuss this case in more detail next.
Before proceeding, we note that in this case although the individual von Neumann entropies $S(A_1^\prime)$ and $S(A_2^\prime)$ will in general differ from $S(A_1)$ and $S(A_2)$, the von Neumann entropy of the union $S(A_1^\prime  A_2^\prime) = S(A_1  A_2) = S^{\rm disc.}(x_1,\ldots,x_4)$, and in fact the minimal surface homologous to $A_1^\prime \cup A_2^\prime$ is the same as the minimal surface homologous to $A_1 \cup A_2$. Thus the bipartite von Neumann entropy is independent of the choice of choosing the disconnected region  given a set of four boundary points.\footnote{The two choices are illustrated in figure \ref{fig:connectedB}.} Moreover, it is independent of the choice of the planar embedding.

Let the logarithm of the cross-ratio $u(x_1,x_2,x_3,x_4)$ be positive; then the boundary anchored geodesics between $x_1$ \& $x_2$ and $x_3$ \& $x_4$ overlap (i.e.\ share non-zero edges on the Bruhat--Tits tree). 
The bulk interpretation of  mutual information $I(x_1,\ldots,x_4)=I(A_1:A_2)$ is that it is given precisely by (twice) the number of edges of overlap between the minimal geodesics homologous to $A_1$ and $A_2$ individually. 
Each such edge corresponds to a bond on the tensor network which extends from a node in $A_1$ to a node of $A_2$, explaining why the combination $I(A_1:A_2)=S(A_1)+S(A_2)-S(A_1A_2)$ is given precisely by twice the number of such edges (up to an overall factor of $\log r$). From the boundary perspective, the mutual information between the path-disjoint intervals $A_1, A_2$ is given by
\eqn{MIbdy}{
I(x_i,x_j,x_k,x_\ell) = I(A_1:A_2) = (2\log r)\: \log_p u(x_i,x_j,x_k,x_\ell)\,,
}
provided $u(x_1,\ldots,x_4)>1$ (which is the same as $u(x_1,\ldots,x_4) \geq p$ since the $p$-adic norm in \eno{uCrossRatio} takes values in $p^\bZ$), with $A_1=(x_i,x_j)$, $A_2=(x_k,x_\ell)$ where $i,j,k,\ell \in \{1,\ldots, \}$ are distinct and chosen such that $A_1, A_2$ are path-disjoint. This result follows from a basic entry in the $p$-adic holographic dictionary between cross-ratios and graph distances on the Bruhat--Tits tree,
\eqn{ugraph}{
\left|{(x-y)(z-w) \over (x-z)(y-w)}\right|_p = p^{-d(a,b)}\,,
}
where $x,y,z,w \in \mathbb{P}^1(\bQ_p)$ such that the bulk geodesics joining $x$ to $z$, and $y$ to $w$ intersect precisely along the path between the bulk points $a$ and $b$ on the Bruhat--Tits tree, and $d(a,b)$ is the graph distance between $a$ and $b$.\footnote{If the bulk geodesics do not intersect along a path on the tree, \eno{ugraph} can still be used after a simple relabelling of the boundary points.}

This possibility is depicted in figure \ref{fig:A1A2}.
We consider the disjoint intervals  $A_1 = (x_1,x_2)$ and $A_2= (x_3,x_4)$ with $A = A_1 \cup A_2$ and $u(x_1,x_2,x_3,x_4) > 1$.
 Applying the contraction rule \eno{ContractRule} to the trace out the complement of $A$ in the state on the left ``splits'' open vertices marked in black, while the dashed bonds on the tensor network turn into cycles. 
Like for figure \ref{fig:A1A4}, the bonds marked in green connect vertices in $A$ to vertices in the complement of $A$,
and the simplified reduced density matrix is given by the subdiagram containing blue vertices, along with black and green bonds. 
Importantly, the reduced density matrix in this case does not split into individual disjoint pieces.
The number of green bonds once again correspond to the (base-$r$ logarithm of the) number of blocks in the Jordan block form of the reduced density matrix, and thus contribute to the von Neumann entropy as explained in section \ref{GENUS0SINGLE}.
It is clear from counting the green bonds that $S^{\rm disc.}(x_1,\ldots,x_4) = S(A_1  A_2) < S(A_1)+S(A_2)$.\footnote{Particularly in figure \ref{fig:A1A2}, $S(A_1)=4, S(A_2)=4$ and $S(A_1A_2)=6$.} The excess on the r.h.s.\ (or equivalently the non-zero mutual information $I(A_1:A_2)$) can precisely be accounted for by the existence of a black bond on the tensor network joining a vertex of $A_1$ with a vertex of $A_2$. 
{\it This establishes the non-negativity of mutual information.}
The edges on the Bruhat--Tits tree corresponding to the green bonds are highlighted in blue, and together they specify the minimal boundary-anchored geodesics homologous to the boundary region $A$.
Thus once again the RT formula holds.
This case is the $p$-adic analog of the situation in figure \ref{fig:realTwoIntB}, where the two regions share mutual information.

We note that while the sample computations presented above for the disconnected interval case considered specific examples (for a specific choice of cutoff and specific value of $p$), the lessons and results obtained here hold in full generality for arbitrarily chosen disconnected regions and arbitrary cutoffs for any prime $p$.
In summary, from the boundary perspective, given path-disjoint  regions $A_1=(x_i,x_j)$ and $A_2=(x_k,x_\ell)$ with $i,j,k,\ell \in \{1,2,3,4\}$ distinct and $A_1, A_2$ path-disjoint, the sign of $\log u(x_i,x_j,x_k,x_\ell)$ fixes whether the mutual information $I(x_i,x_j,x_k,x_\ell)=I(A_1:A_2)$ is positive or vanishing. 
Combining the cases described above, we write
\eqn{MIbdyGen}{
I(x_i,x_j,x_k,x_\ell) &= (2\log r)\: \max\{ 0, \log_p u(x_i,x_j,x_k,x_\ell) \} \cr 
&= (2 \log r)\: \gamma_p\!\left({x_{i\ell}x_{jk} \over x_{ij}x_{k\ell}} \right) \log_p \left|{x_{ij}x_{k\ell} \over x_{i\ell}x_{jk}} \right|_p \cr 
&\geq 0,
}
where $\gamma_p(x)$ is the characteristic function on $\bZ_p$, that is $\gamma_p(x) = 1$ if $x \in \bZ_p$ (equivalently $|x|_p \leq 1$), and zero otherwise, and we emphasize the non-negativity of mutual information in the final inequality.
From the bulk perspective, mutual information equals twice the number of shared edges between the minimal surfaces homologous to the individual regions $A_1$ and $A_2$ (or equivalently twice the number of bonds in the tensor network which start in $A_1$ and end in $A_2$) up to an overall factor of $\log r$.
The entropy of the disconnected region, $S(A)$ equals the entropy of any disconnected region built out of boundary points $x_1,\ldots,x_4$, thus we alternately denote the bipartite entropy as $S(x_1,\ldots,x_4)$. Using the results from this  and the previous subsections, we write
\eqn{Sdisconnect}{
& S^{\rm disc.}(x_1,x_2,x_3,x_4) \cr 
    &= S(x_i,x_j) + S(x_k,x_\ell) - I(x_i,x_j,x_k,x_\ell) \cr 
    &= 2\left( \log_p { |\mathfrak{B}(x_i,x_j)|_{\rm PS} \over |\epsilon|_p} + \log_p { |\mathfrak{B}(x_k,x_\ell)|_{\rm PS} \over |\epsilon|_p} -  \gamma_p\!\left({x_{i\ell}x_{jk} \over x_{ij}x_{k\ell}} \right) \log_p \left|{x_{ij}x_{k\ell} \over x_{i\ell}x_{jk}} \right|_p \right)\log r  \cr
    &=  \big( \delta(x_i \to x_j,x_i \to x_j) + \delta(x_k \to x_\ell,x_k \to x_\ell) - 2| \delta(x_i \to x_j,x_k \to x_\ell)| \big) \log r
}
where we used \eno{SACAB}, \eno{CABlength}, \eno{lengthP1Qp} and \eno{lengthxy} for the first two terms in the last equality above,\footnote{Moreover, we made convenient choices for the arbitrary node $C$ in \eno{lengthxy} to obtain the simplified forms in \eno{Sdisconnect}.} while the third term in the last equality explicitly counts the number of shared edges between the minimal surfaces $x_i \to x_j$ and $x_k \to x_\ell$ (for regions $A_1$ and $A_2$ respectively) which appear in the bulk interpretation of mutual information. 
Recall that the indices $i,j,k,\ell \in \{1,2,3,4\}$ are chosen such that the subregions $(x_i,x_j)$ and $(x_k,x_\ell)$ are path-disjoint. Thus the three terms on the r.h.s.\ in the first equality of \eno{Sdisconnect} depend on the initial choice of the dual graph tensor network (i.e.\ the choice of planar embedding). 
The second and third equalities show explicitly the functional form of the three terms, entirely in terms of the boundary coordinates (as well as the UV cut off). 
We now show that $S^{\rm disc.}(x_1,x_2,x_3,x_4)$ as given by the particular combination  in \eno{Sdisconnect} is {\it independent} of the choice of a planar embedding. 

\begin{figure}[!t]
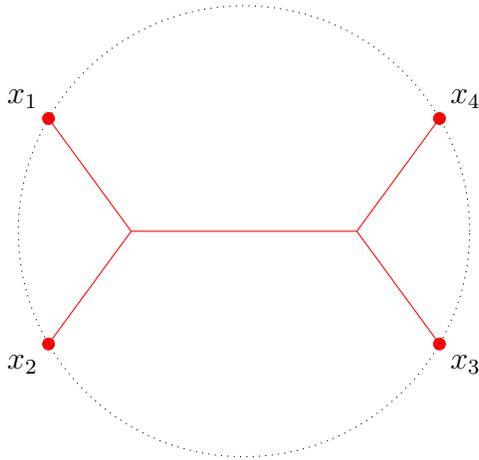

    \centering
    \[
    \musepic{\uconfig}
    \]
    \caption{Boundary anchored bulk geodesics on the Bruhat--Tits tree and the boundary point configuration such that $u(x_1,x_2,x_3,x_4) \leq 1$.}
    \label{fig:config}
\end{figure}

Assume, without loss of generality, $u(x_1,x_2,x_3,x_4) \leq 1$ (the associated configuration on the Bruhat--Tits tree is shown in figure \ref{fig:config}). (If not, we relabel the boundary points to ensure the inequality holds.) The choice of the path-disjoint intervals $A_1=(x_i,x_j)$ and $A_2=(x_k,x_\ell)$ depends on the choice of the planar embedding. Depending on the chosen planar embedding, there are in all three inequivalent possibilities:\footnote{Technically, one needs to be careful about the orientation of the interval to ensure there is no overlap, but we will assume  proper orientations have already been chosen to ensure the subregions are path-disjoint. This simply amounts to being careful about the order of the boundary points in specifying the intervals $A_1, A_2$; however the argument in the following is insensitive to this ordering as long as we assume the intervals are path-disjoint.} 
\begin{itemize}
    \item $x_i=x_1, x_j=x_2, x_k=x_3, x_\ell = x_4$, or
    \item $x_i=x_1, x_j=x_3, x_k=x_2, x_\ell = x_4$, or
    \item $x_i=x_1, x_j=x_4, x_k=x_2, x_\ell = x_3$.
\end{itemize}
In each of the three cases, the graph-theoretic quantity in the third line of \eno{Sdisconnect} is easily computed to explicitly verify that $S^{\rm disc.}(x_1,x_2,x_3,x_4)$ is identical in all cases, and in fact equals the {\it minimal} length of the geodesics homologous to the intervals $A_1$ and $A_2$.\footnote{This observation is also illustrated in figure \ref{fig:realTwoInt} where the two cases shown have the same minimal surface homologous to the disconnected regions.}
Thus the bipartite entanglement for a disconnected region is given exactly by the RT formula. 
The bipartite entanglement $S^{\rm disc.}(x_1,\ldots,x_4)$ as defined above is not only $\PGL(2,\bQ_p)$ invariant but also independent of the choice of planar embedding (equivalently the choice of a valid tensor network associated with the bulk geometry).

We are now in a position to show that the Araki--Lieb inequality~\cite{Araki1970} is satisfied as well. In our setup, this corresponds to showing that the inequality
\eqn{AL}{
S(x_1,x_2,x_3,x_4) \geq |S(x_i,x_j) - S(x_k,x_\ell)|
}
holds, where the various terms are defined below \eno{MIx}. Once again, without loss of generality, we assume the boundary point configuration of figure \ref{fig:config}. 
As already discussed, in this case, $S(x_1,\ldots,x_4)$ is proportional to the sum of the lengths of the unique geodesics joining $x_1$ to $x_2$ and $x_3$ to $x_4$. 
On the other hand, the entropies of the path-disjoint intervals, $S(x_i,x_j)$ and $S(x_k,x_\ell)$ are proportional to lengths of the unique geodesics joining $x_i$ to $x_j$ and $x_k$ to $x_\ell$, respectively. 
Comparing lengths of geodesics in figure \ref{fig:config} it is immediately clear that for all possible choices of $i,j,k,\ell$ (subject to the requirements specified below \eno{MIx}), the inequality \eno{AL} reduces to checking whether $a+b \geq |a-b|$ for positive real numbers $a,b$. This is clearly true, and thus establishes the Araki--Lieb inequality for path-disjoint intervals. 
The simplicity in comparison of lengths of geodesics such as the ones in figure \ref{fig:config} is a direct consequence of ultrametricity of the $p$-adic numbers (or equivalently, the simplifying tree structure of the bulk geometry).

Together, the Araki--Lieb inequality and the non-negativity of mutual information, which we have now established for path-disjoint intervals, are referred to as the subadditivity property of entropy. 
In the next subsection, we define path-adjoining intervals, and the proofs presented here  extend easily to this case (we leave them as trivial exercises for the reader), thus establishing subadditivity in full generality in our setup.

\subsection{More entropy inequalities: SSA and MMI}
\label{ssec:Inequalities}

So far we have shown that the $p$-adic bipartite entropy satisfies an RT-like formula, as well as  subadditivity of entropy. One should  also expect strong subadditivity (SSA) and  monogamy of mutual information (MMI)~\cite{Hayden:2011ag} to hold.  Indeed in this section we establish these inequalities holographically. 

Given three regions $A_1, A_2$ and $A_3$, SSA is the statement that~\cite{Lieb:1973zz,Lieb:1973cp}
\eqn{SSA123}{
S(A_1 A_2) + S(A_2 A_3) \geq S(A_1A_2A_3) + S(A_2)\,,
}
or equivalently\footnote{The inequality \eno{SSA123alt} can be obtained from \eno{SSA123} (and vice versa) by first purifying the system $\rho_{A_1A_2A_3}$ by formally adding a fourth region $A_4$.}
\eqn{SSA123alt}{
S(A_1 A_2) + S(A_2 A_3) \geq S(A_1) + S(A_3)\,.
}
In the previous subsection, we  discussed the bipartite entropy of unions of path-disjoint regions. However, here we will focus on regions $A_i$ such that in a given planar embedding, they are disjoint (i.e.\ they do not share any nodes on the tensor network) but are ``adjoining'', that is they share end-points on the Bruhat--Tits tree. We refer to them as ``path-adjoining'' regions.
\begin{definition}
Given a planar embedding, two regions $A_1$ and $A_2$ are {\it path-adjoining} if they are disjoint as sets of nodes on the tensor network, but there exists exactly one ``shortest bond'' on the network which contracts a vertex in $A_1$ with a vertex in $A_2$.
\end{definition}
A consequence of this definition is that if two regions are path-adjoining, then written as a set of nodes ``in between'' two boundary points, the two regions share a common boundary point. (This notion is also $\PGL(2,\mathbb{Q}_p)$ invariant.) The converse of this statement is not always true.

Given four boundary points $x_1,\ldots,x_4$ and any choice of a planar embedding, we will assume that regions $A_1$ and $A_2$ are path-adjoining as well as $A_2$ and $A_3$ are path-adjoining.\footnote{\label{fn:pathdisjoint}In this paper, we will not discuss the case where the regions $A_1, A_2$ and $A_3$ are path-disjoint from each other, although we expect SSA to hold here as well. This case requires an input data of six distinct boundary points. The notion of bipartite entropy presented in section \ref{GENUS0DOUBLE} given a set of four boundary points should generalize in a systematic way to the case of six (and higher) boundary points, but we leave this for future work.} 
Without loss of generality, we take $A_1=(x_i,x_j), A_2=(x_j,x_k)$ and $x_3=(x_k,x_\ell)$, where $i,j,k,\ell$ are distinct indices from the set $\{1,2,3,4\}$ chosen such that $A_1$ and $A_2$ are path-adjoining, and similarly for $A_2$ and $A_3$. 
This setup might be familiar to the reader from the holographic proof of strong subadditivity over the reals~\cite{Headrick:2007km}. The proof presented here is similar in spirit  but has a distinct $p$-adic flavor as will be apparent shortly.

Under the hypotheses of the previous paragraph, we can write
\eqn{Sindiv}{
S(A_1) &= S(x_i,x_j) \qquad\quad S(A_2) = S(x_j,x_k) \qquad\qquad S(A_3) = S(x_k,x_\ell) \cr 
S(A_1 A_2) &= S(x_i,x_k) \qquad S(A_2 A_3) = S(x_j,x_\ell) \qquad S(A_1A_2A_3) = S(x_i,x_\ell)\,,
}
where the two-point bipartite entropy $S(x,y)$ is the entropy of a connected region $(x,y)$, discussed previously in section \ref{GENUSZERORESULTS}. Consequently all terms in \eno{SSA123} (and \eno{SSA123alt}) turn into bipartite entropies of connected regions. Thus to check SSA, we need to show
\eqn{ToShowSSA}{
S(x_i,x_k) + S(x_j,x_\ell) \stackrel{!}{\geq} S(x_i,x_\ell) + S(x_j,x_k) \qquad S(x_i,x_k) + S(x_j,x_\ell) \stackrel{!}{\geq} S(x_i,x_j) + S(x_k,x_\ell)\,.
}
To be concrete, (without loss of generality) we label the given boundary points $x_1,\ldots,x_4$ such that the pairwise boundary anchored bulk geodesics intersect as shown in figure \ref{fig:config}. Now recall from section \ref{GENUSZERO} that in a genus zero background,  $S(x,y)$ is given simply by the unique minimal geodesic joining boundary points $x$ and $y$. Thus the inequalities in \eno{ToShowSSA} turn into (trivial) statements about lengths of various boundary anchored geodesics in figure \ref{fig:config}. Further they can be related them directly to the conformal cross-ratios as we now explain.

For example, suppose in a planar embedding we can choose $i=1, j=2, k=3, \ell=4$. Then it is clear from comparing lengths of minimal geodesics in figure \ref{fig:config} that the first inequality in \eno{ToShowSSA} is saturated, while the second one is obeyed in the strict sense. In fact, the equality $S(x_1,x_3)+S(x_2,x_4) = S(x_1,x_4) + S(x_2,x_3)$ of the lengths of geodesics has the same content as the triviality of the cross-ratio, $u(x_1,x_3,x_2,x_4) = |(x_{13} x_{24})/ (x_{14}x_{23})|_p = 1$. Similarly, the inequality $S(x_1,x_3)+S(x_2,x_4) > S(x_1,x_2)+S(x_3,x_4)$ has identical content as the inequality $u(x_1,x_2,x_4,x_3) = |(x_{12} x_{34})/ (x_{13}x_{24})|_p < 1$.\footnote{Refer to the discussion around \eno{ugraph}.}
The 23 other permutations of assignments for $i,j,k,\ell$ which could possibly be made over all possible planar embeddings admit  identical analysis so we do not repeat it here. This confirms SSA.

To summarize, in each case one of the two inequalities in \eno{ToShowSSA} is saturated,\footnote{In the case of path-disjoint intervals $A_1, A_2$ and $A_3$ (see the comment in footnote 
\ref{fn:pathdisjoint}), we do not expect such a saturation of one of the inequalities to hold in general.}  while the other remains an inequality.\footnote{The inequality is obeyed strictly unless the boundary points $x_1,\ldots,x_4$ are such that the geodesics connecting them in the bulk meet at a single bulk point. This corresponds in figure \ref{fig:config} to the collapse of the internal bulk geodesic to a single point.} SSA is interpreted as an inequality between lengths of geodesics and  admits a dual description in terms of boundary cross-ratios. 
From the boundary perspective, SSA (for path-adjoining regions) has the same content as the following statement about cross-ratios: Given four boundary points $x_1,\ldots,x_4$, up to a relabelling of coordinates one always has $|(x_{12}x_{34})/(x_{13}x_{24})|_p \leq 1$ with $|(x_{14}x_{23})/(x_{13}x_{24})|_p = 1$, which follows from the ultrametric nature of $p$-adic numbers.

\begin{figure}[!t]
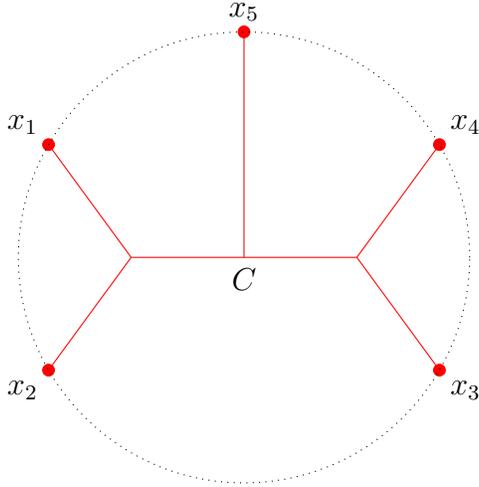

    \centering
    \[
    \musepic{\uvconfig}
    \]
    \caption{Boundary anchored bulk geodesics on the Bruhat--Tits tree and the boundary point configuration such that $u(x_1,x_2,x_4,x_5) \leq 1$ and $u(x_3,x_4,x_1,x_5) \leq 1$. Up to relabelling, this is the most general configuration of five boundary points at the terminus of the Bruhat--Tits tree.}
    \label{fig:5ptconfig}
\end{figure}

Next let's turn to MMI (also referred to as the negativity of tripartite information)~\cite{Hayden:2011ag}. Given three disjoint intervals $A, B$ and $C$ (in our terminology in a given planar embedding, they can be either path-disjoint or path-adjoining or a mix of both such that no two intervals overlap), MMI is the following inequality obeyed by mutual information,
\eqn{MMI}{
I(A:BC) \geq I(A:B) + I(A:C)\,
}
or equivalently
\eqn{MMIalt}{
I(A:B:C) \equiv I(A:B) + I(A:C) - I(A:BC) \leq 0\,.
}
Such an inequality does not hold in general for arbitrary quantum mechanical states, but is special to quantum states which admit a holographic dual. The inequality makes sense even for adjoining intervals since the divergences in the individual mutual information pieces  cancel out among the various terms.
We will now prove \eno{MMI} holds in the $p$-adic tensor network setting for (connected) intervals $A,B$ and $C$ chosen in an arbitrary planar embedding such that they are either path-adjoining or path-disjoint but never overlapping. 
In fact we show the inequality is always saturated.
We will first prove this in the case these intervals are specified in terms of given  set of {\it five} boundary points (see figure \ref{fig:5ptconfig}), and then extend this to full generality.

Fix a planar embedding. Let's first consider the case where $B$ and $C$ are chosen such that they are path-adjoining but $B \cup C$ is path-disjoint from $A$. There are then three inequivalent choices of intervals in figure \ref{fig:5ptconfig}:\footnote{We will suppress keeping track of orientation of intervals and simply assume the intervals are chosen with the correct orientation such that they are path-adjoining or path-disjoint as desired. Keeping track of orientations simply adds an extra layer of detail without changing the basic analysis.}
\begin{itemize}
    \item $A=(x_1,x_2), B=(x_3,x_4), C=(x_4,x_5),$ or
    \item $A=(x_5,x_1), B=(x_2,x_3), C=(x_3,x_4),$ or
    \item $A=(x_2,x_3), B=(x_4,x_5), C=(x_5,x_1).$
\end{itemize}
In each of these cases, the mutual information measures $I(A:BC), I(A:B)$ and $I(A:C)$ take the form of $I(x_i,x_j,x_k,x_\ell)$ for appropriately chosen $i,j,k,\ell$ and can be determined simply by considering the overlap of minimal geodesics for given intervals $A, B, C$ and $B\cup C$, as discussed in detail in section \ref{GENUS0DOUBLE}. We immediately see that the inequality \eno{MMI} is saturated in all three cases -- in the first case each of the individual mutual information measures are identically zero, while in the second and third cases there is non-trivial overlap of minimal geodesics so not all $I$'s vanish.

The remaining cases involve choosing $A, B$ and $C$ such that $A$ is path-adjoining to either $B$ or $C$, where at the same time $B$ and $C$ may be path-adjoining or path-disjoint to each other. In all such cases, some of the mutual information measures will diverge, but the divergences still cancel out on both sides of \eno{MMI}. We leave it as a simple exercise to the reader to consider the finitely many inequivalent cases to consider and verify that in each case, \eno{MMI} is satisfied, and in fact saturated. Thus we conclude that in the case of five points, \eno{MMI} is saturated in the $p$-adic setting. 

The restriction to five boundary points  allowed us to prove \eno{MMI} in the $p$-adic setting in almost full generality. The only case remaining is when the intervals $A,B$ and $C$ are pairwise path-disjoint, in which case we need six boundary points to specify the intervals. Once again there are a small number of cases to individually consider, with the analysis identical to the previously studied cases. We find \eno{MMI} is saturated here as well. In all, we conclude in the $p$-adic tensor network considered in this paper, $I(A:B:C) = 0$, that is
\eqn{MMIpadic}{
I(A:BC) = I(A:B) + I(A:C)\,.
}
It is interesting to compare this observation with the result over real CFTs where it was shown that the inequality is saturated for a massless fermion in two dimensions (i.e.\ mutual information is exactly extensive), but not for instance for the massless scalar~\cite{Hayden:2011ag,Casini:2004bw,Casini:2005rm}.

\subsection{Black hole backgrounds}
\label{GENUS1BTZ}

We now change gears and present some of the computational methods and results for black hole entropy and a Ryu--Takayanagi like formula for minimal geodesics in the black hole background. This discussion is essentially an extension of sections~\ref{BTZResults} and \ref{RTBTZresults}; we refer to these for the basic setup of the tensor network geometry. We will define the integer length of the horizon to be $\tau = \log_p|q|_p^{-1}$ with $q$ our uniformizing parameter; we also assume a nondegenerate geometry so $\tau > 1$. As before, the cutoff is $\Lambda$ (defined now from the horizon) and the rank is $r+1$, where we assume $(r+1)/2 \geq 2\Lambda + 1$ and $(r+1)/2 \geq \tau$. The first condition ensures there are enough boundary dangling legs for minimal surfaces to extend into the bulk and the second condition ensures a similar property for the center vertex.

As in the genus 0 case, our first task is to calculate the norm of the black hole boundary state. This quantity is obtained by contracting all legs, including those at the center vertex behind the horizon. Denoting the boundary state of a black hole of size $\tau$ as $|\psi_{\tau} \rangle$, we may compute $\langle \psi_{\tau} | \psi_{\tau} \rangle$ graphically using techniques of the previous sections. This involves counting the number of each type of tensor, where again the type is the number of legs contracted with other tensors (in the network), $v_c$. The number of dangling legs is $v_d$, and $v_c + v_d = r+1$ for every tensor.

In the black hole geometry, one can see all vertices with $v_c$ in the set $\{2, 4, \ldots, 2\Lambda,  2\Lambda +1\}$ will appear, as well as the central vertex which always has $v_c = \tau$. The counting is similar to the genus $0$ case, though all numbers explicitly depend on $\tau$. The multiplicity of each type of vertex is found in table \ref{tb:BTZvertices}, where we have singled out the center vertex multiplicity, even though it may coincide with one of the other types of tensors.
\begin{table}[h!]
\centering 
\begin{tabular}{|c|c|c|}
\hline
$v_c$ & $v_d$ & multiplicity $M$\\
\hline \hline 
$\tau$ & $(r+1) - \tau$ & $1$ \\
$2\Lambda + 1$ & $(r+1) - (2\Lambda+1)$ & $\tau$ \\
$2\Lambda$ & $(r+1) - 2\Lambda$ & $(p-2)\tau$ \\
$2\Lambda - 2$ & $(r+1) - (2\Lambda -2)$ & $(p-1)(p^1- p^0) \tau$\\
$2\Lambda - 4$ & $(r+1) - (2\Lambda -4)$ & $(p-1)(p^2- p^1) \tau$\\
\vdots & \vdots & \vdots \\
4 & $(r+1) - 4$ & $(p-1)(p^{\Lambda-2}- p^{\Lambda-3}) \tau$\\
2 & $(r+1) - 2$ & $(p-1)(p^{\Lambda-1}- p^{\Lambda-2}) \tau$ \\
\hline
\end{tabular}
\caption{Types of vertices in a black hole holographic state $|\psi_{\tau} \rangle$.}
\label{tb:BTZvertices}
\end{table}

As in previous sections, we contract each vertex and obtain splits which contribute an overall factor of $r^{(v_d-v_c)/2}$ for each vertex as prescribed by \eno{ContractRule}. After, splitting we must now count the number of new cycles created, where each new cycle contributes a factor of $r$. The number of cycles is the number of internal lines in the tensor network. These considerations mean that the norm of the state will go as a power of $r$, and we find:
\eqn{BTZpsiNorm}{
\log \langle \psi_{\tau} | \psi_{\tau} \rangle = \frac{1}{2}\sum_{{\rm type\ }i} \frac{{ M^{(i)} v_d^{(i)}}}{2}\log r = \left( \frac{r+1}{2} + \frac{\tau}{2}p^{\Lambda - 1}((p-1)r-(p+1))\right )\log r \,.
}

\begin{figure}[t!]
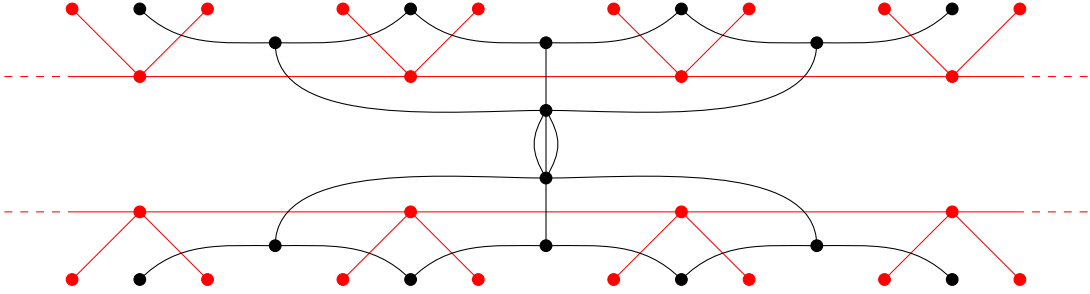

\[
\musepic{\gluehole}
\]
    \caption{The density matrix associated to the thermal state on the boundary of a BTZ black hole tensor network.}
    \label{fig:BTZglued}
\end{figure}

Having found the norm of the state, we may now compute the density matrix and entropy which comes from tracing out the central vertex behind the horizon, as shown in figure \ref{fig:BTZglued}. The intuition is that these degrees of freedom cannot be associated to any boundary state, so we should trace them out of the Hilbert space. The result is a mixed density matrix describing only the boundary qudits. As we are only tracing out one vertex, the result is surprisingly simple and parallels the computation of the entanglement entropy at genus $0$.

Applying our rule for tensor contractions to the center vertex as in figure \ref{fig:BTZglued}, where the two sides denote $| \psi_{\tau} \rangle$ and $\langle \psi_{\tau} |$, we see that in general we have a mixed density matrix with $\tau$ bonds stretched across the two sides. This is somewhat reminiscent of a two-sided BTZ black hole, as depicted in figure \ref{fig:BTZ2sided}. Based on the computation for the entanglement entropy, one would expect the thermodynamic entropy of this state to be proportional to $\tau$, and this can be supported by explicit analytic computation. 

Performing the split of the center vertex gives a factor of $r^{(v_d-v_c)/2} = r^{(r+1)/2 - \tau}$, and the general density matrix has a form similar to \eno{RhoAInterval}, with $r^{\tau}$ blocks of all $1$'s of size $r^{\sigma}$:
\eqn{RhoBTZ}{
 \rho_{BH} = { r^{\frac{(r+1)}{2} - \tau} \over \langle \psi_{\tau}|\psi_{\tau} \rangle} \begin{pmatrix}
 	\begin{bmatrix} 1 & \cdots & 1 \\
    \vdots & \ddots & \vdots \\
    1 & \cdots & 1
    \end{bmatrix} r^{\sigma} &
     \vspace{-1 mm} \\  \hspace{-3.5 mm} r^{\sigma} \\
    & \ddots &  \\
    & & \vspace{-1mm} \hspace{3.5 mm} r^{\sigma} \\  &  & r^{\sigma} \begin{bmatrix} 1 & \cdots & 1 \\
    \vdots & \ddots & \vdots \\
    1 & \cdots & 1
    \end{bmatrix}
 \end{pmatrix}_{r^{\sigma +\tau} \times r^{\sigma +\tau}}
}
We may fix the value of $\sigma$ by diagrammatic computation, but it is easier to simply impose the unit trace condition. Using the value of \eno{BTZpsiNorm}, we find $\sigma$ is given by the second piece, 
\eqn{BTZsigmasize}{
\sigma = \frac{\tau}{2}p^{\Lambda - 1}((p-1)r-(p+1)) \,.
}
We now compute the von Neumann entropy of this state, which corresponds to the black hole entropy.
\eqn{BTZvNent}{
S_{BH} = - \Tr \rho_{BH} \log \rho_{BH}\,,
}
where each block of $\rho_{BH}$ may be diagonalized before taking the trace. This gives a sum of identical terms with the $\sigma$ dependence cancelling,
\begin{equation}
S_{BH} = - \sum^{r^{\tau}} r^{-\tau} \log r^{-\tau}\,,
\end{equation}
or the main result of this section,
\eqn{BTZEntropy}{
S_{BH} = \tau \log r = (\log r) \log_p|q|_p^{-1}  \,.
}
We see that the von Neumann entropy of the boundary state is large and directly proportional to the perimeter of the event horizon. 


We now briefly discuss the entanglement entropy between an interval and its complement in the thermal background, dual to minimal geodesics in the black hole geometry. We have already explained the results of these computations in section~\ref{RTBTZresults}, which match the expectations of real AdS/CFT and the cut rule for perfect tensors. The computation of specific examples is straightforward using the rules we have used throughout this section, but the general formula is cumbersome to present in detail, so we elect to explain the basic geometry and the result of the contractions.

\begin{figure}[t]
\[
\begin{tikzpicture}[scale=0.9]
\draw[color=white,use as bounding box] (-2.2,-2.2) rectangle (2.2,2.2);
\tikzstyle{vred}=[draw,scale=0.4,color=red,fill=red,circle]
\foreach \x in {0,1,...,11} {
	\coordinate (b\x) at (\x*360/12 - 30 :2);
	};	
\draw[thick,blue] (b3) arc (-45:-135:1.4); 	
\draw[thick] (b3) arc (60:120:2);
\draw[dotted] (0,0) circle[radius=2];
\draw[fill=gray!50!white,pattern=north east lines] (0,0) circle[radius=0.5];
\draw (b3) node[anchor=south west] {$x_1$};
\draw (b5) node[anchor=south east] {$x_2$};
\draw (b4) node[anchor=south] {$A$};
\end{tikzpicture}
\quad
\begin{tikzpicture}[scale=0.9]
\draw[color=white,use as bounding box] (-2.2,-2.2) rectangle (2.2,2.2);
\tikzstyle{vred}=[draw,scale=0.4,color=red,fill=red,circle]
\foreach \x in {0,1,...,11} {
	\coordinate (b\x) at (\x*360/12 - 30 :2);
	};	
\draw[thick] (b8) arc (210:-30:2);
\draw[dotted] (0,0) circle[radius=2];
\draw[fill=gray!50!white,pattern=north east lines] (0,0) circle[radius=0.5];
\coordinate (a1) at (180:0.65);
\coordinate (a2) at (0:0.65);
\draw[thick,blue] (a1) arc (180:0:0.65); 	
\draw[thick,blue] (b8) to[out=45,in=-90,looseness=1] (a1);
\draw[thick,blue] (b0) to[out=135,in=-90,looseness=1] (a2);
\draw (b8) node[anchor=north east] {$x_1$};
\draw (b0) node[anchor=north west] {$x_2$};
\draw (b4) node[anchor=south] {$A$};
\end{tikzpicture}
\quad
\begin{tikzpicture}[scale=0.9]
\draw[color=white,use as bounding box] (-2.2,-2.2) rectangle (2.2,2.2);
\tikzstyle{vred}=[draw,scale=0.4,color=red,fill=red,circle]
\foreach \x in {0,1,...,11} {
	\coordinate (b\x) at (\x*360/12 - 30 :2);
	};
\draw[thick,blue] (b11) arc (45:135:1.4); 	
\draw[thick] (b9) arc (240:-60:2);
\draw[dotted] (0,0) circle[radius=2];
\draw[thick,blue] (0,0) circle[radius=0.65];
\draw[fill=gray!50!white,pattern=north east lines] (0,0) circle[radius=0.5];
\draw (b9) node[anchor=north east] {$x_1$};
\draw (b11) node[anchor=north west] {$x_2$};
\draw (b4) node[anchor=south] {$A$};
\end{tikzpicture}
\]
\caption{A schematic depiction of the three topologically distinct cases of minimal geodesics for a connected region $A$.}
\label{fig:realBHInt}
\end{figure}
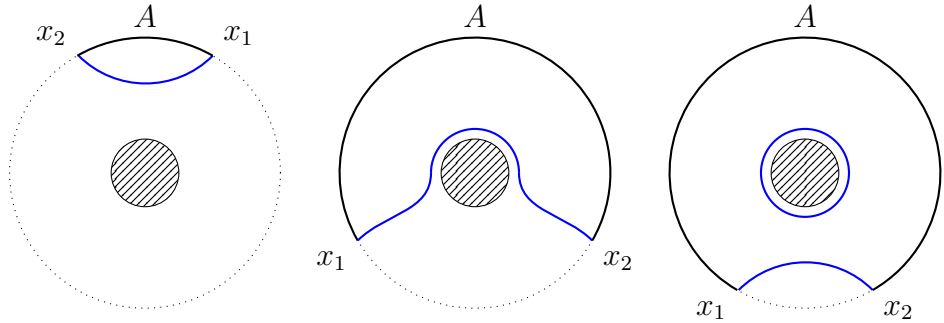

Tracing out a boundary region in the black hole background is obtained by combining the two graphical rules for reduced density matrices so far. 
The mixed density matrix is constructed  by gluing the state and its dual along the boundary interval to be traced out (implementing the partial trace), as well  as  along the center black hole vertex. Physically, these two gluings are two separate effects, but the resulting mixed state has an entropy which is sensitive both to the black hole horizon size and to the interval size. 
As before, we apply the rules of splits and cycles to the perfect tensors which are contracted, and from this point of view we treat the various contractions on equal footing to ultimately obtain the correct entropy. 
Several cases are possible, as the entanglement entropy of a region is no longer equal to that of the complement due to the presence of the black hole. There are essentially three possible cases which are schematically depicted in figure \ref{fig:realBHInt}:
\begin{itemize}
    \item Given a cutoff, if the region to be traced out is sufficiently small such that the entanglement wedge does not approach near the horizon of the black hole as in the first picture,  the resulting entropy will be completely insensitive to the presence of the horizon. As in the genus 0 case, this entropy is proportional to the log of the interval size and can be graphically computed by counting the number of bonds shared across the traced out region after performing all contractions. (These are exactly the bonds which cut the minimal surface on the genus $1$ tree geometry.)
    \item For a larger region, the graphical rules imply that bonds crossing the horizon interfere with those for the traced out region. The suspended bonds between the state and its dual now use up some of the bonds which were originally part of the horizon contraction. This is interpreted as the minimal surface wrapping around the horizon, and a computation reveals the entropy is given by exactly this length. This schematically looks like the middle figure.
    \item For a sufficiently large region, the available bonds inside the black hole become exhausted. The entanglement is now given by the number of remaining bonds across the state and the dual, which corresponds to a minimal surface which wraps the other side and includes the black hole; this is show in the final figure. The entropy is given by the sum of the horizon area and the length of the minimal surface.
\end{itemize}
In each case the minimal geodesic is homologous to the boundary region as desired.
 From these basic geometric rules, the results of section~\ref{RTBTZresults} follow, as one can see by direct though non-trivial calculation. The minimal surfaces we find closely resemble their archimedean counterparts, but are distinctly discrete and ultrametric.

We take the success of this network as a prediction for entanglement in thermal $p$-adic AdS/CFT. We also suspect the methods we have described for single intervals in black hole backgrounds generalize, possibly to higher genus black holes and more intervals. 

\section{Geometric Properties of the Tensor Networks}
\label{GEOMETRIC}

In this section we discuss in more detail some aspects of the geometry of the tensor
networks introduced above. In particular we discuss more in detail the symmetries and
the dependence in the choice of embedding. We show that the construction can be carried
out in a purely $p$-adic setting, where the tensor network lives on the Drinfeld $p$-adic plane
and is determined by a choice of sections of the projection from the Drinfeld plane to the
Bruhat--Tits tree. We also show a similar construction of the tensor network for the genus one case
on a fundamental domain for the action of the Schottky group on the Drinfeld plane. We also
discuss measures on the $p$-adic Tate--Mumford curve induced by different restrictions of
the Patterson--Sullivan measure on the projective line. Finally we
discuss the limit of the density matrices when the entire infinite tree is considered,
interpreted as states on an approximately finite dimensional von Neumann algebra. 

\subsection{Tensor networks: symmetries and embeddings}
\label{TNPROPS}

Recall from  section \ref{ssec:padic}
 that we can label the nodes on the Bruhat--Tits tree as cosets $G= \PGL(2,\mathbb{Q}_p) =\bigcup_{i=1}^\infty g_i H$ where $H = \PGL(2,\bZ_p)$ and $g_i \in G$.
Suppose we pick  a particular planar embedding, or in other words, make a choice of all incidence relations among bonds on the dual graph consistent with  definition \ref{def:dualgraph}.  Further, focus on the particular bond in the dual graph specified by the corresponding edge between the nodes $g_1 H$ and $g_2 H$ on the Bruhat--Tits tree. Let it be incident with a bond in the dual graph  corresponding to the edge between the nodes $g_3 H$ and $g_4 H$. After an isometric $G$ transformation, suppose the nodes on the Bruhat--Tits tree go to the cosets $g_1^\prime H, g_2^\prime H, g_3^\prime H$ and $g_4^\prime H$ respectively. Then the $G$ transformation sends the edge on the Bruhat--Tits tree between $g_1 H$ and $g_2 H$ to the edge between $g_1^\prime H$ and $g_2^\prime H$ (and similarly for the other edge). Correspondingly, the bonds on the dual graph transform as well, in such a way that all incidence relations are preserved on the dual graph. In other words, $G$ acts as an isometry on the dual graph. The point of intersection of the two bonds on the dual graph before the $G$ transformation was a node on the dual graph where the two bonds corresponding to cosets $g_1 H$ \& $g_2H$ and $g_3H$ \& $g_4H$ met. After the transformation, the intersection node on the dual graph is mapped to the node which is the point of intersection of the bonds specified by the cosets $g_1^\prime H$ \& $g_2^\prime H$ and $g_3^\prime H$ \& $g_4^\prime H$ respectively. 

Other choices for the dual graph can be obtained as follows. Starting with the Bruhat--Tits tree where the nodes are specified via the cosets $g_i H$ for $i=0,1,\ldots, \infty$, we specialize to a particular planar embedding.
Now we perform an isometric $G$ transformation, which transforms $g_i H \to g_i^\prime H$ for all $i$. The original dual graph incidence relations, given in terms of cosets $g_iH$  transform to the ones in terms of the transformed cosets $g_i^\prime H$ as explained in the previous paragraph. However, we can construct a {\it different} dual graph, whose bond incidence relations are the original incidence relations (in terms of original labelling of the cosets $g_iH$) but the Bruhat--Tits tree nodes are given in terms of the transformed cosets $g_i^\prime H$. (Here we are using the fact that $G=\bigcup_{i=1}^\infty g_i H = \bigcup_{i=0}^\infty g_i^\prime H$.) 
From this we conclude there are at least as many possible dual graphs as elements of the isometry group $G$.
In fact there exist more choices for the dual graph. One can act with any element of the automorphism group of the Bruhat--Tits (which is still an isometry) and obtain a different planar embedding. All such planar embeddings are allowed but there is no preferred choice among them. 
For the purposes of computation, we usually picked a particular choice of a dual graph (i.e.\ a particular  planar embedding); however, the final physical result of the computations was always independent of this choice as discussed in previous sections.

\subsection{Drinfeld plane and the dual graph}\label{ssec:Drinfeld}

In our previous discussion of the tensor networks on the dual graph, we
have constructed such dual graph by realizing the tree (or a finite portion
of the tree) embedded inside an ordinary plane. This can look
at first very artificial: the Bruhat--Tits tree is a nonarchimedean $p$-adic 
object hence a natural construction of associated tensor networks should 
exist entirely inside the $p$-adic world and should not depend on the
choice of an embedding in an archimedean space like the plane.

Indeed, we show in the following that it is in fact possible to realize the
tensor networks described in the previous sections entirely in the
$p$-adic setting, on the $p$-adic Drinfeld plane. Thus, while we continue
to draw them in the ordinary plane for graphical convenience and simplicity,
one should really think of these tensor networks as living on the Drinfeld plane.

More precisely, we describe here a notion of ``dual graph" to an embedding of the
Bruhat-Tits tree as a $1$-skeleton in the Drinfeld $p$-adic upper half plane, given by
a choice of a lift of the natural projection $\Upsilon: \Omega \to T$ of the $p$-adic
plane to the tree. We first describe a toy model based on a tubular neighbourhood of
a tree in ordinary $3$-space, and then we explain how this model adapts to the
case of the Drinfeld $p$-adic upper half plane.

\subsubsection{An archimedean toy model}
\label{ssec:toymodel}

We discuss first a toy model in a simpler archimedean setting, where we
consider a homogeneous tree $T$ of valence $q+1$ embedded in a $3$-dimensional
Euclidean space and a $2$-dimensional surface ${\mathcal S}$ given by the boundary of a small 
tubular neighbourhood 
${\mathcal N}(T)$ of the tree, ${\mathcal S}=\partial {\mathcal N}$.  
In this section we will take $q$ to be a positive integral power of $p$.
We identify ${\mathcal N}(T)$ with the disk bundle of the
normal bundle and we denote by $\Pi: {\mathcal S} \to T$ the projection restricted to ${\mathcal S}$. 
Then, for almost all points $x$ in the tree $T$ the preimage $\Pi^{-1}(x)$ is a circle,
while the image of the star $S(v)$ of half-edges around a vertex $v$ is a ``pair of pants" figure
with $q+1$ holes.

Choose two lifts of the projection map $\Pi: {\mathcal S} \to T$, so that their images
give two disjoint embeddings of the tree $T$ in ${\mathcal S}$. For example, take the
sections so that the two trees cut each circle in the fiber of $\Pi$ in antipodal points.
We call the two images $T$ and $T'$. Also fix a sufficiently small $\epsilon>0$
and a tubular neighbourhood ${\mathcal N}'_\epsilon(T')$ of size $\epsilon$
of the tree $T'$ inside ${\mathcal S}$. The chosen
epsilon should be small enough that the distance on ${\mathcal S}$ between $\partial {\mathcal N}'_\epsilon(T')$ and
$T$ is bounded below by a nonzero quantity, say greater than $3\epsilon$. 
Let ${\mathcal C}$ denote the countable collection of curves 
on ${\mathcal S}$ given by ${\mathcal C} = \partial {\mathcal N}'_\epsilon(T')$.

Choose then a sequence of $\epsilon_i>0$ with the property that the
series converges with $\sum_i \epsilon_i <\epsilon$. We consider
tubular neighbourhood ${\mathcal N}'_{\eta}(T')$ for each $\eta=\epsilon+\epsilon_i$.
Let ${\mathcal C}_{\eta} =\partial {\mathcal N}'_{\eta}(T')$. All these curves are at distance
at least $\epsilon$ from $T$.

Also fix a real number $t$ with $0<t<1$, and for each edge $e$ of $T$, identified
with the set of points $x_t(e)=(1-t)v + t v'$ with $v,v'$ the endpoint vertices, consider the
circle $\Pi^{-1}(x_t(e))$ in ${\mathcal S}$. Note that it is possible to find such a $t$ so that all
$\Pi^{-1}(x_t(e))$ are indeed circles.

We construct a ``dual graph" to the copy $T$ of the tree embedded in ${\mathcal S}$ as
follows. Fix a base vertex $v_0$ in the tree $T$. Let $e_0, \ldots, e_q$ be the
edges of $T$ adjacent to $v_0$. For each of these edges $e_i$ let $x_i$ be the
point of intersection with the circle $S^1_i=\Pi^{-1}(x_t(e_i))$ constructed as above.

Consider the path $\ell_i$ given by the two arcs $\gamma_{i,1}, \gamma_{i,2}$ of $S^1_i$ between the
point $x_i$ and the two points $y_{i,1}, y_{i,2}$ of intersection between $S^1_i$ and ${\mathcal C}=\partial{\mathcal N}'_\epsilon(T')$ together with the two infinite curves $C_{i,1}, C_{i,2}$ in ${\mathcal C}$ that start at the points 
$y_{i,1}, y_{i,2}$ pointing in the direction of $e_i$, that is, 
$\ell_i =\gamma_{i,1}\cup \gamma_{i,2}\cup C_{i,1} \cup C_{i,2}$.

We proceed in a similar way for an arbitrary edge $e$ in $T$. Let $e$ be an
edge that is at distance $N$ from the root vertex $v_0$. Then consider the
curve $\ell_e =\gamma_{e,1,}\cup \gamma_{e,2}\cup C_{e,1} \cup C_{e,2}$,
where $\gamma_{e,j}$ are the two arcs along $S^1_e =\Pi^{-1}(x_t(e))$
connecting the point $x_e$ of intersection between $e$ and $S^1_e$ and
the two points $y_{e,j}$ of intersection between $S^1_e$ and the ${\mathcal C}_\eta$
for $\eta=\epsilon+\epsilon_N$, and $C_{e,j}$ the infinite paths along ${\mathcal C}_\eta$
that start from $y_{e,j}$ in the direction of $e$.
\begin{figure}[t]
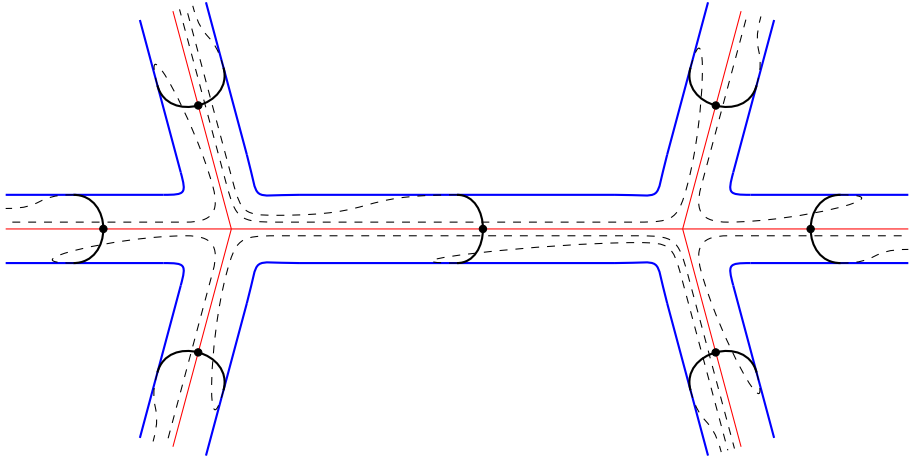

\centering
\[
\musepic{\dualDrinfeld}
\]
\caption{The dual graph locus in the Drinfeld plane.}
\label{fig:drinfelddual}
\end{figure}

Our ``dual graph" ${\mathcal D}(T)$ of $T$ in ${\mathcal S}$ consists of the collection of edges $\ell_e$
with vertices given by their endpoints at infinity, modulo an equivalence relation: 
if two subarcs $C_{e,j}$ $C_{e',j'}$ of two curves 
$\ell_e$ and $\ell_{e'}$ are always at a distance less than $\epsilon$ outside of 
a compact region in ${\mathcal S}$, then their endpoints at infinity are identified. 
This is shown in figure \ref{fig:drinfelddual}.

\subsubsection{The $p$-adic plane}
\label{ssec:pplane}

The toy model considered above explains the heuristics of what we
would like to construct on the Drinfeld plane. Indeed the Drinfeld
plane $\Omega$, introduced in \cite{Dri}, can be thought of as a $p$-adic analog 
of the tubular neighbourhood ${\mathcal S}=\partial {\mathcal N}(T)$ of the Bruhat-Tits tree $T$.

For ${\mathbb K}$ a finite extension of ${\mathbb Q}_p$ with residue field ${\mathbb F}_q$, for some $q=p^r$, 
the Drinfeld plane $\Omega$ can be identified with ${\mathbb P}^1({\mathbb C}_p) \smallsetminus {\mathbb P}^1({\mathbb K})$,
or equivalently with the set of homothety classes of invertible ${\mathbb K}$-linear maps $\varphi: {\mathbb K}^2 \to {\mathbb C}_p$,
$\varphi: (x,y) \mapsto x \zeta_0 + y \zeta_1$ for $(\zeta_0:\zeta_1)\in {\mathbb P}^1({\mathbb C}_p) \smallsetminus {\mathbb P}^1({\mathbb K})$. 
It is endowed with a projection map $\Upsilon: \Omega \to T$ to the Bruhat--Tits tree of ${\mathbb K}$.
Given two adjacent vertices $v,v'$ connected by an edge $e$ in $T$, parameterized by
$e_t=(1-t)v +t v'$, for $0\leq t \leq 1$, the projection map satisfies (see section~2 of \cite{BoutCar})
\eqn{}{
\Upsilon^{-1}(v) &= \{ \zeta \in {\mathbb C}_p \,:\, |\zeta|\leq 1\} \smallsetminus \bigcup_{a\in {\mathcal O}_{\mathbb K}/\pi {\mathcal O}_{\mathbb K}} \{ \zeta \in {\mathbb C}_p \,:\,  
|\zeta - a| < 1 \} \cr 
 \Upsilon^{-1}(v') &=\{ \zeta \in {\mathbb C}_p \,:\, |\zeta|\leq q^{-1} \} \smallsetminus \bigcup_{b \in \pi {\mathcal O}_{\mathbb K}/\pi^2 {\mathcal O}_{\mathbb K}} 
\{ \zeta \in {\mathbb C}_p \,:\,   |\zeta - b| < q^{-1} \} 
}
where $v=[M]$, $v'=[M']$ with $\pi M \subset M' \subset M$, and for $e_t = (1-t) v + t v'$, for $0<t<1$, along the edge $e$
\eqn{}{ 
\Upsilon^{-1}(e_t) =\{ \zeta \in {\mathbb C}_p \,:\, |\zeta|\leq q^{-t} \}\,. 
}
One can therefore view it as an analog of the surface ${\mathcal S}$ with its decomposition into
a collection of ``pairs of pants", as discussed previously.  More extensive discussions of
the geometry of the Drinfeld plane can be found in \cite{BoutCar} and \cite{DasTeit}.

Given the star $S(v)$ of a vertex $v$ (given by the vertex together with all its adjacent edges) 
in the Bruhat--Tits tree, consider the regions $\Sigma(v)=\Upsilon^{-1}(S(v))$ in the Drinfeld 
plane $\Omega$. The sets $\Sigma(v)$ are a covering of $\Omega$ with nerve $T$.
A cell $\tau$ in the Bruhat-Tits tree is given by an edge together with its
two adjacent vertices. Given a cell $\tau$ corresponding to an edge $e$ we denote by \
$\Sigma(\tau):=\Sigma(v) \cap \Sigma(v') =\Upsilon^{-1}(\tau)$, with $\partial(e)=\{ v,v' \}$, 
given by
\eqn{}{ 
\Sigma(\tau) = \{ \zeta \in {\mathbb C}_p \,:\, |\zeta|\leq 1\} \smallsetminus \bigcup_{a\in ({\mathcal O}_{\mathbb K} \smallsetminus \pi {\mathcal O}_{\mathbb K})/\pi {\mathcal O}_{\mathbb K}} \{ \zeta \in {\mathbb C}_p \,:\,  |\zeta - a| < 1 \} \cr 
\smallsetminus \bigcup_{b \in \pi {\mathcal O}_{\mathbb K}/\pi^2 {\mathcal O}_{\mathbb K}} 
\{ \zeta \in {\mathbb C}_p \,:\,   |\zeta - b| < q^{-1} \}\,. 
}

It is known by \cite{SchStu} and \cite{DeSha} that the de Rham cohomology 
of the Drinfeld plane can be computed in terms of certain combinatorial 
harmonic forms on the Bruhat--Tits tree, with the map realizing this 
identification given by a $\PGL(2,{\mathbb K})$-equivariant residue map. 
The results of \cite{SchStu} and \cite{DeSha} in fact holds more generally for 
harmonic forms on higher rank Bruhat--Tits buildings and de Rham 
cohomology of higher rank Drinfeld symmetric spaces (the complement in
${\mathbb P}^n({\mathbb C}_p)$ of the ${\mathbb K}$-rational hyperplanes).

In fact, the case of rank two,
involving the $p$-adic plane and the Bruhat--Tits tree of $\PGL(2,{\mathbb K})$, was
also discussed in \cite{Sch} (see also \cite{vdP}), with an extension to the case of quotients by
$p$-adic Schottky groups, Mumford curves and quotients of the Bruhat--Tits tree.
In the setting discussed in \cite{vdP} one identifies the holomorphic
one-forms on the Drinfeld plane with currents on the Bruhat--Tits tree, via  a
residue map.

A current on an oriented locally finite graph ${\mathcal G}$ is a map $\mu:  E({\mathcal G}) \to {\mathbb Z}$ from the
oriented edges of ${\mathcal G}$ to the integers satisfying the conditions
\eqn{}{ 
\mu(\bar e)=-\mu(e)\,,
}
where $\bar e$ denotes the edge with the reverse orientation, and
\eqn{}{ 
\sum_{e: s(e)=v} \mu(e) =0 \,.
}
Currents  form an abelian group, denoted by ${\mathcal C}({\mathcal G})$.
One can also consider currents on a locally finite directed graph ${\mathcal G}$ with values in 
a field ${\mathbb K}$ of characteristic zero, by taking ${\mathcal C}({\mathcal G},{\mathbb K})={\mathcal C}({\mathcal G})\otimes_{\mathbb Z} {\mathbb K}$.

There is an algebra of $p$-adic holomorphic functions on the Drinfeld upper half plane
(see e.g.~\cite{Man}), which we denote by ${\mathcal O}(\Omega)$. There is a short exact 
sequence (Corollary 2.1.2 of \cite{vdP} and \cite{Sch} p.~225) relating currents ${\mathcal C}(T_{\mathbb K},{\mathbb K})$ 
on the Bruhat-Tits tree and $p$-adic holomorphic $1$-forms on the Drinfeld upper half plane 
\begin{equation}\label{Omegaseq}
0 \to {\mathcal O}(\Omega) \stackrel{d}{\to} \Omega^1(\Omega) \to {\mathcal C}(T_{\mathbb K},{\mathbb K}) \to 0\,.
\end{equation}
Thus, ${\mathbb K}$-valued currents on the Bruhat-Tits tree provide a combinatorial way of
describing holomorphic $1$-forms modulo exact forms. The map that assigns a
current on $T$ to a 1-form on $\Omega$ is the residue map 
\eqn{}{ 
\omega \mapsto \sum_i {\rm res}_{\partial D_i}(\omega) 
}
where the $D_i$ are a collection of finitely many disjoint disks in ${\mathbb P}^1({\mathbb C}_p)$ 
such that their union contains ${\mathbb P}^1({\mathbb K})$.

The group of currents ${\mathcal C}(T_{\mathbb K},{\mathbb K})$ can be identified with the group of finitely additive 
${\mathbb K}$-valued measures on ${\mathbb P}^1({\mathbb K})=\partial T_{\mathbb K}$ with zero total mass $\mu({\mathbb P}^1({\mathbb K}))=0$,
by the identification $\mu(U(e)):=\mu(e)$, for any edge $e$ in $T_{\mathbb K}$ with $U(e)\subset {\mathbb P}^1({\mathbb K})$
the clopen set consisting of ends of half infinite paths in the tree starting with $e$. In turn we can
identify the set of these measures with 
\eqn{}{ 
{\rm Ker} (\Phi: {\rm Hom}_{\mathbb Z} ({\mathcal C}^\infty({\mathbb P}^1({\mathbb K})),{\mathbb Z}) \to {\mathbb K}) 
}
by identifying $\mathbb{K}$-valued measures as functionals acting by integration on locally constant functions
with the zero mass condition represented by the vanishing of $\Phi: \mu \mapsto \mu(1)$.
This gives the case of rank two of \cite{SchStu} and \cite{DeSha} with the identification of
the first de Rham cohomology of the Drinfeld plane with
\eqn{}{ 
H^1_{dR}(\Omega) \cong  {\rm Ker} (\Phi: {\rm Hom}_{\mathbb Z} ({\mathcal C}^\infty({\mathbb P}^1({\mathbb K})),{\mathbb Z})\to {\mathbb K})\,.
}

\subsubsection{Other models of $p$-adic planes}

There are other possible models of $p$-adic plane, where one can directly use
metric properties. The version considered in \cite{Gel} section~2.2.8 has the advantage
that it does have a hyperbolic metric and geodesics behaving in many ways
(e.g.~the trace formula in \cite{Yas}) like the usual hyperbolic plane, but it does not
have a nice relation to the Bruhat--Tits tree, while the version of \cite{Guill} has
a projection to the Bruhat--Tits tree and is defined so as to have metric properties
but it does not generalize to higher ranks, unlike the Drinfeld plane. For these
reasons, especially in view of developing higher rank generalizations of the
$p$-adic AdS/CFT correspondence based on Bruhat--Tits buildings, we prefer
to use the Drinfeld plane model of a $p$-adic plane. 

\subsubsection{The dual locus in the Drinfeld plane}\label{ssec:dualpplane}

Due to the topological nature of $p$-adic spaces, we cannot quite literally 
perform the same construction described in the previous toy model case, but
we aim at identifying a locus in $\Omega$ that has similar properties to the
dual graph ${\mathcal D}(T)$ described in the previous case, although it will not
be a graph.

As in the previous case, consider two sections $s,s': T \to \Omega$ that lift
the projection $\Upsilon: \Omega \to T$, with disjoint images $s(T)$ and $T'=s'(T)$.
Also consider a collection of nested sets in $\Omega$ (with strict inclusions) 
\eqn{}{ 
T'\subset {\mathcal N}_0^-\subset {\mathcal N}_0^+ \subset {\mathcal N}_1^- \subset {\mathcal N}_1^+ \subset \cdots \subset {\mathcal N}_k^- \subset {\mathcal N}_k^+ \subset \cdots 
}
and such that ${\mathcal N}=\cup_{k,\pm} {\mathcal N}_k^\pm$ is disjoint from $s(T)$. 
We require that the sets ${\mathcal N}_k^\pm$ have compatible
projection maps $\Pi_k^\pm$ to $T'$ with $\Upsilon \circ \Pi_k^\pm = \Upsilon$ and 
$\Pi_k^- |_{{\mathcal N}^+_{k-1}}=\Pi_{k-1}^+$ and $\Pi_k^+|_{{\mathcal N}_k^-}=\Pi_k^-$. 
We also require that each ${\mathcal N}_k^\pm$ has trivial cohomology.
The explicit description of \cite{Sch,SchStu} recalled above of de Rham cohomology of 
$\Omega$ in terms of residues and harmonic forms on the Bruhat-Tits tree shows 
that such regions ${\mathcal N}_k^\pm$ can be constructed, by showing that the restriction of the
holomorphic forms in $\Omega^1(\Omega)$ to ${\mathcal N}$ have trivial residues.   
Since the residues of the holomorphic forms on $\Omega$
are all supported along circles $\partial D$ that are boundaries of the ``pairs of pants" regions,
it suffices to ensure that the region ${\mathcal N}$ does not contain any of these circles.
For instance, one can construct the ${\mathcal N}_k^\pm$ by choosing nested sets $S_k^\pm$ of sections
of the projection $\Upsilon: \Omega \to T$ containing the section $s'$ with 
$s\notin \cup_{k,\pm} S^\pm_k$.

Given an edge $e$ in $T$ and a chosen base vertex $v_0$, orient all the edges of $T$ away 
from $v_0$ and let $T(e)\subset T$ be the subtree with root vertex $v=s(e)$, consisting of all
vertices and edges of $T$ that are reachable from $s(e)$ along an oriented path. Let  
$\Omega(e):=\Upsilon^{-1}(T(e))$.

Fix a $t$ with $0<t<1$ and let $\Sigma_t =\Upsilon^{-1}(e_t)$ for the point $e_t=(1-t)v + t v$ on the edge $e$.
Then to an edge $e$ in the Bruhat-Tits tree at a distance $N$ from a fixed root vertex $v_0$ 
we associate a region $L_e$ obtained as the union $L_e = \Gamma_e \cup {\mathcal C}_{N,e}$, where
$\Gamma_e$ is the region of $\Omega$ given by 
\eqn{}{ 
\Gamma_e = \Sigma_t \smallsetminus (\Sigma_t \cap {\mathcal N}_N^-) 
}
and ${\mathcal C}_{N,e}={\mathcal N}_{N,e}^+\smallsetminus {\mathcal N}_{N,e}^-$, where ${\mathcal N}_{k,e}^\pm$
is the region given by 
\eqn{}{ 
{\mathcal N}_{k,e}^\pm={\mathcal N}_k^\pm \cap \Omega(e)\,.
}
The $L_e$ are mutually disjoint regions in $\Omega$, with endpoints at the
boundary at infinity ${\mathbb P}^1(K)$ of $\Omega$. 
We define the analog of the ``dual graph" of our Archimedean toy model to be
the region 
\eqn{}{ 
{\mathcal D}(T) := \cup_e L_e \subset \Omega\,.
}
This depends on the choice of the sections $s,s'$, of $t$, and of the regions ${\mathcal N}_k^\pm$.

As in the Archimedean setting we think of tensor networks supported on the
dual graph, here we consider a tensor networks supported in the region ${\mathcal D}(T)$,
with ``bonds" along the loci $L_e$. A geodesic in the  tree $s(T)$ cuts a number of 
such bonds equal to its length in number of edges. We place vertices of the ``dual graph"
${\mathcal D}(T)$ at its limit points at infinity in ${\mathbb P}^1({\mathbb K})$. These are also limit points of
the geodesics in the Bruhat--Tits tree, by construction.

\subsection{Genus one case: Tate--Mumford elliptic curves} 
\label{ssec:geometryTATE}

As discussed earlier,
in the genus one case, which gives the $p$-adic BTZ black hole, we consider a rank 
one $p$-adic Schottky group $\Gamma \subset \PGL(2,{\mathbb K})$, generated by 
a single hyperbolic element $\gamma$ with two fixed points
in the boundary ${\mathbb P}^1({\mathbb K})$. We can always identify the
endpoints with the points $\{ 0, \infty \}$. 
Instead of the Bruhat--Tits tree we then consider the quotient graph
$T/\Gamma$. This consists of a polygonal ring with infinite trees attached to
the vertices, as illustrated in figure \ref{fig:padicBTZ}. The Mumford curve $X_\Gamma({\mathbb K}) = \partial T/\Gamma$
is a Mumford--Tate $p$-adic elliptic curve with Tate uniformization
$X_\Gamma({\mathbb K}) =({\mathbb P}^1({\mathbb K})\smallsetminus \{ 0, \infty \})/\Gamma$.

The Schottky group $\Gamma=\gamma^{\mathbb Z}$ also acts on the Drinfeld plane
$\Omega={\mathbb P}^1({\mathbb C}_p)\smallsetminus {\mathbb P}^1({\mathbb K})$
and we can consider the quotient $\Omega/\Gamma$. Since the projection map
$\Upsilon: \Omega \to T$ is equivariant with respect to the $\PGL(2,{\mathbb K})$
action, hence with respect to $\Gamma$, we obtain an induced projection
$\Upsilon: \Omega/\Gamma \to T/\Gamma$. By choosing a lift of this projection we
can embed a copy of the graph $T/\Gamma$ inside $\Omega/\Gamma$ as a $1$-skeleton,
with boundary at infinity given by the ${\mathbb K}$-rational points of the Mumford--Tate
elliptic curve, $\partial T/\Gamma = \partial \Omega/\Gamma = X_\Gamma ({\mathbb K})$. 

\begin{figure}
\centering
\begin{tikzpicture}[scale=1.4]
\tikzstyle{vertex}=[draw,scale=0.25,fill=black,circle]
\tikzstyle{ver2}=[draw,scale=0.4,fill=red,circle]
\tikzstyle{ver3}=[];

\newcommand{\wpar}{0.2}
\newcommand{\widthpar}{0.65ex}

\foreach \x in {0,...,5} {
	\coordinate (a\x) at (\x*60:1);
	\foreach \z in {-1,1} {
		\coordinate (b\x_\z) at ($ (a\x) + (\x*60 + \z*30:1) $) ;
		\draw[thin, color=red, name path = tree] (b\x_\z) -- (a\x);
		\draw[thick, dotted] 
			($ ($ (b\x_\z)!.3!(a\x) $) !\widthpar!90:(a\x)$) 
			to[out=\x*60 + \z*30 , in = \x*60 + \z*30 , looseness = 1.3] 
			($ ($ (b\x_\z)!.3!(a\x) $) !\widthpar!-90:(a\x)$) ;
		\draw[thick, name path=circ] 
			($ ($ (b\x_\z)!.3!(a\x) $) !\widthpar!90:(a\x)$) 
			to[out=\x*60 + \z*30 + 180 , in = \x*60 + \z*30 + 180 , looseness = 1.3] 
			($ ($ (b\x_\z)!.3!(a\x) $) !\widthpar!-90:(a\x)$) ;
		\path [name intersections={of=circ and tree}];
		\draw (intersection-1) node[vertex] {};
		};
	};

\foreach \x/\y in {0/1,1/2,2/3,3/4,4/5,5/0} {
	\draw[thin,color=red, name path = loop] (a\x) -- (a\y);
	\coordinate (a\x\y) at ($ (a\x)!.2!(a\y) $);
	\coordinate (a\y\x) at ($ (a\x)!.8!(a\y) $);
  	\coordinate (U\x\y) at ($ (a\x\y)!\widthpar!90:(a\x) $);
	\coordinate (U\y\x) at ($ (a\y\x)!\widthpar!-90:(a\y) $);
  	\coordinate (L\x\y) at ($ (a\x\y)!\widthpar!-90:(a\x) $);
	\coordinate (L\y\x) at ($ (a\y\x)!\widthpar!90:(a\y) $);
	\draw[thick, name path=circ] 
		($ (a\x)!.5!(a\y) !\widthpar!90:(a\y)$) 
		to[out=\y*60 + 60 , in = \y*60 + 60, looseness = 1.3] 
		($ ($ (a\x)!.5!(a\y) $) !\widthpar!90:(a\x) $);
	\draw[thick, dotted] 
		($ (a\x)!.5!(a\y) !\widthpar!90:(a\y)$) 
		to[out=\y*60 - 120, in = \y*60 - 120, looseness = 1.3] 
		($ ($ (a\x)!.5!(a\y) $) !\widthpar!90:(a\x) $);
	\path [name intersections={of=circ and loop}];
	\draw (intersection-1) node[vertex] {};
	\foreach\z in {-1,1} {
		\coordinate (V\x_\z) at ($ ($ (a\x)!.35!(b\x_\z) $) !\widthpar!90*\z:(a\x) $);
		\coordinate (O\x_\z) at ($ ($ (a\x)!.2!(b\x_\z) $) !\widthpar!-90*\z:(a\x) $);
		\coordinate (Q\x_\z) at ($ ($ (a\x)!.8!(b\x_\z) $) !\widthpar!-90*\z:(a\x) $);
		\coordinate (W\x_\z) at ($ ($ (a\x)!.8!(b\x_\z) $) !\widthpar!90*\z:(a\x) $);
		\draw[thick] (V\x_\z) -- (W\x_\z);
		\draw[thick] (O\x_\z) -- (Q\x_\z);
		};
	};
\foreach \x/\y/\z in {0/1/2,1/2/3,2/3/4,3/4/5,4/5/0,5/0/1} {
	\draw[thick] (V\x_1) to[in = \x*60 - 30 + 180, out = \x*60 + 30 + 180, looseness = 1.5] (V\x_-1);
	\draw[thick] (L\x\y) -- (L\y\x);
	\draw[thick] (U\x\y) -- (U\y\x);
	\draw[thick] (U\x\y) to[out = \x*60 - 60, in = \x*60 - 150, looseness = 1.5] (O\x_1);
	\draw[thick] (O\y_-1) to[out = \y*60 + 150, in = \y*60 + 60, looseness = 1.5] (U\y\x);
	\draw[thick] (L\y\x) to[out = \y*60 + 60, in = \y*60 - 60, looseness = 1.5] (L\y\z);
	};

\foreach \n in {0, 1, 2, 3, 4, 5} {
	\foreach \z in {-1,1} {
		\foreach \x in {-1,0,1} {
			\coordinate (d\n\x\z) at ($ (b\n_\z) + (\n*60 + \z*15 + \x*30:1) $) ;
			\draw[thin,color=red,name path= branch] (d\n\x\z) -- (b\n_\z);
			\draw[thin, color=red,name path=outer] (d\n\x\z) -- ($ (b\n_\z) + (\n*60 + \z*15 + \x*30:1.15) $);
			\draw[thick,name path= circ]
				($ (d\n\x\z)!\widthpar!90:(b\n_\z) $)
				to[out=\n*60 + \z*15 + \x*30 + 180, in = \n*60 + \z*15 + \x*30 + 180, looseness=1.25] 
				($ (d\n\x\z)!\widthpar!-90:(b\n_\z) $) ;
			\draw[thick,name path=undercirc]
				($ (d\n\x\z)!\widthpar!90:(b\n_\z) $)
				to[out=\n*60 + \z*15 + \x*30, in = \n*60 + \z*15 + \x*30, looseness=1.25] 
				($ (d\n\x\z)!\widthpar!-90:(b\n_\z) $) ;		
			\path [name intersections={of=circ and branch,by=Bl}];
			\draw (Bl) node[vertex] {};
			\path [name intersections={of=outer and undercirc,by=Wh}];
			\draw (Wh) node[scale=0.3,fill=white,circle] {};
			\draw[thin, color=red] (d\n\x\z) -- ($ (b\n_\z) + (\n*60 + \z*15 + \x*30:1.15) $);
			\draw[thin, color=red, dotted] ($ (b\n_\z) + (\n*60 + \z*15 + \x*30:1.15) $) -- ($ (b\n_\z) + (\n*60 + \z*15 + \x*30:1.3) $);
		};
		\foreach \x in {-1,1} {
			\draw[thick] 
				($ (d\n\x\z)!\widthpar!90*\x:(b\n_\z) $)
				to[out=\n*60 + \z*15 + \x*30 + 180, in = \n*60 + \z*15 + 180, looseness=5.5] 
				($ (d\n0\z)!\widthpar!-90*\x:(b\n_\z) $) ;
			\draw[thick]
				($ (d\n\x\z)!\widthpar!-90*\x:(b\n_\z) $)
				--
				($  ($(d\n\x\z)!.8!(b\n_\z)$) !\widthpar!-90*\x:(b\n_\z) $);
			\draw[thick]
				($ ($ (b\n_\z)!0.2!(a\n) $) !\widthpar!-90*\x:(a\n) $)
				to[in=  \n*60 + \z*15 + \x*30 + 180, out = + \n*60 + \z*30, looseness = 1.1]
				($  ($(d\n\x\z)!.8!(b\n_\z)$) !\widthpar!-90*\x:(b\n_\z) $);
		};
	};
};
\end{tikzpicture}
    \caption{A sketch of the quotient of the Drinfeld plane associated to a genus-one Mumford curve. The details of the bonds of the dual graph are not shown; the reader should compare figure~\ref{fig:drinfelddual}.}
    \label{fig:mumforddual}
\end{figure}
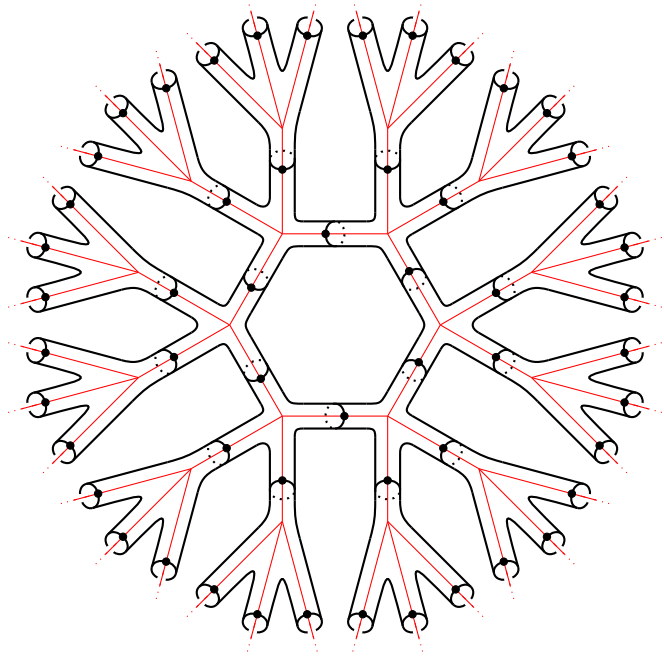

\subsubsection{Dual locus for the Tate--Mumford curve} 

The same construction described above of a ``dual locus" in the Drinfeld plane 
to a lift of the projection to the Bruhat--Tits tree, realizing a copy of $T$ as a 
$1$-skeleton in the $p$-adic plane, can be adapted to the genus one case.
We illustrate here how the construction changes using the toy model of a 
tubular neighbourhood of a tree in $3$-space, which is easier to show visually.
The corresponding construction on the $p$-adic Drinfeld plane itself then
proceeds as in the previous case following the model of the tubular 
neighbourhood of the tree. In this case we consider the tubular neighbourhood of
the geodesic $L_{\{0,\infty\}}$ in the tree, and we fix a choice of a base point
on this tubular neighbourhood, away from both chosen lifts of the tree via 
disjoint preimages of the projection. We construct the dual graph by considering
a family of loops from the chosen base point around each loop $\Pi^{-1}(x_t(e_i))$
for edges along $L_{\{0,\infty\}}$, while for all other edges we repeat the construction
as in the genus zero case. This gives a collection of curves whose image in the quotient lies on the surface illustrated in figure \ref{fig:mumforddual}.
Consider a fundamental domain ${\mathcal F}$ of the action of the group $\Gamma$ that contains 
the chosen base point, and a collection of curves contained in this fundamental domain, so that the quotient looks like the curves in figure \ref{fig:drinfelddual} drawn on the surface of figure \ref{fig:mumforddual}.
These determine the ``dual graph" on
the quotient by $\Gamma$.

\subsubsection{A measure on the Tate--Mumford curve} 

In the genus-zero case, the boundary ${\mathbb P}^1({\mathbb K})$ of the Bruhat--Tits tree
$T_{\mathbb K}$ is a finite extension ${\mathbb K}$ of ${\mathbb Q}_p$ carries a measure
that is the Patterson--Sullivan measure for the action of $\PGL(2,{\mathbb K})$ on
$T_{\mathbb K}$, which has the full boundary ${\mathbb P}^1({\mathbb K})$ as limit set. 
It is known by the general construction of \cite{Coor} that any Gromov-hyperbolic space
with a proper discontinuous action of an isometry group determines a Patterson--Sullivan 
measure on the hyperbolic boundary, with support on the limit set of the group, and 
quasi conformal of dimension equal to the Hausdorff dimension of the limit set. In particular,
in the case of a $p$-adic Schottky group $\Gamma$ of rank at least two acting on the Bruhat-Tits tree
and its boundary, one obtains in this way a Patterson--Sullivan measure supported on
the limit set $\Lambda_\Gamma\subset {\mathbb P}^1({\mathbb K})$ of the Schottky group. The
properties of this Patterson--Sullivan measure are used to prove rigidity results for
Mumford curves, \cite{CornKool}. However, notice that the Patterson--Sullivan measure
lives on the limit set $\Lambda_\Gamma$, which is the complement of the boundary region
that determines the Mumford curve $X_\Gamma({\mathbb K}) = ( {\mathbb P}^1({\mathbb K}) \smallsetminus \Lambda_\Gamma )/\Gamma$.  Thus, unlike the genus zero case, the natural construction of a
Patterson--Sullivan measure does not produce a measure on the Mumford curve, but only
a measure on the limit set. Moreover, in the particular case of genus one, even this measure
on the limit set would be uninteresting, since in the genus one case the limit set only consists
of two points (which we can always assume to be $0$ and $\infty$), rather than a Cantor
set type object as in the higher genus cases. One can also see that the other interesting group
action that is present in the case of Mumford curves, namely the action of the automorphism
group of the curve, also fails to give rise to an interesting Patterson--Sullivan 
measure (except in genus zero where the automorphism group of the projective line is $\PGL(2,{\mathbb K})$ and
one obtains again the Patterson--Sullivan measure on ${\mathbb P}^1({\mathbb K})$). Indeed,
in the case with genus at least two the automorphism group ${\rm Aut}(X)$ of the Mumford curve $X$
is a finite group hence the limit set is empty, hence we do not have a Patterson--Sullivan measure
supported on the Mumford curve $X$ itself.

In the case of genus one (BTZ black hole) the automorphism
group of the elliptic curve is a semidirect product of the elliptic curve $E$ itself as a group
(acting on itself by translations) by the automorphism group $Aut(E)$ of this group. In particular,
the action on the Bruhat--Tits tree of arbitrary translations along the geodesic with
endpoints $\{0,\infty \}$ induces automorphisms of the Mumford curve, which act
on the infinite graph $T/\Gamma$ and its boundary $\partial T/\Gamma=X_\Gamma$ as
rotations of the central polygonal ring and of all the outgoing trees attached to it. The
change of orientation that exchanges the endpoints $\{0,\infty \}$ also induces a
self map of $T/\Gamma$ and its boundary $X_\Gamma$. 
Again we do not obtain a non-trivial limit set on the boundary  Mumford
curve $X({\mathbb K})=({\mathbb P}^1({\mathbb K})\smallsetminus \{ 0, \infty \})/\Gamma$, hence
we cannot just replace the Patterson--Sullivan measure on ${\mathbb P}^1({\mathbb K})$
with a similar Patterson--Sullivan measure on the Mumford curves $X_\Gamma({\mathbb K})$
of genus at least one.

However, for the genus one case of Tate--Mumford elliptic curves that we are mainly interested in here, it is possible to define a measure on $X_\Gamma({\mathbb K})$ induced by
the Patterson--Sullivan measure on ${\mathbb P}^1({\mathbb K})$. Consider the geodesic 
$L_{\{ 0, \infty \}}$ in the Bruhat--Tits tree $T$ that connects the fixed points $\{ 0, \infty \}$ of 
the Schottky group. Fix a fundamental domain ${\mathcal F}_\Gamma$ of the action of the
Schottky group $\Gamma \simeq {\mathbb Z}$ on $T$. The intersection 
${\mathcal F}_\Gamma\cap L_{\{ 0, \infty \}}$ consists of a finite set of vertices in bijective
correspondence with the vertices of the central polygon in the graph $T/\Gamma$.

There are then two main choices for how to construct a measure on $X_\Gamma({\mathbb K})$ 
using the Patterson--Sullivan measure on ${\mathbb P}^1({\mathbb K})$. The first choice
generates a measure on $X_\Gamma({\mathbb K})$ that is invariant under the automorphisms
of $X_\Gamma$ induced by arbitrary translations along $L_{\{ 0, \infty \}}$, while the second
one does not have this invariant property.

For the first construction, fix 
a choice of a root vertex $v_0$ in the tree $T$ contained in ${\mathcal F}_\Gamma\cap L_{\{ 0, \infty \}}$,
and 
consider the tree $T_0$ stemming from $v_0$ with first edges the $q-1$ directions at $v_0$ that are not
along $L_{\{ 0, \infty \}}$. Let $\Omega_0({\mathbb K}) \subset {\mathbb P}^1({\mathbb K})$ be
the boundary region $\Omega_0({\mathbb K})=\partial T_0$, endowed with the restriction $\mu_0 = \mu |_{\Omega_0}$ of the Patterson--Sullivan measure $\mu$ on ${\mathbb P}^1({\mathbb K})$. 
Every other subtree of the Bruhat--Tits tree that has root vertex on $L_{\{ 0, \infty \}}$ and first edges 
not in the direction of $L_{\{ 0, \infty \}}$ is obtained from $T_0$ via the action of a translation along 
$L_{\{ 0, \infty \}}$. We can endow the boundary region of these trees with copies of the same
measure $\mu_0$. In this way we obtain a measure on ${\mathbb P}^1({\mathbb K})\smallsetminus \{ 0,\infty\}$
that has infinite total mass and that is invariant under arbitrary translations along $L_{\{ 0, \infty \}}$.
Since it is in particular invariant under the action of the rank one Schottky group $\Gamma$ with
limit set $\{ 0, \infty \}$ it descends to a measure on $X_\Gamma=
({\mathbb P}^1({\mathbb K})\smallsetminus \{ 0,\infty\})/\Gamma$. This measure on $X_\Gamma$
has finite total mass, since it consists of finitely many copies of $\mu_0$ (one for each tree
stemming from one of the vertices of the central polygon of $T/\Gamma$), hence we can
normalize it to a probability measure on $X_\Gamma$ which is invariant under the automorphisms
induced by translations along $L_{\{ 0, \infty \}}$ and also by orientation reversal.

The second construction is similar, but instead of considering the tree $T_0$ stemming from
the root vertex $v_0$ along the directions complementary to $L_{\{ 0, \infty \}}$, we consider
now the forest $T_{{\mathcal F}}$ which is the disjoint union of the trees $T_v$ stemming from the
vertices in ${\mathcal F}_\Gamma\cap L_{\{ 0, \infty \}}$. We denote by $\Omega_{\mathcal F}$
the corresponding boundary region $\Omega_{\mathcal F}=\partial T_{{\mathcal F}}\subset {\mathbb P}^1({\mathbb K})$. The normalized restriction $\mu_{{\mathcal F}}$ of the Patterson--Sullivan 
measure $\mu$ on ${\mathbb P}^1({\mathbb K})$ to the region $\Omega_{\mathcal F}$ induces a
$\Gamma$-invariant measure on ${\mathbb P}^1({\mathbb K}) \smallsetminus \{ 0, \infty \}$ of infinite
total mass, and a probability measure on the quotient $X_\Gamma({\mathbb K})=
({\mathbb P}^1({\mathbb K})\smallsetminus \{ 0,\infty\})/\Gamma$.

While the first construction gives a more ``symmetric" measure on $X_\Gamma({\mathbb K})$,
the symmetry under arbitrary translations along $L_{\{ 0, \infty \}}$ has the disadvantage that
the boundary measure no longer keeps track of geodesic paths along the central polygonal
graph in $T/\Gamma$. The second measure instead is more useful for our purposes: while
invariance under translations in $\Gamma$ means that the measure descends to the
quotient $X_\Gamma({\mathbb K})$, hence it does not detect the number of times that
a path in the bulk $T/\Gamma$ wraps around the central polygon, it still does distinguish the
number of polygon edges along the polygon modulo its total length.

\subsection{AF algebras and limits of density matrices}
\label{AFALGEBRA}

Our construction of density matrices using the dual graph, as described in
sections~\ref{GENUSZERO}-\ref{COMPUTATION}, is based on fixing a level in the Bruhat--Tits tree, namely considering only the
vertices that are at distance at most $n$ steps from a fixed root vertex (which is related to the UV cutoff parameter $\Lambda$ in definition \ref{def:cutoff}). This determines
in turn the rank of the tensors in the tensor network and the number of dangling legs at 
the vertices of the dual graph. In order to consider the entire Bruhat--Tits tree, we need to
perform a limiting procedure over this construction at finite levels. This means considering
limits, in the appropriate sense, of density matrices of increasing ranks. This limiting
procedure can be made precise in the setting of AF-algebras and states. We now describe
this briefly.

A Bratteli diagram \cite{Bratteli} is an infinite directed graph with vertex set $V=\cup_{n=0}^\infty V_n$
and edge set $E=\cup_{k=1}^\infty E_n$ where edges $e\in E_n$ have source and target
$s(e)\in V_{n-1}$ and $t(e)\in V_n$. We assume here that each $V_n$ and $E_n$ is a finite
set and that each vertex $v\in V_n$ emits at least one edge and when $n\geq 1$ also receives
at least one edge. Each vertex $v\in V$ is labelled by a positive integer $N_v \in {\mathbb N}$,
with the property that the number of edges $N_{v,v'}=\# \{ e\in E_n \,:\, s(e)=v,\, t(e)=v'\}$, for 
given $v\in V_{n-1}$ and $v'\in V_n$, satisfies the estimate $N_v \cdot N_{v,v'}\leq N_{v'}$. 
Equivalently, we can consider only diagrams for which there is at most a single
edge $e$ between two given vertices $v,v'$, decorated with a multiplicity $N_{v,v'}\in {\mathbb N}$.
We will work with diagrams without multiple edges and with both
vertex and edge multiplicities $N_v$, $N_{v,v'}$.

A finite dimensional complex $C^*$-algebra is a direct sum $\oplus_i M_{N_i}({\mathbb C})$ of
matrix algebras (Wedderburn theorem) and $C^*$-algebra homomorphisms between them are
completely specified by assigning multiplicities on the matrix algebra components. Thus, a
direct system $(A_n, \varphi_n)$ of finite dimensional $C^*$-algebras and injective homomorphisms 
$\varphi_n: A_{n-1}\to A_n$ between them can be completely described by a Bratteli diagram with 
$V_n$ the set of matrix components and $N_v$ the mutliplicities, 
$A_n=\oplus_{v\in V_n} M_{N_v}({\mathbb C})$. The edges $e_{v,v'}\in E_n$ 
and their multiplicities $N_{v,v'}$  then uniquely specify the
injective map $\varphi_n : A_{n-1}\to A_n$ by letting $N_{v,v'}$ be the multiplicity of
$M_{N_v}({\mathbb C})$ into $M_{N_{v'}}({\mathbb C})$.  Thus, Bratteli diagrams provide
a very convenient graphical way of describing direct limits $A=\varinjlim_n (A_n,\varphi_n)$ of
finite dimensional $C^*$-algebras.  The $C^*$-algebras that can be obtained as such
limits are called AF-algebras.

A particular case of AF-algebras is given by the uniformly hyperfinite algebras, or
UHF-algebras. These are direct limits of sequences $(A_n,\varphi_n)$ where the
morphisms $\varphi_n$ are unit-preserving. This in particular implies that
the restrictions to matrix blocks $M_{N_v}({\mathbb C}) \to M_{N_{v'}}({\mathbb C})$
satisfy $N_v\cdot N_{v,v'}=N_{v'}$, namely the block $M_{N_v}({\mathbb C})$ is
mapped into $M_{N_{v'}}({\mathbb C})$ with multiplicity $N_{v'}/N_v$.

A state on $C^*$-algebra $A$ is a continuous linear functional 
$\omega: A \to {\mathbb C}$ which satisfies
positivity $\omega(a^* a)\geq 0$ for all $a\in A$, and is normalized $\omega(1)=1$ if
the algebra is unital. In other words, it is the noncommutative analog of a measure. 
When $A$ is finite dimensional, states are given by density matrices $\rho$
with $\omega(a)={\rm Tr}(\rho\, a)$. 
If a $C^*$-algebra $A$ is an AF-algebra, obtained as a direct limit $A=\varinjlim_n (A_n,\varphi_n)$
corresponding to a Bratteli diagram ${\mathbb B}={\mathbb B}(A_n,\varphi_n)$, in general we can describe 
states on $A$ in terms of density matrices $\rho_n$ on the finite dimensional algebras $A_n$ and 
a compatibility condition on the diagram ${\mathbb B}$, \cite{Evans}.

We consider here the case of 
a direct system of finite dimensional algebras $(A_n,\varphi_n)$ associated to
the Bratteli diagram ${\mathbb B}$, with $A_n=\oplus_{v\in V_n} M_{N_v}({\mathbb C})$. The associated
convex set of density matrices ${\mathcal M}(A_n)$ describing states on $A_n$ can be described as 
\eqn{}{ 
{\mathcal M}(A_n)=\left\{ \sum_{v\in V_n} \lambda_v \rho_v\,:\, \lambda=(\lambda_v)\in \Sigma_{V_n},\,\,
\rho_v \in {\mathcal M}_{N_v} \right\} 
}
where $\Sigma_{V_n}=\{ \lambda=(\lambda_v)\,:\, \lambda_v\geq 0, \, \sum_v \lambda_v=1\}$ 
is the simplex on the set $V_n$ and 
${\mathcal M}_N=\{ \rho\in M_N({\mathcal C})\,:\, \rho=\rho^*, \, \rho\geq 0,\,  {\rm Tr}(\rho)=1\}$ is
the set of density matrices of rank $N$.  The density matrices $\rho\in {\mathcal M}(A_n)$ 
have the same matrix block decomposition as the elements of $A_n$.  Let ${\mathcal M}(A_n)^0$
denote the set of $\omega \in {\mathcal M}(A_n)$ such that $\omega(e_n)\neq 0$, for the idempotent
$e_n=\varphi_n(1)$. In the case of a UHF-algebra $e_n=1$ hence this condition is always satisfied. 
One can then define maps $R_n: {\mathcal M}(A_n)^0 \to {\mathcal M}(A_{n-1})^0$ by setting
\eqn{}{ 
R_n(\omega)(a_{n-1})=\frac{\omega(\varphi_n(a_{n-1}))}{\omega(e_n)}\,.
}
This determines a projective system $({\mathcal M}(A_n)^0, R_n)$ with ${\mathcal M}(A)^0$
the inverse limit. For $\psi_n : A_n \to A$ the maps to the direct limit, one has maps $\tilde R_n: {\mathcal M}(A)^0\to {\mathcal M}(A_n)^0$ given by $\tilde R_n (\omega)(a_n)=\omega(\psi_n(a_n))/\omega(\psi_n(1))$. 
Thus, we can identify elements in the projective limit ${\mathcal M}(A)^0$ with those sequences 
$\{ \omega_n \}_{n\in {\mathbb N}}$ with $\omega_n\in {\mathcal M}(A_n)$ that have the property
that $\omega_{n-1}=R_n(\omega_n)$.  A state $\omega$ on the AF-algebra $A$ defines such a
sequence by setting $\omega_n(a_n)=\omega(\psi_n(a_n))/\omega(\psi_n(a))$ and conversely
every such sequence determines a state $\omega(a)=\omega_n(\psi_n(a_n))/\omega_n(\psi_n(1))$
for $a=\psi_n(a_n)$, which is well defined because of the compatibility $\omega_{n-1}=R_n(\omega_n)$.

Thus, in order to obtain a limit of the density matrices $\rho_n$ associated to the boundary of the tensor network
on the dual graph of the Bruhat--Tits tree truncated at level $n$, we need to show that they give 
rise to a sequence of states $\omega_n(a_n)={\rm Tr}(\rho_n a_n)$ for $a_n \in A_n$ that satisfy 
the compatibility $\omega_{n-1}=R_n(\omega_n)$.

In the case we are considering here, the AF-algebra is constructed in the following way, with
a Bratteli diagram whose underlying graph is the Bruhat--Tits tree.
As in the previous sections, we denote by $\Lambda\in {\mathbb N}$ the cutoff on the Bruhat--Tits 
tree of ${\mathbb Q}_p$. Thus, at level $\Lambda$, we consider a finite tree with $(p+1) 
p^{\Lambda -1}$ leaves. As before, we assume the choice of a fixed planar embedding of the
Bruhat--Tits tree. Let $A,B$ be two complementary regions in the  boundary of the 
finite tree at level $\Lambda$, determined by the choice of two boundary points $x,y$.
We associate to the tree, the level $\Lambda$, and the choice of the regions $A$ and $B=A^c$, 
a finite dimensional algebra of the form 
$A_\Lambda = M_{r^\sigma}({\mathbb C})^{\oplus r^{C_{AB}}}$, a direct sum of $r^{C_{AB}}$
copies of the complex algebra of $r^\sigma \times r^\sigma$ matrices,
where both $C_{AB}$ and $\sigma$ depend on $\Lambda$ and are defined as in the
previous sections. The explicit expression for $\sigma=\sigma(\Lambda)$ is
obtained through the vanishing condition of \eqref{TraceCond}. We write 
$\Delta\sigma(\Lambda)=\sigma(\Lambda+1)-\sigma(\Lambda)$. The explicit expression
for $\Delta\sigma(\Lambda)$ can also be computed directly from \eqref{TraceCond} and 
from table~\ref{tb:vertices}. The quantity $C_{AB}$, which measures the normalized geodesic
length in the cutoff tree connecting $x$ to $y$, changes by $2$ when both points are pushed one
step forward towards the boundary of the Bruhat--Tits tree when the cutoff $\Lambda$ is
increased to $\Lambda +1$.  The embeddings 
$\varphi_{\Lambda+1}: A_\Lambda \hookrightarrow A_{\Lambda+1}$ is obtained by
mapping each $r^{\sigma(\Lambda)} \times r^{\sigma(\Lambda)}$ block of $A_\Lambda$ into an 
$r^{\sigma(\Lambda+1)} \times r^{\sigma(\Lambda+1)}$ block by repeating the same block
$r^{\Delta\sigma(\Lambda)}$ times, and then repeating the resulting configuration of $r^{C_{AB}(\Lambda)}$
blocks of size $r^{\sigma(\Lambda+1)} \times r^{\sigma(\Lambda+1)}$ for $r^2$ times. This gives a
matrix consisting of $r^{C_{AB}(\Lambda+1)}$ blocks of size $r^{\sigma(\Lambda+1)} \times r^{\sigma(\Lambda+1)}$, which is an element of $A_{\Lambda+1}$. The map $\varphi_{\Lambda+1}: A_\Lambda \hookrightarrow A_{\Lambda+1}$ constructed in this way is unital, hence the resulting AF-algebra
$A=\varinjlim_\Lambda A_\Lambda$ is a UHF-algebra.  Thus, to show that the density matrices
$\rho_\Lambda$ of the form specified in \eqref{RhoAInterval} determine a state on the limit AF-algebra,
we need to check that they satisfy the compatibility condition $\omega_\Lambda=R_{\Lambda+1}(\omega_{\Lambda+1})$ where $\omega_\Lambda(a)={\rm Tr}(\rho_\Lambda a)$ is the state on
the algebra $A_\Lambda$ determined by the density matrix $\rho_\Lambda$. This condition means
that, for all $a\in A_\Lambda$,
${\rm Tr}(\rho_\Lambda \, a) = {\rm Tr}(\rho_{\Lambda+1}\, \varphi_{\Lambda+1}(a))$.

The density matrix $\rho_\Lambda$ consists of $r^{C_{AB}(\Lambda)}$ blocks of
size $r^{\sigma(\Lambda)}\times r^{\sigma(\Lambda)}$ where all the entries in each
of these blocks are equal to $1$, with an overall normalization factor equal to 
$r^{-(\sigma(\Lambda)+C_{AB}(\Lambda))}$ that makes ${\rm Tr}(\rho_\Lambda)=1$
(see \eqref{RhoAInterval}). An element $a\in A_\Lambda$ is a matrix of the same size
that also consists of $r^{C_{AB}(\Lambda)}$ blocks of size 
$r^{\sigma(\Lambda)}\times r^{\sigma(\Lambda)}$, with each block given by an
arbitrary matrix in $M_{r^{\sigma(\Lambda)}}({\mathbb C})$. Thus, the evaluation of
${\rm Tr}(\rho_\Lambda \, a)$ just yields the sum of the entries of $a$ normalized by
the factor $r^{-(\sigma(\Lambda)+C_{AB}(\Lambda))}$. Under the map
$\varphi_{\Lambda+1}: A_\Lambda \hookrightarrow A_{\Lambda+1}$ the matrix $a$ is
mapped to $r^{\Delta\sigma(\Lambda)}$ copies of each block and $r^2$ copies of
the resulting matrix.  Since all the nonzero entries of this resulting matrix 
$\varphi_{\Lambda+1}(a)$ fall inside one of the blocks where all the entries 
of the density matrix $\rho_{\Lambda+1}$ are equal to $1$, the evaluation of 
${\rm Tr}(\rho_{\Lambda+1}\, \varphi_{\Lambda+1}(a))$ gives the sum of the entries of $a$
repeated as many times as each block of $a$ is repeated in $\varphi_{\Lambda+1}(a)$,
normalized by $r^{-(\sigma(\Lambda+1)+ C_{AB}(\Lambda+1))}$. This gives
\eqn{}{ 
{\rm Tr}(\rho_{\Lambda+1}\, \varphi_{\Lambda+1}(a)) &= r^{-(\sigma(\Lambda+1)+ C_{AB}(\Lambda+1))} \cdot (\sum_{ij} a_{ij}) \cdot r^{\Delta\sigma(\Lambda)} \cdot r^2 \cr  
&= r^{-(\sigma(\Lambda)+C_{AB}(\Lambda))}
\cdot (\sum_{ij} a_{ij}) \cr  
&= {\rm Tr}(\rho_\Lambda \, a) \,, 
}
hence the compatibility condition is satisfied and the density matrices $\rho_\Lambda$ determine
a state on the UHF-algebra $A=\varinjlim_\Lambda A_\Lambda$.

\section{Outlook} 
\label{DISCUSSION}

The study of holography over nonarchimedean fields such as the $p$-adics is still a very young area, and there is much more to be learned both from the study of models in the continuum and from the relation to tensor network constructions such as we have pursued here. In these paragraphs, we summarize a few questions and directions that seem worthy of further investigation. 

One lesson of our computations is that, in this $p$-adic setting, it is more natural to think of the entropy as a function of boundary points and configurations of points, rather than as a function of boundary regions (intervals). This is in spite of the fact that our computations always rely on a choice of region in the dual tensor network. Thus, our results are consistent with an interpretation in a continuum $p$-adic field theory, for example in terms of correlation functions of twist operators (as used in real two-dimensional CFT computations by~\cite{Holzhey,Calabrese:2004eu}). This is also consistent with the physical intuition that the main contributions to entanglement entropy should arise from UV modes localized near the entangling surface. In our scenario, as in CFT$_2$, the entangling surfaces are just points, and in particular live at the boundary of the Bruhat--Tits tree itself, rather than being associated to the dual tensor network. It would be interesting to find a calculational framework depending only on the positions of entangling surfaces that would work in parallel fashion in real and $p$-adic field theories.

One can interpret our results as giving predictions for entanglement entropies in certain continuum $p$-adic field theories, which we expect to be valid up to certain overall theory-dependent factors (such as the overall normalization). For pure states, these predictions include the connected interval (two point) entropy in~\eno{RTgenus0} and~\eno{RTgenus0circle}, disconnected-interval (four point) entropy~\eno{Sdisconnect},  and mutual information~\eno{MIbdyGen}; when considering the thermal state dual to a $p$-adic BTZ black hole, we give the form of the connected interval result in~\eno{lengthBTZ0} and~\eno{lengthBTZ}. Furthermore, our proofs of entropy inequalities are evidence in support of such results---such as subadditivity and strong subadditivity---in continuum $p$-adic field theory. Extending these results to holographic codes~\cite{Pastawski:2015qua} which include bulk logical inputs would be a natural next step. 

It would also be interesting to investigate the recently conjectured duality between entanglement of purification and entanglement wedge  cross-section~\cite{Terhal,Takayanagi:2017knl,Nguyen:2017yqw} and its extensions~\cite{Bao:2017nhh,Espindola:2018ozt,Bao:2018gck,Umemoto:2018jpc}  in this setup, as well as other measures of entanglement for mixed states, such as entanglement negativity~\cite{Rangamani:2014ywa} and the conjectured bulk interpretation (see e.g.~\cite{Chaturvedi:2016rcn,Chaturvedi:2016rft,Jain:2017uhe,Kudler-Flam:2018qjo}).
The simplifying features of the tensor networks studied here provide an effective computational framework to explore such questions. 

Along these lines, we have shown that many aspects of the bulk $p$-adic geometries closely parallel the situation in real AdS/CFT. Even so, it is comparably simpler to work with the discrete geometries, and this specific network provided a model in which we could efficiently compute many holographic entropy quantities. One might hope this trend will continue, and we expect that more complicated holographic quantities will be computationally easier to study in the $p$-adic setting. A major goal of this program is to reconstruct bulk quantities in smooth AdS from knowledge of the corresponding $p$-adic quantities for all $p$. We hope to return to the reconstruction of real AdS quantities from $p$-adic in future work. 

As mentioned above, it would furthermore be interesting to study the possibility of gluing together the Hamiltonians of section~\ref{CRSS} to give a general semiclassical construction (over finite fields) of spin systems whose vacua are constructed from networks of perfect tensors. One could imagine that such a construction could produce either a spin system, with a Hamiltonian possibly of commuting-projector type, or a system described by a path integral over discrete ($\FF_p$-valued) classical degrees of freedom; either version of the construction would be interesting, and would lead to a unification of the tensor network perspective with a full-fledged quantum system evolving dynamically in time (see e.g.~\cite{Osborne:2017woa} for a different take on this). One could also examine if our setup allows insights into the connection between holographic correlators and entropy measures. 

We note that as also familiar from the tensor networks literature, the tensor network dual to the Bruhat--Tits tree does not account for sub-AdS effects. However, we showed that these tensor networks can be embedded in the Drinfeld $p$-adic plane.  Although the geometry of the Drinfeld plane did not play a role in our entropy computations, it may become relevant in the investigation of sub-AdS holography (see~\cite{Bao:2018pvs} for another perspective on this). 

Finally, generalizations of the BTZ black hole given by higher genus Mumford curves (quotients by higher rank Schottky groups) and higher dimensional models based on higher rank buildings may also exhibit more intricate relations between entanglement entropy on the boundary $p$-adic varieties and the geometry of the bulk regions, and may help identify a covariant generalization of the RT formula in the context of $p$-adic theories.

We look forward to returning to many of these questions in future work.

\section*{Acknowledgments}

I.A.S.\ thanks D.~Aasen, J.~Keating, and J.~Walcher for conversations, and the Kavli Institute for Theoretical Physics in Santa Barbara for hospitality as this manuscript was being completed; he also gratefully acknowledges partial support by the  Deutsche Forschungsgemeinschaft, within the
framework of the Exzellenzinitiative an der Universit\"at Heidelberg.
M.H.\ and S.P.\ thank Perimeter Institute for their kind hospitality while this work was in its early stages. 
The work of M.H.\ and S.P.\ was supported in part by Perimeter Institute for Theoretical Physics. 
M.H.\ would like to thank S.S.~Gubser and Princeton University for their hospitality while this work was being completed, and work done at Princeton was supported in part by the Department of Energy under Grant No.~DE-FG02-91ER40671, and by the Simons Foundation, Grant 511167 (SSG).
M.H.\ is also partially supported by the U.S. Department of Energy, Office of Science, Office of High Energy Physics, under Award Number DE-SC0011632.
M.M.\ is partially supported by  
NSF grant DMS-1707882, by NSERC Discovery Grant RGPIN-2018-04937 and Accelerator Supplement grant RGPAS-2018-522593, and by the Perimeter Institute for Theoretical Physics. Research at Perimeter Institute is supported by the Government of Canada through the Department of Innovation, Science and Economic Development and by the Province of Ontario through the Ministry of Research, Innovation and Science. 
Research at the Kavli Institute is supported in part by the National Science
Foundation under Grant No.~PHY-1748958.
\appendix

\section{The Three-Qutrit Code} 
\label{THREEQ} 

In this appendix, based on the discussion in section \ref{CLASSQUANTCODES}, we give an explicit example of a quantum Reed-Solomon code which has quantum error-correcting and `perfectness' properties. 
We start with the classical Reed--Solomon code $[n,k,n-k+1]_{q^2}$ (note the $q^2$ in the subscript) and choose parameters $n=q=3$, $k=(q-1)/2 = 1$, and $X = \mathbb{P}^1(\mathbb{F}_3) \smallsetminus \{\infty\} = \{[1:0],[1:1],[1:2]\}$. The $[3,1,3]_{3^2}$-code, which will serve as an example of a $D$-type code from section \ref{ALGEBCODES} (see discussion around \eno{HermInProd}), takes the form $D = \{ (f_a(1,0),f_a(1,1),f_a(1,2)): a=(a_0) \in \mathbb{F}_{3^2}, f_a \in \mathbb{F}_{3^2}[u,v]\}$. Since $k=1$, the homogeneous polynomial takes a particularly simple form, $f_a(u,v) = a_0 \in \mathbb{F}_{3^2}$. Thus the $[3,1,3]_{3^2}$-code becomes, $D=\{ (a_0,a_0,a_0): a_0 \in \mathbb{F}_{3^2}\}$.

It is easy to check that $D$ is self-orthogonal with respect to the Hermitian inner product \eno{HermInProd}. Take $a=(a_0,a_0,a_0), b=(b_0,b_0,b_0) \in D$, where $a_0,b_0 \in \mathbb{F}_{3^2}$, then
\eqn{DSOcheck}{
a * b = 3 a_0 b_0^3 = 0\,.
}
One can also check self-orthogonality by constructing the dual code $D^\perp$. It follows from the definition of the dual code in footnote \ref{fn:DualCode} that if $b = (b_1,b_2,b_3) \in D^\perp$, then for all $a = (a_0,a_0,a_0) \in D$,
\eqn{DdualCond}{
a * b = a_0 (b_1^3 + b_2^3 + b_3^3) \stackrel{!}{=} 0\,.
}
Since \eno{DdualCond} must hold for all $a_0 \in \mathbb{F}_{3^2}$, we must have $b_1^3+b_2^3+b_3^3 = 0$. Writing $b_i = \sum_{j=1}^2 b_{i,j}\gamma_j$ where $b_{i,j} \in \mathbb{F}_3$ and $\{ \gamma_1, \gamma_2 \} = \{1,x \}$ form a basis for $\mathbb{F}_{3^2}$ as an $\mathbb{F}_3$-vector space, we have the result, $b_i^3 = b_{i,1}  - b_{i,2} x$. So the condition of duality becomes
\eqn{DdualCondAgain}{
& (b_{1,1}+b_{2,1}+b_{3,1}) - (b_{1,2}+b_{2,2}+b_{3,2})x= 0 \quad
\Rightarrow \quad b_{1,j} = 2 b_{2,j}+2 b_{3,j} \qquad j=1,2\,,
}
or, in other words, $D^\perp = \{ (2b_2 + 2b_3, b_2, b_3): b_2, b_3 \in \mathbb{F}_{3^2}\}$. One checks that $D \subset D^\perp$, thus $D$ is self-orthogonal. While establishing self-orthogonality via the dual code may seem a bit roundabout, the construction of the dual code helps determine $d_Q$, the minimum distance of the quantum code. It is straightforward to show that in this example, $d_Q = \min \{ {\rm wt}(v): v \in D^\perp \smallsetminus D\} = 2$.

Now consider the classical Reed--Solomon $[2n,2k,2n-2k+1]_q$-code $C$, with $n=q=3$ and $k=(q-1)/2=1$ as before. 
The input $2$-tuple now becomes $a=(a_0,b_0) \in \mathbb{F}_{3}^2$, and the code takes the form, $C = \left\{ \big( (f_{a_0}(1,0),f_{b_0}(1,0)),  (f_{a_0}(1,1),f_{b_0}(1,1)),  (f_{a_0}(1,2),f_{b_0}(1,2)) \big)\right\}$ where $f_a[u,v] \in \mathbb{F}_{3}$ is the homogeneous polynomial given by \eno{homPol}. 
Like before, setting $k=1$ leads to a significant simplification and results in $C = \left\{ \big( (a_0,b_0),  (a_0,b_0),  (a_0,b_0) \big) : (a_0,b_0) \in \mathbb{F}_3^2\right\}$.

To establish the self-orthogonality of $C$, we use the inner product given in \eno{TrInProd} or equivalently \eno{TrInProdEqui}. Given $(a,b) = \big( (a_0,b_0),  (a_0,b_0),  (a_0,b_0) \big) \in C \subset \mathbb{F}_3^{2n}$ and similarly $(a^\prime, b^\prime) \in C$,  using \eno{TrInProd} we have 
\eqn{CdualCond}{
(a,b) * (a^\prime, b^\prime) = \Tr ( 3 a_0 b_0^\prime - 3 a_0^\prime b_0) = 0\,,
}
which tells us $C$ is self-orthogonal.\footnote{One can also construct $C^\perp$ \emph{ a la} our construction of $D^\perp$ outlined above, and verify self-orthogonality and check that $d_Q = \min \{ {\rm wt}(v,w): (v,w) \in C^\perp \smallsetminus C\} = 2$.} Alternatively, we can use the inner product in \eno{TrInProdEqui}. First note that since $q=3$, the parameter $r=1$ (recall that $q=p^r$), and consequently $\varphi$ is the trivial identity automorphism. This is because $\varphi(a) = (\varphi(a_1), \varphi(a_2), \varphi(a_3))$ where $a \in \mathbb{F}_3^n$, but since $r=1$, we have $\varphi(a_1) = a_1 \in \mathbb{F}_3$. Then, given $(a,b), (a^\prime,b^\prime) \in C$ as before and using \eno{TrInProdEqui} along with the fact that $\varphi$ is the identity transformation, we have
\eqn{CdualCondEqui}{
(a,b) * (a^\prime, b^\prime) = \langle a,b^\prime \rangle - \langle a^\prime, b \rangle = 3a_0 b_0^\prime - 3a_0^\prime b_0 = 0\,,
}
which confirms self-orthogonality. 

The trivial action of $\varphi$ is due to the fact that we have chosen $q$ to be a prime number, rather than a power of a prime. This makes the construction of the corresponding abelian subgroup  especially simple: For every element $(a,b) \in C$ (where $a, b \in \mathbb{F}_3^n$), the corresponding group elements  are given by $\xi^{i} E_{a,b}$ for $0 \leq i \leq 2$.  The cardinality of $C$, $|C| = 9$ is small enough that we may explicitly list all its elements. They are
\eqn{C625}{
C = \{ ((0,0)^3, (0,1)^3, (0,2)^3, (1,0)^3, (1,1)^3, (1,2)^3, (2,0)^3, (2,1)^3, (2,2)^3\} \,,
}
where $(i,j)^3 \equiv \big( (i,j),(i,j),(i,j) \big) \in \mathbb{F}_3^{2 \times 3}$. The operators in the corresponding subgroup are given by
\eqn{S625}{
{\cal S} = \{ \xi^{i} E_{0^3,0^3}, \xi^i E_{0^3,1^3}, \xi^i E_{0^3,2^3}, \xi^{i} E_{1^3,0^3}, \xi^i E_{1^3,1^3}, \xi^i E_{1^3,2^3}, \xi^{i} E_{2^3,0^3}, \xi^i E_{2^3,1^3}, \xi^i E_{2^3,2^3} : 0\leq i\leq 2 \} \,,
}
 where $j^3 \equiv (j,j,j) \in \mathbb{F}_3^3$. In constructing the simultaneous eigenspaces of these operators, we can ignore the overall scalar factors $\xi^i$ in \eno{S625}, since they come from the center of the Heisenberg group. Now using \eno{EabDef} we conclude that if $(a,b) \in C$, then $E_{a,b} = T_{a_0} R_{b_0} \otimes T_{a_0} R_{b_0} \otimes T_{a_0} R_{b_0}$. The matrices $T_{a_0} R_{b_0}$ where $a_0, b_0 \in \mathbb{F}_3$ are given by
\eqn{TaRbMatq3}{
T_0 R_0 &= \begin{pmatrix} 1 & & \\ & 1 &  \\ & & 1 \end{pmatrix}  \quad T_0 R_1 = \begin{pmatrix} 1 & & \\ & \xi &  \\ & & \xi^2 \end{pmatrix} \quad T_0 R_2 = \begin{pmatrix} 1 & & \\ & \xi^2 &  \\ & & \xi \end{pmatrix} \cr
T_1 R_0 &= \begin{pmatrix} 0 & 1 & 0  \\ 0 & 0 & 1  \\ 1 & 0 & 0 \end{pmatrix} \quad T_1 R_1 = \begin{pmatrix} 0 & \xi & 0  \\ 0 & 0 & \xi^2  \\ 1 & 0 & 0 \end{pmatrix} \quad T_1 R_2 = \begin{pmatrix} 0 & \xi^2 & 0  \\ 0 & 0 & \xi  \\ 1 & 0 & 0 \end{pmatrix} \cr 
T_2 R_0 &= \begin{pmatrix} 0 & 0 & 1  \\ 1 & 0 & 0  \\ 0 & 1 & 0 \end{pmatrix} \quad T_2 R_1 = \begin{pmatrix} 0 & 0 & \xi^2  \\ 1 & 0 & 0  \\ 0 & \xi & 0 \end{pmatrix}  \quad T_2 R_2 = \begin{pmatrix} 0 & 0 & \xi  \\ 1 & 0 & 0  \\ 0 & \xi^2 & 0 \end{pmatrix}\,.
}
Recalling that the orthonormal basis qubits $|a_0 \rangle$ with $a_0 \in \mathbb{F}_3$ are given by
\eqn{q3Qubits}{
|0 \rangle = \begin{pmatrix} 1 \\ 0 \\ 0 \end{pmatrix} \qquad |1 \rangle = \begin{pmatrix} 0 \\ 0 \\ 1 \end{pmatrix} \qquad |2 \rangle = \begin{pmatrix} 0 \\ 1 \\ 0 \end{pmatrix}\,,
}
it is straightforward to check that the operators $E_{a,b}$ have the following common eigenvectors:
\eqn{q3Eigenvecs}{
|A \rangle &= | 000 \rangle + |111 \rangle + | 222 \rangle \quad  |B \rangle = | 012 \rangle + |120 \rangle + | 201 \rangle \quad |C \rangle = | 021 \rangle + |210 \rangle + | 102 \rangle \cr
|D \rangle &= | 001 \rangle + |112 \rangle + | 220 \rangle \quad  |E \rangle = | 010 \rangle + |121 \rangle + | 202 \rangle \quad |F \rangle = | 100 \rangle + |211 \rangle + | 022 \rangle \cr
|G \rangle &= | 002 \rangle + |110 \rangle + | 221 \rangle \quad  |H \rangle = | 020 \rangle + |101 \rangle + | 212 \rangle \quad |I \rangle = | 200 \rangle + |011 \rangle + | 122 \rangle\,,
}
where $|ijk \rangle \equiv |i \rangle \otimes |j \rangle \otimes |k \rangle \in \left(\mathbb{C}^3\right)^{\otimes n}$. For completeness, we tabulate the eigenvalues of each eigenvector under the operators $E_{a,b} \in {\cal S}$:
\eqn{q3Eigenvals}{
  \begin{tabular}{c||c|c|c|c|c|c|c|c|c}
    & $E_{0^3,0^3}$ & $E_{0^3,1^3}$ & $E_{0^3,2^3}$ & $E_{1^3,0^3}$ & $E_{1^3,1^3}$ & $E_{1^3,2^3}$ & $E_{2^3,0^3}$ & $E_{2^3,1^3}$ & $E_{2^3,2^3}$ \\ \hline\hline
   $|A\rangle $ & $1$ & $1$ & $1$ & $1$ & $1$ & $1$ & $1$ & $1$ & $1$ \\[1pt] \hline
   $|B\rangle $ & $1$ & $1$ & $1$ & $1$ & $1$ & $1$ & $1$ & $1$ & $1$ \\[1pt] \hline
   $|C\rangle $ & $1$ & $1$ & $1$ & $1$ & $1$ & $1$ & $1$ & $1$ & $1$ \\[1pt] \hline
   $|D\rangle $ & $1$ & $\xi^2$ & $\xi$ & $1$ & $\xi^2$ & $\xi$ & $1$ & $\xi^2$ & $\xi$ \\[1pt] \hline
   $|E\rangle $ & $1$ & $\xi^2$ & $\xi$ & $1$ & $\xi^2$ & $\xi$ & $1$ & $\xi^2$ & $\xi$ \\[1pt] \hline
   $|F\rangle $ & $1$ & $\xi^2$ & $\xi$ & $1$ & $\xi^2$ & $\xi$ & $1$ & $\xi^2$ & $\xi$ \\[1pt] \hline
   $|G\rangle $ & $1$ & $\xi$ & $\xi^2$ & $1$ & $\xi$ & $\xi^2$ & $1$ & $\xi$ & $\xi^2$ \\[1pt] \hline
   $|H\rangle $ & $1$ & $\xi$ & $\xi^2$ & $1$ & $\xi$ & $\xi^2$ & $1$ & $\xi$ & $\xi^2$ \\[1pt] \hline
   $|I\rangle $ & $1$ & $\xi$ & $\xi^2$ & $1$ & $\xi$ & $\xi^2$ & $1$ & $\xi$ & $\xi^2$ 
  \end{tabular}\,.
 }
 From this table, one can see that the invariant subspace of the abelian subgroup corresponding to $C$---which is the code subspace of the corresponding quantum CRSS code---is spanned by~$\ket{A}$, $\ket{B}$, and~$\ket{C}$. Other equivalent choices of code subspace are the other mutual eigenspaces of the operators in~$C$, corresponding to $\ket{D}$, $\ket{E}$, and~$\ket{F}$ or to~$\ket{G}$, $\ket{H}$, and~$\ket{I}$. The resulting quantum code is the quantum $[\![3,1,2]\!]_3$ Reed--Solomon/three-qutrit code, corresponding to the simplest example of a four-index perfect tensor given in~\cite{Pastawski:2015qua}; it maps basis states according to the rule
\deq{
\ket{0} \mapsto \frac{1}{\sqrt{3}}\ket{A}, \quad \ket{1} \mapsto \frac{1}{\sqrt{3}}\ket{B}, \quad \ket{2} \mapsto \frac{1}{\sqrt{3}}\ket{C}.
}

{\small
\bibliographystyle{ssg}
\bibliography{draft}
}

\end{document}